\documentclass[final,1p,times]{elsarticle}

\usepackage{graphicx}
\usepackage{amssymb}

\newcommand{\Lya}{Ly$\alpha$~}
\newcommand{\beq}{\begin{equation}}
\newcommand{\eeq}{\end{equation}}

\journal{Physics Reports}

\begin{document}

\def\Omm{{\Omega_m}}
\def\Ommz{{\Omega_m^{\,z}}}
\def\Omr{{\Omega_r}}
\def\Omk{{\Omega_k}}
\def\Oml{{\Omega_{\Lambda}}}
\def\Omb{{\Omega_b}}
\def\tot{{\rm tot}}
\def\toteff{{\rm tot,eff}}
\def\xb{{\bf x}}
\def\kb{{\bf k}}
\def\yb{{\bf y}}
\def\rb{{\bf r}}
\def\vb{{\bf v}}
\def\ub{{\bf u}}
\def\cN{c_{\rm N}}
\def\del{\delta}

\begin{frontmatter}

  \title{The Rise of the First Stars: Supersonic Streaming, Radiative
    Feedback, and 21-cm Cosmology}

  \author[a,b,c,d]{Rennan Barkana} \ead{barkana@tau.ac.il}
  \address[a]{Raymond and Beverly Sackler School of Physics and
    Astronomy, Tel Aviv University, Tel Aviv 69978, Israel}
  \address[b]{Sorbonne Universit\'{e}s, Institut Lagrange de Paris
    (ILP), Institut d'Astrophysique de Paris,\\ UPMC Univ Paris 06/CNRS,
    France} \address[c]{Department of Astrophysics, University of
    Oxford, Denys Wilkinson Building, Keble Road, Oxford OX1 3RH, UK}
  \address[d]{Perimeter Institute for Theoretical Physics, 31 Caroline
    St N., Waterloo, ON N2L 2Y5, Canada}

\begin{abstract}

  Understanding the formation and evolution of the first stars and
  galaxies represents one of the most exciting frontiers in astronomy.
  Since the universe was filled with hydrogen atoms at early times,
  the most promising method for observing the epoch of the first stars
  is to use the prominent 21-cm spectral line of hydrogen. Current
  observational efforts are focused on the cosmic reionization era,
  but observations of the pre-reionization cosmic dawn are also
  beginning and promise exciting discoveries. While observationally
  unexplored, theoretical studies predict a rich variety of
  observational signatures from the astrophysics of the early galaxies
  that formed during cosmic dawn. As the first stars formed, their
  radiation (plus that from stellar remnants) produced feedback that
  radically affected both the intergalactic medium and the character
  of newly-forming stars. Lyman-$\alpha$ radiation from stars
  generated a strong 21-cm absorption signal, observation of which is
  currently the only feasible method of detecting the dominant
  population of galaxies at redshifts as early as $z \sim 25$. Another
  major player is cosmic heating; if due to soft X-rays, then it
  occurred fairly early $(z \sim 15$) and produced the strongest
  pre-reionization signal, while if it is due to hard X-rays, as now
  seems more likely, then it occurred later and may have dramatically
  affected the 21-cm sky even during reionization. In terms of
  analysis, much focus has gone to studying the angle-averaged power
  spectrum of 21-cm fluctuations, a rich dataset that can be used to
  reconstruct the astrophysical information of greatest interest. This
  does not, however, diminish the importance of finding additional
  probes that are complementary or amenable to a more
  model-independent analysis. Examples include the global
  (sky-averaged) 21-cm spectrum, and the line-of-sight anisotropy of
  the 21-cm power spectrum. Another striking feature may result from a
  recently recognized effect of a supersonic relative velocity between
  the dark matter and gas. This effect enhanced large-scale
  clustering and, if early 21-cm fluctuations were dominated by small
  galactic halos, it produced a prominent pattern on 100~Mpc scales.
  Work in this field, focused on understanding the whole era of
  reionization and cosmic dawn with analytical models and numerical
  simulations, is likely to grow in intensity and importance, as the
  theoretical predictions are finally expected to confront 21-cm
  observations in the coming years.

\end{abstract}


\end{frontmatter}

\newpage
\tableofcontents
\newpage

\section{Introduction and Overview}
\label{intro21}

Galaxies around us have been mapped systematically out to a redshift
$z \sim 0.3$ by recent large surveys \cite{sdss,2df}. The observed
galaxy distribution shows a large-scale filament-dominated ``cosmic
web'' pattern that is reproduced by cosmological numerical simulations
\cite{mill}. This structure is well-understood theoretically
\cite{bond} as arising from the distribution of the primordial density
fluctuations, which drove hierarchical structure formation in the
early universe. Recent observations have been pushing a new frontier
of early cosmic epochs, with individual bright galaxies detected
reliably from as early as $z = 11.1$ \cite{z11p1}, which corresponds
to $t \sim 400$~Myr after the Big Bang. However, it is thought that
the bulk of the early stars formed in a large number of very small
galactic units, which will be difficult to observe individually. In
particular, high-resolution numerical simulations show that the truly
earliest stars formed within $\sim 10^6 M_{\odot}$ dark matter halos
\cite{Bromm,Abel}. These simulations can only follow small cosmic
volumes, and thus begin to form stars much later than in the real
universe, but analytical methods show that the very first such stars
within our light cone must have formed at $z \sim 65$ (age $t \sim 35$
Myr) \cite{first,anastasia}.

\begin{figure}[tbp]
\includegraphics[width=\textwidth]{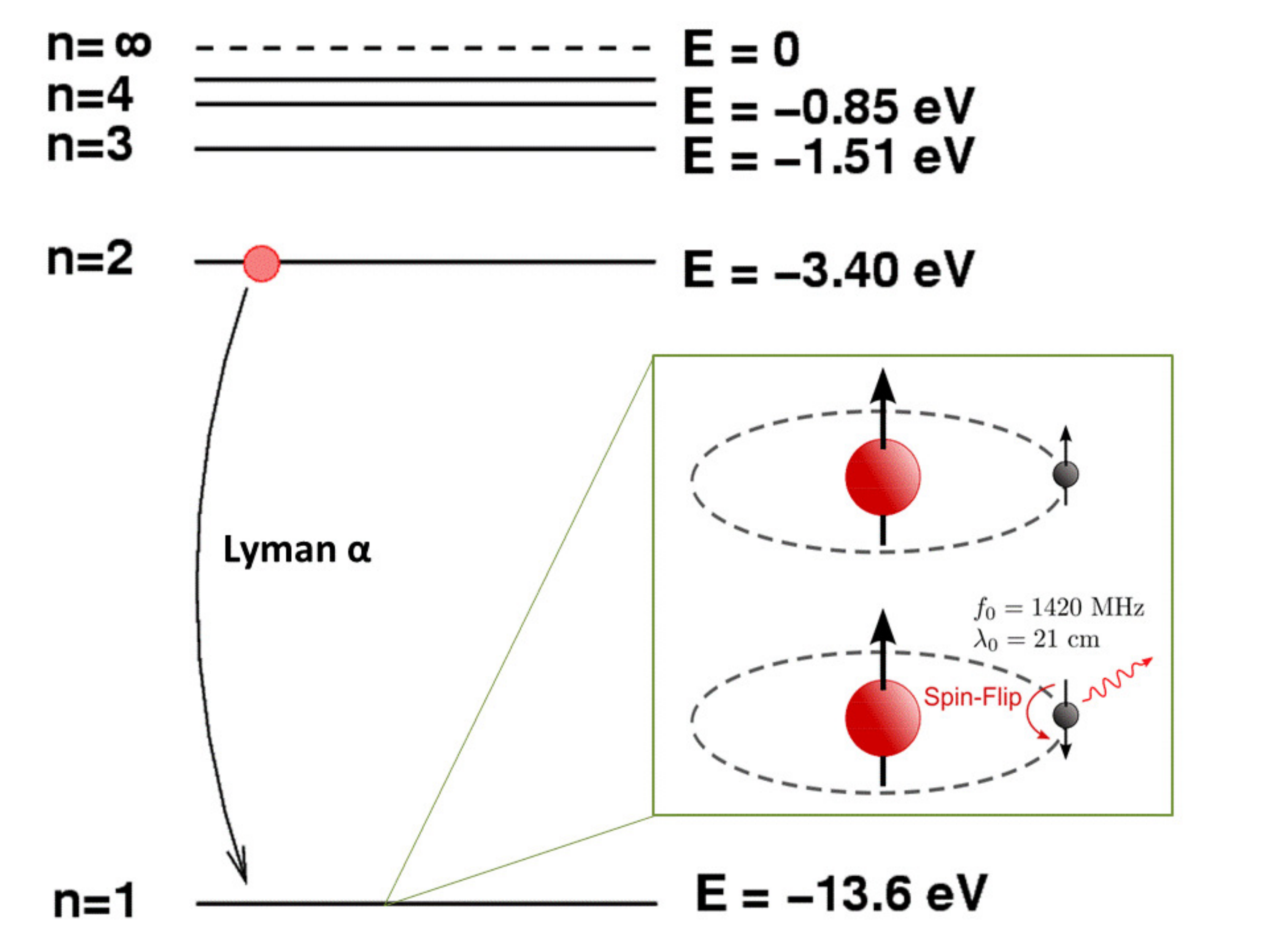}
\caption{The energy levels of hydrogen. The ionization energy of
  hydrogen is 13.6~eV, and the Lyman-$\alpha$ (Ly$\alpha$) line
  (10.2~eV) corresponds to the $n=2$ to $n=1$ transition. The
  spin-flip transition (inset on the right) is a much lower energy
  splitting ($5.87 \times 10^{-6}$~eV) within the ground state ($n=1$)
  of hydrogen, corresponding to a wavelength of 21~cm and a frequency
  of 1420~MHz.  Credits: Main portion: Michael Richmond
  (http://spiff.rit.edu/classes/phys301/lectures/spec\_lines/spec\_lines.html);
  Inset: Tiltec via Wikimedia Commons.}
\label{f:Basic21cm}
\end{figure}

The best hope of observing the bulk population of early stars is via
the cosmic radiation fields that they produced. The mean radiation
level traces the cosmic star formation rate, while spatial
fluctuations reflect the clustering of the underlying sources, and
thus the masses of their host halos. In particular, the hyperfine
spin-flip transition of neutral hydrogen (H~I) at a wavelength of
21~cm (Figure~\ref{f:Basic21cm}) is potentially the most promising
probe of the gas and stars at early times. Observations of this line
at a wavelength of $21\times (1+z)$ cm can be used to slice the
universe as a function of redshift $z$ (or, equivalently, distance
along the line of sight), just like any atomic resonance line in
combination with the cosmological redshift. Together with the other
two dimensions (angular position on the sky), 21-cm cosmology can thus
be used to obtain a three-dimensional map of the diffuse cosmic H~I
distribution \cite{hogan}, in the previously unexplored era of
redshifts $\sim 7 - 200$.

Absorption or emission by the gas along a given line of sight changes
the 21-cm brightness temperature $T_b$, measured relative to the
temperature of the background source, which here is the cosmic
microwave background (CMB) \cite{Madau}. The observed $T_b$ is
determined by the spin temperature $T_S$, an effective temperature
that describes the relative abundance of hydrogen atoms in the excited
hyperfine level compared to the ground state. Primordial density
inhomogeneities imprinted a three-dimensional power spectrum of 21-cm
intensity fluctuations on scales down to $\sim 10$~kpc (all sizes
henceforth are comoving unless indicated otherwise), making it the
richest dataset on the sky \cite{Loeb04}.  The potential yield of
21-cm observations is further increased by the expected anisotropy of
the 21-cm power spectrum \cite{BL05a,Nusser,APindian,MeAP}.

The 21-cm signal vanished at redshifts above $z \sim 200$, when the
gas kinetic temperature, $T_k$, was close to the CMB temperature,
$T_{\rm CMB}$, making the gas invisible with respect to the CMB
background. Subsequently, the gas cooled adiabatically, faster than
the CMB, and atomic collisions kept the spin temperature $T_S$ of the
hyperfine level population below $T_{\rm CMB}$, so that the gas
appeared in 21-cm absorption \cite{scott}. As the Hubble expansion
continued to rarefy the gas, radiative coupling of $T_S$ to $T_{\rm
  CMB}$ started to dominate over collisional coupling of $T_S$ to
$T_k$ and the 21-cm signal began to diminish.

\begin{figure}[tbp]
\includegraphics[width=\textwidth]{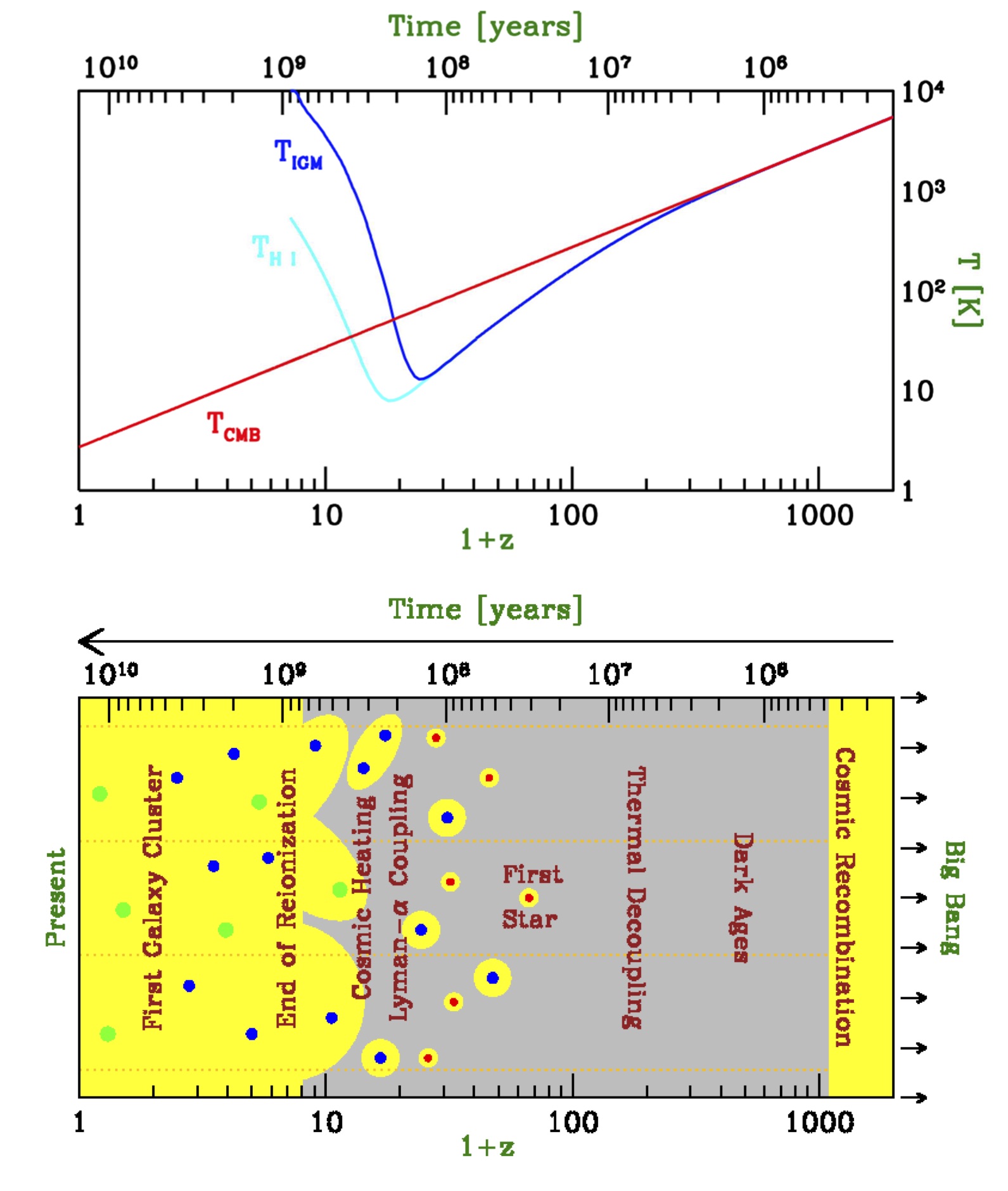}
\caption{Overview of cosmic history, with the age of the universe
  shown on the top axis and the corresponding redshift (plus one) on
  the bottom axis. Bottom panel: Yellow represents ionized hydrogen,
  and gray is neutral. Observers probe the cosmic gas using the
  absorption of background light (dotted lines) by atomic hydrogen.
  Stars formed in halos whose typical size continually grew with time,
  going from the first generation that formed through
  molecular-hydrogen cooling (red dots), to the larger galaxies that
  formed through atomic cooling and likely dominated cosmic
  reionization (blue dots), all the way to galaxies as massive as the
  Milky Way (green dots). Top panel: Corresponding sketch of the
  cosmic mean gas temperature within the IGM, including neutral
  regions only (cyan) or also ionized regions (blue) assumed to be at
  10,000~K; these are compared to the CMB temperature (red curve). The
  gas was initially thermally coupled to the CMB, until it
  adiabatically cooled more rapidly, and was then heated first by
  X-ray heating and subsequently by reionization. Updated and expanded
  version of a Figure from \cite{BSc}.}
\label{f:history}
\end{figure}

Once stars began to form, their radiation produced feedback on the
intergalactic medium (IGM) and on other newly-forming stars, and
substantially affected the 21-cm radiation. The first feedback came
from the ultraviolet (UV) photons produced by stars between the \Lya
and Lyman limit wavelengths (i.e., energies in the range of
$10.2-13.6$~eV). These photons propagated freely through the Universe
and some of them redshifted or scattered into the \Lya resonance, and
coupled $T_S$ to $T_k$ once again \cite{Madau} through the
Wouthuysen-Field \cite{Wout, Field} effect by which the two hyperfine
states are mixed through the absorption and re-emission of a \Lya
photon. Meanwhile, Lyman-Werner (LW) photons in nearly the same energy
range ($11.2-13.6$~eV) dissociated molecular hydrogen and eventually
ended the era of primordial star formation driven by molecular cooling
\cite{haiman}, leading to the dominance of larger halos. X-ray photons
also propagated far from the emitting sources and began early on to
heat the gas \cite{Madau}. Once $T_S$ grew larger than $T_{\rm CMB}$,
the gas appeared in 21-cm emission over the CMB level. Emission of UV
photons above the Lyman limit by the same galaxies initiated the
process of cosmic reionization, creating ionized bubbles in the
neutral gas around these galaxies. Figure~\ref{f:history} 
shows a brief summary of early cosmic history, and Table~\ref{Tab:age}
lists the observed frequency corresponding to 21-cm radiation from
various redshifts, as well as the age of the Universe.

\begin{table*}
\begin{center}
\begin{tabular}{|l |l |l |}
\hline
$1+z$ & Observed 21-cm Frequency [MHz] & Cosmic Age [Myr]\\
\hline
1 & 1.42 GHz & 13.8 Gyr \\
2 & 710 & 5.88 Gyr \\
3 & 473 & 3.29 Gyr \\
4 & 355 & 2.15 Gyr \\
7 & 203 & 934 \\
10 & 142 & 547 \\
15 & 94.7 & 297 \\
20 & 71.0 & 192 \\
25 & 56.8 & 137 \\
30 & 47.3 & 104 \\
40 & 35.5 & 67.5 \\
50 & 28.4 & 48.2 \\
60 & 23.7 & 36.5 \\
  \hline
\end{tabular}
\caption{The observed frequency corresponding to 21-cm radiation from
  a source at redshift $z$, and the age of the Universe, listed versus
  $1+z$. Units are as in the column labels except where indicated
  otherwise.}
\end{center}
\label{Tab:age}
\end{table*} 

The subject of cosmic reionization began in earnest after Gunn \&
Peterson (1965) \cite{GP} used a just-discovered quasar to show that
the Universe around it was highly ionized. This led to much
theoretical work on how the Universe might have been reionized. The
subject of 21-cm cosmology is a more recent one. Hogan \& Rees (1979)
\cite{hogan} worked out the basic ideas and noted that 21-cm
observations could probe the properties of cosmic gas including its
density, temperature, and spin temperature (which, they noted, could
be different from the kinetic temperature). Scott \& Rees (1990)
\cite{scott} revisited the subject, now in the modern context of
galaxy formation in a Universe dominated by cold dark matter; they
were the first to note that 21-cm cosmology could probe reionization.
Madau et al.\ (1997) \cite{Madau} first considered 21-cm radiation
during cosmic dawn, before the epoch of reionization
(EOR)\footnote{Two notes on common usage: The era/epoch of
  reionization is often denoted ``EOR'' in the literature; ``Cosmic
  dawn'' usually refers to the period between the formation of the
  first stars until the beginning of the EOR, although sometimes it is
  used as a general name for the entire period including the EOR.},
and highlighted the eras of \Lya coupling and of early cosmic heating.
However, 21-cm cosmology was relatively slow in developing. For
example, in a major review of the field in 2001 \cite{Barkana:2001},
we devoted 3 pages out of 114 to this topic, which at the time was
considered only one of many promising avenues in the field.

Cosmic reionization remained the dominant subject in the field of the
first stars for some time longer. After several years of confusion
about the basic character of reionization (see \S~\ref{s:reion}), the
now-standard paradigm was developed. Barkana \& Loeb (2004)
\cite{BL04} showed that the surprisingly strong clustering of
high-redshift halos leads to large ionized bubbles due to groups of
clustered galaxies, and to an inside-out topology (with high-density
regions reionizing early, leaving the voids for last). Furlanetto et
al.\ (2004) \cite{fzh04} created a quantitative analytical model that
yielded the first predictions of the distribution of H~II bubble
sizes, showing that 10~Mpc (comoving) is typical for the central stage
of reionization. This picture of reionization based on semi-analytic
models \cite{BL04,fzh04} was then verified by large-scale numerical
simulations, starting with Iliev et al.\ (2006) \cite{Iliev}. The
theoretical expectation that the bubbles of reionization were large
provides a critical boost for observational efforts to discover the
resulting 21-cm fluctuations, since if higher angular resolution were
required, this would make it harder to reach the sensitivity needed to
detect the faint cosmic signal.

Cosmic reionization was initially thought to be the only source of
fluctuations available for 21-cm interferometers (other than
primordial density and temperature fluctuations, which create
significantly smaller signals than those driven by galaxies and their
strongly enhanced clustering). The earlier 21-cm events of cosmic dawn
pointed out by Madau et al.\ (1997) \cite{Madau} were considered to be
highly uniform, since the photons that drove them (\Lya photons in the
case of \Lya coupling, and X-ray photons in the case of cosmic
heating) can travel $\sim 100$~Mpc in a neutral Universe before
interacting. Unless rare objects such as quasars dominated, this
seemed to imply a uniform cosmic transition that could only be seen
with global 21-cm measurements that track the sky-averaged spectrum
\cite{tozzi}. Cosmic dawn was opened up to interferometric
observations when Barkana \& Loeb (2005) \cite{BL05b} applied the same
idea of unusually large fluctuations in the abundance of early
galaxies, which had helped understand reionization, to earlier epochs.
Spatial fluctuations in the \Lya intensity were shown to have led, in
fact, to rather large 21-cm fluctuations from the \Lya coupling era.
The same idea was then applied by Pritchard \& Furlanetto (2007)
\cite{Jonathan07} to the X-ray background during the cosmic heating
era, showing that a large signal of 21-cm fluctuations should be
expected in this case as well.

The entire story of 21-cm cosmology as described thus far is at the
moment purely theoretical, but a great international effort is
underway to open up the observational field of 21-cm cosmology.  In
particular, several arrays of low-frequency radio telescopes have been
constructed (and are now operating) in order to detect the 21-cm
fluctuations from cosmic reionization (and beyond). Current efforts
include the Murchison Wide-field Array (MWA \cite{MWAref}), the Low
Frequency Array (LOFAR \cite{LOFARref}), the Giant Metrewave Radio
Telescope (GMRT \cite{GMRT}), and the Precision Array to Probe the
Epoch of Reionization (PAPER \cite{PAPER}), and future plans have been
made for the Hydrogen Epoch of Reionization Array (HERA;
http://reionization.org/) and the Square Kilometer Array (SKA;
https://www.skatelescope.org/); a 21-cm cosmology pathfinder of the
latter is the New Extension in Nan\c{c}ay Upgrading LOFAR (NenuFAR).
Although the expected foregrounds (dominated by Galactic synchrotron)
are much brighter than the 21-cm signal, they are not expected to
include sharp spectral features.  Thus, although ongoing experiments
are expected to yield noisy maps, the prospects for extraction of the
21-cm signal (and from it the reionization history) are quite
promising, using the key statistic of the 21-cm power spectrum
\cite{Bowman,McQuinn,Sources} as well as other statistics
\cite{BiP,CM03,Fur04b,Ichikawa}. A different approach is to measure
the total sky spectrum and detect the global reionization signal
arising from the overall disappearance of atomic hydrogen
\cite{Fur06,PL,nono}; current and future efforts (some also targeting
eras before reionization) include the Experiment to Detect the Global
EOR Step (EDGES \cite{bowman2}), the Large Aperture Experiment to
Detect the Dark Ages (LEDA; http://www.cfa.harvard.edu/LEDA/), and the
lunar-orbiting Dark Ages Radio Explorer (DARE;
http://lunar.colorado.edu/dare/).

A novel effect that was only noticed recently is the supersonic
velocity difference between the gas and the dark matter \cite{TH10}.
This intriguing effect (often called the streaming velocity) is
predicted to have influenced the gas distribution at high redshift as
well as early galactic halos.

The plan for this review is to first lay out the theoretical
groundwork for galaxy formation, in general (\S~\ref{s:gform}) and at
high redshift in particular (\S~\ref{s:gformz}), followed by the
basics of 21-cm cosmology (\S~\ref{s:21physics}). What follows is a
detailed discussion of the velocity streaming effect and its
consequences (\S~\ref{s:stream}). Next, we discuss in detail the
milestones of radiative feedback during early cosmic history
(\S~\ref{s:milestone}), and then outline their 21-cm signatures
(\S~\ref{s:Final21cm}).  Finally, we summarize the review and conclude
with a general outlook on the field (\S~\ref{s:fin}).

\section{Galaxy Formation: Basic Theory}

\label{s:gform}

The fundamental theoretical understanding of galaxy formation as
related to the earliest generations of stars and galaxies has been
reviewed extensively \cite{Barkana:2001,LoebBook}. Here we provide a
brief updated overview and summarize some useful results and formulas.

\subsection{Cosmological background}

\label{s:cosmo}

In General Relativity, the metric for a space which is spatially
homogeneous and isotropic is the Robertson-Walker metric, which can be
written in the form \beq \label{RW}
ds^2=dt^2-a^2(t)\left[\frac{dR^2}{1-k\,R^2}+R^2
  \left(d\theta^2+\sin^2\theta\,d\phi^2\right)\right]\ , \eeq where
$a(t)$ is the cosmic scale factor which describes expansion in time,
and $(R,\theta,\phi)$ are spherical comoving coordinates. The constant
$k$ determines the geometry of the metric; it is positive in a closed
universe, zero in a flat universe, and negative in an open universe.
Observers at rest at fixed $(R,\theta,\phi)$ remain at rest, with
their physical separation increasing with time in proportion to
$a(t)$. A given observer sees a nearby observer at physical distance
$D$ receding at the Hubble velocity $H(t)D$, where the Hubble constant
at time $t$ is $H(t)=d\ln a(t)/dt$. Light emitted by a source at time
$t$ is observed at the present time $t_0$ with a redshift
$z=1/a(t)-1$, where we set $a(t_0) \equiv 1$.

The Einstein field equations of General Relativity yield the Friedmann
equation \cite{Weinberg1972,KT1990} \beq H^2(t)=\frac{8 \pi
  G}{3}\rho-\frac{k}{a^2}\ ,\eeq which relates the expansion of the
universe to its matter-energy content. For each component of the
energy density $\rho$, with an equation of state $p=p(\rho)$, the
density $\rho$ varies with $a(t)$ according to the equation of energy
conservation \beq d (\rho R^3)=-p d(R^3)\ . \eeq With the critical
density \beq \rho_C(t) \equiv \frac{3 H^2(t)}{8 \pi G} \eeq defined as
the density needed for $k=0$, we define the ratio of the total density
to the critical density as \beq \Omega \equiv \frac{\rho}{\rho_C}\ .
\eeq With $\Omm$, $\Oml$, and $\Omr$ denoting the present
contributions to $\Omega$ from matter (including cold dark matter as
well as a contribution $\Omega_b$ from baryons), vacuum density
(cosmological constant), and radiation, respectively, the Friedmann
equation becomes \beq \frac{H(t)}{H_0}= \left[ \frac{\Omm} {a^3}+
  \Oml+ \frac{\Omr}{a^4}+ \frac{\Omk}{a^2}\right]^{1/2}\ , \eeq where
we define $H_0$ and $\Omega_0=\Omm+\Oml+\Omr$ to be the present values
of $H$ and $\Omega$, respectively, and we let \beq \Omk \equiv
-\frac{k}{H_0^2}=1-\Omega_0\ . \eeq In the particularly simple
Einstein-de Sitter (EdS) model ($\Omm=1$, $\Oml=\Omr=\Omk=0$), the
scale factor varies as $a(t) \propto t^{2/3}$. Even models with
non-zero $\Oml$ or $\Omk$ approach the Einstein-de Sitter behavior at
high redshifts, i.e., when \beq (1+z)\gg {\rm max} \left [\Omk /
  \Omega_m, (\Omega_\Lambda/\Omega_m)^{1/3}\right]\ , \label{highz1}
\eeq as long as we do not reach extremely early times at which $\Omr$
cannot be neglected. The approach to EdS is particularly rapid in
practice given that current observations imply $\Omk \approx 0$. In
this EdS regime (which we will also refer to as the high-$z$ regime),
$H(t) \approx 2/(3t)$.

We assume the best-fit cosmological parameters of the $\Lambda$CDM
(cosmological constant $\Lambda$ plus cold dark matter) model, based
on the first-year data of the Planck satellite \cite{Planck}:
$h=0.678$ (where the present Hubble constant is $H_0=100\, h\mbox{ km
  s}^{-1}\, \mbox{Mpc}^{-1}$), a flat Universe with matter density
parameter $\Omm = 0.307$ and the rest (adding up to unity) in a
cosmological constant, and $\Omb = 0.0482$. In convenient units, the
comoving cosmic mean density of matter in the Universe is: \beq
\bar{\rho}_m = 3.91 \times 10^{10} \left( \frac{\Omega_m h^2} {0.141}
\right) \frac{M_\odot} {\rm Mpc^3}\ , \eeq where the physical density
at redshift $z$ is higher by a factor of $(1+z)^3$.  Also, in the
high-$z$ regime, \beq H(z) \approx H_0 \frac{\sqrt{\Omm}} {a^{3/2}}\ ,
\eeq and the age of the universe is
\begin{equation}
  t\approx {2\over 3\, H_0 \sqrt{\Omega_m}} \left(1+z\right)^{-3/2} =
  5.49 \times 10^8 \left( \frac{\Omega_m h^2} {0.141} \right)^{-1/2} 
  \left(\frac{1+z}{10}\right)^{-3/2}{\rm yr}\ .
\label{highz2}
\end{equation}
Another useful quantity is the comoving (or particle) horizon, the
maximum distance from which light could have traveled to an observer
in the age of the universe. It thus represents a causal horizon for
physical influences (other than gravity). Its general expression, and
its approximate value in the high-redshift regime defined above (not
including an early period of inflation), are: \beq \eta = \int_0^{\,
  t} \frac{c\,dt'}{a(t')} \approx 5.05 \left( \frac{\Omega_m h^2}
  {0.141} \right)^{-1/2} \left(\frac{1+z}{10}\right)^{-1/2}{\rm Gpc}\
.
\label{e:hor}
\eeq We note that cosmologists often explicitly take out the Hubble
constant in expressions, e.g., distances in cosmology are expressed in
units of $h^{-1}$~Mpc (and wavenumbers in $h$~Mpc$^{-1}$).  This is to
some degree a remnant of an earlier time when $h$ was uncertain by
nearly a factor of two. Now that $h$ has been determined to equal 0.7
up to a few percent, it may be preferable to simply use units of Mpc
in papers, and specify the assumed $h$ in case it is needed for
precise comparisons.

In the standard hot Big Bang model, the universe is initially hot and
the energy density is dominated by radiation. The transition to matter
domination occurs at $z \sim 3400$, but the universe remains hot
enough that the gas is ionized, and electron-photon scattering
effectively couples the baryonic matter and the radiation. At $z \sim
1100$ the temperature drops below $\sim 3000$ K and protons and
electrons recombine to form neutral hydrogen. The photons then
decouple and travel freely until the present, when they are observed
as the CMB.

\subsection{Linear perturbation theory}

\label{s:lin}

Observations of the CMB (e.g., \cite{Planck}) show that the universe
at recombination was extremely uniform, but with large-scale
fluctuations in the energy density and gravitational potential of
roughly one part in $10^5$. Such small fluctuations, generated in the
early universe, grew over time due to gravitational instability, and
eventually led to the formation of galaxies and the large-scale
structure observed in the present universe.

We distinguish here between physical/proper and comoving coordinates.
Using vector notation, the physical coordinate ${\bf r}$ corresponds
to a comoving position $\xb=\rb/a$. In a homogeneous universe with
density $\rho$, we describe the cosmological expansion in terms of an
ideal pressure-less fluid of particles each of which is at fixed
$\xb$, expanding with the Hubble flow $\vb=H(t) \rb$ where
$\vb=d\rb/dt$.  Onto this uniform expansion we impose small
perturbations, given by a relative density perturbation \beq
\delta(\xb) \equiv \frac{\rho(\rb)}{\bar{\rho}}-1\ , \eeq where the
mean fluid density is $\bar{\rho}$, with a corresponding peculiar
velocity $\ub \equiv \vb - H \rb$. Then the fluid is described by the
continuity and Euler equations in comoving coordinates
\cite{Peebles1980,Peebles1993}. The gravitational potential $\phi$ is
given in turn by the Poisson equation, in terms of the density
perturbation. This fluid description is valid for describing the
evolution of collisionless cold dark matter particles until different
particle streams cross.  After such ``shell-crossing'', the individual
particle trajectories must in general be followed, but this typically
occurs only after perturbations have grown to become non-linear.
Similarly, baryons can be described as a pressure-less fluid as long
as their temperature is negligibly small, but non-linear collapse
leads to the formation of shocks in the gas.

For small perturbations $\delta \ll 1$, the fluid equations can be
linearized and combined to yield \beq \frac{\partial^2\delta}{\partial
  t^2}+2 H \frac{\partial\delta}{\partial t}=4 \pi G \bar{\rho}
\delta\ . \eeq This linear equation has in general two independent
solutions, the so-called growing and decaying modes. Starting with
random initial conditions, the growing mode comes to dominate the
density evolution. Thus, until it becomes non-linear, the density
perturbation maintains its shape in comoving coordinates and grows in
proportion to a growth factor $D(t)$. In the Einstein-de Sitter model
(or, at high redshift, in other models as well) the growth factor is
simply proportional to $a(t)$. Given the standard normalization of
$D(a=1)=1$, in EdS we would simply have $D(a)=a$, while in
$\Lambda$CDM with our standard parameters, at high redshift we have
$D(a) \approx 1.28 a$. In other words, the recent dominance by the
cosmological constant in $\Lambda$CDM suppresses the linear growth of
structure down to the present by a factor of 1.28 compared to a
universe that continued to follow the EdS model.

More generally, there is also a decaying mode that declines with time
rapidly after matter-radiation equality, as $a^{-3/2}$ in EdS. More
importantly, in the presence of baryons, the difference between the
distribution of dark matter and baryons persists for much longer (see
\S~\ref{sec:bary}).

The spatial form of the initial density fluctuations can be described
in Fourier space, in terms of Fourier components $\delta_\kb$,
where\beq \delta_\kb = \int d^3x\, \delta(\xb) e^{-i \kb \cdot \xb}\ ;
\ \ \ \delta(\xb) = \int \frac{d^3k}{(2 \pi)^3}\, \delta_\kb e^{i \kb
  \cdot \xb}\ .  \eeq We note that the $(2 \pi)^3$ factor is sometimes
switched (or split) between these two equations, so care must be taken
when comparing results that use different conventions for this factor
within the definitions of the Fourier transform and its inverse. Here
we have introduced the comoving wavevector $\kb$, whose magnitude $k$
is the comoving wavenumber which is equal to $2\pi$ divided by the
wavelength.

The Fourier description is particularly simple for fluctuations
generated by cosmic inflation \cite{KT1990}. Inflation generates
perturbations given by a Gaussian random field, in which different
$\kb$-modes are statistically independent, each with a random phase.
The statistical properties of the fluctuations are determined by the
variance of the different $\kb$-modes, and the variance is described
in terms of the power spectrum $P(k)$ as follows: \beq
\left<\delta_{\kb} \delta_{{\bf k'}}^{*}\right>=\left(2 \pi\right)^3
P(k)\, \delta^{(3)}_D \left(\kb-{\bf k'}\right)\ , \eeq where
$\delta^{(3)}_D$ is the three-dimensional Dirac delta function. Note
that $P(k)$ has units of volume, or, more generally, the power
spectrum of some quantity has units of volume times the square of the
units of that quantity.

In standard models, inflation produces a primordial power-law spectrum
$P(k) \propto k^n$ with $n \sim 1$. Perturbation growth in the
radiation-dominated and then matter-dominated universe results in a
modified final power spectrum, characterized by a turnover at a scale
of order the horizon $cH^{-1}$ at matter-radiation equality, and a
small-scale asymptotic shape of $P(k) \propto k^{n-4}$. On large
scales the power spectrum evolves in proportion to the square of the
growth factor, and this simple evolution is termed linear evolution.
On small scales, the power spectrum changes shape due to the
additional non-linear gravitational growth of perturbations, yielding
the non-linear (also called ``full'') power spectrum. The overall
amplitude of the power spectrum is not specified by current models of
inflation, and it is usually set observationally using the CMB
temperature fluctuations or local measures of large-scale structure.

Since density fluctuations may exist on all scales, in order to
determine the formation of objects of a given size or mass it is
useful to consider the statistical distribution of the smoothed
density field.  Using a window function $W(\yb)$ normalized so that
$\int d^3\yb\, W(\yb)=1$, the smoothed density perturbation field,
$\int d^3\yb \delta(\xb+\yb) W(\yb)$, itself follows a Gaussian
distribution with zero mean. For the particular choice of a spherical
top-hat, in which $W$ is constant within a sphere of radius $R$ and is
zero outside it, the smoothed perturbation field measures the
fluctuations in the mass $M$ in spheres of radius $R$. Indeed, a halo
of mass $M$ forms out of an initial (i.e., when $\delta \rightarrow
0$) region of comoving radius $R$, where \beq M = \frac{4}{3} \pi
\bar{\rho}_0 R^3 = 1.64 \times 10^{8} \left( \frac{\Omega_m h^2}
  {0.141} \right) \left( \frac{R}{100\ {\rm kpc}} \right)^3 M_\odot\ .
\eeq The inverse relation is: \beq R = 84.8 \left( \frac{\Omega_m
    h^2}{0.141} \right)^{-1/3} \left( \frac{M} {10^8
    M_\odot}\right)^{1/3} {\rm kpc}\ .  \eeq

The normalization of the present power spectrum is often specified by
the value of $\sigma_8 \equiv \sigma(R=8 h^{-1} {\rm Mpc})$. For the
top-hat, the smoothed perturbation field is denoted $\delta_R$ or
$\delta_M$ (in reference to the equivalent mass $M$). The variance
$\langle \delta_M \rangle^2$ is \beq S(M)=\sigma^2(M)= \sigma^2(R)=
\int_0^{\infty}\frac{dk}{2 \pi^2} \,k^2 P(k) \left[\frac{3
    j_1(kR)}{kR} \right]^2\ ,\label{eqsigM}\eeq where $j_1(x)=(\sin
x-x \cos x)/x^2$. The function $\sigma(M)$ plays a crucial role in
estimates of the abundance of collapsed objects, as described below.
We note that Eq.~\ref{eqsigM} in the limit of no smoothing (i.e., $R
\rightarrow 0$) shows that the contribution of power at wavenumber $k$
per $\log k$ to the variance at a point is $k^3 P(k)/(2 \pi^2)$. The
relative (dimensionless) fluctuation level at $k$ is defined as the
root mean square, i.e., the square root of this contribution to the
variance, and in 21-cm cosmology, if the cosmic mean 21-cm brightness
temperature at some redshift is $\langle T_{\rm b} \rangle$, then the
21-cm fluctuation level $\delta T_{\rm b}$ at $k$ (usually in units of
mK) is defined as: \beq \delta T_{\rm b} = \langle T_{\rm b} \rangle
\sqrt{ \frac{k^3 P(k)}{2 \pi^2}}\ , \eeq in terms of the dimensionless
21-cm power spectrum $P(k)$ (i.e., the power spectrum of relative
fluctuations in the 21-cm brightness temperature).

\subsection{Non-linear collapse}

\label{sec:NL}

The small density fluctuations evidenced in the CMB grow over time as
described in the previous subsection, until the perturbation $\delta$
becomes of order unity, and the full non-linear gravitational problem
must be considered. The dynamical collapse of a dark matter halo can
be solved analytically only in cases of particular symmetry, with the
simplest case being that of spherical symmetry. Although this model is
restricted in its direct applicability, the results of spherical
collapse have turned out to be surprisingly useful in understanding
the properties and distribution of halos in models based on cold dark
matter.

In spherical collapse, at the moment when the top-hat collapses to a
point, the extrapolated overdensity as predicted by linear theory is
\cite{Peebles1980} $\delta_L\, = 1.686$ in the Einstein-de Sitter
model, with a weak dependence on $\Omm$ and $\Oml$ in the more general
case. Thus, a top-hat collapses at redshift $z$ if its linear
overdensity extrapolated to the present day (also termed the critical
density of collapse) is \beq \delta_{\rm crit}(z)=\frac{1.686}{D(z)}\
, \label{deltac} \eeq where we again set $D(z=0)=1$.

Even a slight violation of the exact symmetry of the initial
perturbation can prevent the top-hat from collapsing to a point.
Instead, the halo reaches a state of virial equilibrium by violent
relaxation (phase mixing). Using the virial theorem $U=-2K$ to relate
the potential energy $U$ to the kinetic energy $K$ in the final state,
the final overdensity relative to the critical density at the collapse
redshift is found to be $\Delta_c=18\pi^2 \simeq 178$ in the
Einstein-de Sitter model. This theoretical value is slightly modified
in $\Lambda$CDM, but conventionally the EdS value (or even the rougher
value of 200) is often used to {\it define}\/ the virial radius
$r_{\rm vir}$ and the virial masses of halos in numerical simulations
and in analyses of observations.

Quantitatively, a halo of mass $M$ collapsing at redshift $z$ (assumed
high enough for the EdS limit) thus has a physical virial radius \beq
r_{\rm vir}=1.51\ \left( \frac{\Omega_m h^2}{0.141} \right)^{-1/3}
\left(\frac{M}{10^8\ M_{\odot} }\right)^{1/3} \left( \frac{\Delta_c}
  {18\pi^2}\right)^{-1/3} \left (\frac{1+z}{10} \right)^{-1}\ {\rm
  kpc}\ ,
\label{rvir}\eeq a corresponding circular velocity, \beq
V_c=\left(\frac{G M}{r_{\rm vir}}\right)^{1/2}= 16.9\ \left(
  \frac{\Omega_m h^2}{0.141} \right)^{1/6} \left( \frac{M}{10^8\
    M_{\odot} }\right)^{1/3} \left( \frac{\Delta_c}
  {18\pi^2}\right)^{1/6} \left( \frac{1+z} {10} \right)^{1/2}\ {\rm
  km\ s}^{-1}\ , \label{Vceqn} \eeq and a virial temperature \beq
\label{tvir} T_{\rm vir}=\frac{\mu m_p V_c^2}{2 k_{\rm B}}=1.03\times
10^4\ \left( \frac{\Omega_m h^2}{0.141} \right)^{1/3}
\left(\frac{\mu}{0.6}\right) \left(\frac{M}{10^8\ M_{\odot}
  }\right)^{2/3} \left( \frac{\Delta_c} {18\pi^2}\right)^{1/3}
\left(\frac{1+z}{10}\right)\ {\rm K} \ , \eeq where $\mu$ is the mean
molecular weight in units of the proton mass $m_p$. Note that the
value of $\mu$ depends on the ionization state of the gas; $\mu=0.59$
for a fully ionized primordial gas, $\mu=0.61$ for a gas with ionized
hydrogen but only singly-ionized helium, and $\mu=1.22$ for neutral
primordial gas.

Although spherical collapse captures some of the physics governing the
formation of halos, structure formation in cold dark matter models
proceeds hierarchically. At early times, most of the dark matter is in
low-mass halos, and these halos continuously accrete and merge to form
high-mass halos. Numerical simulations of hierarchical halo formation
indicate a roughly universal spherically-averaged density profile for
the resulting halos (Navarro, Frenk, \& White 1997 \cite{NFW},
hereafter NFW), though with considerable scatter among different
halos. The NFW profile has the form \beq \rho(r) \propto \frac{1} {\cN
  x (1+\cN x)^2}\ , \label{NFW} \eeq where $x=r/r_{\rm vir}$ and $\cN$
is the concentration parameter.

In addition to characterizing the properties of individual halos, a
critical prediction of any theory of structure formation is the
abundance of halos, i.e., the number density of halos as a function of
mass, at any redshift. While the number density of halos can be
measured for particular cosmologies in numerical simulations, an
analytical model helps us gain physical understanding, can be used to
explore the dependence of the halo abundance on the cosmological
parameters, and can be extrapolated to regimes that cannot be reached
by current simulations. It is also a starting point towards building
models for the abundances of galaxies and galaxy clusters.

A simple analytical model that has become the basis for work in this
field was developed by Press \& Schechter (1974) \cite{ps74}. The
model is based on the ideas of a Gaussian random field of density
perturbations, linear gravitational growth, and spherical collapse. To
determine the abundance of halos at a redshift $z$, in this model we
use $\delta_M$, the density field smoothed on a mass scale $M$, as
defined in the previous subsection. Although the model is based on the
initial conditions, it is usually expressed in terms of redshift-zero
quantities. Thus, we use the linearly-extrapolated density field,
i.e., the initial density field at high redshift extrapolated to the
present by simple multiplication by the relative linear growth factor.
A useful entity is the 'present power spectrum', which refers to the
initial power spectrum, linearly-extrapolated to the present without
including non-linear evolution. Since $\delta_M$ is distributed as a
Gaussian variable with zero mean and standard deviation $\sigma(M)$
[which is determined by Eq.~(\ref{eqsigM}) from the present power
spectrum], the probability that $\delta_M$ is greater than some
$\delta$ equals $(1/2)\, {\rm erfc} \left[ \delta/ \left(\sqrt{2}
    \,\sigma \right) \right]$. The fundamental ansatz is to identify
this probability with the fraction of dark matter particles that are
part of collapsed halos of mass greater than $M$, at redshift $z$.
There are two additional ingredients: First, the value used for
$\delta$ is $\delta_{\rm crit}(z)$ given in Eq.~(\ref{deltac}),
which is the critical density of collapse found for a spherical
top-hat (extrapolated to the present since $\sigma(M)$ is calculated
using the present power spectrum); and second, the fraction of dark
matter in halos above $M$ is multiplied by an additional factor of 2
in order to ensure that every particle ends up as part of some halo
with $M>0$. Thus, the final formula for the mass fraction in halos
above $M$ at redshift $z$ is \beq
\label{PSerfc} F(>M | z)={\rm erfc}\left[ \frac{\delta_{\rm crit}(z)}
  {\sqrt{2}\,\sigma(M) } \right]\ . \eeq

This ad-hoc factor of 2 is necessary, since otherwise only positive
fluctuations of $\delta_M$ would be included. Bond et al.\ (1991)
\cite{bond91} found a more satisfactory derivation of this correction
factor, using a different ansatz. In their derivation, the factor of 2
originates from the so-called ``cloud-in-cloud'' problem: For a given
mass $M$, even if $\delta_M$ is smaller than $\delta_{\rm crit}(z)$,
it is possible that the corresponding region lies inside a region of
some larger mass $M_L>M$, with $\delta_{M_L}>\delta_{\rm crit}(z)$. In
this case the original region should be counted as belonging to a halo
of mass $M_L$. Bond et al.\ (1991) \cite{bond91} showed that, under
certain assumptions, the additional contribution results precisely in
a factor of 2 correction. We note that this work is the basis of the
extended Press-Schechter model, which is mentioned later in this
review.

The halo abundance (or halo mass function) is \beq \frac{dn}{dM} =
\frac{\bar{\rho}_m}{M} \left|\frac{d S}{d M} \right| f(\del_{\rm
  crit}(z),S)\ ,
\label{eq:abundance} \eeq where $dn$ is the comoving number density of
halos with masses in the range $M$ to $M+dM$, $S=\sigma^2(M)$ is the
variance on scale $M$, and $f(\del_{\rm crit}(z),S)\, dS$ is defined
to be the mass fraction contained at $z$ within halos with mass in the
range corresponding to $S$ to $S+d S$. In the Press-Schechter model
\cite{ps74}, \beq f_{\rm PS}(\del_{\rm crit}(z),S) = \frac{1} {\sqrt{2
    \pi}}\, \frac{\nu }{S} \exp\left[-\frac{\nu^2}{2} \right]\ ,
\label{eq:PS} \eeq where $\nu=\del_{\rm crit}(z)/\sqrt{S}$ is the
number of standard deviations that the critical collapse overdensity
at $z$ represents on the mass scale $M$ corresponding to the variance
$S$.

The classic Press-Schechter \cite{ps74} model for the halo mass
function fits numerical simulations only roughly, and in particular it
substantially underestimates the abundance of the rarest, most massive
halos. The halo mass function of Sheth \& Tormen (1999)
\cite{shetht99}, with modified best-fit parameters \cite{shetht02},
fits numerical simulations much more accurately \cite{jenkins}. It is
given by: \beq f_{\rm ST}(\del_{\rm crit}(z),S) = A' \frac{\nu }{S}
\sqrt{\frac{a'} {2 \pi}} \left[ 1+\frac{1}{(a' \nu^2)^{q'}} \right]
\exp\left[-\frac{a' \nu^2}{2} \right]\ , \label{eq:ST} \eeq with
best-fit parameters $a'=0.75$ and $q'=0.3$, and where normalization to
unity is ensured by taking $A'=0.322$.

In addition to the overall, mean abundance of halos, another key
question in cosmology and galaxy formation is the spatial distribution
of the halo number density. In particular, since halos form due to
gravity, massive halos should form in larger numbers in regions of
high overall density than in low-density voids. Thus, density
fluctuations are expected to lead to spatial fluctuations in the halo
number density. This leads to the concept of halo (or galaxy) bias, a
now-standard concept in galaxy formation
\cite{ps74,k84,b86,ck89,bond91,mw96}. Particularly simple is the case
of linear bias, i.e., when the distribution of galaxies is a
proportionally amplified (``biased'') version of that of the
underlying density of matter. Mathematically this means that the
relative fluctuations in the number density of galactic halos
($\delta_{\rm g}$) are proportional to the relative fluctuations in
the underlying density of matter $\delta$: \beq \delta_{\rm g} = b\,
\delta\ ,
\label{e:bias} \eeq where $b$ is the linear bias factor (or simply
"the bias").

This simple result is expected to be reasonably accurate when looking
at fluctuations on large scales $s$ (usually tens of comoving Mpc or
more). Several conditions must be satisfied for this to be true. The
scale $s$ must be much larger than the spatial scales involved in
forming the individual galactic halos whose clustering is being
considered; this allows a separation of scales that is the basis of a
simple approximation called a peak-background split \cite{ck89}.
Also, in order to avoid non-linear effects, the scale $s$ must be
large enough that typically $\delta \ll 1$, i.e., the variance is
small when the density field is averaged on the scale $s$. Similarly,
$\delta_{\rm g} \ll 1$ on that scale is advisable. Finally, gravity
must dominate galaxy formation, or at least, any other effects (such
as astrophysical feedbacks) must operate on much smaller scales than
$s$. Of these conditions, the first two tend to be more favorable at
high redshifts, since the galaxies are typically small and thus
associated with small formation scales, and density fluctuations on
large scales are still relatively small. However, the last two
conditions become more problematic, as discussed in great detail in
the rest of this review.  High-redshift galaxies are highly biased, so
their fluctuations are much larger than those in the underlying
density (section~\ref{s:unusually}); and since early galaxies were
typically small, they were susceptible to a variety of external
feedbacks that operate on scales of order 100~Mpc, including the
supersonic streaming velocity (section~\ref{s:stream}) and various
radiative feedbacks (section~\ref{s:milestone}).

\subsection{Baryons: linear evolution, pressure, and cooling}

\label{sec:bary}

Baryons play a major role in cosmology. On the largest scales, their
coupling to the photons in the early universe leaves them clustered
differently from the dark matter, with the difference decaying away
only gradually. On smaller scales, the baryonic pressure suppresses
gravitational growth. Most directly, of course, the baryons are
important since stars form out of baryons that cool and collapse to
high density. To get started, we note a useful number: the cosmic mean
number density of hydrogen (including both neutral and ionized forms)
is \beq \bar{n}_H(z) = 1.89 \times 10^{-7} \left( \frac{\Omb h^2}
  {0.0221} \right) (1+z)^3\, {\rm cm}^{-3}\ , \eeq assuming that
$76\%$ of the baryon mass density is in hydrogen.  The number density
of helium is smaller by a factor of 12.7.

As noted in \S~\ref{s:lin}, in the presence of dark matter only, the
linear perturbations are dominated by a growing mode that is $\propto
a$ in EdS, as the decaying mode drops rapidly, $\propto a^{-3/2}$ in
EdS.  On large scales, baryons also respond to gravity only (after
cosmic recombination), but their initial conditions are different, as
their strong coupling to the CMB photons up to recombination
suppresses their sub-horizon fluctuations. Thus, cosmic recombination
begins a period of baryonic infall, during which the baryons gradually
catch up with the dark matter perturbations \cite{Peebles1980}.
Specifically, if we denote the perturbations of the dark matter and
baryon density $\delta_{\rm dm}$ and $\delta_{\rm b}$, respectively,
and their mass fractions within the total matter density $f_{\rm
  dm}=(\Omm-\Omb)/\Omm$ and $f_{\rm b}=\Omb/\Omm$, then it is useful
to work with the perturbation of the total density, $\delta_{\rm
  tot}=f_{\rm dm} \delta_{\rm dm}+f_{\rm b} \delta_{\rm b}$, and the
difference $\delta_{\rm diff} = \delta_{\rm b} - \delta_{\rm tot}$.
In linear perturbation theory, $\delta_{\rm tot}$ has the usual
growing and decaying modes (i.e., $\propto a$ and $\propto a^{-3/2}$
in EdS), while the two solutions for $\delta_{\rm diff}$ are constant
($\propto 1$) and $\propto a^{-1/2}$ in EdS. Thus, the baryon
perturbation $\delta_{\rm b}$ approaches $\delta_{\rm tot}$ gradually
from below. This approach can be described through their relative
difference. If we approximately include only the dominant modes, this
key quantity decays as \cite{NBhalos} \beq r_{\rm LSS} \equiv
\frac{\delta_{\rm diff}} {\delta_{\rm tot}} \approx -\frac{0.3 \%} {a}
\ .  \eeq This decay is slow enough that the gradual baryonic infall
is in principle observable in high-redshift 21-cm measurements
\cite{BL05c} and perhaps also in the distribution of galaxies at low
redshift \cite{BL11,Maayane}.

During this era of baryonic infall, and before cosmic heating from
radiative astrophysical sources, the gas cooled adiabatically with the
expansion. Traditional calculations
\cite{Peebles1980,MaBert1995,CMBFAST} assumed a uniform speed of sound
for the gas at each redshift, but a more careful consideration of the
combination of adiabatic cooling and Compton heating substantially
modifies the temperature perturbations on all scales
\cite{Yamamoto1,Yamamoto2,BL05c,NB05}.

On small scales, the density evolution is no longer purely
gravitational, as the gas pressure suppresses baryon perturbations.
The relative force balance at a given time can be characterized by the
Jeans scale, which is the minimum scale at which a small perturbation
will grow due to gravity overcoming pressure gradients. If the gas
has a uniform sound speed $c_{\rm s}$, then the comoving Jeans
wavenumber is \beq k_{\rm J} = \frac{a}{c_{\rm s}} \sqrt{4 \pi G
  \bar{\rho}_m}\ .  \eeq In the simple limit where the gas cools
adiabatically (after thermally decoupling from the CMB at $z \sim
150$), this gives a characteristic Jeans mass (defined in terms of a
sphere of diameter equal to the Jeans wavelength) \cite{Barkana:2001}
\beq M_{\rm J}\equiv {4\pi\over 3} \left({\pi \over k_{\rm
      J}}\right)^3 \bar{\rho}_m = 5.89\times
10^3\left({\Omega_mh^2\over 0.141}\right)^{-1/2} \left({\Omega_b
    h^2\over 0.0221}\right)^{-3/5} \left({1+z\over
    10}\right)^{3/2}~M_\odot\ .
\label{eq:m_j}
\eeq

The Jeans mass, however, is not the whole story, since it is related
only to the evolution of perturbations at a given time. When the Jeans
mass itself varies with time, the overall suppression of the growth of
perturbations depends on a time-averaged Jeans mass, the filtering
mass \cite{GH98}. To define it, we start from the regime of
large-scale structure (i.e., scales too large to be affected by
pressure, but much smaller than the horizon and the scale of baryon
acoustic oscillations), where, as noted above, $r_{\rm LSS}$ does not
depend on $k$, and is simply a function of redshift. On smaller
scales, the next-order term describing the difference between the
baryons and dark matter is the $k^2$ term \cite{GH98}, and the
filtering wavenumber $k_{\rm F}$ and corresponding mass scale $M_{\rm
  F}$ are defined through \cite{NBhalos} \beq \frac{ \delta_{\rm b}}
{\delta_{\rm tot}} = 1 + r_{\rm LSS} - \frac{k^2} {k_{\rm F}^2}\ ;\ \
\ M_{\rm F}\equiv {4\pi\over 3} \left({\pi \over k_{\rm F}}\right)^3
\bar{\rho}_m\ .  \eeq This filtering mass scale captures how the whole
history of the evolving Jeans mass affects the final baryon
perturbations that result at a given time. Starting at early times,
since the baryon fluctuations are very small before cosmic
recombination, the gas pressure (which depends on $\delta_{\rm b}$)
starts out small, so the filtering mass starts from low values and
rises with time up to a value of $\sim 3 \times 10^4 M_\odot$
\cite{NBhalos} around redshift 30. It then drops due to the cooling
cosmic gas, but the drop is very gradual (reaching $\sim 2 \times 10^4
M_\odot$ at $z=10$ in the absence of cosmic heating or reionization)
due to the remaining after-effects of the suppression of gas infall at
higher redshifts.  This behavior is significantly different from the
Jeans mass, which declines rapidly with time (as in Eq.~\ref{eq:m_j})
and drops below $10^4 M_\odot$ at $z=13$.

What makes the filtering mass even more useful is that it seems to
offer in many situations a good estimate of the minimum halo mass that
manages to accrete a significant amount of gas (e.g., $50\%$ of the
cosmic baryon fraction). It is natural to expect some relation between
this characteristic, minimum halo mass and the filtering mass, since
the gas fraction in a collapsing halo reflects the total amount of gas
that was able to accumulate in the collapsing region during the
entire, extended collapse process. For example, a sudden change in gas
temperature immediately begins to affect gas motions (through the
pressure-gradient force), but has only a gradual, time-integrated
effect on the overall amount of gas in a given region. In this way,
the minimum accreting mass is analogous to the linear-theory filtering
mass. However, the former is defined within the deeply non-linear
regime, so the two masses may not necessarily agree quantitatively.

Gnedin \cite{G00b} first compared the filtering mass to the
characteristic mass in numerical simulations, suggesting that they are
approximately equal in the post-reionization universe in which the IGM
is hot and ionized. However, he used a non-standard definition of the
filtering mass that equals 8 times the standard definition given
above. Subsequently, higher-resolution simulations did not find a
clear relation between the theoretically calculated filtering mass and
the characteristic mass measured in post-reionization simulations
\cite{Hoeft,Okamoto}. However, the heating within simulations of
inhomogeneous reionization is complex, and thus the filtering mass
(which depends on the thermal history) is difficult to compute
directly. The filtering mass has been shown to agree to within a
factor of $\sim 1.5$ with the characteristic mass measured in
simulations at higher redshifts, throughout the era prior to
significant cosmic heating or reionization, as well as after a
controlled, sudden heating \cite{NaozMF1,NaozMF2}. Thus, the issue of
the possible usefulness of the filtering mass after reionization has
not been settled, but alternative models have been recently proposed
to fit results from post-reionization simulations
\cite{MesingerMF,McQuinnMF}.

The conclusion is that prior to cosmic heating and reionization, gas
is expected to accumulate significantly in dark matter minihalos down
to a mass of $\sim 3 \times 10^4 M_\odot$ \cite{NaozMF2}. This minimum
accretion mass later rises during cosmic heating and even more rapidly
within ionized regions during cosmic reionization. In addition, even
at the highest redshifts, the minimum mass is boosted in regions of
significant streaming velocity (see \S~\ref{s:stream} below).

\begin{figure}[tbp]
\includegraphics[width=\textwidth]{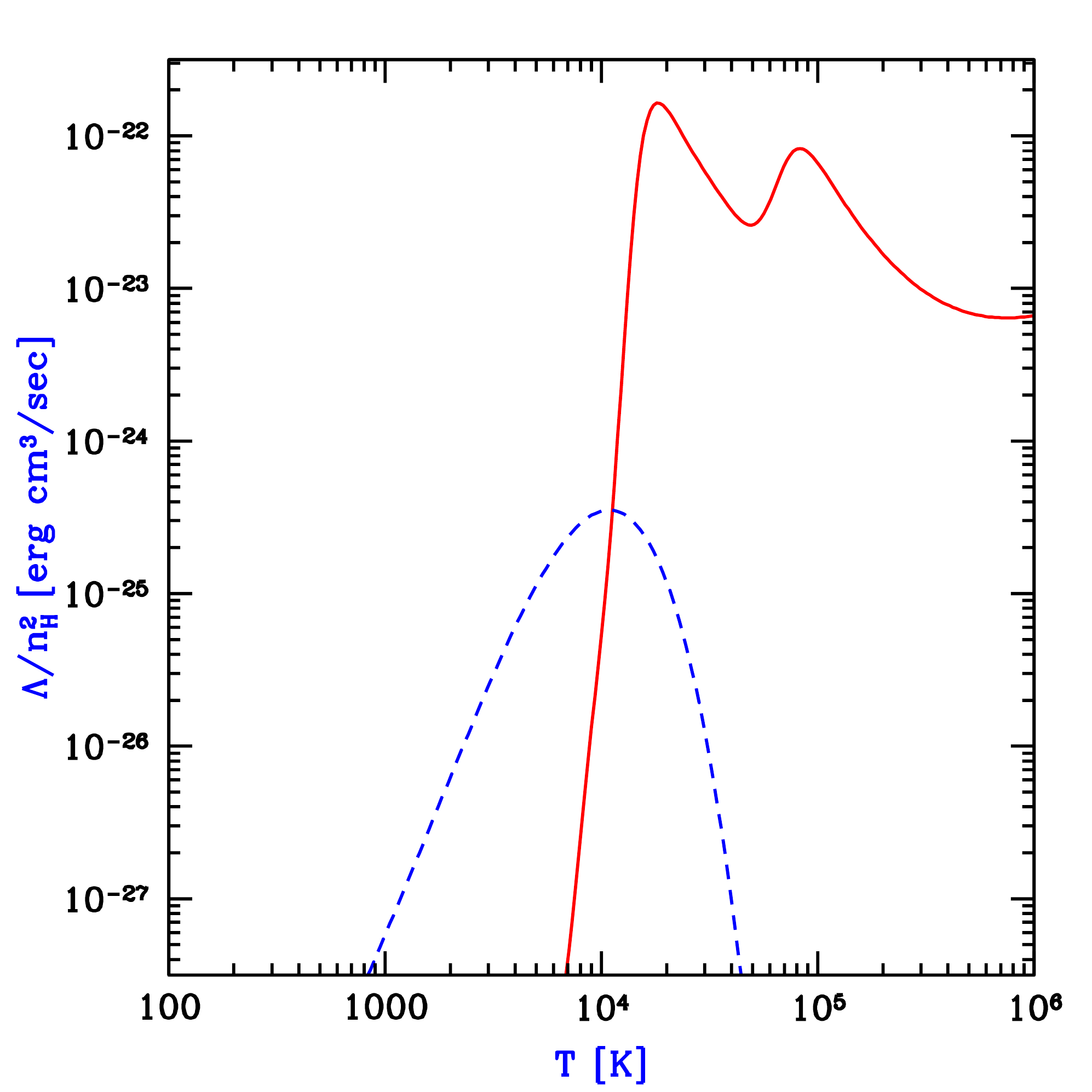}
\caption{Cooling rates as a function of temperature for a primordial
  gas composed of atomic hydrogen and helium, as well as molecular
  hydrogen, in the absence of any metals or external radiation. The
  plotted quantity $\Lambda/n_H^2$ is roughly independent of density
  (unless $n_H \gg 10\ {\rm cm}^{-3}$), where $\Lambda$ is the volume
  cooling rate (in erg/sec/cm$^3$). The solid line shows the cooling
  curve for an atomic gas, with the characteristic peaks due to
  collisional excitation of H~I and He~II. The dashed line shows the
  additional contribution of molecular cooling, assuming a molecular
  abundance equal to $0.1\%$ of $n_H$. From \cite{Barkana:2001}.}
\label{f:cooling}
\end{figure}

We end this section with a brief summary of cooling.
Figure~\ref{f:cooling} shows the cooling curve for primordial gas,
prior to metal enrichment. Primordial {\it atomic}\/ gas can radiate
energy only once hydrogen or helium are significantly ionized, so such
cooling is limited to gas at temperatures above $\sim 10^4$~K. At high
redshifts, most of the gas is in halos with relatively low masses, so
that even if the accreted gas is shocked and heated to the virial
temperature (Eq.~\ref{tvir}), it is unable to cool. However, in the
presence of even a small ionized hydrogen fraction, molecular hydrogen
can acquire sufficient abundance to provide significant cooling
\cite{1967H2}, and its rotational and vibrational transitions allow
cooling down to below $10^3$~K. Further details about primordial gas
cooling are reviewed elsewhere \cite{Barkana:2001}.

\section{Galaxy Formation: High-Redshift Highlights}

\label{s:gformz}

In this section we expand on several features of galaxy formation that
are particularly important at high redshifts. We first discuss
fluctuations in the density of galaxies, which are generally important
in cosmology, but at high redshift the fluctuations become unusually
large and this has significant consequences that reverberate
throughout this review. We then discuss various challenges of
numerical simulations and approaches to deal with them. While
simulations have become an indispensable tool in cosmology, it is
important to bear in mind that they have fundamental limitations, some
of them specific to, or worsening at, high redshifts. For example,
while simulations at low redshift can be continually tested by and
improved based on astronomical observations, this is not currently
possible (or is at least far more limited) at high redshift. Finally,
we discuss the formation of the very first stars, obviously a subject
of great theoretical and numerical interest, hopefully with
significant observational traces as well.

\subsection{Large fluctuations in the galaxy number density}

\label{s:unusually}

A broad, common thread runs through much of the recent theoretical
development of cosmic reionization and 21-cm cosmology: The density of
galaxies (or stars) varies spatially, with the fluctuations becoming
surprisingly large at high redshift, even on quite large cosmological
scales \cite{BL04}. This can be understood from the standard theory of
galaxy formation as due to the fact that the first galaxies
represented rare peaks in the cosmic density field.

\begin{figure}[tbp]
\includegraphics[width=\textwidth]{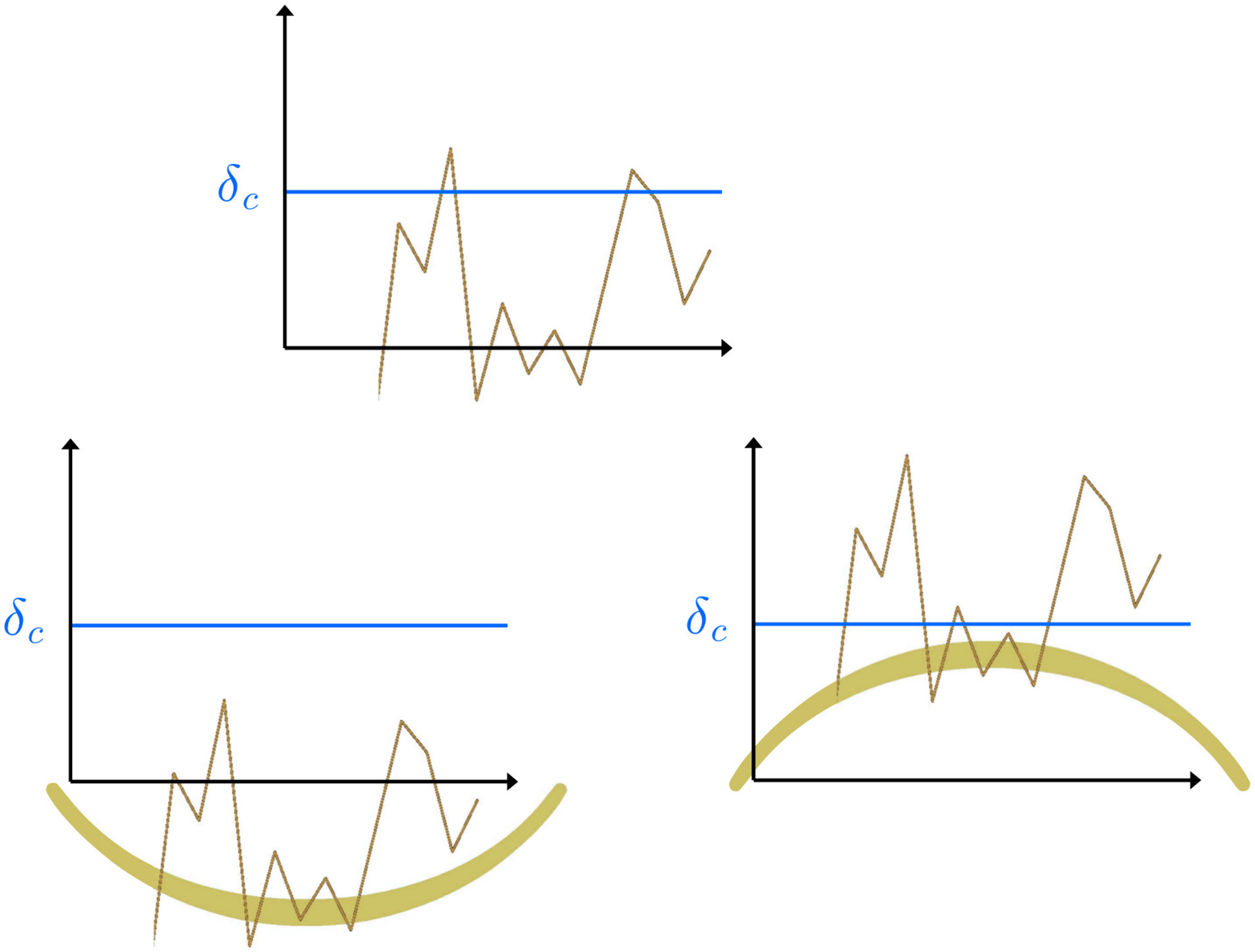}
\caption{Simple illustration of the large bias of high-redshift
  galaxies, which is the main idea driving the character of
  reionization \cite{BL04} and the 21-cm fluctuations during cosmic
  dawn \cite{BL05b}. To form a halo, the total (linearly-extrapolated)
  density fluctuation must reach a value $\delta_c$ [denoted
  $\delta_{\rm crit}(z)$ in Eq.~(\ref{deltac})], from the sum of
  large-scale and small-scale density fluctuations. Thus, a
  large-scale void (bottom left) might have no halos, a typical region
  (top) a couple small halos, while a region with a large-scale
  overdensity (bottom right) will have many halos, both small and
  large. See text for additional explanation.}
\label{f:ePS}
\end{figure}

As an analogy, imagine searching on Earth for mountain peaks above
5000 meters. The 200 such peaks are not at all distributed uniformly
but instead are found in a few distinct clusters on top of large
mountain ranges. Similarly, in order to find the early galaxies, one
must first locate a region with a large-scale density enhancement, and
then galaxies will be found there in abundance. For a more detailed
argument, note that galactic halos form roughly in regions where the
(linearly extrapolated) density perturbation reaches above a fixed
threshold value $\delta_{\rm crit}(z)$ (see section~\ref{sec:NL}).
Now, the total density at a point is the sum of contributions from
density fluctuations on various scales (Figure~\ref{f:ePS}). For
initial perturbations from inflation (which follow the statistics of a
Gaussian random field), the fluctuations on different scales are
statistically independent. Thus, the same small-scale density
fluctuations are added, in different regions, to various
long-wavelength density fluctuations. In an over-dense region on large
scales, the small-scale fluctuations only need to supply the missing
amount needed to reach $\delta_{\rm crit}(z)$, while in a large-scale
void, the same small-scale fluctuations must supply a total density of
$\delta_{\rm crit}(z)$ plus the extra density missing within the void.
This means that a larger fraction of the volume within the over-dense
region will reach above $\delta_{\rm crit}(z)$ in total density, and
thus more halos will form there. Now, at high redshift, when density
fluctuations had not yet had time for much gravitational growth, the
effective threshold value $\delta_{\rm crit}(z)$ is many times larger
than the typical density fluctuation on the scales that form galactic
halos. In other words, each halo represents a many-$\sigma$
fluctuation. Under Gaussian statistics, the fraction of points above
$t\sigma$ changes rapidly with $t$, once $t$ is $2-3$ or higher.
Thus, the abundance of halos in a given region changes rapidly with
small changes of the mean density in the region (and this mean density
is set by large-scale density modes). The density of star formation is
thus expected to show strongly biased (i.e., amplified) fluctuations
on large scales \cite{BL04}. These large-scale fluctuations at high
redshift, and their great observational importance, had for a long
time been underestimated, in part because the limited range of scales
available to numerical simulations put these fluctuations mostly out
of their reach.

\begin{figure}[tbp]
\includegraphics[width=\textwidth]{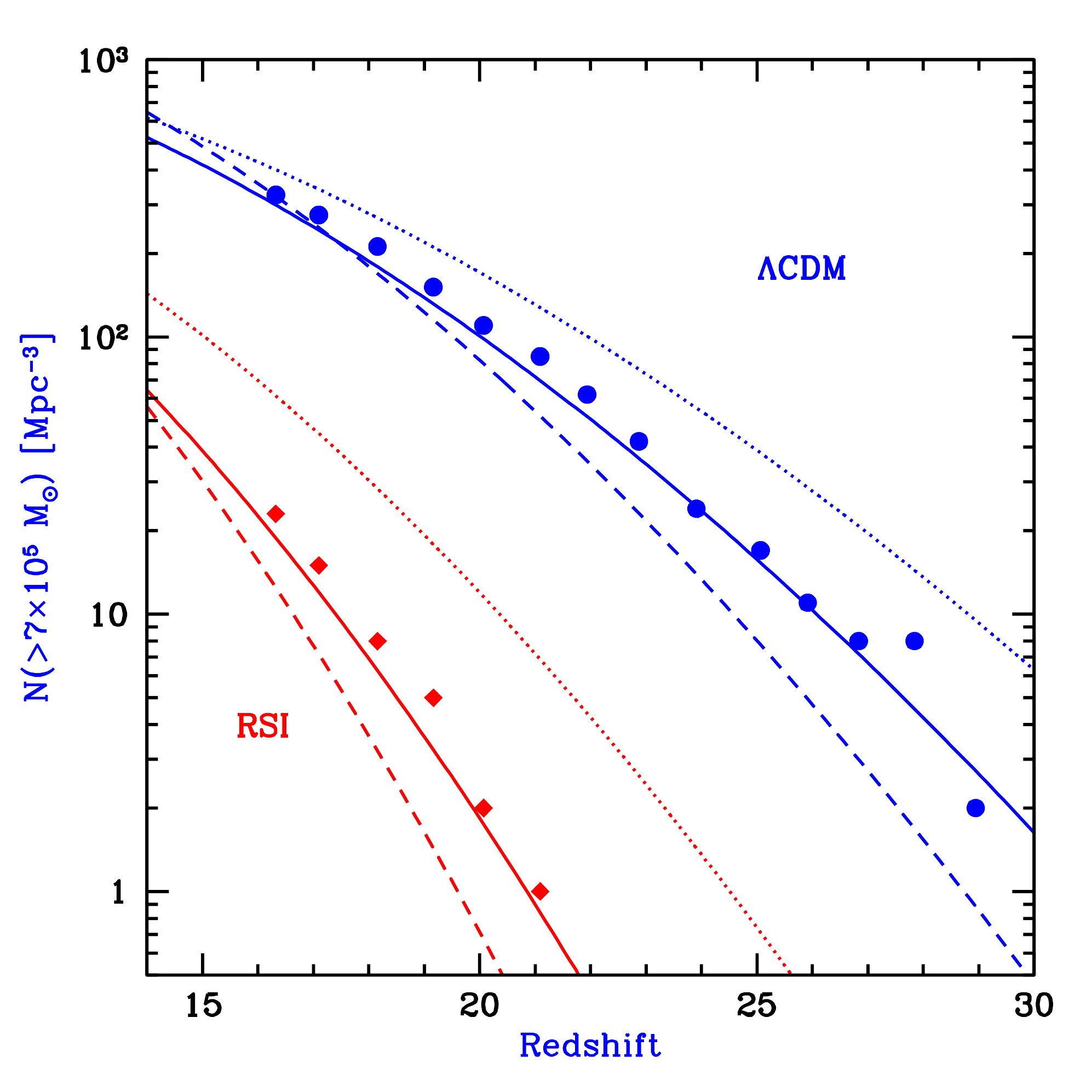}
\caption{Halo mass function at high redshift in a 1 Mpc box at the
  cosmic mean density. Data points show the number of halos above mass
  $7 \times 10^5 M_{\odot}$ as measured in simulations (from Figure~5
  of \cite{yoshida}) with two different sets of cosmological
  parameters, the scale-invariant $\Lambda$CDM model of \cite{yoshida}
  (upper curves), and their running scalar index (RSI) model (lower
  curves). Each data set is compared with three theoretically
  predicted curves. The simulated values are well below the cosmic
  mean of the halo mass function (dotted lines). However, the
  prediction of the Barkana \& Loeb (2004) \cite{BL04} hybrid model
  (solid lines) takes into account the periodic boundary conditions of
  the small simulation box and matches the simulation results fairly
  well. The pure extended Press-Schechter model (dashed lines) is too
  low. From \cite{BL04}.}
\label{f:Nbody}
\end{figure}

Figure~\ref{f:Nbody} illustrates a further effect, which is that the
limited box size of simulations leads to a delay of halo formation, or
equivalently, an underestimate of the abundance of halos (and stars)
at any given time.  The reason is that the periodic boundary
conditions within the finite simulation box artificially set the
amplitude of large-scale modes (above the box size) to zero. There are
many such volumes in the real Universe, with various mean densities
(that follow a Gaussian distribution, within linear perturbation
theory). Since galaxies (especially at high redshift) are highly
biased, most of them form in those volumes that have an unusually high
mean density. Thus, a simulated volume at the cosmic mean density is
not representative of the locations of stars.

This limitation of simulations is most acute for the very first star
in the Universe, a challenge of special interest for simulators
because it represents in principle a perfectly clean problem, before
the first entrance of the complexities of astrophysical feedback from
prior star formation. The very first star formed in a very rare
high-density region. Indeed, the large size of the real Universe
allowed such a rare fluctuation to be found somewhere by chance, but
it is unlikely to be found within a small simulation box, even if the
simulation has the right abundance of galaxies (while real
simulations, in addition, artificially lower this abundance when
setting the mean density in the box to the cosmic mean density). For
example, one of the first high-resolution ``first star'' simulations
formed its first star only at redshift 18.2 \cite{abn02}, while
analytical methods show that the first stars must have formed at $z
\sim 65$ \cite{first,anastasia} within our past light cone (i.e., so
that we can in principle see them as they formed), or a further
$\Delta z \sim 6$ earlier \cite{avihabit} within the entire volume of
the observable Universe (so that we can see them or their remnants
after they formed). On this point, we note that there were some early,
rough analytical estimates of the formation redshift of the very first
stars \cite{miralda03,wise05}.

More generally, Barkana \& Loeb (2004) \cite{BL04} developed a hybrid
model that allows one to predict the modified halo mass function in
regions of various sizes and various average densities. As noted in
section~\ref{sec:NL}, for the cosmic mean halo abundance, the classic
Press-Schechter \cite{ps74} model works only roughly, while the halo
mass function of Sheth \& Tormen (1999) \cite{shetht99} (with modified
best-fit parameters \cite{shetht02}) fits numerical simulations much
more accurately \cite{jenkins}.  Now, a generalization of the
Press-Schechter model known as the extended Press-Schechter model
\cite{bond91} allows the prediction of the halo mass function in a
given volume (of given size and mean density) compared to the cosmic
mean mass function. No simple generalization of this type is known for
the Sheth-Tormen mass function, but Barkana \& Loeb \cite{BL04}
pointed out that this problem can be overcome since the prediction of
the extended Press-Schechter model for the change {\it relative}\/ to
the cosmic mean mass function has been shown to provide a good fit to
numerical simulations over a wide range of parameters
\cite{mw96,casas02,shetht02}. Thus, the Barkana \& Loeb \cite{BL04}
hybrid model starts with the Sheth-Tormen mass function and applies a
correction based on the extended Press-Schechter model. The model
gives a good match to simulations even in volumes that strongly
deviate from the cosmic mean halo function (Figure~\ref{f:Nbody}).

The idea of unusually large fluctuations in the abundance of early
galaxies first made a major impact on studies of cosmic reionization,
leading to the conclusion that reionization occurs inside-out, with
typical H~II bubbles that are larger and thus easier to observe than
previously thought \cite{BL04} (see \S~\ref{s:reion}). The same idea
soon found another important application in a different regime,
leading to the prediction of 21-cm fluctuations from earlier times
during cosmic dawn. The study of fluctuations in the intensity of
early cosmic radiation fields began with \Lya radiation \cite{BL05b}
(see \S~\ref{s:Lya}) and continued to other fields including the
X-rays responsible for early cosmic heating \cite{Jonathan07} (see
\S~\ref{s:heating}). These are all sources of 21-cm fluctuations, and
are thus the main targets for 21-cm radio interferometers. Clearly,
the idea of substantial large-scale fluctuations in galaxy numbers is
a driver of much of the current theoretical and observational interest
in 21-cm cosmology as a way to probe the era of early galaxy
formation. The recent discovery of the streaming velocity (see
\S~\ref{s:stream}) has added a new flavor to this general theme.

\subsection{Simulations at high redshift: challenges and approaches}

\label{s:numerical}

In this subsection we discuss several aspects of simulations of the
high-redshift Universe. First, we discuss some challenges and
limitations of current numerical simulations, particularly when
applied to early galaxy formation at high redshifts. Some of the
issues we discuss can be addressed with additional study (e.g.,
setting the initial conditions accurately), while other difficulties
are likely to remain for the foreseeable future (such as uncertainties
related to star formation and stellar feedback). We then briefly
discuss other approaches: analytical models and semi-numerical
simulations.

\begin{figure}[tbp]
\includegraphics[width=\textwidth]{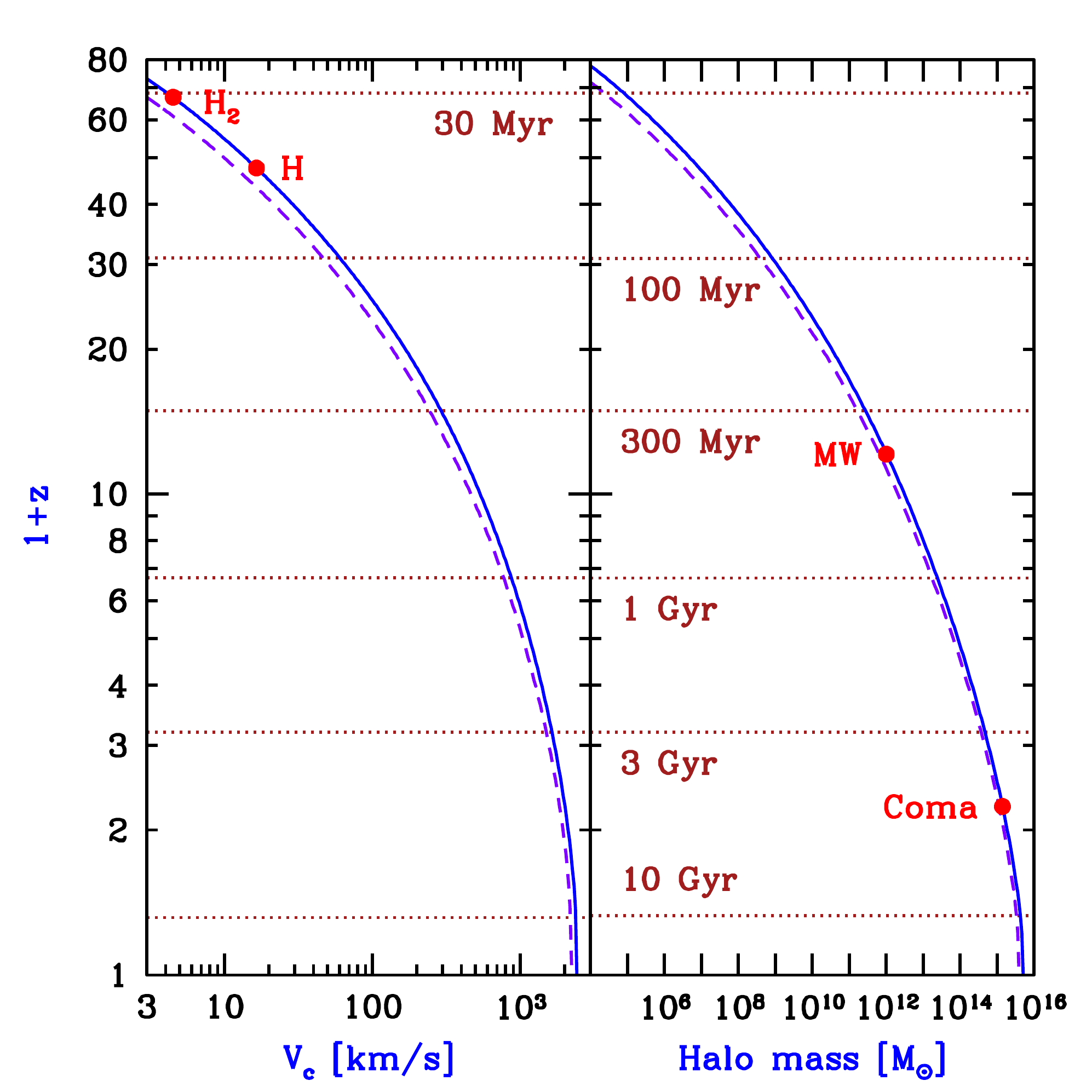}
\caption{The median redshift for the first appearance (in our past
  light cone) of various populations of halos: either halos above a
  minimum circular velocity (left panel) or a minimum mass (right
  panel). Dots indicate in particular the first star-forming halo in
  which $H_2$ allows the gas to cool, the first galaxy that forms via
  atomic cooling (H), as well as the first galaxy as massive as our
  own Milky Way and the first cluster as massive as Coma. The
  horizontal lines indicate the elapsed time since the Big Bang. The
  results from two sets of cosmological parameters (solid curves
  \cite{Spergel07} and dashed curves \cite{Viel06}) illustrate the
  systematic error due to current uncertainties in the values of the
  cosmological parameters. From \cite{first}.}
\label{f:first}
\end{figure}

We begin with a number of challenges that are important to recognize
when evaluating the results of numerical simulations. As explained in
the previous subsection, the large size of the real Universe implies
that stars began to form very early.  More generally, halos of various
masses (or circular velocities) are predicted to have begun to form
much earlier than the typical redshifts we are accustomed to, both
from current numerical simulations and current observations.
Figure~\ref{f:first} shows that while the very first star formed (in
our past light cone) via molecular cooling at $z \sim 65$, the first
generation of more massive atomic-cooling halos formed at $z \sim 47$
\cite{first}. While the Milky Way halo mass is fairly typical in
today's Universe, the very first such halo formed at $z \sim 11$, and
the first Coma cluster halo at $z \sim 1.2$.

A direct simulation of the entire observable universe out to the
spherical shell at redshift 70 would require a simulated box of length
25,000~Mpc on a side. Actual simulations, which often form a ``first
star'' at redshift 20 or 30, effectively explore a very different
environment from $z \sim 65$, in terms of the CMB temperature, the
cosmic and virial halo densities (of both the dark matter and gas),
the halo merger history, and high-redshift effects such as the
difference between the power spectra of baryons and dark matter
(discussed further below). Even if simulations do not attempt to
approach the very first star, critical physical effects at high
redshift push simulations towards the requirement of large boxes. The
fact that the typical bubble scale of cosmic reionization is tens of
Mpc (see \S~\ref{s:reion}) already implies a minimum box size of $\sim
100$~Mpc for this era.  However, the streaming velocity
(\S~\ref{s:stream}), which is important early on, has a typical
coherence scale of $\sim 100$~Mpc, and the radiation fields
responsible for early feedback (\S~\ref{s:milestone}) -- \Lya
coupling, Lyman-Werner feedback, and cosmic heating -- fluctuate
significantly on a similar scale. In particular, hard X-rays heat from
afar and can extend the heating era into cosmic reionization
(\S~\ref{s:heating} and \S~\ref{s:late}).

A significant presence of any one of these effects is enough to force
any reasonable simulation during these epochs to a minimum box size of
$\sim 400$~Mpc. Another consideration that pulls in the same direction
is that observations of the 21-cm signal are easier (and currently
only possible) on large scales. The sensitivity of a radio
interferometer is degraded as the angular resolution is increased
[Eq.~(\ref{e:Tthermal})]. Thus, numerical simulations are squeezed
between the need to cover a huge volume, on the one hand, and the need
to adequately resolve each halo, on the other hand.  This becomes
especially demanding at early times, when most of the star formation
occurs in very low-mass halos. Consider, for example, an N-body
simulation of a 400~Mpc box in which $10^6 M_\odot$ halos are resolved
into 500 dark matter particles each. Extensive tests \cite{converge}
show that this resolution is necessary in order to determine the
overall properties of an individual halo (such as halo mass) just
crudely, to within a factor of two; for better accuracy or to
determine properties such as star formation, more particles are
required.  Even with just 500, this would require a simulation with a
total of $10^{15}$ particles, much higher than numbers that are
currently feasible. Truly resolving star formation within these halos
also requires hydrodynamics and radiative transfer at very high
resolution. 

\begin{figure}[tbp]
\includegraphics[width=\textwidth]{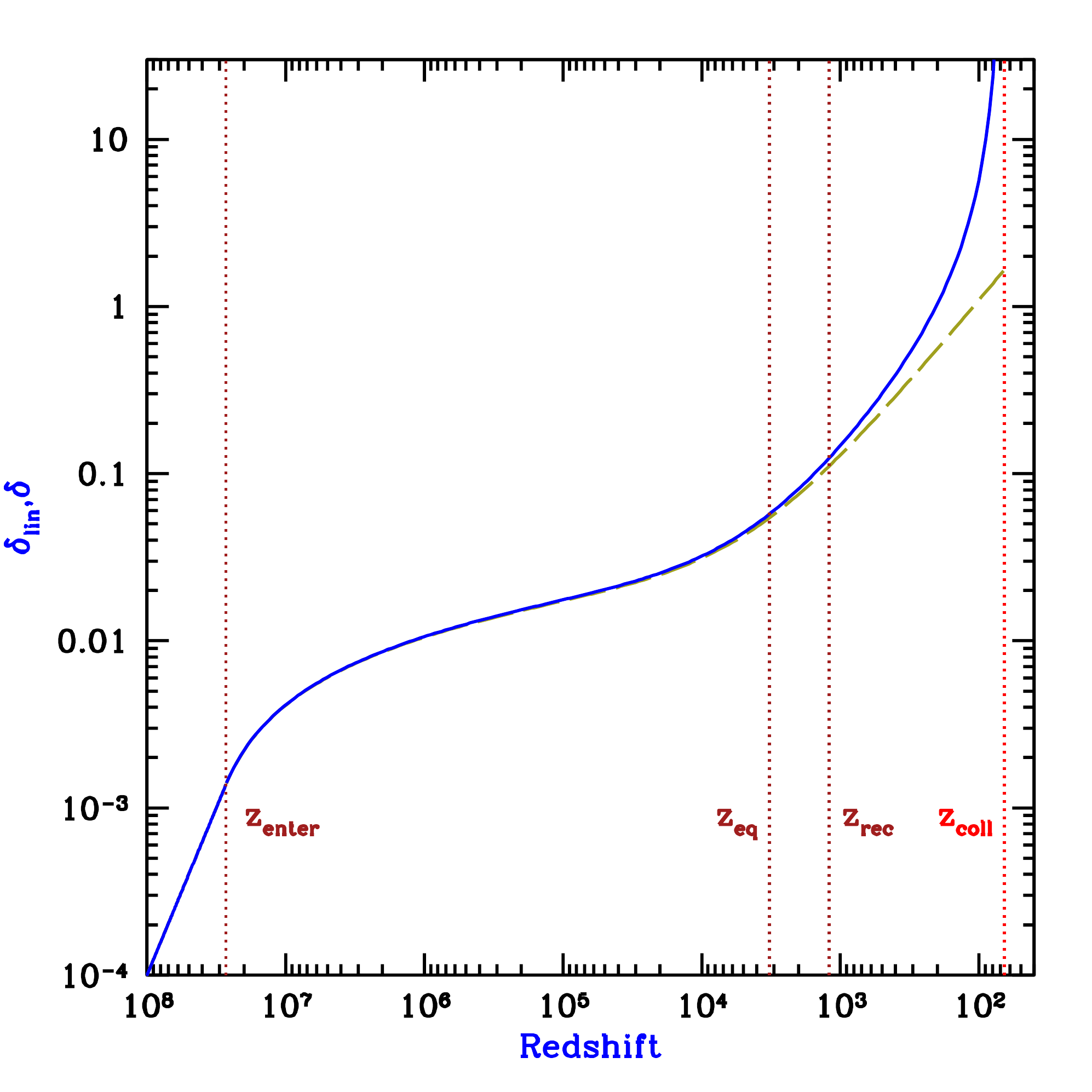}
\caption{Evolution of the fractional overdensity $\delta$ for a
  spherical region containing $10^5 M_{\odot}$ that collapses at
  $z=66$ (approximately corresponding to the host halo of the very
  first star in our past light cone). We show the fully non-linear
  $\delta$ (solid curve) and the linearly-extrapolated $\delta$
  (dashed curve).  We indicate the redshifts of halo collapse ($z_{\rm
    coll}$), cosmic recombination ($z_{\rm rec}$), matter-radiation
  equality ($z_{\rm eq}$), and entry into the horizon ($z_{\rm
    enter}$). Note that the overdensity shown here corresponds to
  synchronous gauge. From \cite{first}.}
\label{f:GRhalo}
\end{figure}

Naoz et al.\ (2006) \cite{first} pointed out another limitation of
current simulations, namely that they do not determine their initial
conditions accurately enough for achieving precise results for
high-redshift halos, especially those hosting the very first stars.
Simulations assume Gaussian random initial fluctuations as might be
generated by a period of cosmic inflation in the early Universe. The
evolution of these fluctuations can be calculated exactly as long as
they are small, with the linearized Einstein-Boltzmann equations. The
need to begin the simulation when fluctuations are still linear forces
numerical simulations of the first star-forming halos to start at very
high redshifts (much higher than starting redshifts in common use that
are often around $z=200$). According to spherical collapse, a halo
forming at redshift $z_{\rm form}$ has an extrapolated linear
overdensity of $\delta = \delta_c \sim 1.7$. Since it grows roughly
with the EdS growing mode, the corresponding perturbation (in the dark
matter) is $\delta \sim 13\%\, [(1+z_{\rm form})/66] $ at cosmic
recombination, and $\delta \sim 6\%\, [(1+z_{\rm form})/66]$ at
matter-radiation equality (see Figure~\ref{f:GRhalo}). The
perturbation reaches $\delta \sim 1\%\, [(1+z_{\rm form})/66] $
extremely early, at $z \sim 10^6$. It re-enters the horizon (after
having left during inflation) when $\delta \sim 0.2\%\, [(1+z_{\rm
  form})/66]$ at $z \sim 3 \times 10^7$; precision at this level would
require setting initial conditions with a non-linear General
Relativistic calculation.

In addition to the problem of non-linearity, there is also the
influence of early cosmic history on the linear and (more
challengingly) non-linear initial conditions. Effects that must be
taken into account include the contribution of the radiation to the
cosmic expansion, suppression of sub-horizon perturbations in the
photon density by the radiation pressure, and the coupling of the
baryons to the photons which suppresses baryon perturbations until
cosmic recombination. Within a spherical collapse calculation, Naoz et
al.\ \cite{first,NBhalos} calculated halo formation including all
these effects (Figure~\ref{f:GRhalo}), and found that they result in
an earlier formation redshift for the first star by $3.3\%$ in $1+z$
(compared to using the standard results from spherical collapse). The
extended period at high redshift when the baryon perturbations remain
suppressed compared to the dark matter is the main cause of this shift
in the formation time, but the contribution of the photons to the
expansion of the universe also makes a significant contribution. A
$3.3\%$ change in $1+z$ at $z \sim 65$ corresponds to a $4.8\%$ change
in the age of the universe, and to an order of magnitude change in the
abundance of halos at a given redshift at $z \sim 65$. The shift in
$1+z$ for the formation of a given halo goes down with time but is
still $1\%$ at $z \sim 20$. In addition, early cosmic history has a
major impact (by factors of two or more) on the amount (and
distribution) of gas that accumulates in the halos that hosted the
first stars \cite{NaozMF2} (see the discussion of the filtering mass
in \S~\ref{sec:bary}); this effect is increased further by the
presence of the streaming velocity (see \S~\ref{s:stream}).
Therefore, even mild precision in numerical simulations of the
formation of the first stars requires a calculation of these effects
on halo formation, in combination with the above-mentioned issue of
non-linearity going back to extremely early times.

Thus, while some processes are calculated with very high precision in
numerical simulations, there are much larger effects that must be
confronted before the results can be considered to be accurate. Even
in the limit of the very first stars, ostensibly a very clean problem
for numerical simulations, the effects just discussed make the problem
difficult, even if all relevant physical processes can someday be
included and numerical convergence fully demonstrated. The current
status of numerical simulation results on the formation of the first
stars is summarized below (\S~\ref{sec:firststars}).

Numerical simulation of galaxy formation {\em beyond}\/ the very first
star (in a given cosmological region) faces even bigger problems,
which can be summarized with one word: feedback. Long-distance
feedback directly from stellar radiation is generated by \Lya photons
(reaching out to $\sim 300$~Mpc), Lyman-Werner photons (out to $\sim
100$~Mpc), and UV ionizing photons (initially absorbed in the
immediate surroundings, but reaching up to $\sim 70$~Mpc by the end of
reionization \cite{WL04}). Some stars have strong stellar winds, and
some explode in supernovae, which deposit thermal and kinetic energy
as well as metals. Stellar remnants such as X-ray binaries produce
X-rays which include hard photons that reach cosmological distances.
Central black holes may also produce thermal and kinetic feedback, as
well as UV and X-ray radiation. Most types of radiation that are
responsible for feedback can be partially absorbed or scattered within
the emitting galaxy or its immediate surroundings, another important
process that depends on the detailed, small-scale distribution of gas
and metals.  Given the basic uncertainties about the detailed physics
even of well-observed present-day astrophysical phenomena such as
magnetic fields, dust, supernovae, the stellar initial mass function,
and central black holes, ab-initio numerical simulations that are
truly self-contained do not seem feasible. Once these various feedback
effects begin to operate, they strongly affect the properties of
subsequent generations of stars and galaxies, so that many observable
predictions become strongly dependent on the generation and results of
feedback.  Numerical simulations can offer increasingly precise
gravity, hydrodynamics, and radiative transfer, but are often limited
by simplistic models of star formation and feedback that are inserted
by hand.  A major issue with astrophysical sources is that truly
simulating their formation process, detailed structure, and feedback
would require resolving length scales that are around 20 orders of
magnitude smaller than the cosmological distances reached by some of
the photons responsible for radiative feedback.  The resulting vast
gulf between the resolution of cosmological simulations and that of
reality means that increasing resolution does not necessarily imply
convergence towards the correct final answer; there could be multiple
regimes of apparent convergence as additional levels of resolution
uncover new physical processes.

On the opposite end from simulations are analytical (or
semi-analytical) models. These models are very flexible, can be easily
used to explore a wide variety of astrophysical possibilities and to
incorporate a range of astrophysical uncertainties, and can be
directly fit to data in order to determine the parameters of well-fit
models. Such models can also be made more quantitatively accurate by
basing them on fits to the results of numerical simulations of early
galaxy formation. However, analytical models are also significantly
limited.  In 21-cm cosmology, perhaps their biggest limitation is that
they usually must assume linear perturbations. While large-scale
density fluctuations are indeed fairly small at early times, the large
bias of high-redshift galaxies (\S~\ref{s:unusually}) leads to quite
non-linear fluctuations in the radiative, astrophysical sources of
21-cm fluctuations. In addition, the highly non-linear fluctuations on
small scales do not completely average out when smoothing on large
scales (as in real observations).  This is due to additional
non-linear relationships in 21-cm cosmology such as the dependence of
21-cm temperature on gas temperature (Eq.~\ref{e:combo2} or
\ref{e:combo3}).  Thus, analytical calculations based on assuming
linear perturbations and linear bias are quite limited in their
accuracy (An important example is the discussion in
\S~\ref{s:anisotropy} of non-linear limits on the accuracy of the
linear result for the anisotropy of the 21-cm power spectrum).

The limitations of both numerical simulations and analytical models
have led to the rise of an intermediate approach that combines some of
the advantages of both. This method is termed hybrid, or
semi-numerical simulation. While there are several specific
approaches, the basic idea is to calculate physical processes directly
on large scales, where everything is relatively simple, and indirectly
on small, highly non-linear scales. On the small scales, halos and
their properties are often adopted from semi-analytical models that
have been fitted to numerical simulation results, or sometimes
directly from the outputs of N-body (i.e., gravity-only) simulations
plus some assumptions about star formation and other astrophysics. On
the large scales, radiation such as X-rays, LW, and \Lya photons can
be directly summed from all sources, albeit with a few approximations
(e.g., the optical depth calculated assuming the cosmic mean density,
and multiple scattering of \Lya photons treated approximately). Also,
for reionization, such codes usually employ an approximation based on
an analytical model for the distribution of H~II bubble sizes
\cite{fzh04} (\S~\ref{s:reion}); fortunately, the resulting ionized
bubble distribution is quite similar to the results of radiative
transfer, except in the fine (small-scale) details (see
Figure~\ref{f:zahn} in \S~\ref{s:reion}). A successful,
publicly-available semi-numerical code in 21-cm cosmology is 21CMFAST
\cite{21cmfast}; results from this code and from the code developed by
the author's group \cite{Cold} are shown in \S~\ref{s:Final21cm}.

To summarize this subsection, numerical simulations of early galaxies
offer the potential advantages of fully realistic source halo
distributions and accurate gravity, hydrodynamics, and radiative
transfer. However, much of the vitality of the field comes from the
major uncertainties associated with the formation of, and feedback
from, astrophysical sources. For example, it is possible that most
early stars were much more massive and thus brighter than modern
stars, or that a relatively large amount of gas collected within
massive mini-quasars in the centers of galaxies. These astrophysical
uncertainties will very likely be resolved only based on direct
observational evidence. As we contemplate the range of possible
observational predictions, it is much easier to explore a wide variety
of astrophysical possibilities with simple analytical models or
semi-numerical hybrid methods that combine processes on a large-scale
grid with a sub-grid model based on numerical simulation results. Once
the observations come in, there will be a need to fit astrophysical
parameters to the data, and this requires a flexible framework and
cannot be done directly with numerical simulations; once a well-fit
model has been found, though, simulations may offer the best way to
compare it in detail with the observations. It is important to note
that discoveries in the field of the first stars and 21-cm cosmology
(as summarized throughout this review) are often driven by large-scale
processes, so due to the limited reach of simulations, many have come
first from analytical or semi-numerical methods.

\subsection{The very first stars}
\label{sec:firststars}

In the previous two subsections we discussed the limitations of
numerical simulations in general, and those of the first stars in
particular.  Still, simulations remain our best tool for trying to
understand and predict the detailed properties of the first stars.
This subject has been extensively reviewed elsewhere
\cite{BrommRev,BrommRev2}, but we briefly summarize it in this
section. In principle, the formation of primordial stars is a clean
numerical problem, as the initial conditions (including the
distribution of the gas and dark matter and the chemical and thermal
history of the gas) are cosmological and not yet affected by
astrophysical feedback. One possible (though still speculative)
complication is the generation and amplification of magnetic fields in
the early universe in time for them to affect the formation of the
first stars \cite{Bfield1,Bfield2,NaozB,BrommRev}.

As mentioned at the end of \S~\ref{sec:bary}, under cosmological
conditions, gas cooling in small early halos is possible only via
molecular hydrogen cooling. Studies of the non-equilibrium chemistry
of $H_2$ formation and destruction \cite{H2a,H2b,H2c,H2d,H2e}
concluded that $H_2$ formation in a collapsing small halo is dominated
by the $H^{-}$ channel, in which the residual free electrons from
cosmic recombination act as catalysts: \beq H + e^- \rightarrow H^- +
\gamma\ ;\ \ \ H^- + H \rightarrow H_2 + e^-\ . \eeq

\begin{figure}[tbp]
\includegraphics[width=\textwidth]{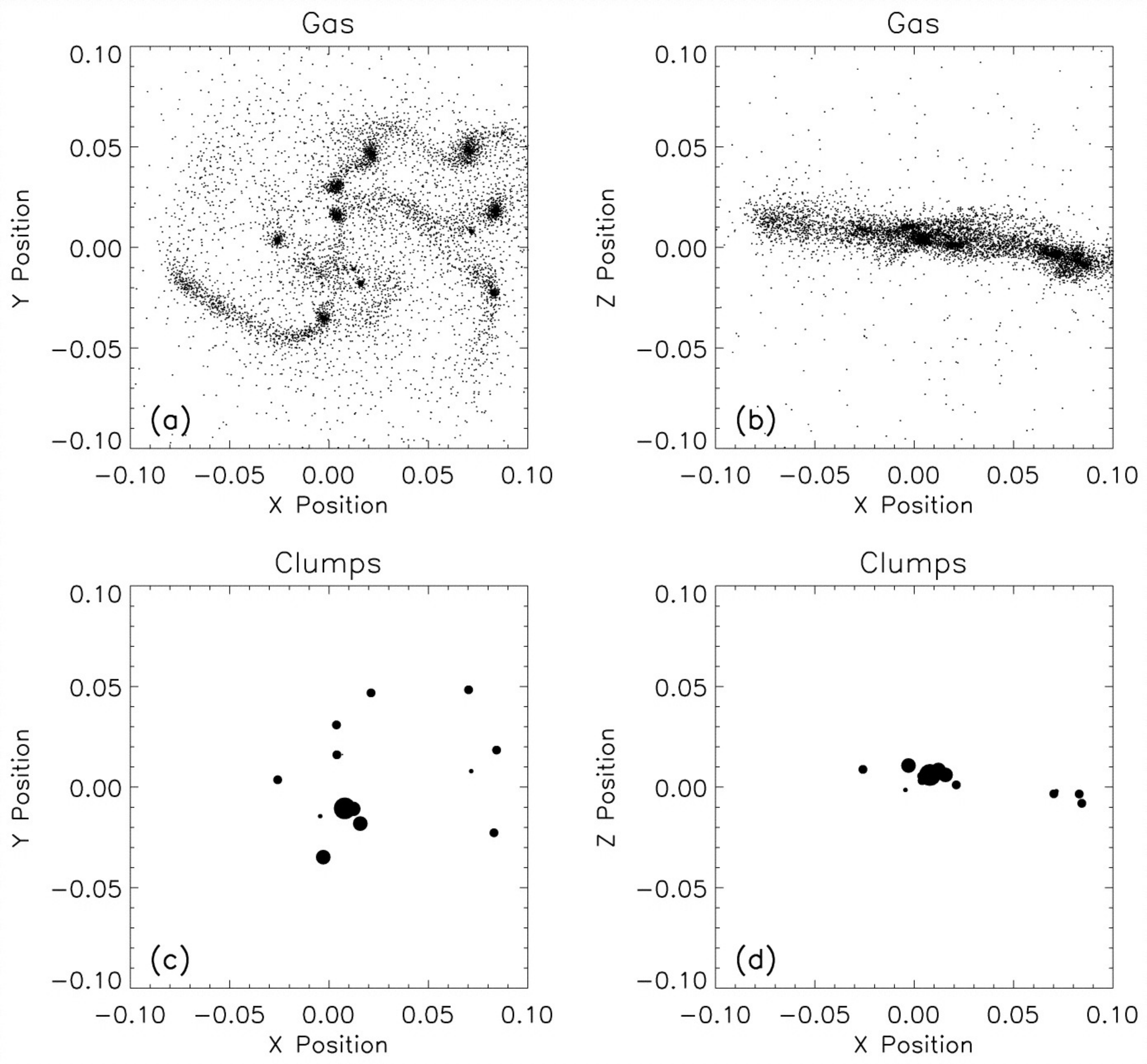}
\caption{Gas and clump morphology at $z = 28$ in the first-star
  simulation of Bromm et al.\ (1999) \cite{Bromm}.  Top row: The
  remaining gas in the diffuse phase. Bottom row: The distribution of
  clumps, where the four increasing dot sizes denote increasing clump
  masses ($>10^2 M_\odot$, $>10^3 M_\odot$, $>5\times 10^3 M_\odot$,
  $>10^4 M_\odot$).  Left panels: Face-on view. Right panels: Edge-on
  view. The length of the box is 30~pc. The gas has settled into a
  flattened configuration with a number of dominant, massive clumps.
  From \cite{Bromm}.}
\label{f:Bromm}
\end{figure}

\begin{figure}[tbp]
\includegraphics[width=\textwidth]{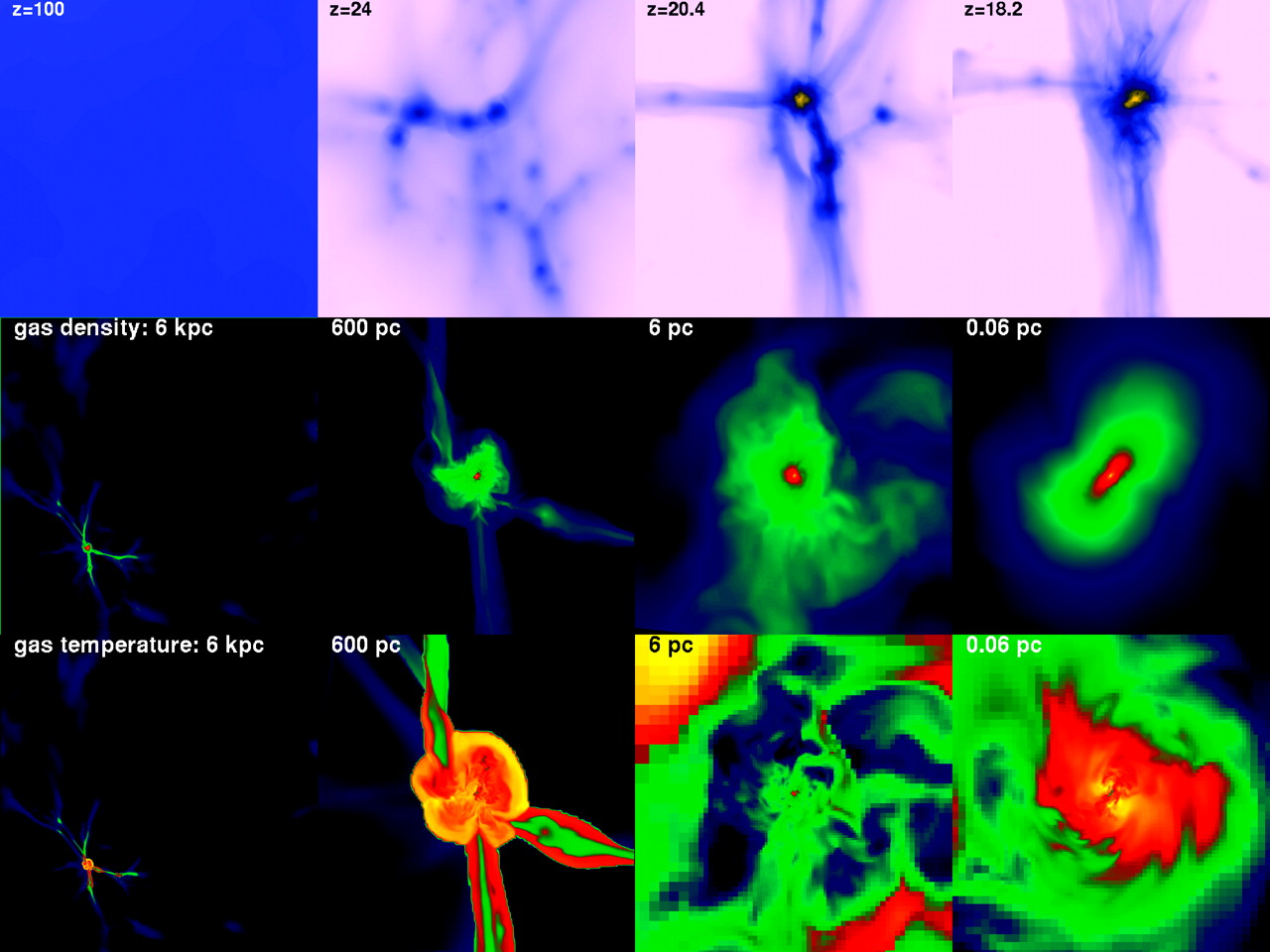}
\caption{The first star in a simulation by Abel et al.\ (2002)
  \cite{Abel}. Top row: Projection of gas density on a 600~pc scale
  (all distances are physical in this Figure), at several redshifts.
  Other two rows: Slices of gas density or temperature on several
  different scales, all at the final redshift of the simulation
  ($z=18.2$). From left to right, the two bottom rows show:
  large-scale filaments; the virial accretion shock; the $H_2$ cooled,
  high-redshift molecular cloud analog; and a warm core containing
  $\sim 100\, M_\odot$ of gas. From \cite{Abel}.}
\label{f:AbelSingle}
\end{figure}

Numerical simulation of the formation of a first (so-called Population
III, or Pop III) star via $H_2$ cooling in a primordial minihalo of
$10^5-10^6 M_{\odot}$ has proven to be a difficult problem, as initial
results that established a prediction of single very massive stars
have recently been replaced by a new paradigm of multiple stellar
systems with a range of masses. Indeed, the first generation of
simulations indicated the formation of massive Pop III stars of $\sim
100 M_\odot$.  Such stars would be short-lived, generate extremely
strong ionizing radiation and stellar winds, and end up producing
massive black hole seeds or pair-instability supernovae. The
expectation of massive stars was consistent between early simulations
evolving an artificial overdensity with a smooth particle
hydrodynamics (SPH) code (Figure~\ref{f:Bromm}) and simulations that
directly employed cosmological initial conditions along with the
impressive resolution of an adaptive mesh refinement (AMR) code
(Figure~\ref{f:AbelSingle}).

\begin{figure}[tbp]
\includegraphics[width=\textwidth]{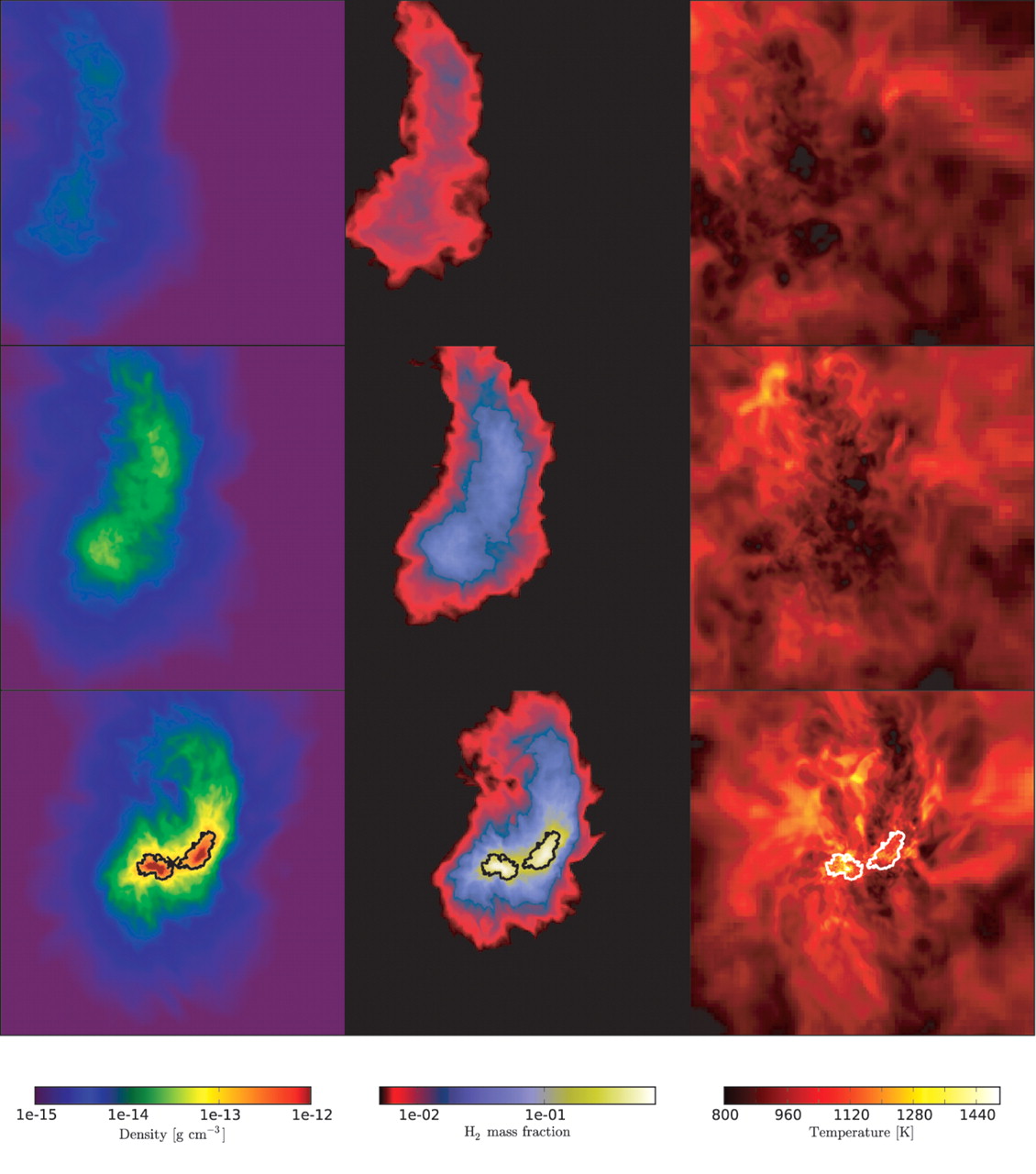}
\caption{The first stars may have been binaries, according to a
  simulation by Turk et al.\ (2009) \cite{AbelBinary}. Shown is the
  average density (left column), $H_2$ mass fraction (middle), and
  temperature (right), projected through a cube 3500~AU on a side. The
  bottom row (in which the two separate gravitationally-bound cores
  are outlined with thick lines) is at the end of the simulation, with
  the other rows showing earlier times by 555 years (middle) or 1146
  years (top). From \cite{AbelBinary}.}
\label{f:AbelBinary}
\end{figure}

\begin{figure}[tbp]
\includegraphics[width=\textwidth]{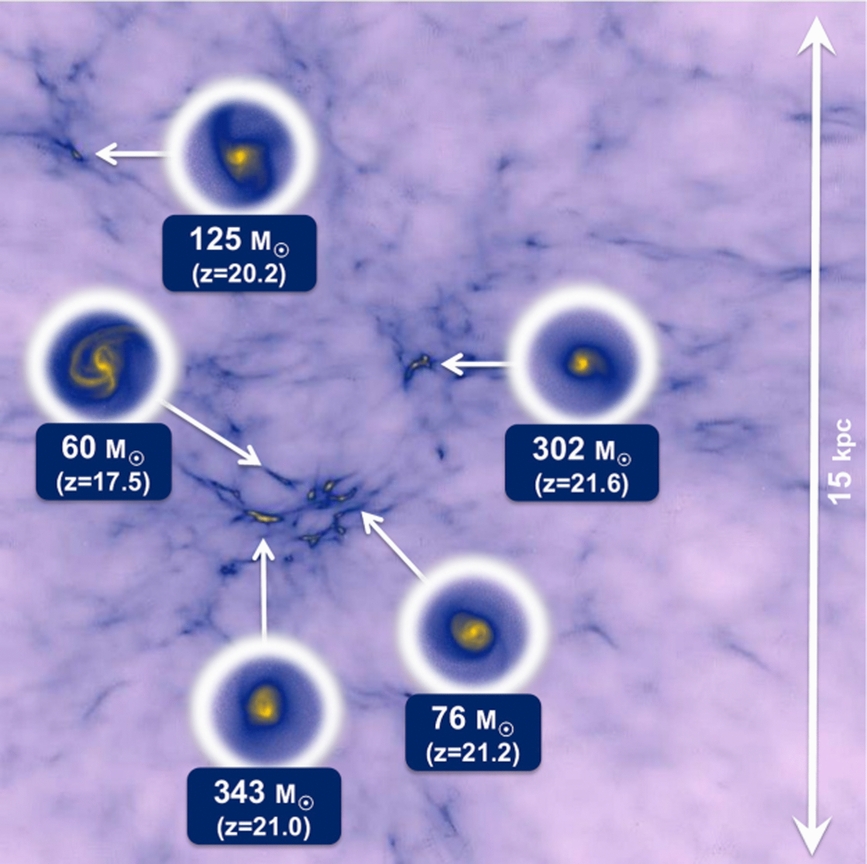}
\caption{The first stars may have had a range of masses, based on a
  simulation by Hirano et al.\ (2014) \cite{YoshidaStars}. The
  projected gas density is shown at $z=25$. Five primordial
  star-forming clouds are highlighted, with each circle showing a
  zoom-in to the central parsec at the formation time of the star; its
  formation redshift and stellar mass are listed. From
  \cite{YoshidaStars}.}
\label{f:YoshidaStars}
\end{figure}

Even for a given set of initial conditions for star formation, the
final properties of the resulting stars depend on a complex process of
proto-stellar evolution. It was initially thought that the rapid
accretion rates characteristic of primordial star-forming regions at
high-redshift would naturally lead to isolated Pop III stars of $100
M_\odot$ or more. However, some simulations \cite{AbelBinary} then
showed the possible formation of binaries (Figure~\ref{f:AbelBinary}),
and further semi-analytical and numerical simulation studies
\cite{PopIIIa,PopIIIb,PopIIIc,PopIIId,YoshidaStars,YStars2} have found
that the clumps have sufficient angular momentum to form a disk, and
that the rapid accretion onto the disk causes it to fragment due to
gravitational instability. While it is too early to draw final
conclusions, the best bet currently is that Pop III stars formed with
a wide range of different masses, but on average were significantly
heavier than later generations of stars (Figure~\ref{f:YoshidaStars}).

\section{21-cm Cosmology}

\label{s:21physics}

An overview of the basic features and early development of 21-cm
cosmology was given in \S~\ref{intro21}. In this section we present
the basic physics in greater detail, then focus on some important
low-temperature corrections, and discuss the important subject of
anisotropy in the 21-cm signal. Finally, we give a brief overview of
the observational aspects of 21-cm cosmology, focusing on the power
spectrum. More details of 21-cm physics and observational techniques
are available in specific reviews of 21-cm cosmology
\cite{F06,MW10,PL12}.

\subsection{Basic physics}

The basic physics of the hydrogen spin transition is determined as
follows. At the low densities typical in cosmological applications,
the gas is far from full thermal equilibrium, and a single temperature
cannot accurately describe the occupancy of various atomic levels.  In
particular, the relative occupancy of the spin levels is usually
described in terms of the hydrogen spin temperature $T_S$, which is an
effective temperature that determines the emission or absorption
properties of the 21-cm line.  Specifically, $T_S$ is defined by \beq
\label{e:T*} \frac{n_1}{n_0}=3\, \exp\left\{-\frac{T_*}{T_S}\right\}\
, \eeq where $n_0$ and $n_1$ are the number densities of the singlet
and triplet hyperfine levels in the atomic ground state ($n=1$),
respectively, and $T_*=0.0682$ K is defined by $k_B T_*=E_{21}$, where
the energy of the 21-cm transition is $E_{21}=5.87 \times 10^{-6}$ eV,
corresponding to a frequency of 1420 MHz (and a precise wavelength of
$\lambda_{21}=21.1$~cm). The factor of 3 in Eq.~(\ref{e:T*}) is the
ratio of statistical weights, i.e., it arises from the degeneracy
factor of the spin 1 excited state. In particular, $T_S \rightarrow
\infty$ would correspond to having the singlet and triplet levels
populated in their statistical 1:3 ratio, $T_S \rightarrow 0$ would
mean an empty excited state, while a population inversion (not
expected in the cosmological 21-cm field) would correspond to negative
$T_S$. Since $T_*$ is such a low temperature, in what follows we make
the standard assumption that all other temperatures (including $T_S$)
are much higher.

A patch of neutral hydrogen at the mean density and with a uniform
$T_S$ produces an optical depth at 21~cm [observed at $21(1+z)$~cm] of
\beq \tau(z) = \frac{3 c \lambda_{21}^2 h_P A_{10} n_{H~I}} {32 \pi
  k_B T_S (1+z) (dv_r/dr)}\ , \label{e:tau} \eeq where $h_P$ is
Planck's constant, $A_{10}=2.85 \times 10^{-15}$ s$^{-1}$ is the
spontaneous decay rate of the hyperfine transition, $n_{H~I}$ is the
number density of hydrogen atoms, and $dv_r/dr$ is the gradient of the
radial velocity along the line of sight, with $v_r$ being the physical
radial velocity and $r$ the comoving distance. In a fully-neutral,
homogeneous universe, $n_{H~I}=\bar{n}_H(z)$ and $dv_r/dr =
H(z)/(1+z)$ in terms of the Hubble parameter $H$. Assuming the
high-redshift (EdS) form for $H(z)$ (see \S~\ref{s:cosmo}), this
yields \beq \tau(z) = 9.85 \times 10^{-3} \left(\frac{T_{\rm CMB}}
  {T_S} \right) \left ( \frac{\Omega_b h} {0.0327} \right)
\left(\frac{\Omm}{0.307}\right)^ {-1/2}
\left(\frac{1+z}{10}\right)^{1/2}\ , \eeq where $T_S$ and $T_{\rm
  CMB}$ are measured at $z$. Since the brightness temperature through
the IGM is $T_{\rm b}^z=T_{\rm CMB} e^{-\tau}+T_S (1-e^{-\tau})$, the
observed mean differential antenna temperature relative to the CMB is
\cite{Madau} \beq T_{\rm b}=(1+z)^{-1} (T_S-T_{\rm CMB}) (1-e^{-\tau})
\simeq 26.8\, {\rm mK}\, \left( \frac{\Omega_b h} {0.0327} \right)
\left(\frac{\Omm}{0.307}\right)^ {-1/2} \left( \frac{1+z} {10}
\right)^{1/2} \left( \frac{T_S-T_{\rm CMB}} {T_S} \right)\ ,
\label{e:Tb}\eeq
where $\tau \ll 1$ is assumed (the relative correction to the linear
term that we kept is $\tau/2$) and $T_{\rm b}$ has been redshifted to
redshift zero. We use here the now standard notation of $T_{\rm b}$
for this final quantity. Note that the brightness temperature is
simply a measure of intensity in equivalent temperature units, defined
in terms of the Rayleigh-Jeans limit of the Planck spectrum: \beq
I_\nu = 2 k_B T_{\rm b} \frac{\nu^2} {c^2}\ .
\label{e:RJ} \eeq Note that in 21-cm cosmology, the CMB is certainly
deep in the Rayleigh-Jeans limit, as its Planck spectrum peaks at a
wavelength of $\sim 2$~mm, while the observed (redshift 0) wavelengths
of relevance to us here are three orders of magnitude larger.

The IGM is observable when $T_S$ differs from $T_{\rm CMB}$, which is
reasonable since $T_S=T_{\rm CMB}$ implies a kind of thermal
equilibrium between the ground-state hyperfine levels of hydrogen and
the CMB background, meaning that the net effect of the gas is neither
absorption nor added emission above the background. The key question
for 21-cm observations is thus the value of the spin temperature. For
intergalactic hydrogen it is determined by three processes. First, by
direct absorption and emission (both spontaneous and stimulated) of
21-cm photons from/into the radio background (which at high redshifts
is simply the CMB), the hyperfine levels of hydrogen tend to
thermalize with the CMB, making the IGM unobservable. If other
processes shift the hyperfine level populations away from such a
thermal equilibrium, then the gas becomes observable against the CMB
in emission or in absorption. In the presence of the CMB alone, the
spin states would reach thermal equilibrium with $T_S=T_{\rm
  CMB}=2.725 (1+z)$ K on a time-scale of $T_*/(T_{\rm CMB} A_{10})
\simeq 3 \times 10^5 (1+z)^{-1}$ yr. This time-scale is much shorter
than the age of the universe at all redshifts after cosmological
recombination.

On the other hand, at high densities the spin temperature comes into
equilibrium with the regular, kinetic temperature $T_K$ that describes
the random velocities of the hydrogen atoms. This equilibrium is
enforced by collisions, which involve energies of order $k_B T_K$, and
drive $T_S$ towards $T_K$ \cite{purcell}. Collisionally-induced
transitions are effective at high redshift, but become less effective
compared to the CMB at low redshift. This may seem surprising given
that as the universe expands, the mean energy density of radiation
decreases faster than that of matter, and the comparison here is
between two-body interactions of the hydrogen atom with either a
photon or a second atom. Part of the explanation is that while the
total radiation energy density goes as $T_{\rm CMB}^4$ (and thus
decreases rapidly with time), the relevant energy density for the
21-cm coupling is that at a fixed physical wavelength of 21~cm; this
is only proportional to $T_{\rm CMB}$ in the Rayleigh-Jeans limit of
the Planck spectrum of the CMB (Eq.~\ref{e:RJ}). In addition, the
collisional rate coefficient (see below) depends strongly on
temperature in the relevant range, and it decreases very rapidly as
the gas cools with time. Thus, if collisions were the only coupling
mechanism of the spin temperature with the kinetic temperature, the
cosmic gas would disappear at 21~cm below $z \sim 30$.

Instead, 21-cm cosmology down to $z \sim 7$ is made possible by a
subtle atomic effect worked out nearly 50 years before its
cosmological significance became widely recognized. This effect is
21-cm coupling as an indirect consequence of the scattering of much
higher-energy \Lya photons \cite{Wout,Field}.  Continuum UV photons
produced by early radiation sources redshift by the Hubble expansion
into the local \Lya line at a lower redshift, or are injected at \Lya
after redshifting and cascading down from higher Lyman lines.  These
photons mix the spin states via the Wouthuysen-Field (hereafter WF)
effect whereby an atom initially in the $n=1$ state absorbs a \Lya
photon (of wavelength $\lambda_\alpha = 1216$~\AA), and the
spontaneous decay that returns it from $n=2$ to $n=1$ can result in a
final spin state that is different from the initial one (These various
energy levels are illustrated in Figure~\ref{f:Basic21cm}). The WF
effect drives $T_S$ to the so-called ``color temperature'' $T_C$,
defined so that the spin-flip transition rates due to \Lya photons
upwards ($P_{01}^{\alpha}$) and downwards ($P_{10}^{\alpha}$) are
related by \cite{Field58}: \beq
\frac{P_{01}^{\alpha}}{P_{10}^{\alpha}} = 3 \left( 1 -
  \frac{T_*}{T_C}\right)\ .  \eeq The color temperature enters since
the $0 \rightarrow 1$ and $1 \rightarrow 0$ scattering events are
caused by photons with slightly different frequencies. It is the
equivalent temperature of a blackbody spectrum that would yield this
transition rate ratio. In general (i.e., including the case of a
non-blackbody radiation background), the color temperature is
determined by the shape of the radiation spectrum near Ly$\alpha$, and
is related to the photon intensity $J$ through \cite{Rybicki} \beq
\frac{h}{k_B T_C} = \frac{2}{\nu} - \frac{d \ln J}{d \nu}\ .  \eeq

Given CMB scattering (which pulls $T_S \rightarrow T_{\rm CMB}$),
atomic collisions ($T_S \rightarrow T_K$), and \Lya scattering
($T_S \rightarrow T_C$), the spin temperature becomes a weighted mean
\cite{Field58}: \beq T_S^{-1}=\frac{T_{\rm CMB}^{-1}+x_c T_K^{-1} +
  x_{\alpha} T_C^{-1}} {1+x_\tot}\ ,
\label{e:TS} \eeq where $x_\tot = x_c + x_{\alpha}$ and the
combination that appears in $T_{\rm b}$ (Eq.~\ref{e:Tb}) is then: \beq
\frac{T_S-T_{\rm CMB}} {T_S} = \frac{x_\tot - T_{\rm CMB} \left( x_c
    T_K^{-1} + x_{\alpha} T_C^{-1} \right)} {1+x_\tot}\ .
\label{e:combo}\eeq Here we have used the notation from
Barkana \& Loeb (2005) \cite{BL05b} in terms of the coupling
coefficients $x_c$ and $x_{\alpha}$ for collisions and \Lya
scattering, respectively. They are given by \cite{Madau} \beq x_c =
{{4 \kappa_{1-0}(T_k)\, n_H T_\star}\over {3 A_{10} T_{\rm CMB}}}\ ,
\eeq where the collisional rate coefficient $\kappa_{1-0}(T_k)$ is
tabulated as a function of $T_k$ \cite{AD,Zyg}, and \beq x_{\alpha} =
\frac{4 P_\alpha T_*} {27 A_{10} T_{\rm CMB}}\ , \eeq in terms of the
\Lya scattering rate $P_\alpha$. Expressed in terms of the
proper \Lya photon intensity $J_{\alpha}$ (defined as the spherical
average of the number of photons hitting a gas element per unit area
per unit time per unit frequency per steradian), \beq x_{\alpha} =
\frac {16 \pi^2 T_\star e^2 f_\alpha} {27 A_{10}\, m_e c\, T_{\rm
    CMB}} J_{\alpha}\ ,
\label{e:xa} \eeq except for a low-temperature correction (see the
next subsection), where $f_\alpha=0.4162$ is the oscillator
strength of the \Lya transition.

The neutral IGM is highly opaque to resonant scattering, which
involves energy transfers between the atomic motion and the photons,
and tends to drive a kind of thermal equilibrium between the photon
energy distribution near \Lya and the kinetic motion of the atoms.
This makes $T_C$ very close to $T_K$ \cite{Field59}, except for
another low-temperature correction (see the next subsection). In the
high-temperature approximation, equations~(\ref{e:TS}) and
(\ref{e:combo}) simplify to: \beq T_S^{-1}=\frac{T_{\rm
    CMB}^{-1}+x_\tot T_K^{-1} } {1+x_\tot}\ ,
\label{e:TS2} \eeq and \beq \frac{T_S-T_{\rm CMB}} {T_S} = \frac{x_\tot}
{1+x_\tot} \left(1 - \frac{T_{\rm CMB}} {T_K} \right)\ .
\label{e:combo2} \eeq

Below $z \sim 200$, the gas is mostly thermally decoupled from the CMB
and $T_K < T_{\rm CMB}$ (until significant X-ray heating), so that
21-cm observations are possible since collisions or \Lya scattering
provide an effective mechanism coupling $T_S$ to $T_K$. While
Eq.~(\ref{e:Tb}) gives the 21-cm brightness temperature in a
fully-neutral, homogeneous universe, in the real Universe $T_{\rm b}$
fluctuates. It is proportional in general to the gas density, and in
partially ionized regions $T_{\rm b}$ is proportional to the neutral
hydrogen fraction. Fluctuations in the velocity gradient term in
Eq.~(\ref{e:tau}) leads to a line-of-sight anisotropy in the 21-cm
signal (\S~\ref{s:anisotropy}).  Also, if $T_S > T_{\rm CMB}$ then the
IGM is observed in emission, and when $T_S \gg T_{\rm CMB}$ the
emission level saturates at a level that is independent of $T_S$.  On
the other hand, if $T_S < T_{\rm CMB}$ then the IGM is observed in
absorption, and if $T_S \ll T_{\rm CMB}$ the absorption strength is a
factor $\sim T_{\rm CMB} / T_S$ larger (in absolute value) than the
saturated emission level. In addition, once the Universe fills up with
\Lya radiation and the WF effect turns on (this is the \Lya coupling
transition, with its peak usually defined as the point when $x_{\rm
  tot}=1$ due mostly to $x_\alpha$), the rapid rise expected during
the early stages of cosmic star formation implies that soon afterwards
$x_\alpha \gg 1$ and $T_{\rm b}$ saturates to a value that no longer
depends on $x_\alpha$.  As a result of these various considerations, a
number of cosmic events (\S~\ref{s:milestone}) are expected to leave
observable signatures in the redshifted 21-cm line
(\S~\ref{s:Final21cm}).

\subsection{Low-temperature corrections}

\label{s:lowT}

There are two corrections to the 21-cm coupling due to \Lya
scattering, which can be important in low-temperature gas. Both arise
from a careful consideration of the multiple scatterings of the
photons near the \Lya resonance with the hydrogen atoms, and how
these scatterings affect the energy distribution of the photons near
the resonance, resulting in a change in the 21-cm coupling. One
correction is due to a difference between the color temperature and
the kinetic temperature of the gas, and the other due to a modified
\Lya scattering rate. We attempt here to clear up confusion in
some of the literature on this subject.

An accurate determination of the \Lya color temperature requires
a careful consideration of radiative scattering including atomic
recoil and energy transfer due to spin exchange. In the limit of a
high optical depth to \Lya scattering (an excellent
approximation in the cosmological context), \beq T_C = T_K \left(
  \frac{1+T_{\rm se}/T_K} {1+T_{\rm se}/T_S}\right)\ , \label{e:TC}
\eeq which differs significantly from $T_K$ once temperatures approach
$T_{\rm se}$, which is given by \beq T_{\rm se} = \frac{m_H c^2} {9
  k_B} \left( \frac{\lambda_\alpha} {\lambda_{21}} \right)^2 = 0.402\
{\rm K}\ , \eeq where $m_H$ is the mass of a hydrogen atom.
Eq.~(\ref{e:TC}) is easily solved simultaneously with
Eq.~(\ref{e:TS}), yielding results that have precisely the same
form as equations~(\ref{e:TS2}) and (\ref{e:combo2}) if we replace
$x_\tot$ by an $x_\toteff$ in which we adopt an effective value
$x_{\alpha,\rm eff}=x_{\alpha}/(1+T_{\rm se}/T_K)$.

The second effect modifies the relation between $J_{\alpha}$ (defined
as the naive \Lya photon intensity, i.e., not including the
modification due to multiple scattering) and the actual \Lya
scattering rate $P_\alpha$. The final result is to multiply
Eq.~(\ref{e:xa}) by an extra factor $S_\alpha$, which depends on
$T_K$ as well as the Gunn-Peterson \cite{GP} optical depth to \Lya
absorption, which for neutral gas at the cosmic mean density is \beq
\tau_{\rm GP}={\pi e^2 f_\alpha \lambda_\alpha n_{\rm H~I} \over m_e c
  H} = 6.62\times 10^5 \left({\Omega_bh\over
    0.0327}\right)\left({\Omega_m\over 0.307}\right)^{-1/2} \left({1+z
    \over 10}\right)^{3/2}\ , \eeq where in the second equality we
used the high-redshift form of the Hubble parameter $H(z)$.

The scattering-rate correction factor $S_\alpha$ is due to the fact
that the H atoms recoil in each scattering, and near the center of the
\Lya line, frequent scatterings with atoms make the photons lose
energy faster. Thus, the number of photons per unit energy at any
instant is smaller than would have been expected without recoil,
leading to a suppression in the scattering rate (i.e., $S_\alpha<1$).
The actual value of $S_\alpha$ is derived from solving the radiative
transfer equation for the photons including scattering and energy
losses. The result is \beq S_\alpha = e^{-0.0128 \left(\tau_{\rm
      GP}/T_K^2 \right)^{1/3}}\ , \eeq with $T_K$ in Kelvin (in this
equation only). Thus, the final results including both low-temperature
corrections are \beq T_S^{-1}=\frac{T_{\rm CMB}^{-1}+x_\toteff
  T_K^{-1} } {1+x_\toteff}\ ,
\label{e:TS3} \eeq and \beq \frac{T_S-T_{\rm CMB}} {T_S} =
\frac{x_\toteff} {1+x_\toteff} \left(1 - \frac{T_{\rm CMB}} {T_K}
\right)\ , \label{e:combo3} \eeq where $x_\toteff = x_c +
x_{\alpha,\rm eff}$, and \beq x_{\alpha,\rm eff} = x_{\alpha} \left(
  1+\frac{T_{\rm se}}{T_K} \right)^{-1} \exp\left[-2.06
  \left({\Omega_bh\over 0.0327}\right)^{1/3}\left({\Omega_m\over
      0.307}\right)^{-1/6} \left({1+z \over 10}\right)^{1/2}
  \left({T_K \over T_{\rm se}}\right)^{-2/3} \right]\ .
\label{e:xaeff3} \eeq Eq.~(\ref{e:combo3}) shows that even with
the low-temperature correction, whether we get 21-cm emission or
absorption is determined solely by whether $T_K$ is larger or smaller
than $T_{\rm CMB}$ (which seems reasonable based on thermodynamics),
while at a given $T_K$, the absolute value of $T_{\rm b}$ increases
monotonically with $x_{\alpha,\rm eff}$. The low-temperature
corrections simply reduce the effective value of $x_{\alpha}$ and thus
reduce the absolute value of $T_{\rm b}$ and delay the onset of \Lya
coupling and \Lya saturation (the latter is when $T_S \rightarrow
T_K$).  Note that we wrote the scattering-rate correction in
Eq.~(\ref{e:xaeff3}) in terms of $T_{\rm se}$ for ease of
comparison with the color-temperature correction.

These results are based on Chuzhoy \& Shapiro (2006) \cite{ChSh06},
who found simple and accurate final expressions based on an
approximate analytical solution (that was also found earlier in a
different context \cite{Grachev}).  The calculation of $S_\alpha$ was
first carried out by Chen \& Miralda-Escud\'e (2004) \cite{Miralda}
(based on a numerical solution to an approximate form of the radiative
transfer equation developed earlier \cite{Basko,Ryb94}), but they made
a numerical error and were off by about a factor of 2. Hirata (2006)
\cite{Hirata} gave complicated fitting formulas to numerical solutions
for both corrections, but the results given above agree with those
formulas to within a relative error of a few percent or better within
the relevant parameter range. Furlanetto \& Pritchard (2006)
\cite{Jonathan06b} developed higher-order analytical solutions and
also compared them to full numerical solutions.  Contrary to
statements in the literature \cite{F06}, no iteration is necessary in
order to include the low-temperature corrections; the results
summarized in this section are accurate at all $T \gtrsim 1$~K, except
at very high temperatures ($>1000$~K) which in the real Universe are
reached only after the \Lya coupling has saturated (and so these
corrections no longer matter). Note also that the scattering
correction factor $S_\alpha$, while calculated slightly differently
for the continuum (redshifting) \Lya photons and the injected (from
higher-level cascades) \Lya photons, has the same value in the two
cases, to high accuracy.

\begin{figure}[tbp]
\includegraphics[width=\textwidth]{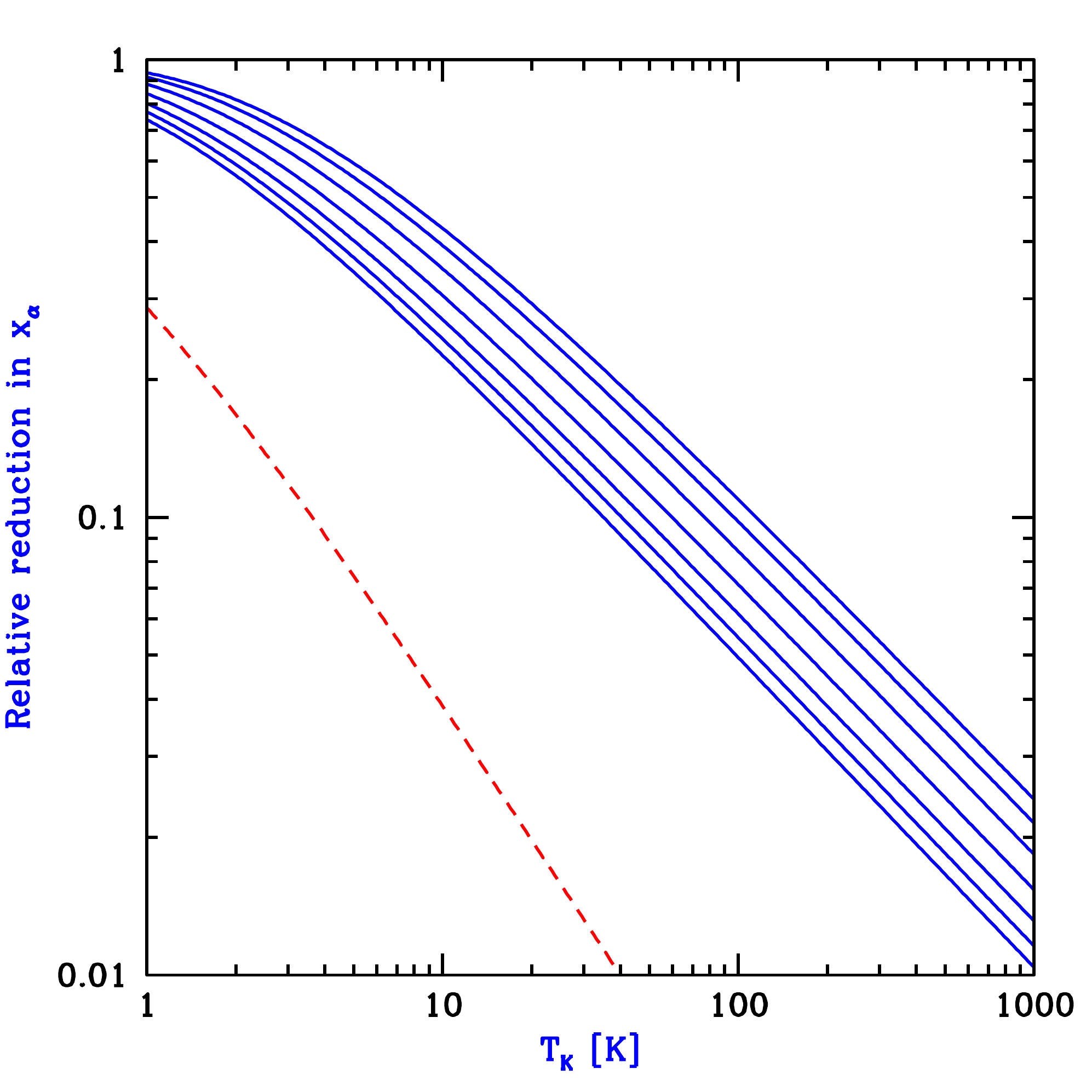}
\caption{Relative reduction in $x_{\alpha}$, i.e., $1-x_{\alpha,\rm
    eff}/x_{\alpha}$, versus the kinetic gas temperature $T_K$.  We
  show the total reduction (solid curves) including both the
  scattering and color-temperature corrections, at redshifts 7, 9, 12,
  17, 25, 35, and 45 (from bottom to top), and the reduction from the
  modified color temperature only (dashed curve). }
\label{f:xalpha}
\end{figure}

The quantitative results are illustrated in Figure~\ref{f:xalpha}. The
scattering correction dominates over the correction from the color
temperature. In practice, the observable effects of the
low-temperature corrections could be important in the real Universe
during the \Lya coupling era.  These corrections affect 21-cm
fluctuations only when the \Lya coupling is significant but has not
yet saturated (since in the saturated limit, 21-cm observations are
independent of $x_\alpha$ and its corrections). As long as the cosmic
gas cools, the strengthening reduction in the effective $x_{\alpha}$
slows the rise of \Lya coupling; once the gas reaches its minimum
temperature and begins to warm up, the declining low-temperature
effect then accelerates \Lya saturation. In realistic models (see
\S~\ref{s:milestone} and \S~\ref{s:Final21cm}), $x_\alpha \sim 1$ is
expected at $z \sim 25$, when the gas has cooled to $\sim 15$~K, while
temperatures as low as $\sim 7-8$~K may be reached at $z \sim 17$
(e.g., in the plausible case of late heating), although $x_\alpha$ is
then expected to already be fairly large. Thus, the low-temperature
corrections may affect $T_{\rm b}$ by up to $\sim 20\%$ within this
redshift range.

\newpage
\subsection{Anisotropy of the 21-cm signal}

\label{s:anisotropy}

As explained previously, the 21-cm signal on the sky is potentially an
extremely rich dataset. This signal is intrinsically three
dimensional, covering the full sky over a wide range of redshifts.
Even if 21-cm fluctuations are only measured statistically in terms of
the isotropically-averaged power spectrum of fluctuations, this power
spectrum versus redshift should yield a powerful dataset that can
probe a wide range of the physics and astrophysics of the first stars
and galaxies (as explored in detail in \S~\ref{s:Final21cm}).

The fluctuations in 21-cm cosmology are potentially even richer, as a
result of a particular form of anisotropy that is expected due to gas
motions along the line of sight \cite{Kaiser87,BA04,BL05a}. This
anisotropy, expected in any measurement of density that is based on a
spectral resonance or on redshift measurements, results from velocity
compression. The point is that spectral absorption is determined
directly by the velocity (along the line of sight) of gas rather than
its position.  As an extreme example, a slab of neutral hydrogen with
no internal motions will all appear to be at the same redshift from an
observer, producing enormous absorption at one particular frequency
and thus appearing like a huge density enhancement at the
corresponding redshift, even though the real, physical density need
not be high (if the slab extends over a long distance along the line
of sight).

More generally, consider a photon traveling along the line of sight
that resonates with absorbing atoms at a particular point. In a
uniform, expanding universe, the absorption optical depth encountered
by this photon probes a particular narrow strip of atoms, since the
expansion of the universe makes all other atoms move with a relative
velocity that takes them outside the narrow frequency width of the
resonance line.  If there is a density peak, however, near the
resonating position, the increased gravity will reduce the expansion
velocities around this point and bring more gas into the resonating
velocity width. Thus, near a density peak, the velocity gradient tends
to increase the 21-cm optical depth above and beyond the direct
increase due to the gas density itself. This effect is sensitive only
to the line-of-sight component of the gradient of the line-of-sight
component of the velocity of the gas, and thus causes an observed
anisotropy in the 21-cm power spectrum even when all physical causes
of the fluctuations are statistically isotropic.  Barkana \& Loeb
(2005) \cite{BL05a} showed that this anisotropy is particularly
important in the case of 21-cm fluctuations. When all fluctuations are
linear, the 21-cm power spectrum takes the form \cite{BL05a} \beq
P_{\rm 21-cm}({\bf k}) = P_{\rm iso}(k) + 2 \mu^2 P_{\rho - {\rm iso}}
(k) + \mu^4 P_{\rho}(k) \ , \label{e:mu4} \eeq where $\mu = \cos
\theta$ in terms of the angle $\theta$ between the wavevector ${\bf
  k}$ of a given Fourier mode and the line of sight, $P_{\rm iso}(k)$
is the isotropic power spectrum that would result from all sources of
21-cm fluctuations without velocity compression, $P_{\rho}(k)$ is the
power spectrum of gas density fluctuations, and $P_{\rho - {\rm iso}}
(k)$ is the Fourier transform of the cross-correlation between the
density and all (isotropic) sources of 21-cm fluctuations. Here the
velocity gradient has led to the appearance of the density power
spectrum due to their simple relationship via the continuity equation.
The three power spectra can more generally be denoted according to the
power of $\mu$ that multiplies each term: \beq P_{\rm 21-cm}({\bf
  k},z) = P_{\mu^0}(k,z) + 3 \mu^2 P_{\mu^2} (k,z) + 5 \mu^4
P_{\mu^4}(k,z)\ , \label{e:muB} \eeq where we have defined the
coefficients according to their angle-averaged size (e.g., $P_{\mu^4}$
is defined accounting for $\langle \mu^4 \rangle = 1/5$), and have
written the redshift dependence explicitly.

Given this anisotropic form, measuring the power spectrum as a
function of $\mu$ should yield three separate power spectra at each
redshift \cite{BL05a}. These probe, in turn, the 21-cm fluctuations
without the velocity gradient term (through the $\mu$-independent
term); basic cosmology (through the intrinsic density power spectrum,
measurable from the $\mu^4$ term even when complex astrophysical
processes contribute to the other terms); and additional information
about the nature and properties of the various sources of 21-cm
fluctuations (through the $\mu^2$ term, which measures the
cross-correlation between density fluctuations and the total isotropic
21-cm fluctuations).

In practice, 21-cm fluctuations on small scales are quite non-linear,
and this non-linearity cannot be completely decoupled from large
scales. In other words, even if the fluctuations are linear on a
particular large scale, the way the fluctuations on that scale are
measured is via a Fourier decomposition of the overall 21-cm
fluctuations, which include non-linear, small-scale fluctuations.
This small-scale averaging may to some degree cancel out, or largely
result in an overall, simple bias factor, but the fact that the
averaging involves non-linearity makes the interpretation of even
large-scale measurements somewhat model-dependent. This is the
double-edged sword of small-scale 21-cm fluctuations: on the one hand,
they make 21-cm cosmology potentially a much larger dataset than CMB
anisotropies \cite{Loeb04}, but on the other hand, they make 21-cm
fluctuations more susceptible to non-linear effects (see the related
discussion in \S~\ref{s:numerical} of non-linear limits on the
accuracy of analytical models).

Numerical investigations during cosmic reionization
\cite{McQuinn,MaoY,Jensen,ShapiroPRL} suggest that indeed, the
decomposition of the line-of-sight anisotropy is more complex than the
simple linear limit. It remains an incontrovertible fact, though, that
the line-of-sight anisotropy makes 21-cm cosmology richer. The
anisotropy allows three separate power spectra to be measured at each
redshift, or more generally, a two-dimensional function of $k$ and
$\mu$. At worst, the interpretation of this large dataset will be
somewhat complicated and will need to be studied numerically, but in
any case the anisotropy makes the 21-cm technique more powerful.
There, are, moreover, two important caveats to these numerical
studies. First, they focused on reionization (dominated by UV
photons), which is a particularly difficult case as it makes the 21-cm
fluctuations intrinsically non-linear on small scales, since the
ionization fraction basically jumps from zero to unity in going from a
neutral region to an H~II bubble. And second, they focused on the
$\mu^4$ term and its promised yield of the primordial power spectrum;
this term, though, is usually the smallest of the three anisotropic
terms (as it does not benefit from the large bias of galaxies which
enhances terms dominated by astrophysical radiation), so it is most
susceptible to non-linear contamination.

\begin{figure}[tbp]
\includegraphics[width=\textwidth]{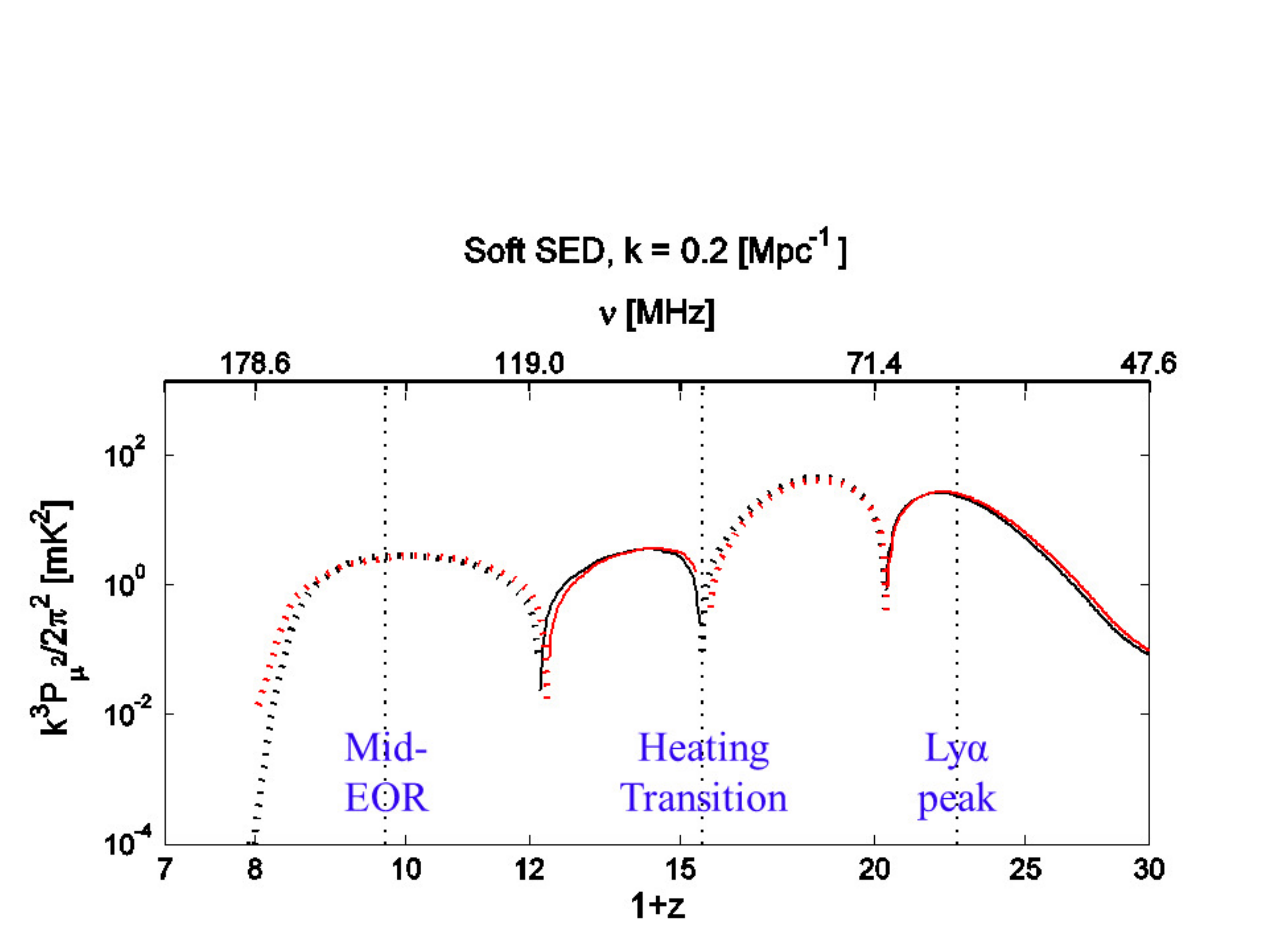}
\caption{The main anisotropic term of the 21-cm power spectrum,
  $P_{\mu^2}(k,z)$, shown in terms of the corresponding variance of
  the 21-cm fluctuations, at wavenumber $k=0.2$~Mpc$^{-1}$. The
  comparison of the actual value [reconstructed by fitting the form of
  Eq.~(\ref{e:muB}) to mock observations] (red) with that from
  assuming perfect linear separation [as in Eq.~(\ref{e:mu4})] (black)
  shows that this quantity withstands non-linearities and can be
  reconstructed accurately.  This quantity, which measures the
  cross-correlation between density fluctuations and 21-cm
  fluctuations, is sometimes positive (solid lines) and sometimes
  negative (dashed lines), as it tracks early history like a cosmic
  clock. It is negative during the EOR as a direct reflection of
  inside-out reionization (\S~\ref{s:reion}): higher density implies
  more galaxies which implies less neutral hydrogen, hence an inverse
  cross-correlation of density and the 21-cm signal.  It is positive
  during the \Lya coupling era, as more galaxies imply stronger \Lya
  radiation and a stronger 21-cm (absorption) signal.  During the
  cosmic heating era, it changes sign at the heating transition (when
  the cosmic H~I gas is first heated above the CMB temperature), the
  point at which heating a gas element switches from reducing the size
  of the 21-cm (absorption) signal to enhancing the size of the
  (emission) signal. The particular model shown here assumes cosmic
  heating by a soft power-law X-ray spectrum (see \S~\ref{s:heating}).
  From \cite{FialkovPRL}.}
\label{f:PRL}
\end{figure}

Recently, Fialkov et al.\ (2015) \cite{FialkovPRL} reconsidered the
anisotropic 21-cm power spectrum using a semi-numerical simulation
that covered a wide period of early cosmic history. Focusing on the
dominant anisotropic term ($P_{\mu^2}$), they showed that the
anisotropy is large and thus potentially measurable at most redshifts,
and it acts as a model-independent cosmic clock that tracks the
evolution of 21-cm fluctuations over various eras (see
Figure~\ref{f:PRL}). Also, they predicted a redshift window during
cosmic heating (at $z \sim 15$) when the anisotropy is small, during
which the shape of the 21-cm power spectrum on large scales is
determined directly by the average radial distribution of the flux
received from X-ray sources at a typical point. This makes possible a
direct and, again, model-independent, reconstruction of the X-ray
spectrum of the earliest sources of cosmic heating.

The velocity gradient anisotropy that we have just discussed is well
known in the context of galaxy redshift surveys \cite{Kaiser87}, where
it is often referred to as ``redshift-space distortions''. In that
case, it is used not as an additional probe of galaxies but of
fundamental cosmology, since it allows a measurement of the amplitude
of the velocity field (a recent example is \cite{Beutler}), which is
related to the rate of change of the growth factor
(section~\ref{s:lin}). A similar velocity gradient anisotropy also
arises in the context of the \Lya forest. In that case, measurements
are mostly one-dimensional (i.e., along the line of sight), so
redshift distortions are more difficult to extract, though they do
affect observations \cite{Seljak}.

An additional source of 21-cm anisotropy is the light-cone anisotropy
\cite{BL06}. While redshift can be converted to position in order to
create three-dimensional cubes for calculating the 21-cm power
spectrum, the line-of-sight direction is intrinsically different from
directions on the sky. The reason is that the look-back time changes
with the radial distance, and the character of the 21-cm fluctuation
sources evolves with time, which results in a line-of-sight effect
that introduces anisotropy. A significant anisotropy can be generated
on large scales near the end of reionization \cite{BL06}, as has been
further studied in numerical simulations
\cite{Datta12,Zawada14,LaPlante14,Datta14}. It is important to clarify
a possible confusing issue here (see, e.g., a clear explanation in
\cite{Datta12}). The light-cone anisotropy refers to 21-cm
fluctuations, which will be observed by radio interferometer
experiments. Interferometers measure the relative fluctuations at each
redshift, and are not sensitive to the mean of the 21-cm intensity at
each redshift. Mathematically, this is equivalent (for a flat sky) to
not being able to measure $\kb$ modes that point directly along the
line of sight ($\mu = 1$). Once the mean at each redshift is properly
removed, the light-cone effect on the power spectrum is then mainly
that the measured power spectrum is a redshift average of the real
power spectrum, since any frequency slice corresponds to a range of
redshifts within our past light cone. Looking towards the future, the
light-cone anisotropy can in principle be reduced as data become
available with improved sensitivity and larger fields of views,
allowing the power spectrum to be measured from thin redshift slices
that minimize the light-cone effect (though the slice should not be
thinner than one wavelength, which implies some remaining averaging
when measuring power on large scales).

Finally, if 21-cm data are analyzed using assumed cosmological
parameters that differ from the true ones, this causes an additional
Alcock-Paczy\'{n}ski \cite{AP} anisotropy that can be used to
constrain cosmological parameters \cite{Nusser,APindian}; in
particular, the technique of Eq.~(\ref{e:mu4}) can be extended, in
principle permitting (in the limit of linear fluctuations) a separate
probe of this anisotropy using the $\mu^6$ term that it induces in the
21-cm power spectrum \cite{MeAP}.

\subsection{Observational aspects}

\label{s:obs}

Attempts to measure the cosmological 21-cm signal must deal with the
much stronger foreground emission, dominated by synchrotron radiation
from electrons in the Milky Way, with other radio sources added on.
Indeed, the brightness temperature of the sky for typical
high-latitude, relatively quite portions of the sky, is \cite{F06}
\beq T_{\rm sky} \sim 180\, \left(\frac{\nu} {180\, {\rm MHz}}
\right)^{-2.6}\, {\rm K}\ . \label{e:Tsky} \eeq This steep increase of
foreground emission with decreasing frequency is the reason that 21-cm
observations become more difficult with increasing redshift;
distortion of the radio signal due to refraction within the Earth's
ionosphere also increases with redshift, down to the critical plasma
frequency of $\nu \sim 20$~MHz below which the ionosphere becomes
opaque. The sky emission in Eq.~(\ref{e:Tsky}) must be compared to the
expected signal of typically a few tens of mK (sky averaged), with
fluctuations of order several mK. The reason that this tiny signal may
be observable, even on top of a foreground that is brighter by at
least a factor of $10^4$, is that the foreground is produced by
synchrotron emission which inherently produces a very smooth frequency
spectrum. 

There are a number of approaches to observing the 21-cm signal from
high redshifts. The simplest, in principle, is measuring the global
21-cm signal, i.e., the sky-averaged, cosmic mean emission as a
function of frequency (i.e., redshift). This can be done with a single
dish (or dipole), but requires a very accurately calibrated instrument
to enable foreground subtraction. Indeed, the sensitivity of a single
dish \cite{shaver} is \beq \Delta T \sim \frac{T_{\rm sys}}
{\sqrt{\Delta \nu \, t_{\rm int}}}\ . \eeq Assuming that the system
temperature is approximately equal to that of the foreground
[Eq.~(\ref{e:Tsky})], and taking a bandwidth of $\Delta \nu = 5$~MHz
centered at $z=10$ ($\nu=129$~MHz), a sensitivity of $\Delta T=10$~mK
only requires an integration time $t_{\rm int}$ of 6 minutes. Thus,
the real issue with global 21-cm experiments is not raw sensitivity,
but the ability to clean out the smooth foreground emission to a
spectral accuracy of one part in $10^4$ or $10^5$. In practice, the
need to subtract out the smoothly-varying foreground implies a
simultaneous removal of the smoothly-varying part of the desired 21-cm
signal. Thus, the absolute level of the global signal likely cannot be
measured, but its variation with frequency may be measurable,
particularly when the frequency gradient of the 21-cm signal is large
during the rises or declines that accompany various milestones of
early cosmic evolution (see section~\ref{s:global}).

The other main approach is to make an interferometric map of the 21-cm
signal. In this case, much more information is available at each
redshift than just a single mean temperature. With a sufficiently high
signal-to-noise ratio, direct tomography/imaging can reveal the full
spatial distribution of the 21-cm signal. However, even if the maps
themselves are noisy, statistical measures such as the 21-cm power
spectrum can be computed with high accuracy, and used to extract many
of the most interesting aspects of cosmic dawn such as the properties
of the galaxies that existed at various times.  In the case of an
interferometer, one basic consideration is the achievable angular
resolution $\theta_D$ (and corresponding comoving spatial resolution
$r_D$), determined by the diffraction limit corresponding to the
longest array baseline $D_{\rm max}$ \cite{F06}: \beq \theta_D \sim
\frac{\lambda} {D_{\rm max}} \sim 7.\mkern-4mu^\prime 3 \left(
  \frac{1+z}{10} \right) \left( \frac{D_{\rm max}}{1\,{\rm km}}
\right)^{-1} \ ; \ \ \ r_D \sim 20 \left(\frac{h}{0.68}\right)^{-1}
\left( \frac{1+z}{10} \right)^{1.2} \left( \frac{D_{\rm max}}{1\,{\rm
      km}} \right)^{-1}\,{\rm Mpc}\ .\eeq For an array of $N$ radio
antennae (or stations), each with an effective collecting area $A_{\rm
  eff}$, the resulting field of view $\Omega_{\rm FoV} = \lambda^2 /
A_{\rm eff}$ corresponds to an angular diameter $\theta_{\rm FoV}$
(and comoving distance $r_{\rm FoV})$: \beq \theta_{\rm FoV} \equiv
\sqrt{ \frac{4 \Omega_{\rm FoV}}{\pi}} = 5.\!\!^\circ1 \left(
  \frac{1+z}{10} \right) \left( \frac{A_{\rm eff}}{700\,{\rm m}^2}
\right)^{-1/2}\ ; \ \ \ r_{\rm FoV} \sim 0.86
\left(\frac{h}{0.68}\right)^{-1} \left( \frac{1+z}{10} \right)^{1.2}
\left( \frac{A_{\rm eff}}{700\,{\rm m}^2} \right)^{-1/2}\,{\rm Gpc}\ .
\eeq In the line-of-sight direction, the comoving length corresponding
to a bandwidth $\Delta \nu$ is \beq r_{\Delta \nu} \sim 18 \left(
  \frac{\Delta \nu} {1\,{\rm MHz}} \right) \left( \frac{1+z}{10}
\right)^{1/2} \left( \frac{\Omega_m h^2} {0.141} \right)^{-1/2}\,{\rm
  Mpc}\ .\eeq Another commonly noted quantity is the total collecting
area of the array \beq A_{\rm coll} = N A_{\rm eff} = 1.8\times10^5
\left(\frac{N}{250} \right) \left( \frac{A_{\rm eff}}{700\,{\rm m}^2}
\right)\,{\rm m}^2\ , \eeq where we have used illustrative values
based roughly on the planned first phase of the Square Kilometer Array
\cite{Koopmans} (though note that $A_{\rm eff}$ is actually expected
to vary with frequency).

A key quantity for interferometric arrays is the sensitivity to power
spectrum measurements. We assume the simple approximation of antennae
distributed over a core area $A_{\rm core}$ in such a way that the
$uv$-density (i.e., the density in visibility space which is
equivalent to a Fourier transform of the sky) is uniform, and a single
beam (i.e., we do not include here the technique of multi-beaming
which can speed up surveys). In this case, the power-spectrum error
due to thermal noise is \cite{Mellema13,Koopmans} \beq \Delta T_{\rm
  PS}^{\,\rm thermal} = \sqrt{\frac{2}{\pi}}\, k^{3/4} \left[D_c^2\,
  \Delta D_c\, \Omega_{\rm FoV} \right]^{1/4} \frac{T_{\rm sys}}
{\sqrt{\Delta \nu \, t_{\rm int}}} \frac{1}{N} \sqrt{\frac{A_{\rm
      core}} {A_{\rm eff}}} \ , \label{e:TthermalA} \eeq which yields
an approximate value of
\begin{eqnarray} \Delta T_{\rm PS}^{\,\rm thermal} & \sim & 0.13\,
  \left( \frac{k}{0.1\,\rm Mpc^{-1}} \right)^{3/4} \left(
    \frac{1+z}{10} \right)^{3.3} \left( \frac{t_{\rm int}} {1000\,\rm
      hr} \right)^{-1/2} \left( \frac{\Delta \nu} {1\,{\rm MHz}}
  \right)^{-1/4} \nonumber \\
  & & \times\, \left( \frac{A_{\rm eff}}{700\,{\rm m}^2}
  \right)^{-3/4} \left( \frac{A_{\rm core}} {3.8 \times 10^5\,{\rm
        m}^2} \right)^{1/2} \left(\frac{N}{250} \right)^{-1} \left(
    \frac{\Omega_m h^2} {0.141} \right)^{-1/8}\, {\rm mK}\ ,
  \label{e:Tthermal} \end{eqnarray} where $D_c$ is the comoving 
distance to redshift $z$, and $\Delta D_c$ equals $r_{\Delta \nu}$
from above. The thermal noise thus increases with $k$, and typically
dominates the expected power spectrum errors on small scales.
Attempting to improve the angular resolution by increasing $D_{\rm
  max}$ would typically imply an increase in $A_{\rm core} \propto 
D^2_{\rm max}$ as well, and thus a worsening power-spectrum
sensitivity at all $k$.

The uncertainty in comparing data to models is usually dominated on
large scales by sample variance (sometimes termed ``cosmic
variance''), which gives a relative error that is roughly proportional
to the inverse square root of the number of modes of wavenumber $k$
that fit into the survey volume. Assuming a cylindrical volume and a
bin width of $\Delta k \sim k$ [assumptions also made in
Eq.~(\ref{e:TthermalA})], this yields \cite{Mellema13} \beq \Delta
T_{\rm PS}^{\,\rm sample} \approx T_{\rm PS}\, \sqrt{ \frac{8 \pi} {
    k^3 r_{\rm FoV}^2\, r_{\Delta \nu}}} \ , \eeq where $T_{\rm PS}$
is the root-mean-square 21-cm brightness temperature fluctuation at
wavenumber $k$. The resulting approximate value is
\begin{eqnarray} 
  \Delta T_{\rm PS}^{\,\rm sample} & \sim & 0.087\, \left(
    \frac{T_{\rm PS}}{2~\rm mK}\right) \left( \frac{k}{0.1\,\rm
      Mpc^{-1}} \right)^{-3/2} \left( \frac{1+z}{10} \right)^{-1.5}
  \left( \frac{\Delta \nu} {1\,{\rm MHz}} \right)^{-1/2} \nonumber \\
  & & \times\, \left( \frac{A_{\rm eff}}{700\,{\rm m}^2} \right)^{1/2}
  \left(\frac{h}{0.68}\right) \left( \frac{\Omega_m h^2} {0.141}
  \right)^{1/4} \, {\rm mK}\ .\end{eqnarray}

We note, though, that these noise estimates (both thermal noise and
sample variance) may in a sense be overestimated, since they are
calculated for a narrow bandwidth at a single redshift (e.g., 1~MHz
around $1+z=20$ corresponds to $\Delta z \sim 0.3$). If a theoretical
model is fit to data covering a wide range of redshifts, then the
model in a sense smoothes the data over the various redshifts,
yielding effectively lower noise overall. Of course, this conclusion
is not model-independent as it relies on the smooth variation with
redshift typically assumed in any model, a smoothness that ties
together, within such a combined fit, the data measured at various
redshifts. A model-independent way to try to reduce the errors would
be to simply average the data over wide redshift bins, but that would
erase some information about the redshift evolution, as well as
features of the power spectrum that may only appear prominently at
particular redshifts. A more direct observational approach is to use
the flexibility available in balancing the amount of integration time
spent per field (with more time leading to lower thermal noise), on
the one hand, and the total number of separate fields of view observed
(with more fields reducing the sample variance), on the other hand.

\section{The Supersonic Streaming Velocity}

\label{s:stream}

Current observational efforts in 21-cm cosmology (and high-redshift
astronomy more generally) are focused on the reionization era
(redshift $z \sim 10$), with earlier times considered more difficult
to observe. However, recent work suggests that at least in the case of
21-cm cosmology, the pre-reionization, $z \sim 20$ era of even earlier
galaxies may produce very interesting signals that make observational
exploration quite promising. One argument for this is based on a
recently noticed effect on early galaxy formation that had been
previously neglected. We discuss here this supersonic streaming
velocity, which has also been reviewed recently in detail
\cite{Fialkov}.

\subsection{Cosmological origins}

Up until recently, studies of early structure formation were based on
initial conditions from linear perturbation theory. However,
Tseliakhovich \& Hirata (2010) \cite{TH10} pointed out an important
effect that had been missing in this treatment. At early times, the
electrons in the ionized gas scattered strongly with the
then-energetic CMB photons, so that the baryons moved together with
the photons in a strongly-coupled fluid. On the other hand, the motion
of the dark matter was determined by gravity, as it did not otherwise
interact with the photons. Thus, the initial inhomogeneities in the
universe led to the gas and dark matter having different velocities.
When the gas recombined at $z \sim 1100$, it was moving relative to
the dark matter, with a relative velocity that varied spatially. The
root-mean-square value at recombination was $\sim 30$ km/s, which was
supersonic (Mach number $\sim 5$). The streaming velocity then
gradually decayed as $\propto 1/a$, like any peculiar velocity, but
remained supersonic (getting down to around Mach 2) until the onset of
cosmic heating. This is true for the root-mean-square value, but the
streaming velocity was lower in some regions, and up to a few times
higher in others.

\begin{figure}[tbp]
\includegraphics[width=\textwidth]{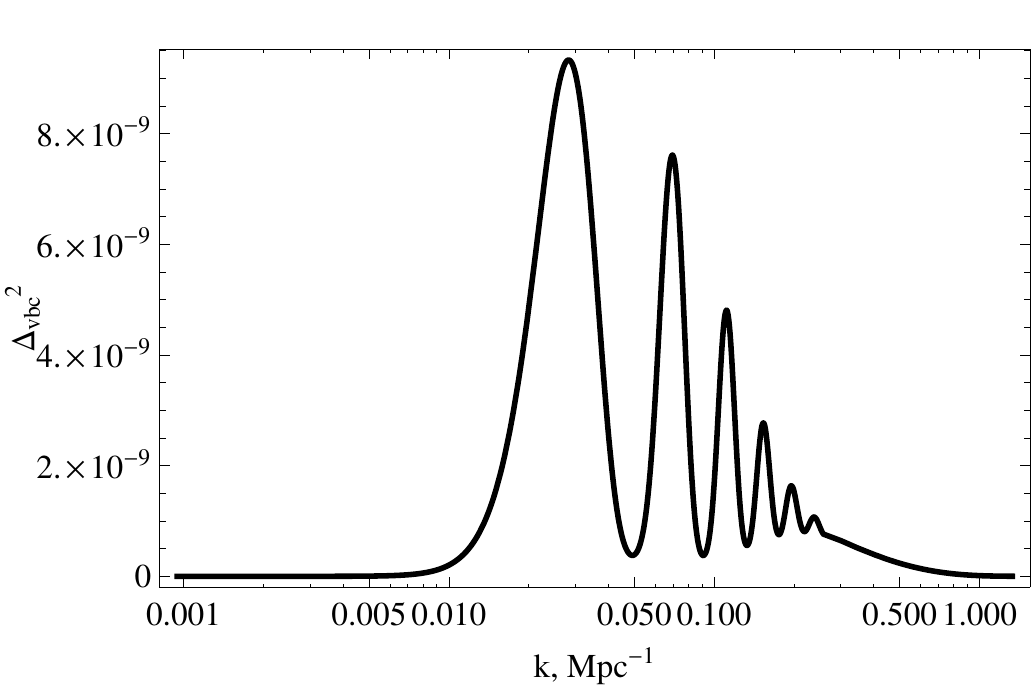}
\caption{The contribution of various scales to the mean squared
  velocity difference between the baryons and dark matter (at the same
  position) at recombination. The contribution per $\log k$ of
  fluctuations at wavenumber $k$ is shown vs.\ $k$. From \cite{TH10}.
}
\label{f:vbc}
\end{figure}

Figure~\ref{f:vbc} shows the contribution of fluctuations on various
scales to the variance of the velocity difference. This highlights two
important properties of this relative motion. First, there is no
contribution from small scales, so that the relative velocity is
uniform in patches up to a few Mpc in size; the velocity is generated
by larger-scale modes, up to $\sim 200$ Mpc in wavelength. The
uniformity on small scales is critical as it allows a separation of
scales between the spatial variation of the velocity (on large scales)
and galaxy formation (on small scales). Each individual high-redshift
mini-galaxy forms out of a small region ($\sim 20$ kpc for a $10^6
M_\odot$ halo) that can be accurately approximated as having a
uniform, local baryonic wind, or a uniform stream of baryons; the
relative velocity is thus also referred to as the ``streaming
velocity''.  The second important feature of Figure~\ref{f:vbc} is the
strong baryon acoustic oscillation (BAO) signature. Arising from the
acoustic oscillations of the photon-baryon fluid before recombination,
this strong BAO signature is a potentially observable fingerprint of
the effect of this relative motion, as is further detailed below.

The relative motion between the dark matter and baryons was not in
itself a surprise (it had been known for decades), but before 2010 it
had not been noticed that this effect was both important and dropped
within the standard approach. The standard initial conditions for both
analytical calculations and numerical simulations had been generated
based on linear perturbation theory, in which each $\kb$ mode evolves
independently. Indeed, the relative velocity is negligible if any
single scale is considered. However, it {\em is}\/ important as an
effect of large scales (which contribute to the velocity difference)
on small scales (which dominate early galaxy formation). Specifically,
the relative motion makes it harder for small-scale overdensities in
the dark matter to gravitationally accrete the streaming gas. Now,
observing such small scales directly would require far higher
resolution than is currently feasible in radio astronomy at high
redshift. Nonetheless, the relative motion is immensely important
because of its effect on star formation. Since stellar radiation
strongly affects 21-cm emission from the surrounding IGM, 21-cm
cosmology offers an indirect probe of the relative velocity effect.

\subsection{Effect on star formation in early halos}

The effect of the streaming velocity on early star formation can be
usefully separated into three effects, both for physical understanding
and for the purposes of analytical modeling. This also tracks the
development of the subject. The first effect of the streaming velocity
on halos to be analyzed was the suppression of the abundance of halos
\cite{TH10}. Since the baryons do not follow the dark matter
perturbations as closely as they would without the velocity effect,
linear fluctuations in the total density are suppressed on small
scales (where the gravitationally-induced velocities are comparable to
or smaller than the relative velocity). According to the standard
theoretical models for understanding the abundance of halos as a
function of mass \cite{ps74,bond91} (\S~\ref{sec:NL}), this should
result in a reduction of the number density of high-redshift halos of
mass up to $\sim 10^6 M_\odot$ \cite{TH10}, a mass range that is
expected to include most of the star-forming halos at early times.

The next effect to be noted \cite{Dalal} was that separately from the
effect on the number of halos that form, the relative velocity also
suppresses the gas content of each halo that does form. It was
initially claimed \cite{Dalal} that this second effect results in
2~mK, large-scale 21-cm fluctuations during \Lya coupling, with a
power spectrum showing a strong BAO signature due to the streaming
velocity effect. These conclusions were qualitatively on the mark but
were later seriously revised quantitatively. In particular, it turned
out \cite{Us,anastasia} that the gas-content effect is a minor one on
star-forming halos, and is mainly important for the lower-mass gas
minihalos that do not form stars.

Meanwhile, many groups began to run small-scale numerical simulations
that followed individual collapsing halos subject to the streaming
velocity
\cite{Maio:2011,Stacy:2011,Greif:2011,mcquinn12,mcquinn12b,naoz1,naoz2}.
In particular, two simulations \cite{Stacy:2011,Greif:2011} first
indicated the presence of a third effect, i.e., that the relative
velocity substantially increases the minimum halo mass for which stars
can form from gas that cools via molecular hydrogen cooling. The
intuitive explanation is that even if a halo does manage to form
(albeit with a reduced gas content), it does not contain the same
dense gas core that it would in the absence of the streaming velocity.
The reason is that the densest part of the halo (which is where stars
first manage to form) comes together well before the rest of the halo,
and is thus strongly disrupted by the streaming velocity (which is
high at early times); thus, after a halo forms in the presence of the
streaming velocity, it is necessary to wait longer for a dense core to
develop and bring about star formation. Given these simulation results
on the increase in the minimum halo mass for star formation, a
physically-motivated fit \cite{anastasia} allowed the development of a
general analytical model of early star formation that includes the
effect of density as well as all three effects of the streaming
velocity on star formation.

\begin{figure}[tbp]
\includegraphics[width=\textwidth]{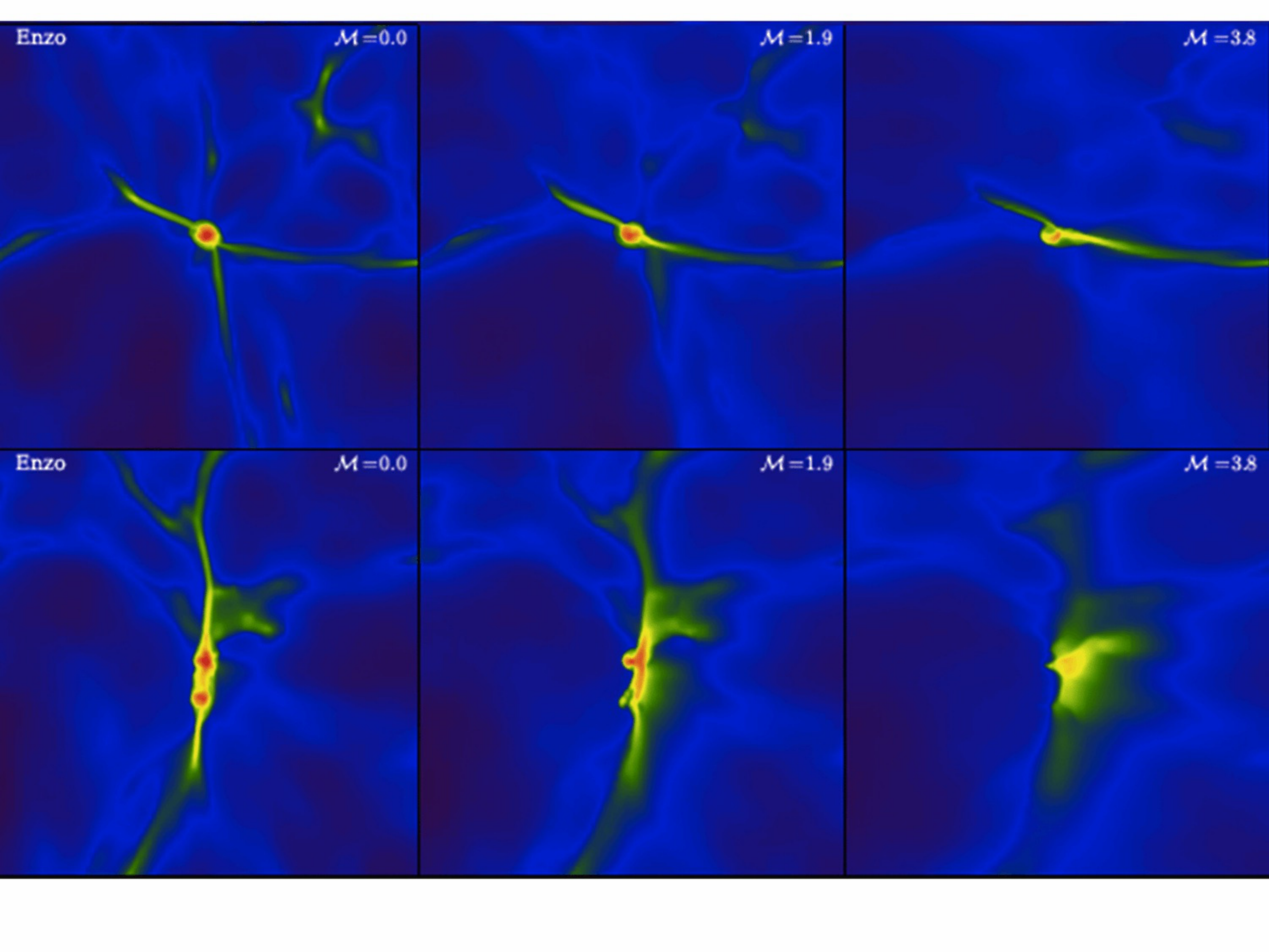}
\caption{The effect of relative velocity on individual halos, from
  numerical simulations (including gravity and hydrodynamics).  The
  colors indicate the gas density, which ranges from $10^{-26} {\rm
    g/cm}^3$ (blue) to $10^{-23} {\rm g/cm}^3$ (red). Two halos are
  shown at $z=20$, with a total halo mass of $2 \times 10^6 M_\odot$
  (top panels) or $8 \times 10^5 M_\odot$ (bottom panels). Panels show
  the result for gas initially moving to the right with a relative
  velocity of 0 (left panels), 1 (middle panels), or 2 (right panels)
  in units of the root-mean-square value of the relative velocity.
  $\mathcal{M}$ indicates the corresponding Mach number at $z=20$.
  From \cite{mcquinn12}.}
\label{f:numsim}
\end{figure}

Figure~\ref{f:numsim} illustrates some of the results of the numerical
simulation studies of the effect of the streaming velocity on galaxy
formation. As expected, a larger velocity suppresses gas accretion
more strongly, in particular reducing the amount of dense gas at the
centers of halos. But beyond just this general trend, the relative
velocity effect gives rise to very interesting dynamics on small
scales. It disrupts gas accretion in an asymmetric way, so that
filaments of accreting gas are disrupted more easily if they are
perpendicular to the local wind direction. In addition, halos that
form in regions of relatively high velocity develop supersonic wakes
as they move through the wind.

\subsection{Consequences}

The immediate major consequence of the streaming velocity effect is
the change in the large-scale distribution of the first stars in the
Universe, and the resulting pattern embedded in the 21-cm sky at very
high redshift. All of this is discussed below, particularly in
\S~\ref{s:2012}, where the distribution of the streaming velocity
field is also shown (Figure~\ref{f:RhoV}). Here we note some other
interesting consequences of the streaming velocity that have been
suggested.

Although the relative velocity only affected low-mass halos at high
redshifts, those halos were the progenitors of later, more massive
galaxies. Thus, the streaming velocity may have indirectly left a mark
on later galaxies through its influence on their star-formation
histories and, thus, on their current luminosity (through their old
stellar content and perhaps its feedback on the formation of younger
stars). This signature may be observable in galaxy surveys, and could
affect probes of dark energy through measurements of BAO positions in
the galaxy power spectrum \cite{UrosBAO1}; indeed, current data imply
an upper limit of $3.3\%$ on the fraction of the stars in luminous red
galaxies that are sensitive to the relative velocity effect
\cite{UrosBAO2}. More directly, the early streaming velocity effect on
star formation in dwarf galaxies could leave remnants in their
properties as measured today, e.g., in the low-mass satellites of the
Milky Way \cite{BovyDvorkin}.

We note, though, that when considering these effects on later
galaxies, it is important to keep in mind the modulation of star
formation by other effects, in particular LW radiation
(\S~\ref{s:Lya}) that suppresses molecular hydrogen cooling, and
reionization, which suppresses gas accretion through photoheating
feedback (\S~\ref{s:reion}). These effects suppressed star formation
in larger halos than the streaming velocity itself, which means that
they affected later progenitors of current galaxies (containing a
larger fraction of the final, present-day stellar content). The
distribution of LW feedback may have reflected in part the initial
relative velocity pattern \cite{FialkovLW}, since the LW radiation
itself was produced by stars in small halos, but reionization occurred
later, likely due to more massive halos (\S~\ref{s:reion}) that were
not affected much by the streaming velocity. Thus, photoheating likely
did not carry a significant signature of the streaming velocity field.

Moving towards higher redshifts, as mentioned, the streaming velocity
likely did not significantly affect the main stages of cosmic
reionization. However, it suppressed the formation of earlier cosmic
populations, perhaps including supermassive black holes at $z > 15$
\cite{Tanaka}. More intriguing (and speculative) are ideas on opposite
effects, whereby a large streaming velocity may have produced a unique
environment that allowed some objects to form. A large relative
velocity may have delayed star formation enough to allow a direct
collapse to a massive black hole \cite{Tanaka2}, or it may have
produced a baryonic density peak that was sufficiently displaced from
dark matter to allow the formation of an early globular cluster
\cite{GCs}. Moving on to the dark ages ($z > 30$), the supersonic
streaming velocity had a number of significant effects on the 21-cm
power spectrum at both large and small scales \cite{Yacine}.

Recently, re-analyses of the streaming velocity efffect point towards
a possible boost of the effect on galaxy formation, due to advection
and the coupling with density \cite{Blazek,Ahn,Schmidt}. Note also
that while the streaming velocity directly affected very small
galaxies, another remnant of early cosmic history, the difference
between the clustering of dark matter and baryons, has affected even
the largest halos down to the present \cite{BL11,Maayane}.

\section{Cosmic milestones of early radiative feedback}
\label{s:milestone}

\subsection{Reionization}
\label{s:reion}

The reionization of the Universe is an old subject. The observation of
transmitted flux short-ward of the \Lya wavelength of quasars
indicated in 1965 that the modern Universe is highly ionized
\cite{GP}. While this led to a gradual growth of literature on the
theoretical development of cosmic reionization (as summarized, e.g.,
in \cite{Barkana:2001}), calculations in the context of modern
cosmological models of hierarchical galaxy formation were first made
in the 1990's. These included the first numerical simulations of
cosmic reionization \cite{CO93,GO97}, and analytical calculations
\cite{FK94,SGB94,KSS94,TSB94,HL97,VS99} that mostly followed the
overall, global progress of reionization, based on counting the
ionizing photons from the rapidly rising star formation while
accounting for recombinations. Exploration of the 21-cm signatures of
reionization began in one of these numerical simulations \cite{GO97}
and in theoretical papers by Shaver et al.\ (1999) and Tozzi et al.\
(2000) \cite{shaver,tozzi}.

There soon began a more detailed discussion of the structure and
character of reionization, important issues for a variety of
observational probes of the era of reionization, especially 21-cm
cosmology. A commonly-assumed simple model was that of instantaneous
reionization, often adopted in calculations of the effect of
reionization on the CMB. This was supported by simulations
\cite{GO97,G00a} that showed a rapid ``overlap'' stage whereby the
transition from individual H~II regions around each galaxy to nearly
full reionization was rapid ($\Delta z \sim 0.1$). Fast reionization
would have made it easier to detect reionization through a sudden jump
in the number of faint \Lya sources \cite{HS99,H02} (given the strong
\Lya absorption due to a neutral IGM).

These same simulations also found that the H~II regions during
reionization were typically quite small, below a tenth of a Mpc for
most of reionization until a sudden sharp rise (to larger than the
simulation box) once only $30\%$ of the hydrogen mass (occupying
$15\%$ of the volume) remained neutral. Predictions made on this basis
\cite{CGO02} were bad news for 21-cm observations, which will find it
difficult to reach the angular resolution required to see such small
features within the cosmological 21-cm signal. Modeling of the effect
of reionization on secondary CMB anisotropies through the kinetic
Sunyaev-Zel'dovich effect (whereby the velocities of free electrons
created by reionization changed the energies of the fraction of CMB
photons that re-scattered) also assumed that the ionized bubble scale
would be very small unless quasars were dominant
\cite{Aghanim96,GruzH98,Santos03}.

Another basic issue about reionization is its structure/topology. At
this time, both analytical models and numerical simulations
\cite{MHR00,G00a} suggested that reionization would be outside-in
(with most ionizing photons leaking to the voids and reionizing them
first, leaving the dense regions for later) rather than inside-out
(which is when the high-density regions around the sources reionize
before the low-density voids).  All of the just-noted conclusions were
based on numerical simulations with box sizes below 10~Mpc. A
simulation of a 15~Mpc box found some ionized regions as large as
3~Mpc \cite{Sokasian}. An even larger, 30~Mpc simulation \cite{Ciardi}
considered a field (average) region and a proto-cluster (i.e., an
overdense region), and found substantial differences between their
reionization histories (thus suggesting fluctuations on quite large
scales), but still supported an outside-in reionization (since the
proto-cluster reionized later than the field region). In hindsight,
most of the results summarized in this and the previous two paragraphs
were incorrect or confusing.

\begin{figure}[tbp]
\includegraphics[width=\textwidth]{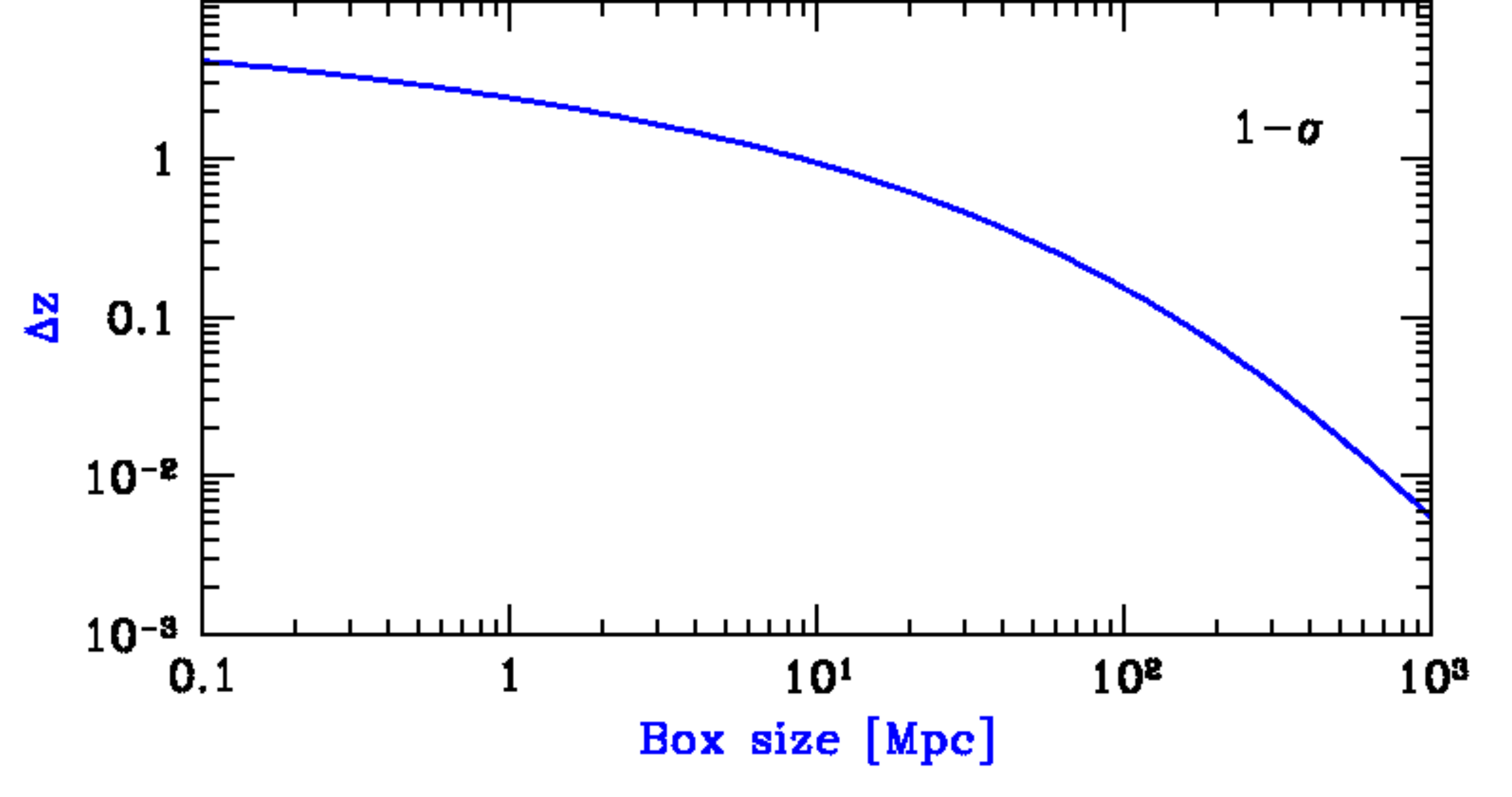}
\caption{Cosmic $1-\sigma$ scatter in the redshift of reionization, or
  any other phenomenon that depends on the fraction of gas in
  galaxies, versus the size of a rectangular region (in the Universe
  or in a simulation). When expressed as a shift in redshift, the
  scatter is predicted to be approximately independent of the typical
  mass of galactic halos. Regions of size 10~Mpc are not
  representative and do not yield an overall picture of reionization,
  since different regions of that size reionize at redshifts that
  differ by a $1-\sigma$ scatter of $\Delta z \sim 1$. One hundred Mpc
  boxes are required in order to decrease $\Delta z$ to well below
  unity $(\sim 0.15)$. From \cite{BL04}. }
\label{f:reionA}
\end{figure}

\begin{figure}[tbp]
\includegraphics[height=0.5\textwidth]{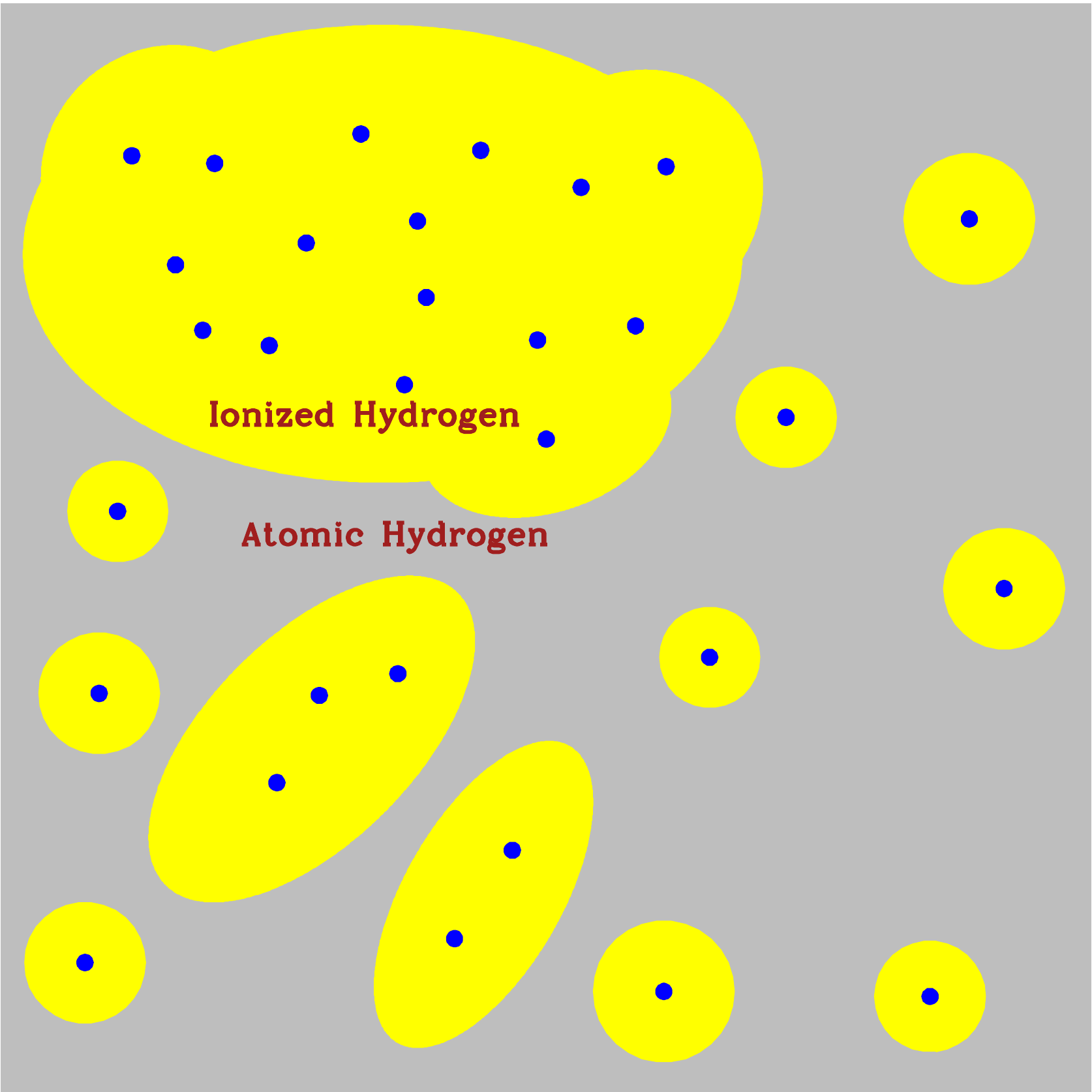}
\includegraphics[height=0.5\textwidth]{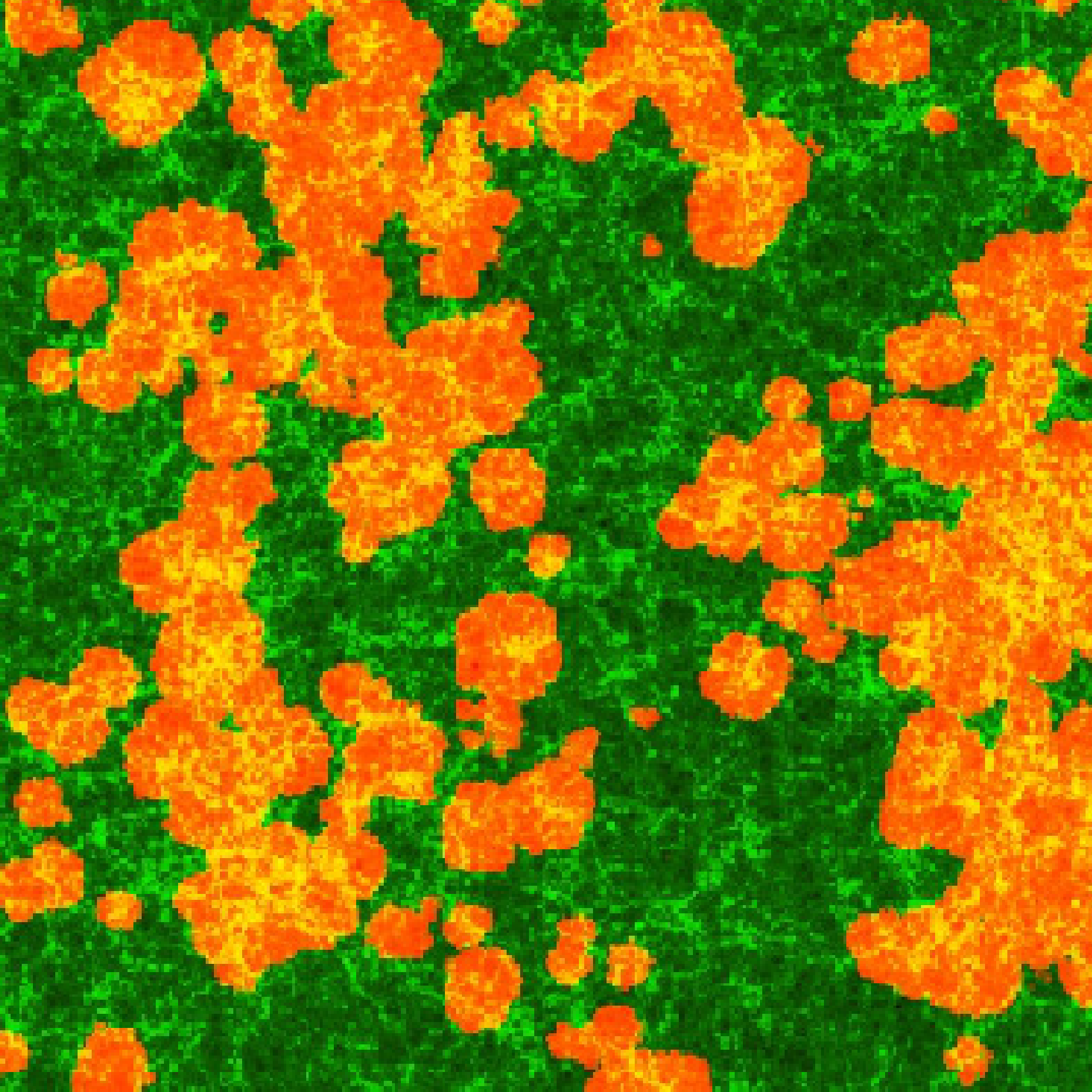}
\caption{During reionization, ionized bubbles were created by
  clustered groups of galaxies \cite{BL04}. The illustration (left
  panel, from \cite{BSc}) shows how regions with large-scale
  overdensities formed large concentrations of galaxies (dots) whose
  ionizing photons produced large ionized bubbles. At the same time,
  other large regions had a low density of galaxies and were still
  mostly neutral. A similar pattern has been confirmed in large-scale
  numerical simulations of reionization (e.g., the right panel shows a
  two-dimensional slice from a 150 Mpc simulation box \cite{mellema}).
  Multiple-source bubbles likely dominated the ionized volume from as
  early as $z \sim 20$ \cite{B08}.}
\label{f:reion}
\end{figure}

The now-accepted paradigm of reionization began to emerge when Barkana
\& Loeb (2004) \cite{BL04} realized that the surprisingly strong
clustering of high-redshift halos (see section \S~\ref{s:unusually})
leads to H~II bubbles driven by multiple clustered galaxies rather
than individual galaxies\footnote{This paper \cite{BL04} was first
  submitted in August 2003 but was only published 11 months later due
  to initial resistance to its novel conclusions.} (see Figures
\ref{f:reionA} and \ref{f:reion}). This clustering is significant even
on scales of tens of Mpc, leading to typical bubble sizes during
reionization that are larger than the total box size of most numerical
simulations of reionization at the time. The strong bias of
high-redshift galaxies also settled the issue of the topology of
reionization \cite{BL04}, showing that it is inside-out; while the
recombination rate was higher in overdense regions because of their
higher gas density, these regions still reionized first, despite the
need to overcome the higher recombination rate, since the number of
ionizing sources in these regions was increased even more strongly as
a result of the strong bias of galaxies\footnote{Quantitatively, the
  number of hydrogen atoms that must be initially ionized in each
  region is proportional to its density, i.e., the effective linear
  bias (Eq.~\ref{e:bias}) for this quantity is unity. The number of
  recombinations that must be overcome goes as density squared, so its
  effective bias is 2. The high-redshift galaxies that are thought to
  have sourced reionization likely had a bias above 2 throughout
  reionization, with a more typical value of 5 or 10.}.  The
outside-in picture, though, is still useful, as it seems likely to
apply to the internal structure of individual H~II bubbles and to the
post-reionization universe.  Another important revision was in the
common view of the effect of reionization on the abundance of dwarf
galaxies in various environments \cite{BL04}.

\begin{figure}[tbp]
\includegraphics[width=\textwidth]{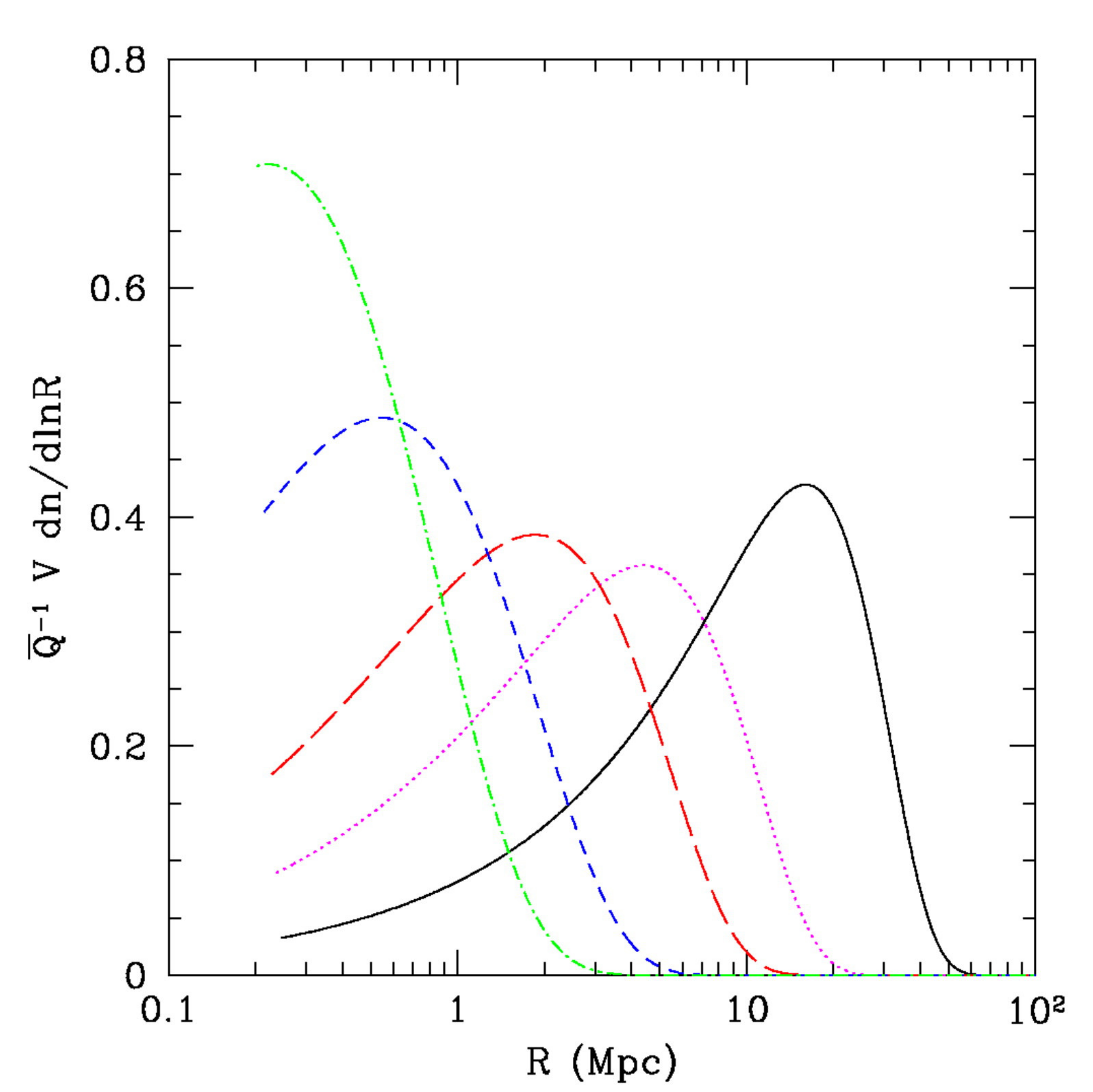}
\caption{Distribution of H~II bubble sizes during reionization. The
  fraction of the ionized volume in bubbles of radius $R$ is shown per
  $\log R$ interval. Dot-dashed, short-dashed, long-dashed, dotted,
  and solid lines are for $z=18$, 16, 14, 13, and 12, respectively, in
  a model in which the cosmic ionized fraction at these times is
  0.037, 0.11, 0.3, 0.5, and 0.74, respectively. From \cite{fzh04}.}
\label{f:fzh}
\end{figure}

The next big step was taken by Furlanetto et al.\ (2004) \cite{fzh04},
who created an analytical model for the distribution of H~II bubble
sizes (Figure~\ref{f:fzh}), based on an ingenious application of the
extended Press-Schechter model \cite{bond91}. This showed how the
typical size rises gradually during reionization, from a few Mpc to
tens of Mpc during the main stages, and allowed an estimate of the
resulting 21-cm power spectrum during reionization.  This picture of
reionization based on semi-analytic models \cite{BL04,fzh04} was then
confirmed by several numerical simulations that reached sufficiently
large scales with boxes of $\sim 100$ Mpc in size (e.g.,
\cite{Iliev,zahn,santos}). The simulations indeed showed the dominance
of large bubbles due to large groups of strongly-clustered galaxies,
though it should be noted that the price of such large boxes was (and
remains) a limited ability to resolve the small galaxies that were
likely the dominant sources of reionization.

This realization, that reionization was characterized by strong
fluctuations on large scales even if the individual galaxies that
caused it were small, has been very important and influential. It has
helped motivate the large number of observational efforts currently
underway in 21-cm cosmology (\S~\ref{intro21}), since large-scale
fluctuations are easier to detect (as they do not require high angular
resolution; see Eq.~(\ref{e:Tthermal})).

Today there remain some major uncertainties about reionization that
will likely only be resolved by 21-cm measurements
(\S~\ref{s:reion21}). In terms of the overall timing, the best current
constraint comes from large-angle polarization measurements of the CMB
which capture the effect of the re-scattering of CMB photons by the
reionized IGM. The latest measured optical depth of $5.5 \pm 0.9\%$
\cite{Planck16} implies a reionization midpoint at a redshift of
$7.5-9$ in realistic models (with reionization completing somewhere in
the range $z=6-8$).  However, the best-fit optical depth has changed
substantially with every new measurement (declining over time), and in
general it is more difficult to constrain small values of the optical
depth since the corresponding reionization signature on the CMB is
then smaller compared to systematic errors. The CMB results do
strongly limit the high-redshift onset of reionization, with a limit
of less than $\sim 10\%$ completion by $z=10$ \cite{Planck16b}. There
have long been hints of a late end to reionization at $z \sim 6$
\cite{D01,Becker01,Fan02,Fan06,z6}, but they have been controversial
due to the expected large fluctuations in the cosmic ionizing
background even after full reionization of the low-density IGM
\cite{B02,N02,L02,MF08,M10}.

As far as the typical halo masses that hosted the dominant sources of
reionization, it is expected that Lyman-Werner radiation dissociated
molecular hydrogen early on \cite{haiman}, so that by the central
stages of reionization star formation required atomic cooling, with a
minimum halo mass for star-formation of $\sim 10^8 M_\odot$. As
reionization proceeded, the hot gas within ionized regions raised the
gas pressure and prevented it from falling into small gravitational
potential wells; this photoheating feedback gradually eliminated star
formation in halos up to a mass of $\sim 3 \times 10^9 M_\odot$, as
has been studied in many calculations and numerical simulations
\cite{R86,E92,H2a,TW96,QKE96,WHK97,GO97,MR98,KI00,G00b,NS00,MesingerMF,McQuinnMF}.
In particular, this means that an era of active star formation in
dwarf galaxies prior to reionization may be observable directly with
next-generation telescopes \cite{BL00,WL06}, or in the star formation
histories of massive high-redshift galaxies \cite{BLhistory}, although
this depends also on the effectiveness of supernova feedback in
small galaxies \cite{DS86,WL13}.

Another interesting issue related to reionization is that of
minihalos, i.e., low-mass halos that collect gas but do not form stars
due to the lack of sufficient cooling. These minihalos formed in large
numbers, clustered strongly around ionizing sources, and contained
enough gas to effectively block most ionizing photons
\cite{HAM01,BL02}. However, the minihalos naturally photoevaporated
once engulfed by H~II regions \cite{BL99,SIR04}, making their effect
on reionization (which they delay) and on 21-cm emission only modest
\cite{Ciardi06,Yue09}. We note that due to their low masses, minihalos
were also strongly affected by the baryon - dark matter streaming
velocity (\S~\ref{s:stream}).

\subsection{\Lya coupling and Lyman-Werner feedback}
\label{s:Lya}

The general course of cosmic history as relevant to 21-cm cosmology
was outlined in \S~\ref{intro21}, and the physics of the 21-cm
transition (including \Lya coupling) was described in detail in
\S~\ref{s:21physics}. Here we briefly summarize \Lya coupling and LW
feedback, as they are among the most important observable events in
early cosmic history.

The IGM can be observed in 21-cm emission or absorption, relative to
the CMB background, only if the hyperfine levels of the hydrogen atom
are not in equilibrium with the CMB. This means that the spin
temperature must differ from the CMB temperature. At the highest
redshifts, atomic collisions overcome the scattering of CMB photons
and drive the spin temperature to the kinetic temperature of the gas.
However, this becomes ineffective at $z \sim 30$, and the spin
temperature then approaches the CMB temperature. Luckily for 21-cm
cosmologists, stellar \Lya photons come to the rescue \cite{Madau},
moving the spin temperature back towards the kinetic temperature
through the indirect Wouthuysen-Field effect \cite{Wout,Field}. The
\Lya coupling era refers to the time during which the \Lya flux
reaches and passes the level needed for effective 21-cm coupling.

Unlike reionization and heating, \Lya coupling and Lyman-Werner
feedback are not cosmic events that change the overall state of the
IGM. \Lya coupling is basically a 21-cm event, and it is important
because of the prospect of detecting 21-cm emission from the early
era ($z \sim 20-30$ \cite{Holzbauer:2011,Complete}) of \Lya coupling.
A 21-cm observation of \Lya coupling (see \S~\ref{s:Lya21} for more
details) is the only currently feasible method of detecting the
dominant population of galaxies from such high redshifts and measuring
their properties, either through a global 21-cm detection of the
strong mean absorption signal or by interferometric measurement of the
substantial 21-cm fluctuations expected from this era \cite{BL05b}.
While still far from the very first stars at $z \sim 65$
\cite{first,anastasia}, this is the highest redshift range currently
envisioned for observing the dominant galaxy population, a feat which
would be very exciting.

\begin{figure}[tbp]
\includegraphics[width=\textwidth]{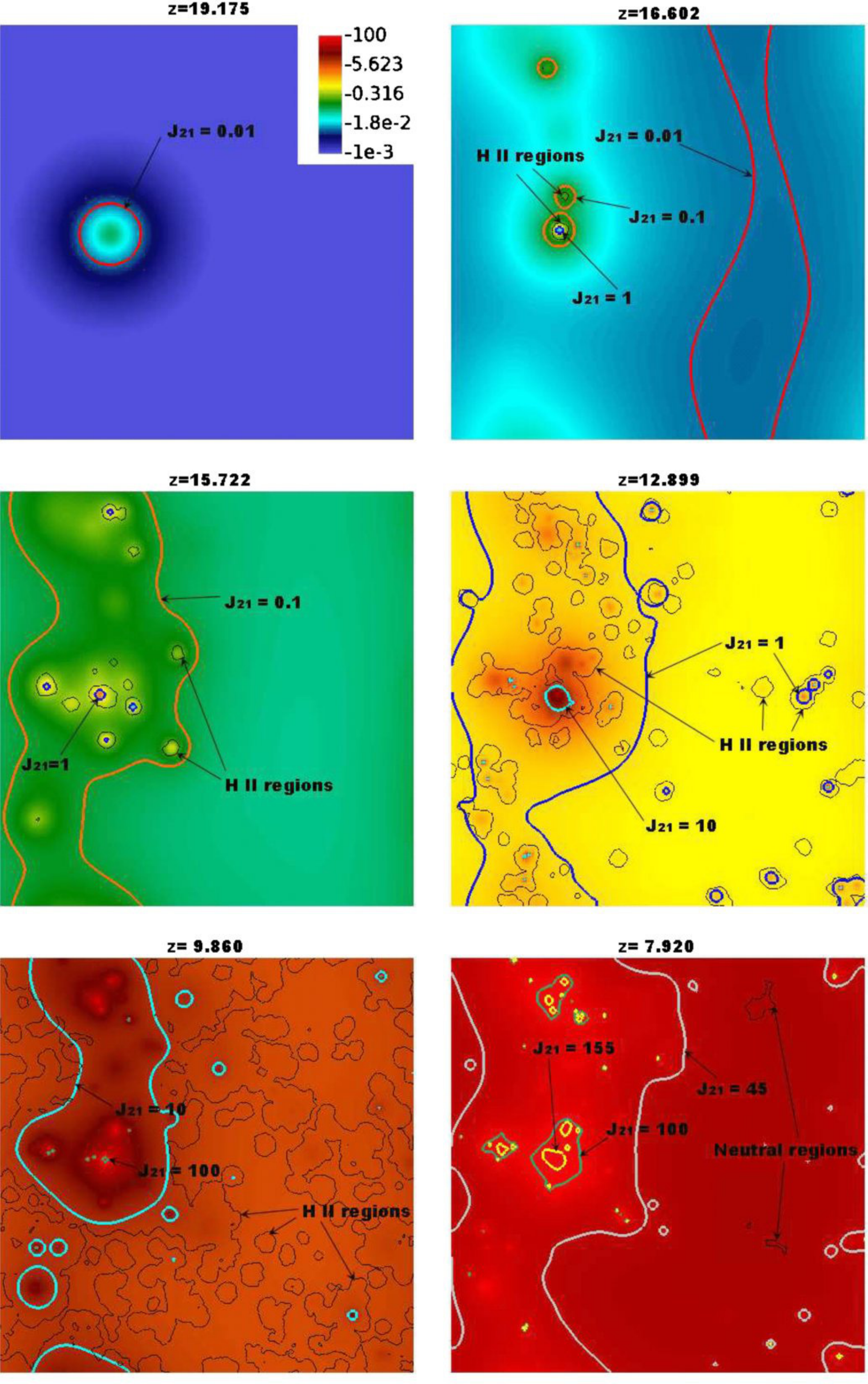}
\caption{Spatial images from a simulation showing the isocontours of
  patchy reionization and the patchy $H_2$-dissociating background on
  a planar slice through a box of volume ($35/h$~Mpc)$^3$ at various
  epochs. The level of $J_{\rm LW,21}$ (the LW photon intensity in
  units of $10^{-21}$~erg~cm$^{-1}$~s$^{-1}$~Hz$^{-1}$~sr$^{-1}$) on
  the grid is depicted by various colors, with the range $[10^{-3} -
  10^2]$ shown on the inset of the top-left panel. On top of each
  $J_{\rm LW,21}$ color map, contours of thick colored lines represent
  different $J_{\rm LW,21}$ levels (red, orange, blue, cyan, and green
  corresponding to $J_{\rm LW,21}=0.01$, 0.1, 1, 10, and 100,
  respectively). The black lines represent ionization fronts. From
  \cite{Ahn:2009}.}
\label{f:LWAhn}
\end{figure}

\begin{figure}[tbp]
\includegraphics[width=\textwidth]{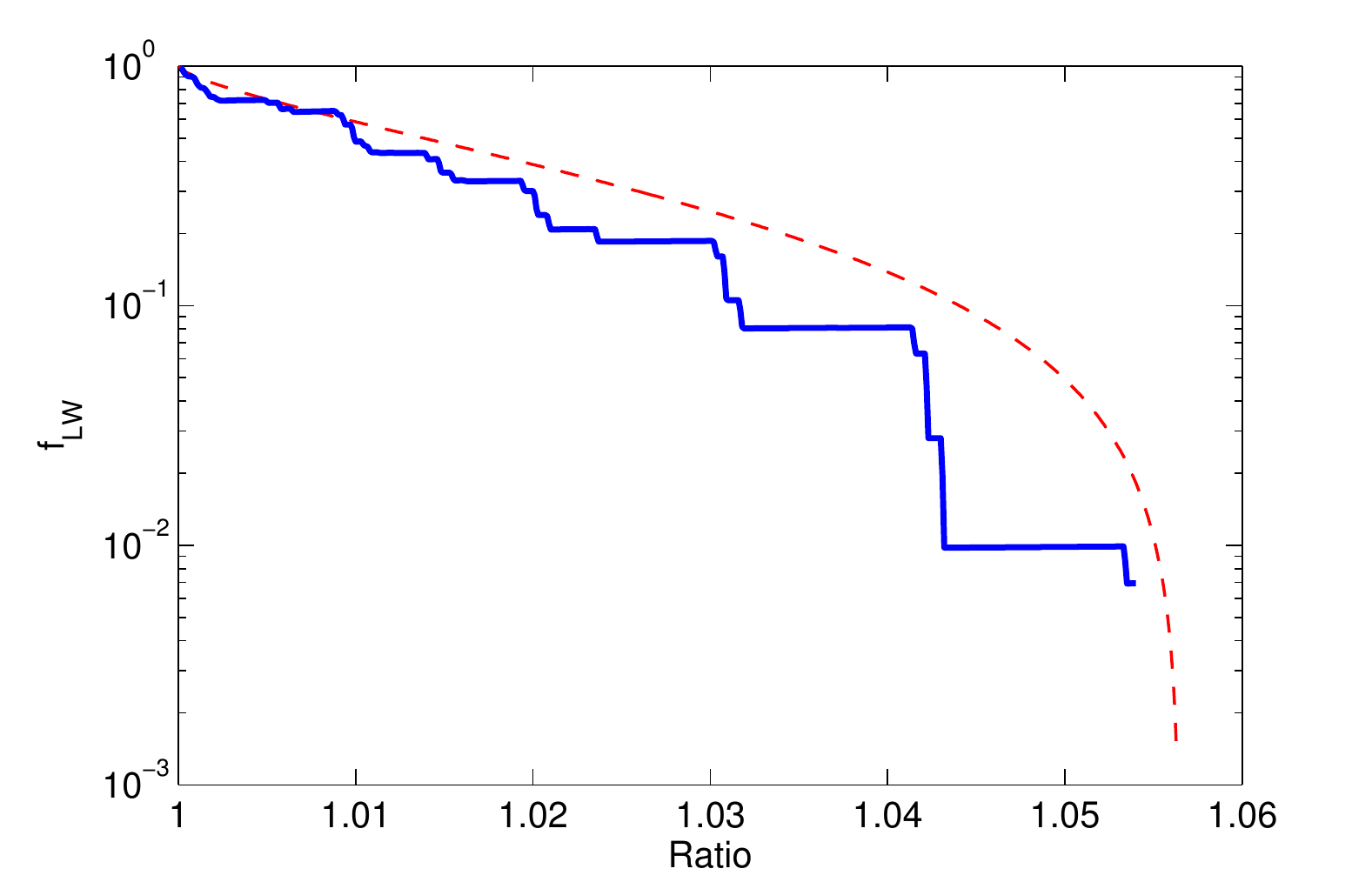}
\caption{The relative effectiveness of causing H$_2$ dissociation in
  an absorber at $z_a$ due to stellar radiation from a source at
  $z_s$, shown versus the ratio $R\equiv (1+z_s)/(1+z_a)$ since in
  this form this function is independent of redshift. The complex
  result (solid curve) incorporates the expected stellar spectrum of
  Population III stars \cite{Bromm:2001,BL05b}, along with the full
  list of 76 relevant LW lines \cite{haiman}. Beyond the max shown
  $R=1.054$ (which corresponds to 104 comoving Mpc at $z=20$), {\bf
    $f_{\rm LW}$} immediately drops by five orders of magnitude. Also
  shown is a commonly used approximation \cite{Ahn:2009} (dashed
  curve) which is based on a flat, averaged LW spectrum. Both
  functions are normalized to unity at $R=1$. From \cite{FialkovLW}.}
\label{f:LW}
\end{figure}

LW feedback is a major feedback effect on the first stars. It
indirectly affects the IGM and the 21-cm sky through its effect on the
radiative output from stars (including Ly$\alpha$, X-ray, and ionizing
radiation). LW feedback dissociates molecular hydrogen and thus it
ended star formation driven by molecular cooling \cite{haiman} in
halos of $\sim 10^6 M_{\odot}$ \cite{Abel,Bromm}. If the overall
(time-averaged) star-formation efficiency in such small, early halos
was significant, then their LW radiation is expected to have produced
significant feedback early on ($z \sim 20-25$)
\cite{haiman,Ahn:2009,Holzbauer:2011,FialkovLW}, at a time when these
halos still dominated the global star formation. This feedback
strengthened gradually as the LW intensity increased, as has been
found in numerical simulations that imposed a LW background on forming
early galaxies (either constant with time
\cite{Machacek:2001,Wise:2007,O'Shea:2008} or increasing more
realistically \cite{VisbalLW}). Because of its gradual rise, LW
feedback did not actually halt or reduce the global star formation,
but it did slow down the otherwise rapid rise of star formation at
high redshifts. Like other inhomogeneous negative feedbacks, LW
feedback increased cosmic equality by first suppressing the sites of
earliest star formation \cite{Ahn:2009,Holzbauer:2011,FialkovLW}
(Figure~\ref{f:LWAhn}).  While some LW photons reached out to a
distance of $\sim 100$~Mpc from each source, the feedback was more
local than that; emission from distant sources was absorbed more
weakly, so that half the effective LW flux seen at a given point came
from sources within $\sim 15$~Mpc away (Figure~\ref{f:LW}).

A discussion of the 21-cm signatures of the \Lya coupling and LW
feedback eras is deferred to \S~\ref{s:Lya21}. We note that in this
topic it is essential to include the baryon - dark matter streaming
velocity (\S~\ref{s:stream}) as well, since it affects the same halos
as the LW feedback, and these same halos may have dominated star
formation during the \Lya coupling era.

\subsection{Cosmic heating}
\label{s:heating}

Before discussing heating in the context of 21-cm cosmology, we begin
with a brief summary of the basic physics of X-ray heating. The
comoving mean free path of an X-ray photon, to photoelectric
absorption in a universe of neutral fraction $x_{\rm H\, I}$,
is\footnote{In Eq.~(\ref{e:mfp}), the power-law dependence of
  $\lambda_X$ on $x_{\rm H\, I}$ is -1; it has sometimes been
  incorrectly listed as +1/3 \cite{F06} or -1/3 \cite{Jonathan07}.}
\cite{F06} \beq \lambda_X \approx 51 \, x_{\rm H\, I}^{-1} \left(
  \frac{1+z}{10} \right)^{-2} \left( \frac{E}{0.5\, \rm keV}
\right)^3\, \rm Mpc\ .
\label{e:mfp} \eeq For photons of energy $E \gtrsim 1$~keV,
$\lambda_X$ becomes a significant fraction of the horizon
(Eq.~\ref{e:hor}), and in that case cosmological redshift effects lead
to a substantial loss of energy between emission and absorption (plus
there is a significant time delay between these two events). Once the
X-rays are absorbed, the resulting (primary) fast electrons then
interact with the surrounding gas through the processes of collisional
excitation, ionization, and electron-electron scattering.  These
secondary processes quickly distribute the original X-ray energy into
ionization (of hydrogen and helium in the IGM), heating (i.e.,
thermalized energy), and excitation (which results in low energy
photons that then escape, so that the energy is effectively lost). The
fraction of energy that goes into heating varies with the ionization
fraction of the background medium, from around a third of the energy
in a neutral medium up to nearly all of the energy in a highly ionized
one \cite{F06,Shull,ChenK,Stoever}.

It has long been known that the Universe was reionized at an early
time (\S~\ref{s:reion}) and thus heated to at least $\sim 10,000$~K by
the ionizing photons. While reionization was a major phase transition
in the IGM, the question of whether the gas had been radiatively
pre-heated prior to reionization is also important. Significant
pre-heating of the IGM directly affects 21-cm observations, and also
produces some photoheating feedback (though much weaker than that due
to reionization).

The dependence of the 21-cm brightness temperature on the kinetic
temperature $T_K$ of the gas takes the form $T_{\rm b} \propto [1 -
T_{\rm CMB}/T_K]$ (Eq.~\ref{e:combo2} or \ref{e:combo3}). Thus, the
midpoint of the heating era, or the central moment of the ``heating
transition'', refers to the moment when the mean gas temperature is
equal to that of the CMB, so that the cosmic mean $T_{\rm b}$ is zero;
actually, the latter would be true in a universe with purely linear
fluctuations, but non-linearities delay the time when $\langle T_{\rm
  b} \rangle=0$ by an extra $\Delta z \sim 0.5$ \cite{FialkovLW}.
Also, clearly $T_{\rm b}$ is more sensitive to cold gas than to hot
gas (relative to the CMB temperature). Indeed, at early times the
21-cm absorption can be very strong (depending on how much the gas
cools), but at late times, once $T_K \gg T_{\rm CMB}$, $T_{\rm b}$
becomes independent of $T_K$ and the 21-cm emission is said to be in
the ``saturated heating'' regime.

\begin{figure}[tbp]
\includegraphics[width=\textwidth]{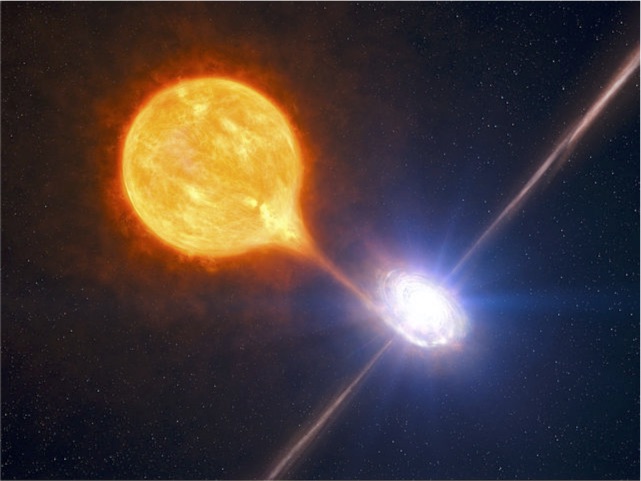}
\caption{The most plausible sources for cosmic heating before
  reionization are early X-ray binaries, dominated by black-hole
  binaries \cite{Frag2} like the one illustrated here, where material
  from a companion spills onto a black hole, resulting in X-ray
  emission from its accretion disk. Credit: ESO drawing from
  http://en.wikipedia.org/wiki/File:A\_stellar\_black\_hole.jpg .}
\label{f:XRB}
\end{figure}

For a long time it was confidently predicted that the universe was
well into the saturated heating regime once cosmic reionization got
significantly underway. The stage for this widespread belief was set
by the landmark paper in 21-cm cosmology by Madau et al.\ (1997)
\cite{Madau}. They considered several possible heating sources, mainly
X-rays from quasars (later observed to disappear rapidly at $z>3$,
e.g., \cite{S08}) and heating from \Lya photons (later shown to be
negligibly small \cite{Miralda,chuzhoy07,ciardiLya}). However, stellar
remnants -- particularly X-ray binaries (Figure~\ref{f:XRB}) -- have
become the most plausible source of cosmic heating. This is the result
of a combination of basic facts: 1) X-rays travel large distances even
through a neutral IGM; 2) Large populations of X-ray binaries should
have formed among the stellar remnants associated with the significant
cosmic star formation that we know must have occurred in order to
reionize the universe; 3) Observations of the local Universe suggest
not only that X-ray binaries form wherever star formation is found,
but that their relative populations increase by an order of magnitude
at the low metallicity expected for high-redshift galaxies
\cite{Mirabel0,Frag2,Mirabel,Basu,Basu2}.

Even with X-ray binaries as the plausible source, the common
expectation of saturated heating before reionization had remained, and
had been assumed in many mock analyses made in preparation for
upcoming data (\cite{Lofar14} is a recent example). A key reason for
this is that until recently, calculations of cosmic X-ray heating
\cite{Fur06,Jonathan07,CLoeb,21cmfast,Mesinger13} had assumed
power-law spectra that place most of the X-ray energy at the
low-energy end, where the mean free path of the soft X-rays is
relatively short. This means that most of the emitted X-rays are
absorbed soon after they are emitted, before much energy is lost due
to cosmological effects. The absorbed energy is then enough to heat
the gas by the time of reionization to $\sim 10$ times the temperature
of the CMB \cite{Cold}.

\begin{figure}[tbp]
\includegraphics[width=\textwidth]{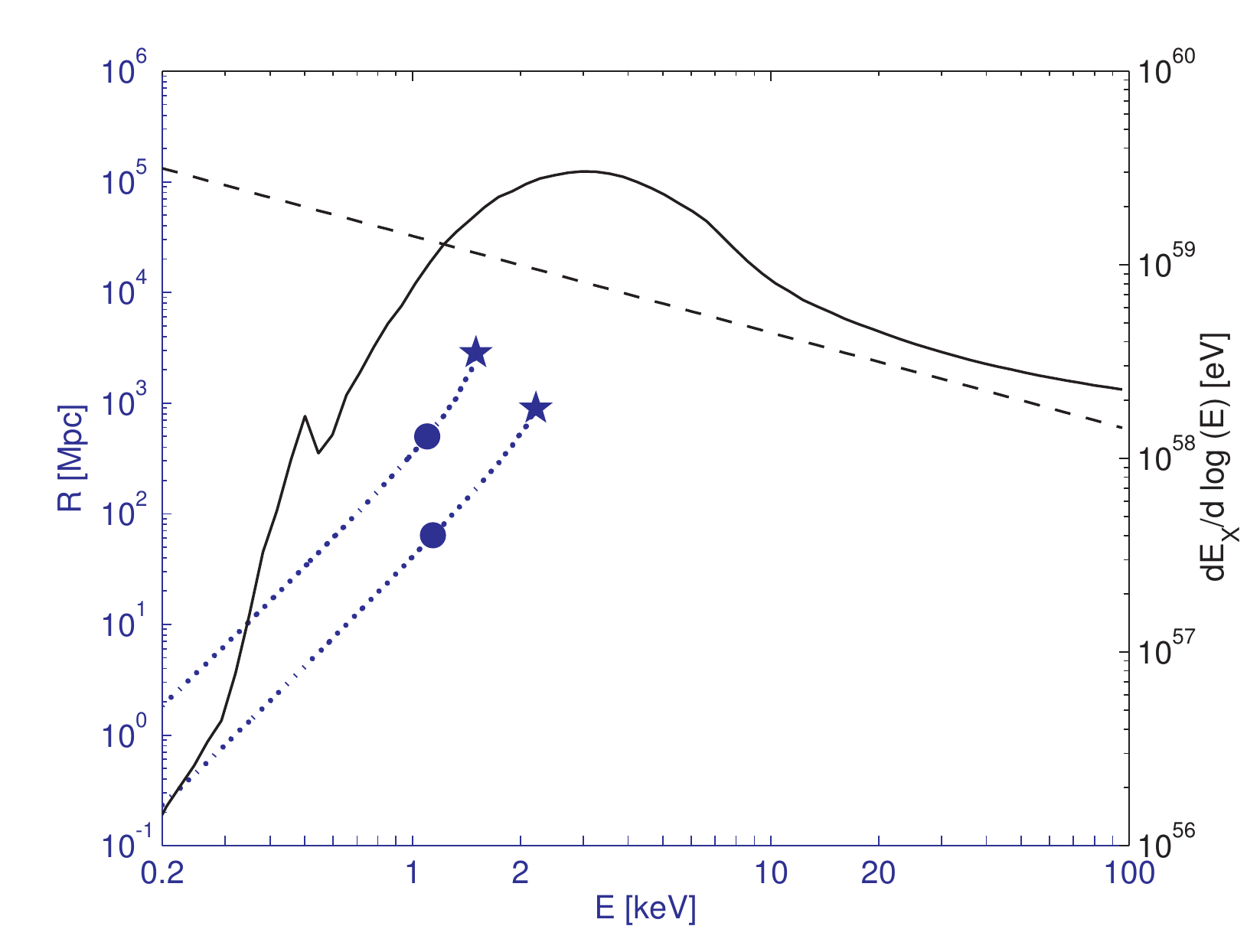}
\caption{X-ray spectra, mean free paths, and horizons. The expected
  spectrum of X-ray binaries at high redshift (solid curve) from
  population synthesis models \cite{Frag1,Frag2} is compared with the
  soft power-law spectrum (dashed curve) adopted until recently
  \cite{Fur06,Jonathan07,CLoeb,21cmfast,Mesinger13}. Both indicate the
  distribution into X-ray photons with energy $E$ of the total X-ray
  energy $E_X$ produced per solar mass of newly-formed stars. The
  X-ray emission of X-ray binaries should be dominated by the most
  massive systems in their high (that is, bright) state \cite{Frag2},
  which is dominated by thermal disk emission, with little emission
  expected or seen \cite{Frag1,spectra,tamura} below 1~keV. Also
  shown are the mean free paths (dotted curves) of X-ray photons
  arriving at $z=10$ (top) or $z=30$ (bottom). For each of these
  redshifts, also indicated are the effective horizon for X-rays
  (defined as a $1/e$ drop-off, like a mean free path) from the
  combined effect of cosmological redshift and time retardation of
  sources ($\bullet$), and the distance to $z=65$ ($\star$), the
  formation redshift of the first star \cite{first,anastasia} (at
  which the mean free path curves are cut off). Note the separate
  $y$~axes that indicate energy content for the spectra (right) or
  comoving distance for the other quantities (left). From
  \cite{Cold}.}
\label{f:coldfig1}
\end{figure}

However, Fialkov et al.\ (2014) \cite{Cold} recognized that the
assumed X-ray spectrum is a critical parameter for both the timing of
cosmic heating and the resulting 21-cm signatures. The average
radiation from X-ray binaries is actually expected to have a much
harder spectrum (Figure~\ref{f:coldfig1}) whose energy content (per
logarithmic frequency interval) peaks at $\sim 3$~keV. Photons above a
(roughly redshift-independent) critical energy of $\sim 1$~keV have
such a long mean free path that by the start of reionization, most of
these photons have not yet been absorbed, and the absorbed ones came
from distant sources that were effectively dimmed due to cosmological
redshift effects. This reduces the fraction of the X-ray energy
absorbed as IGM heat by about a factor of 5, enough to push the moment
of the heating transition into the expected redshift range of cosmic
reionization (and thus, we will refer to this case as late heating).
For this and other reasons, the spectrum of the X-ray heating sources
is a key parameter for 21-cm cosmology, as further discussed in
\S~\ref{s:2012}.

Based on low-redshift observations, other potential X-ray sources
appear sub-dominant compared to X-ray binaries. One such source is
thermal emission from hot gas in galaxies, which has a relatively soft
X-ray spectrum. Its X-ray luminosity in local galaxies \cite{mineo} is
(for a given star-formation rate) about a third of that of X-ray
binaries. Given the above-mentioned order-of-magnitude increase
expected in the emission from X-ray binaries at high redshift, the
thermal gas would have to be highly efficient at high redshift in
order to contribute significantly. Also, some theoretical arguments
suggest that X-rays produced via Compton emission from relativistic
electrons in galaxies could be significant at high redshift
\cite{oh01}, though again the increase would have to be very large
compared to such emission in low-redshift sources; the expected
spectrum in this case (flat from $\sim 100$~eV to $\sim 100$~GeV)
would deliver most of the energy above 1~keV and thus count as a hard
spectrum in terms of 21-cm signatures.  Another possible heating
source, large-scale structure shocks, is likely ineffective
\cite{shocks,shocks2,mcquinn12b}.

A natural X-ray source to consider is the population of bright
quasars. As noted above, while quasars are believed to dominate the
X-ray background at low redshift \cite{vasudevan}, their rapid decline
beyond $z \sim 3$ \cite{S08} suggests that their total X-ray
luminosity (including an extrapolation of their observed luminosity
function) is sub-dominant compared to X-ray binaries during and prior
to reionization \cite{Frag2}. The rarity of quasars at early times is
natural since they seem to be hosted mainly by halos comparable in
mass to our own Milky Way; the \Lya absorption signature of gas infall
provides direct evidence for this \cite{BL03}.  More promising for
early heating, perhaps, is the possibility of a population of
mini-quasars, i.e., central black holes in early star-forming halos.
This must be considered speculative, since the early halos were so
small compared to galactic halos in the present universe that the
corresponding black-hole masses are expected to fall in a very
different range from observed quasars, specifically within the
intermediate black-hole range ($10^2 - 10^4 M_\odot$) that local
observations have probed only to a limited extent \cite{imbh}. Thus,
the properties of these mini-quasars are highly uncertain, and various
assumptions can allow them to produce either early or late heating
\cite{tanakaX,ciardiLya}.  Local observations can be used to try to
estimate the possible importance of mini-quasars. An internal feedback
model that is consistent with observations of local black-hole masses
as well as high-redshift quasar luminosity functions \cite{WL03}
indicates a mini-quasar contribution that is somewhat lower than X-ray
binaries \cite{Cold}, though the uncertainties are large. Regarding
the spectrum, standard models of accretion disks \cite{ss} around
black holes predict that the X-ray spectrum of mini-quasars
\cite{tanakaX} should peak at $1-5$ keV, making it a hard spectrum for
cosmic heating that is quite similar to that of X-ray binaries.

Regardless of the source of X-rays, an important parameter is the
degree of absorption in high-redshift halos compared to locally
observed galaxies. If we assume that the gas density in high-redshift
halos increases proportionally with the cosmic mean density, then the
column density through gas (within a galaxy or a halo) is proportional
to $(1+z)^2 M_{\rm halo}^{1/3}$. This simple relation suggests that
absorption of X-rays should increase at high redshift, since the
redshift dependence should have a stronger effect than the decrease of
the typical halo mass. However, complex astrophysics could
substantially affect this conclusion, since the lower binding energy
of the gas in low-mass halos could make it easier to clear out more of
the blockading gas. Given the large uncertainty in internal absorption
(on top of the other uncertainties in source properties), it is likely
that only 21-cm observations will determine the precise
characteristics of the high-redshift sources of cosmic heating.

\section{21-cm Signatures of the First Stars}

\label{s:Final21cm}

Ongoing and planned interferometric observations in 21-cosmology hope
to reach a sub-mK sensitivity level \cite{McQuinn,F06} (see also
\S~\ref{intro21}). The best current observational upper limit is from
PAPER \cite{PAPER}: 22.4~mK at a wavenumber range of $k=0.1 -
0.35$~Mpc$^{-1}$ at $z=8.4$, around an order of magnitude away from
plausible predictions (or two orders of magnitude in terms of the
power spectrum). Global 21-cm experiments (measuring the total sky
spectrum) are also being pursued, with the best result thus far (from
the EDGES experiment) \cite{bowman2} being a lower limit of $\Delta z
> 0.06$ for the duration of the reionization epoch. In the next few
subsections we focus on 21-cm fluctuations, and consider global
experiments separately in \S~\ref{s:global}.

\subsection{21-cm signatures of reionization}

\label{s:reion21}

\begin{figure}[tbp]
\includegraphics[width=\textwidth]{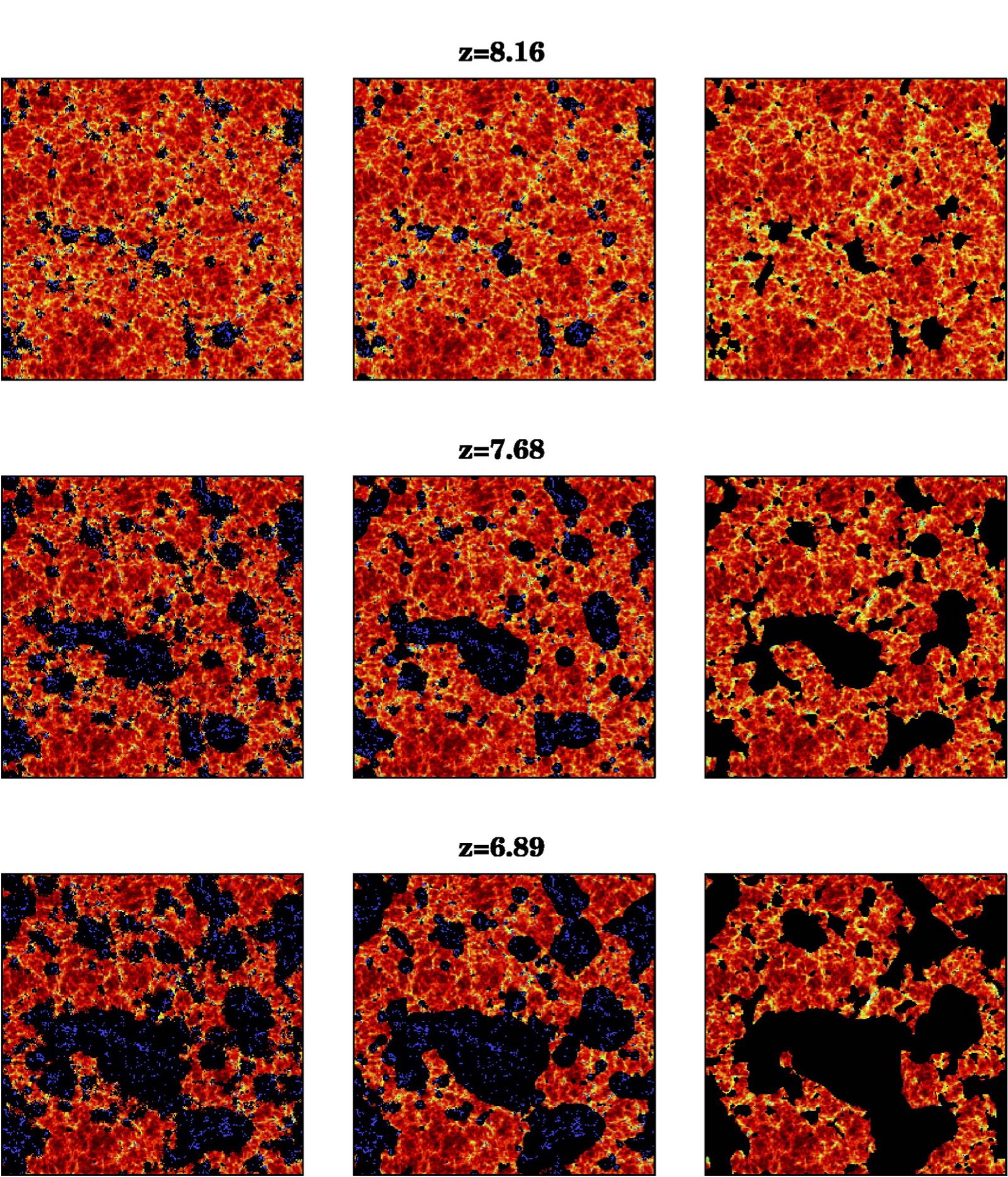}
\caption{21-cm maps during reionization, in radiative transfer
  simulations versus a semi-numeric approach. Each map is 94~Mpc on a
  side and 0.36~Mpc deep. The ionized fractions are 0.11, 0.33, and
  0.52 for $z=8.16$, 7.26, and 6.89, respectively. Left column:
  Radiative transfer calculation with ionizing sources (blue dots).
  Middle column: Halo smoothing procedure with sources from the N-body
  simulation. Right column: Matching semi-numerical model based on
  \cite{fzh04} and using the initial, linear dark matter overdensity.
  From \cite{zahn}.}
\label{f:zahn}
\end{figure}

\begin{figure}[tbp]
\includegraphics[width=0.498\textwidth]{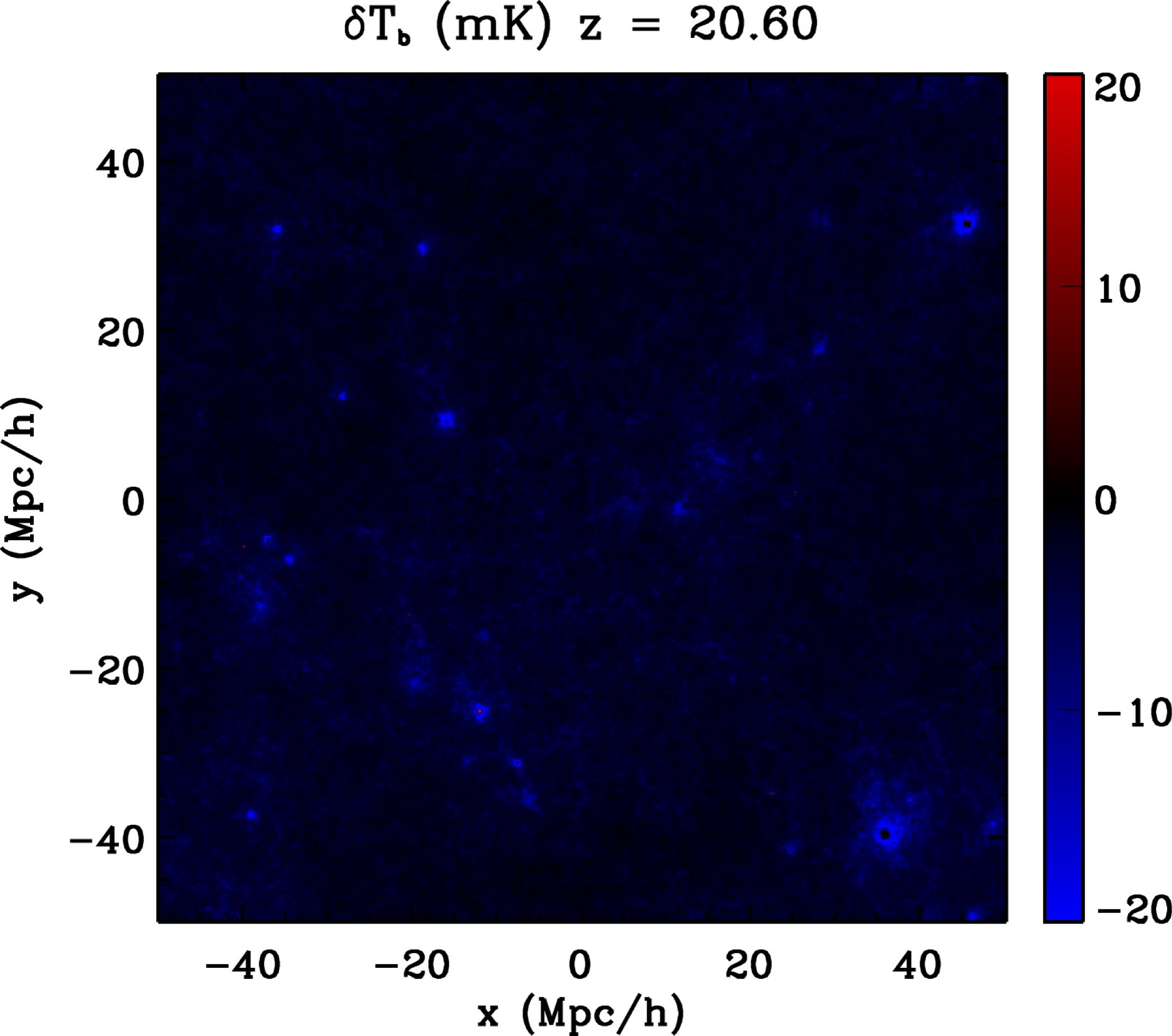}
\includegraphics[width=0.498\textwidth]{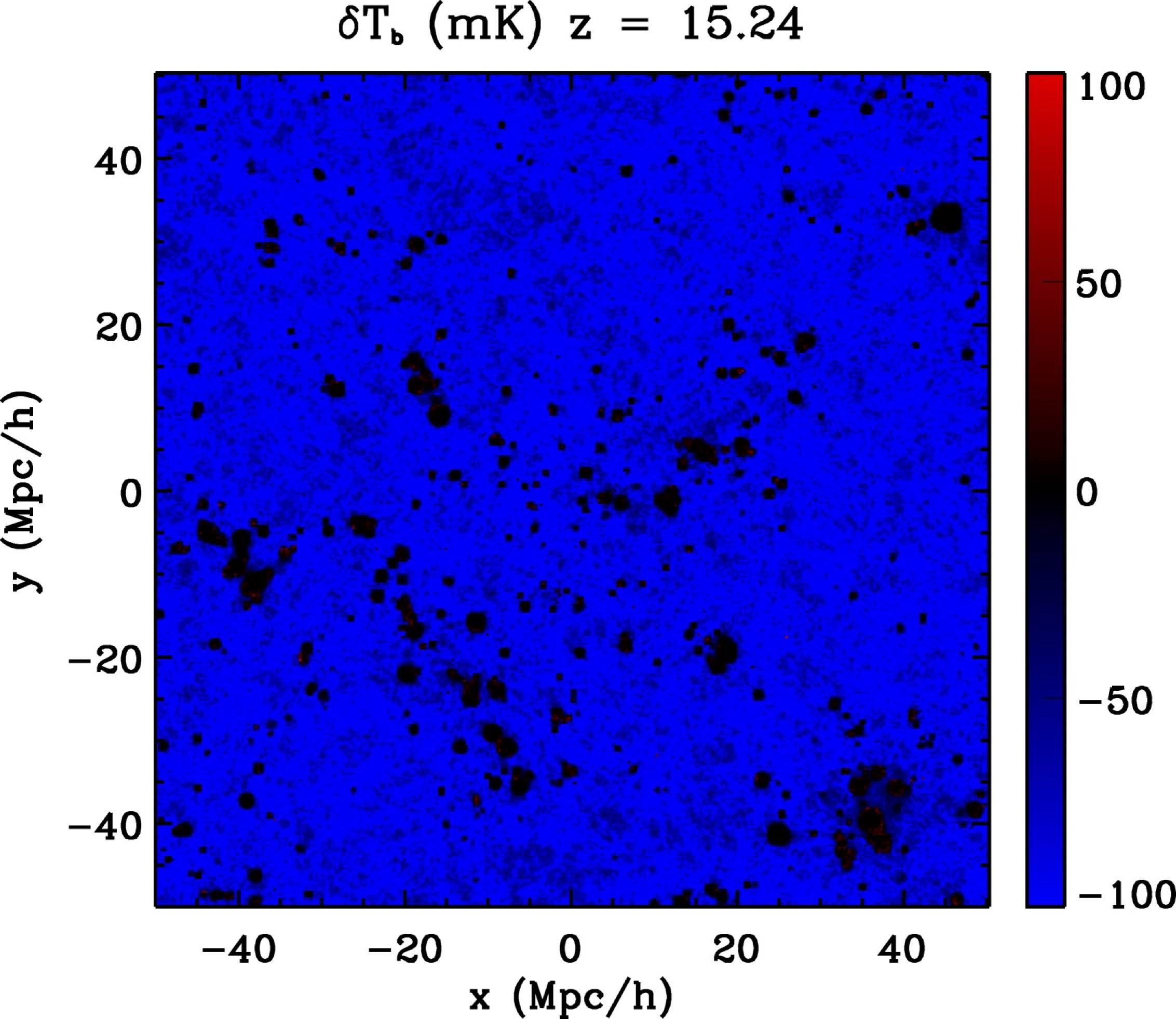}
\includegraphics[width=0.498\textwidth]{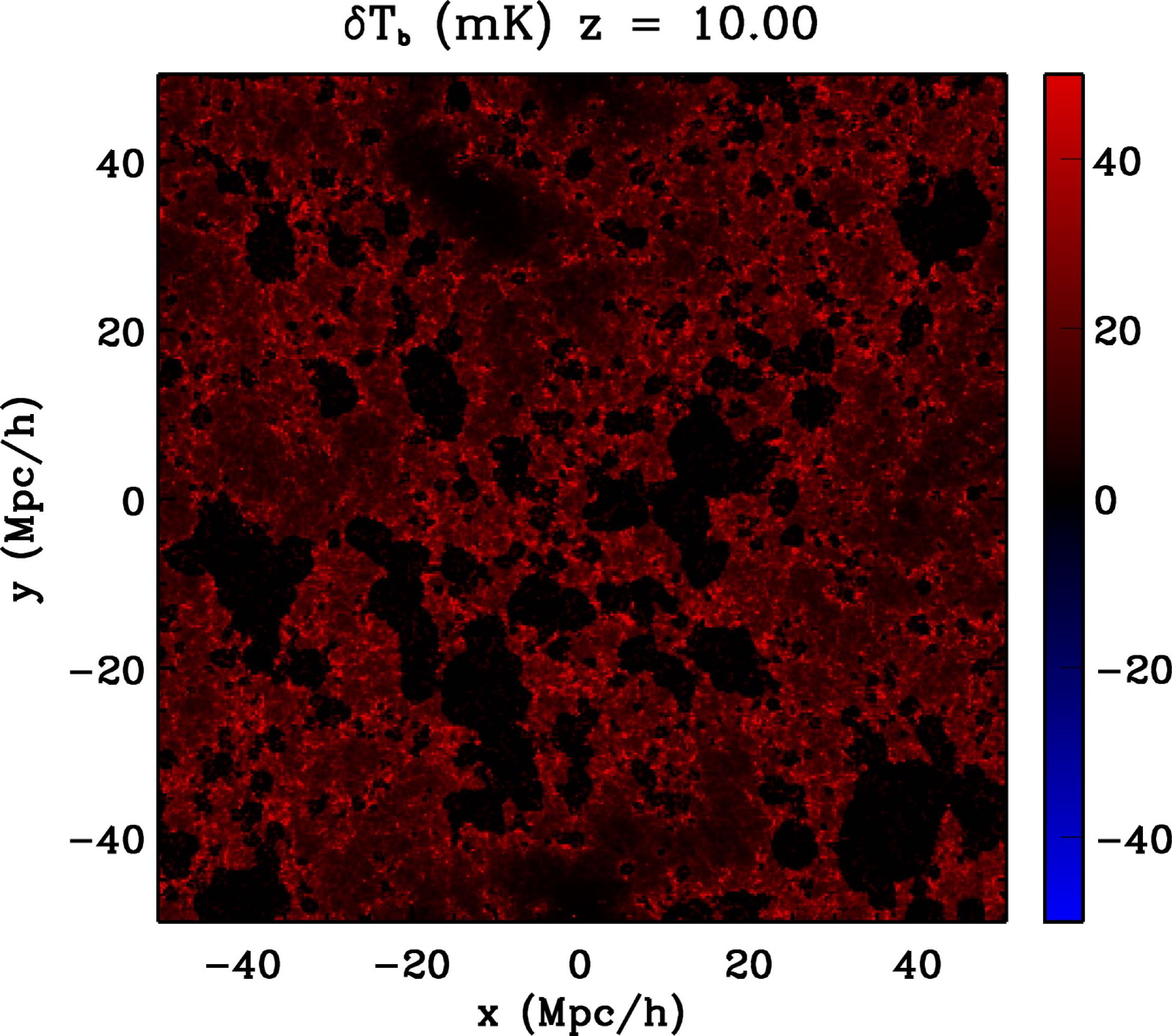}
\includegraphics[width=0.498\textwidth]{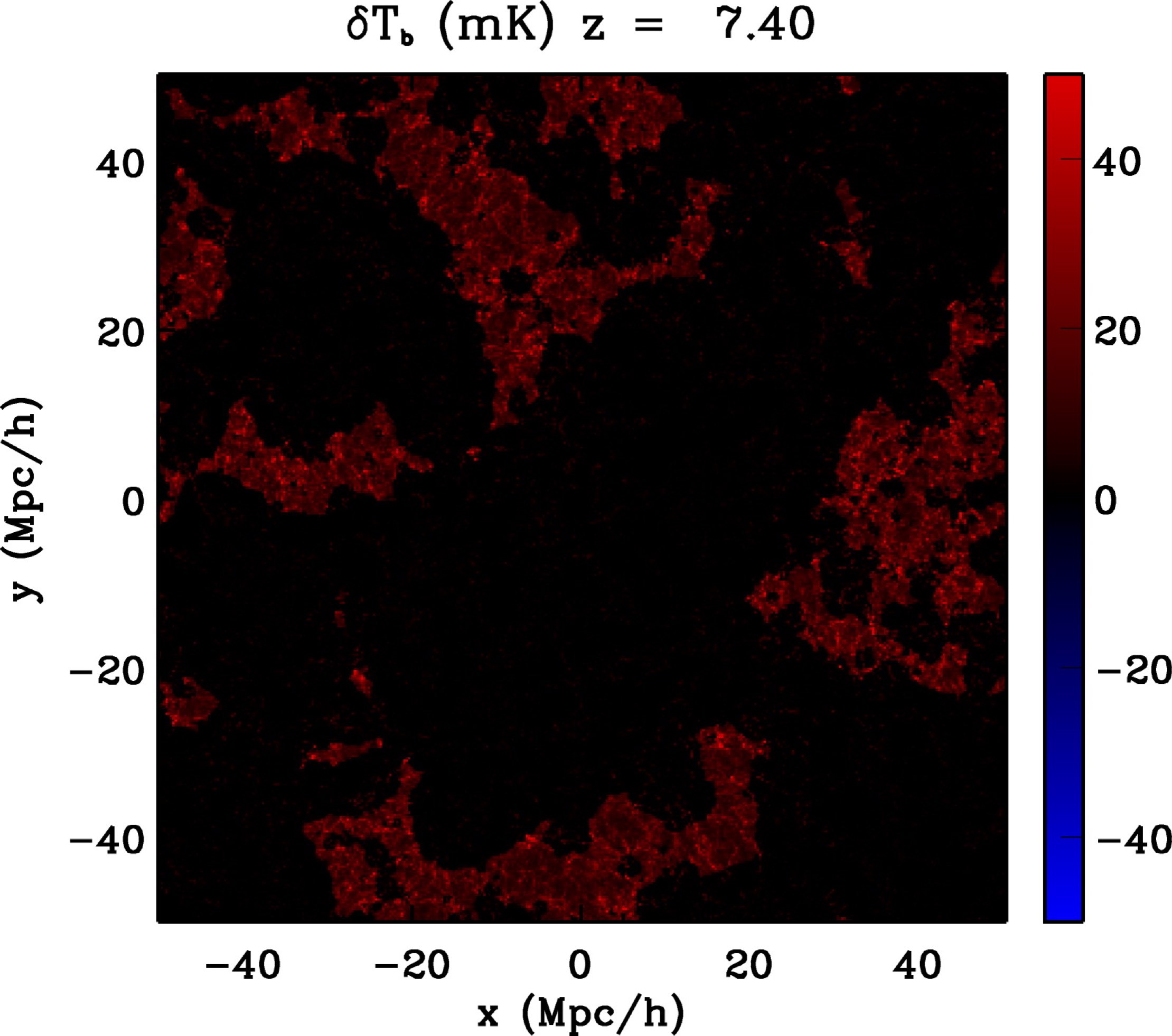}
\caption{21-cm maps from a $100/h$~Mpc simulation box that includes
  inhomogeneous \Lya and X-ray radiation fields, in addition to
  reionization. The ionized fractions are 0.0002, 0.03, 0.35, and 0.84
  for $z=20.60$, 15.24, 10.00, and 7.40, respectively. From
  \cite{santos}.}
\label{f:santos}
\end{figure}

In \S~\ref{s:reion} we discussed the important realizations that
reionization was driven by groups of galaxies, the early galaxies were
strongly clustered on large scales, and reionization had an inside-out
topology. These features of reionization should all be observable with
21-cm cosmology. Figure~\ref{f:zahn} shows an example of 21-cm maps
during reionization, as predicted by numerical simulations; a
semi-numerical model gives a quite similar reionization field though
it differs in the fine details. Another example is shown in
Figure~\ref{f:santos}, which is from a simulation that computes the
ionization, Ly$\alpha$, and X-ray fields. 

\begin{figure}[tbp]
\includegraphics[width=0.498\textwidth]{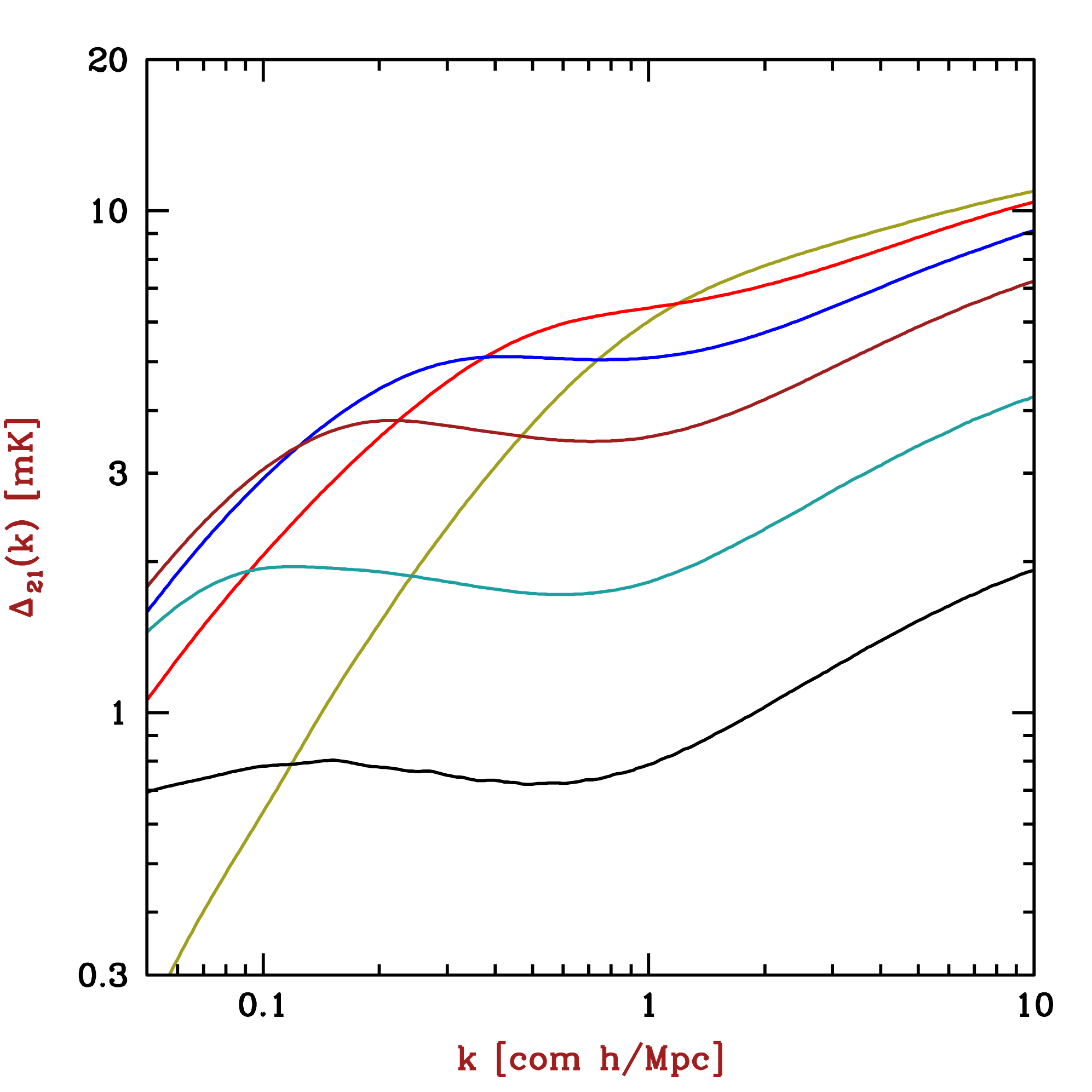}
\includegraphics[width=0.498\textwidth]{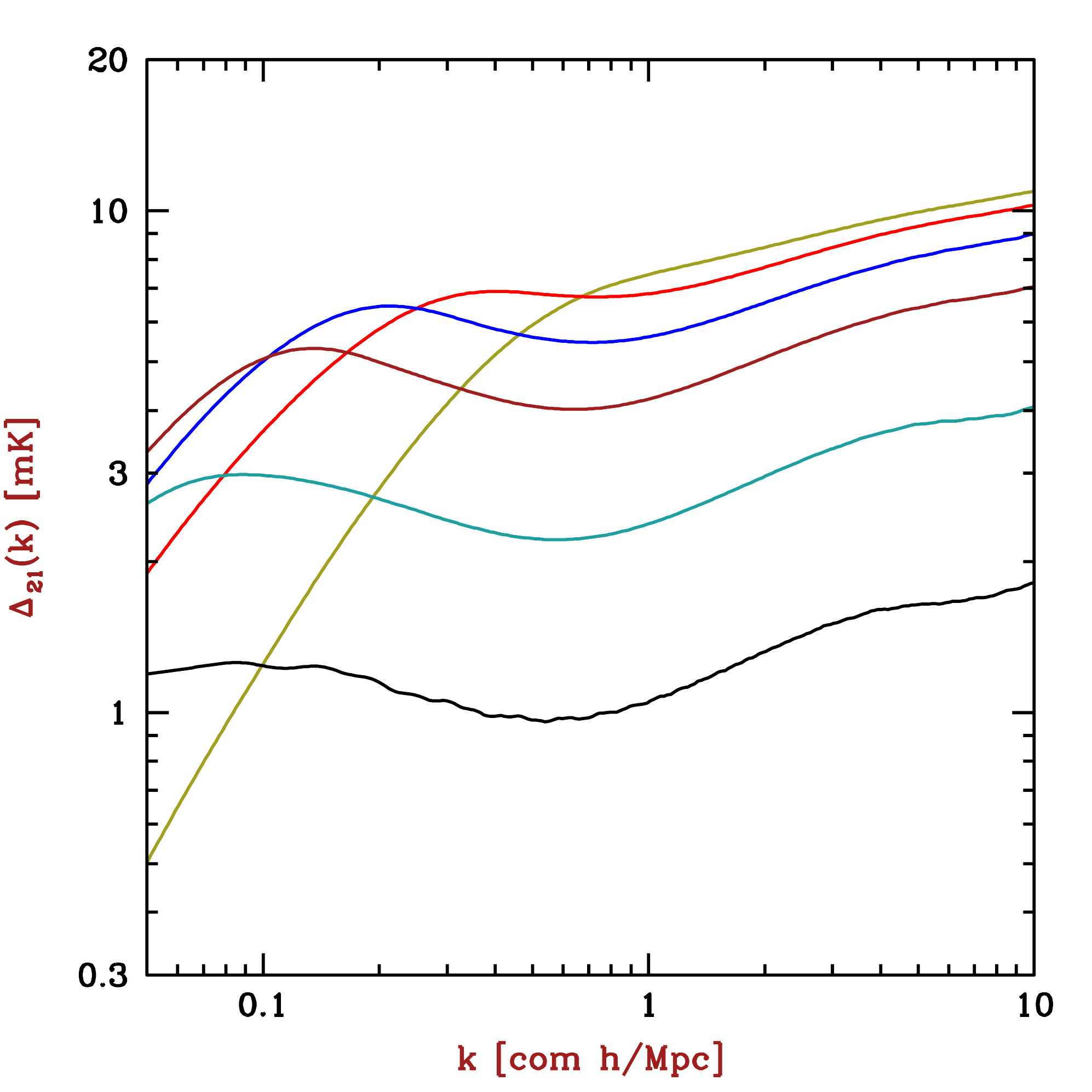}
\caption{Evolution of the 21-cm power spectrum throughout
  reionization, for a model that sets the cosmic mean ionized fraction
  $\bar{x}_i=98\%$ at $z=6.5$.  Shown are times when $\bar{x}_i=10\%$,
  $30\%$, $50\%$, $70\%$, $90\%$, and $98\%$ (from top to bottom at
  large $k$). At the very end of reionization, atomic hydrogen remains
  only within galaxies (this gas is not included in these plots). The
  panels show two different possibilities for the masses of galactic
  halos, assuming a minimum circular velocity for star formation of
  $V_{\rm c} = 35$~km/s (left panel) or 100~km/s (right panel). From
  \cite{Sources}.}
\label{f:scan1}
\end{figure}

The typical evolution of the 21-cm power spectrum during cosmic
reionization is illustrated in Figure~\ref{f:scan1}, using an
analytical model \cite{Sources} that was shown to be in reasonable
agreement with numerical simulations. Early on, when the cosmic
ionized fraction is $\sim 10\%$, the 21-cm power spectrum simply
traces the baryon density power spectrum (assuming here the limit of
saturated \Lya coupling and saturated heating). As reionization
advances, H~II bubbles form around individual sources and begin to
overlap between nearby sources, giving the 21-cm power spectrum an
extra hump on large scales, with the corresponding $k$ gradually
decreasing as the typical size of the bubbles increases. At the final
stages of reionization, the 21-cm intensity probes the distribution of
remaining neutral gas in large-scale underdensities, and at the very
end, atomic hydrogen remains only within galaxies.
Figure~\ref{f:scan1} also illustrates how the 21-cm power spectrum can
be used to probe the properties of the galaxies that are the sources
of reionization. By artificially setting various values for the
minimum circular velocity (or mass) of halos that dominate star
formation, it is possible to simulate cases where small galaxies
dominate or where large galaxies do (the latter case illustrating a
situation where internal feedback is highly effective within small
galaxies). Placing a fixed total amount of ionizing intensity within a
smaller number of more massive halos has a number of effects on the
21-cm power spectrum; large halos are rarer and more strongly
biased/clustered, leading to a higher power spectrum (in amplitude), a
more prominent H~II bubble bump that extends to somewhat larger
scales, and a more rapid reionization process (in terms of the
corresponding redshift range).

\begin{figure}[tbp]
\includegraphics[width=\textwidth]{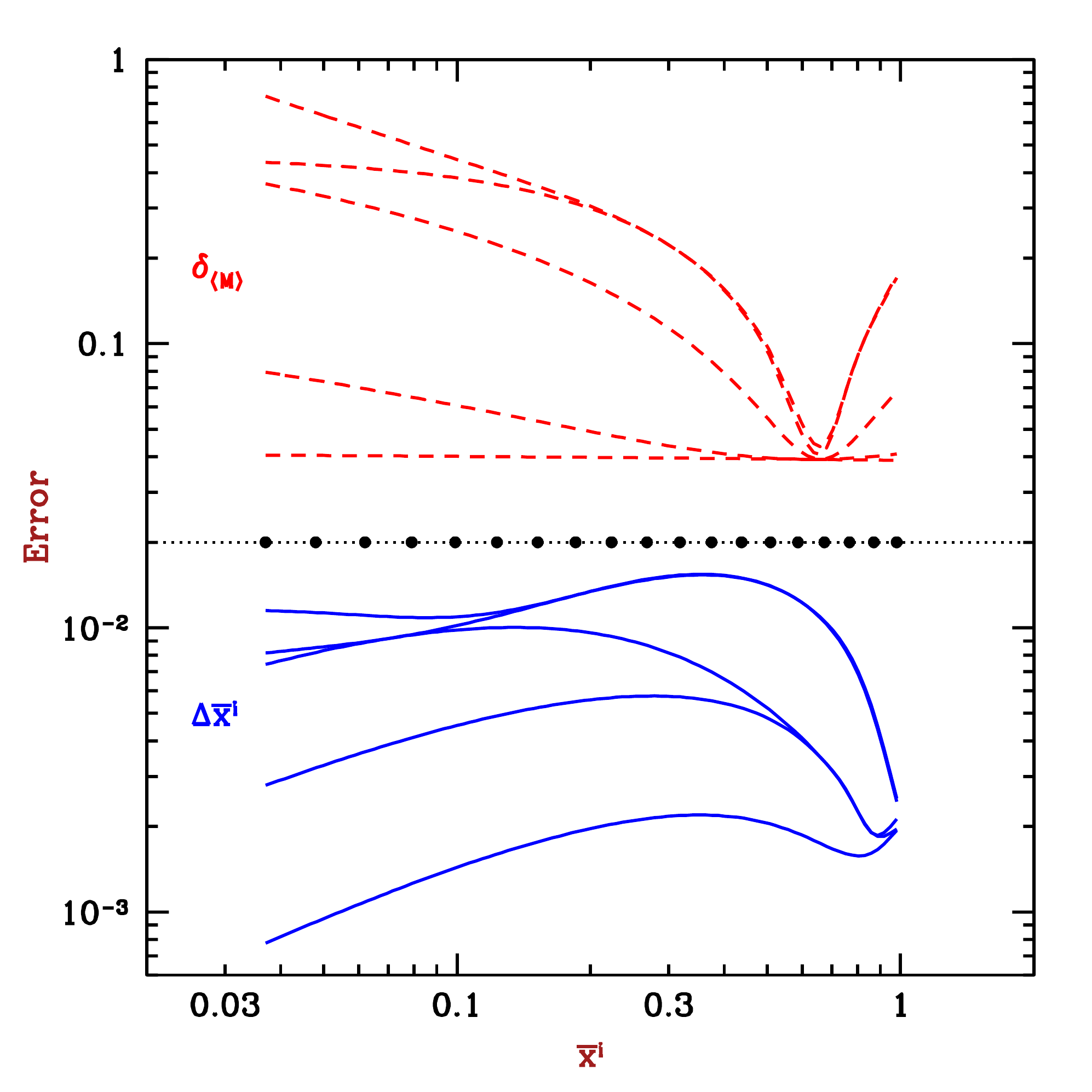}
\caption{Expected reconstruction errors throughout cosmic
  reionization, from fitting models to mock data of the 21-cm power
  spectrum (with the expected errors of a first-generation
  experiment). The models are based on an analytical model
  \cite{Evan,B07} that is in reasonable agreement \cite{Sources} with
  numerical simulations of reionization. The $x$-axis shows the stage
  of reionization, i.e., the fraction of the IGM that has been
  reionized ($\bar{x}_i$).  Models of varying degrees of flexibility
  are considered, with 2--6 free parameters (bottom to top in each set
  of curves). The input model of the mock universe sets the end of
  reionization (defined as $98\%$ of the IGM being ionized) at
  $z=6.5$, with galactic halos assumed to have a minimum circular
  velocity (Eq.~\ref{Vceqn}) $V_c = 35$ km/s. A horizontal dashed line
  separates the two areas of the plot that show the expected relative
  error in the intensity-weighted mean mass of galactic halos (top)
  and the {\it absolute}\/ error in the ionized fraction (bottom).
  Dots on the horizontal line show the values of $\bar{x}_i$
  corresponding to the 19 assumed observed redshifts (in the range
  $z=6.5 - 12$). From \cite{Sources}.}
\label{f:scan21cm}
\end{figure}

\begin{figure}[tbp]
\includegraphics[width=\textwidth]{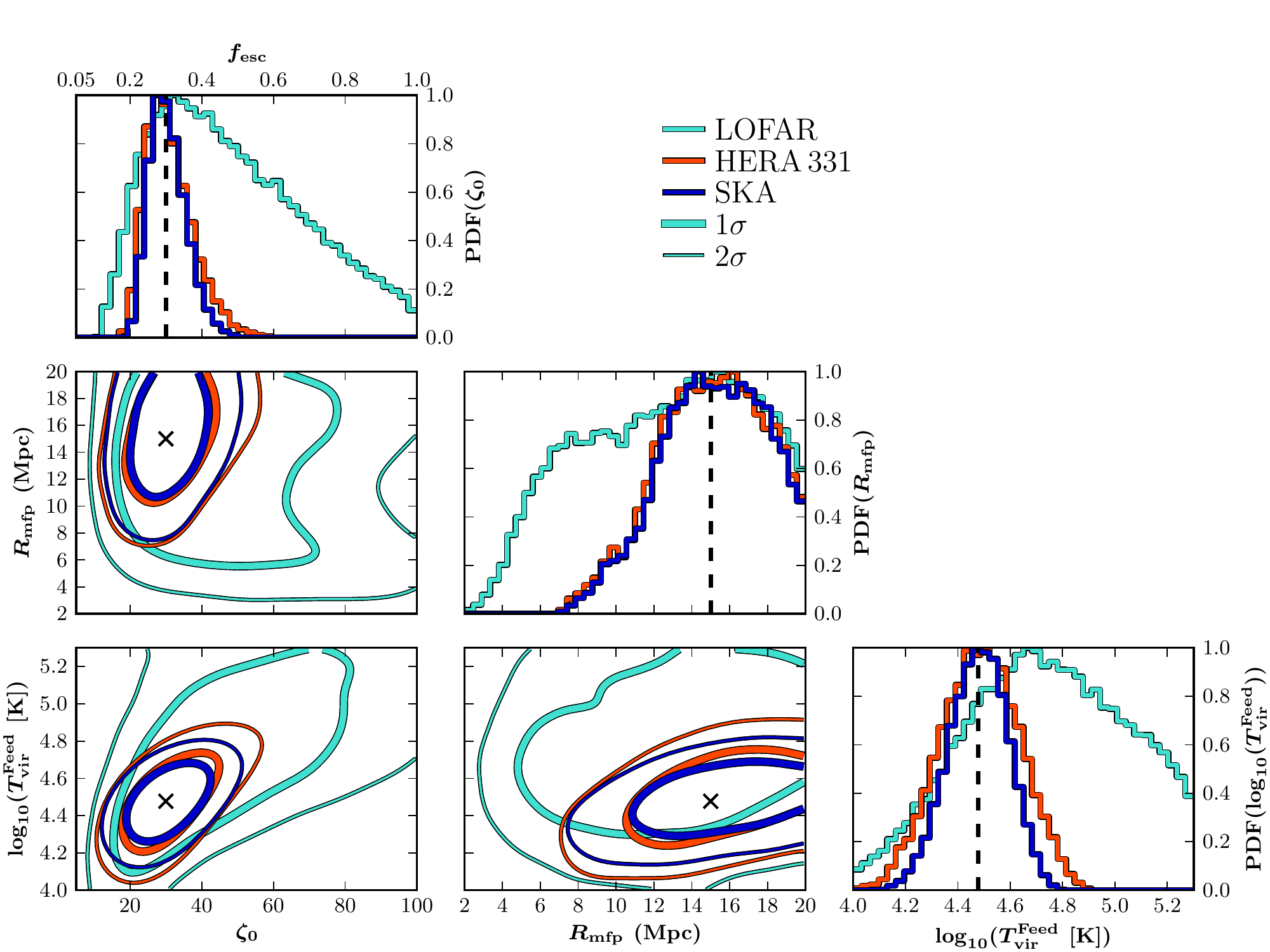}
\caption{The recovered constraints from 21CMMC on various reionization
  model parameters from combining three independent ($z = 8$, 9 and
  10) 1000~hr observations of the 21~cm power spectrum. Three
  different telescope arrays are compared: the current LOFAR
  (turquoise), and the future HERA (red) and SKA phase 1 (blue).
  Across the diagonal panels, the 1D marginalized PDFs are shown for
  the recovered reionization parameters [the ionizing efficiency
  $\zeta_0$, showing the corresponding escape fraction of ionizing
  photons $f_{\rm esc}$ on the top; $R_{\rm mfp}$, the mean free path
  of ionizing photons within ionized regions; and $\log_{10}(T_{\rm
    vir}^{\rm Feed})$, where $T_{\rm vir}^{\rm Feed}$ is the minimum
  virial temperature of star-forming halos], with the input model
  parameter value indicated by a vertical dashed line. In the three
  panels below the diagonal, 2D joint marginalized likelihood contours
  are shown for various pairs out of the three reionization
  parameters.  The $1\sigma$ (thick) and $2\sigma$ (thin) contours are
  shown, with crosses marking the input parameter values.  From
  \cite{21CMMC}.}
\label{f:21CMMC}
\end{figure}

An important question is how to fit the 21-cm data that are expected
soon from the cosmic reionization era. In general, the 21-cm power
spectrum during reionization is a complex superposition of the
fluctuations in density and ionization (and possibly heating: see
\S~\ref{s:late}); in order to interpret it quantitatively and
reconstruct the history of reionization and of early galaxy formation,
a flexible model is needed. Fitting to data cannot be done directly
with numerical simulations, and is difficult even with a
faster-running semi-numerical code. Thus, the first maximum likelihood
fitting of mock data \cite{Sources} was done with the analytical model
noted above. The computational efficiency of this approach made it
possible to employ a flexible six-parameter model that parameterized
the uncertainties in the properties of high-redshift galaxies;
specifically, the parameters were the coefficients of quadratic
polynomial approximations to the redshift evolution of two parameters:
the minimum circular velocity of galactic halos, and the overall
efficiency of ionizing photon production within galaxies. The
conclusion (see Figure~\ref{f:scan21cm}) was that observations with a
first-generation experiment should measure the cosmic ionized fraction
to $\sim 1\%$ accuracy at the very end of reionization, and a few
percent accuracy around the mid-point of reionization. The mean halo
mass hosting the ionizing sources should be measurable to better than
$10\%$ accuracy when reionization is 2/3 of the way through, and to
$20\%$ accuracy throughout the central stage of reionization
\cite{Sources}. Recently the semi-numerical code 21CMFAST
\cite{21cmfast}, in a sped-up version that employs some
approximations, has been incorporated directly within 21CMMC, a Monte
Carlo Markov Chain statistical analysis code. One result derived with
this code (see Figure~\ref{f:21CMMC}) is that combining three
observations (at $z = 8$, 9 and 10) of the 21-cm power spectrum will
allow upcoming 21-cm arrays to accurately constrain the basic
parameters of reionization \cite{21CMMC}.

\subsection{21-cm signatures of \Lya coupling and LW feedback}

\label{s:Lya21}

\begin{figure}[tbp]
\includegraphics[width=\textwidth]{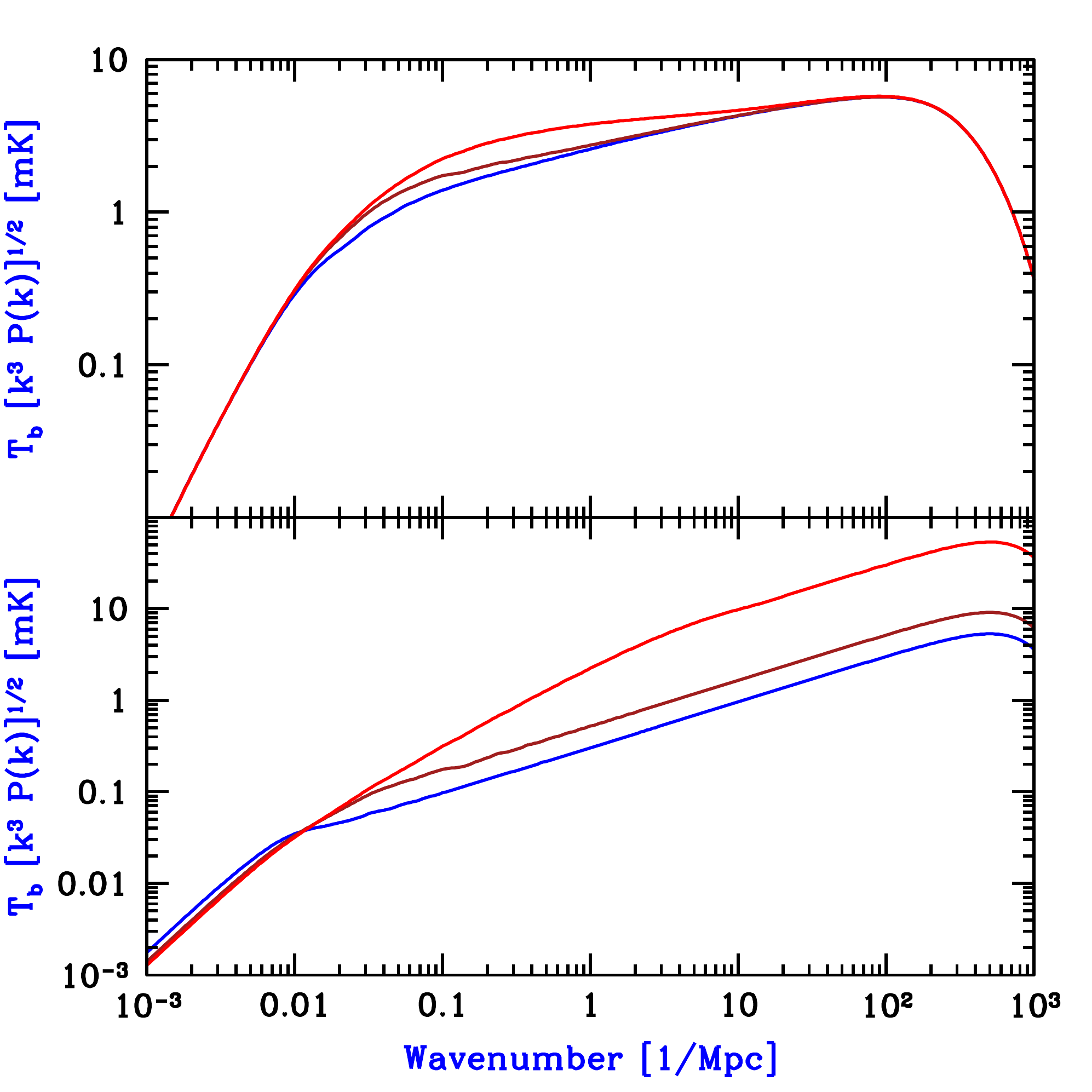}
\caption{Power spectra of 21-cm brightness fluctuations (in mK units)
  from \Lya fluctuations, plotted versus (comoving) wavenumber. Shown
  are two components of the power spectrum that in principle can be
  separated (in the limit of linear perturbations) based on the
  line-of-sight anisotropy of the 21-cm fluctuations \cite{BL05a}
  (\S~\ref{s:anisotropy}): $P_{\mu^2}$ (top panel) contains
  contributions directly from density fluctuations and from the
  density-induced fluctuations in galaxy density and therefore in \Lya
  flux, and $P_{\rm un-\delta}$ (bottom panel) is due to \Lya
  fluctuations from Poisson fluctuations in galaxy numbers. These
  results are for galaxies formed via atomic cooling in halos at
  $z=20$, with a star formation efficiency set to produce the \Lya
  coupling transition at this redshift. They also assume linear
  fluctuations, and that the IGM gas cooled adiabatically down to this
  redshift. Each set of solid curves includes, from bottom to top at
  $k=0.1$ Mpc$^{-1}$, stellar radiation emitted up to Ly$\beta$,
  Ly$\delta$, or full Lyman-band emission, all assuming Pop III stars.
  Note that the results shown here from the first such prediction
  \cite{BL05b} were later updated (Figure~\ref{f:warp}).}
\label{f:zCut21cm}
\end{figure}

As previously discussed, the idea of unusually large fluctuations in
the abundance of early galaxies (\S~\ref{s:unusually}) first made a
major impact on studies of cosmic reionization (\S~\ref{s:reion}). The
same idea was also key in opening up cosmic dawn, prior to
reionization, to interferometric 21-cm observations, by launching the
study of fluctuations in the intensity of early cosmic radiation
fields. The fact that fluctuations in the galaxy number density cause
fluctuations even in the intensity of long-range radiation was first
shown, specifically for the \Lya radiation background, by Barkana \&
Loeb (2005) \cite{BL05b}. The spin temperature of hydrogen atoms in
the IGM is coupled to the gas temperature indirectly through the
Wouthuysen-Field effect \cite{Wout,Field}, which involves the
absorption of \Lya photons (\S~\ref{s:21physics}). While it had been
previously known \cite{Madau, Miralda} that this \Lya coupling likely
occurred in the IGM due to \Lya photons emitted by early stars at $z
\sim 20-30$, this radiation background had been assumed to be uniform.
This intuition was based on the fact that each atom sees \Lya
radiation from sources as far away as $\sim 300$ Mpc. However, it
turns out that relatively large, potentially observable, 21-cm
fluctuations are generated during the era of initial \Lya coupling,
for two reasons: fluctuations in the number density of the (highly
biased) early galaxies are significant even on scales of order
100~Mpc, and also a significant fraction of the \Lya flux received by
each atom comes from sources at smaller distances. Since relatively
few galaxies contribute most of the flux seen at any given point,
Poisson fluctuations can be significant as well, producing correlated
21-cm fluctuations (since a single galaxy contributes \Lya flux to
many surrounding points in the IGM).  If observed, the \Lya
fluctuation signal would not only constitute the first detection of
these early galaxies, but the shape and amplitude of the resulting
21-cm power spectrum would also probe their average properties
\cite{BL05b} (Figure~\ref{f:zCut21cm}).

\begin{figure}[tbp]
\includegraphics[width=\textwidth]{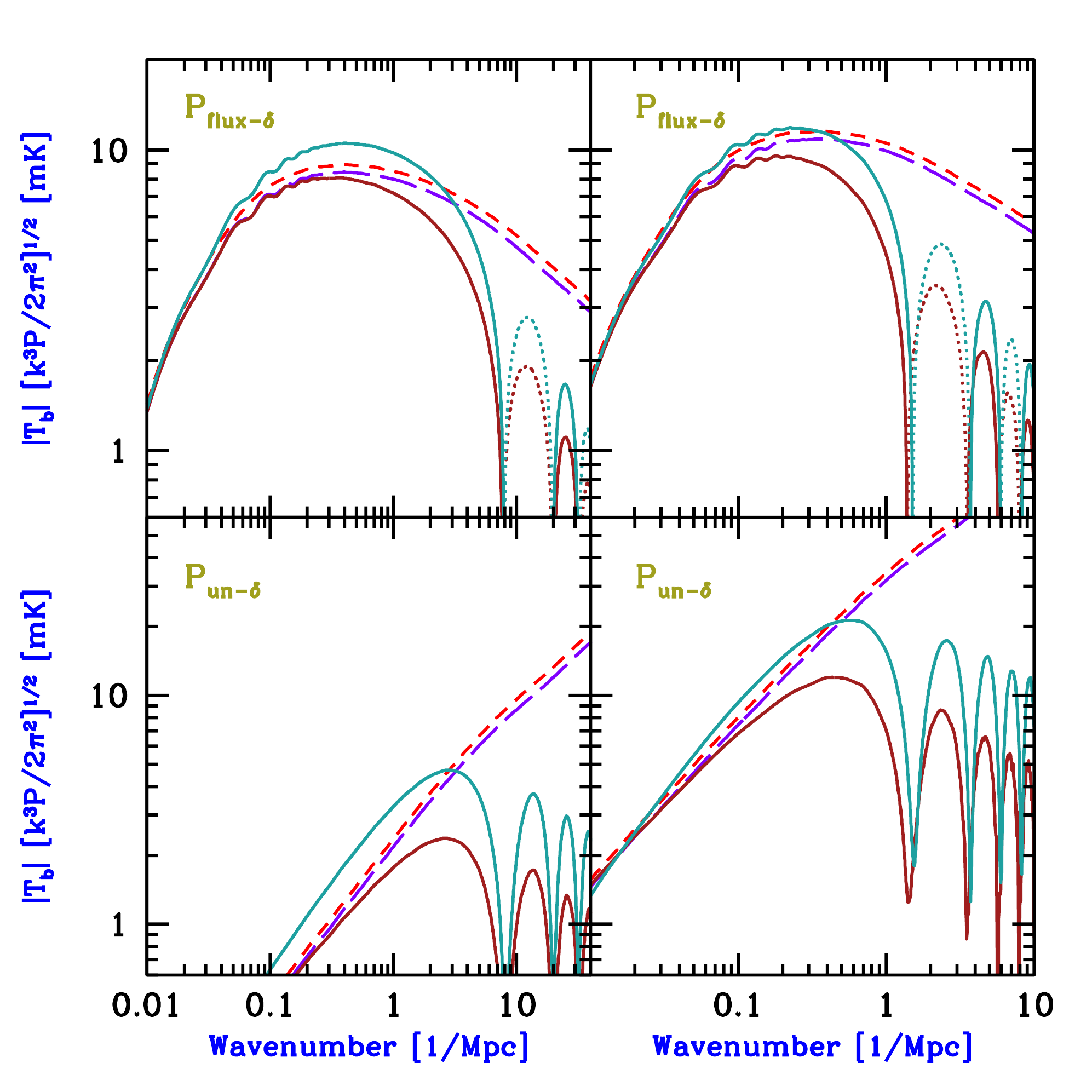}
\caption{21-cm power spectrum $P$ (in terms of the brightness
  temperature fluctuation at wavenumber $k$) as a function of $k$.
  Shown are two components of the power spectrum that in principle can
  be separated (in the limit of linear perturbations) based on the
  line-of-sight anisotropy of the 21-cm fluctuations \cite{BL05a}
  (\S~\ref{s:anisotropy}): $P_{\rm flux-\delta}$ (top panels) contains
  the contribution of density-induced \Lya fluctuations and $P_{\rm
    un-\delta}$ (bottom panels) is due to \Lya fluctuations from
  Poisson fluctuations in galaxy numbers. Compared here are the
  earlier result from \cite{BL05b} (including the correction from
  \cite{Hirata,Jonathan06a}) (short-dashed curves), the result
  corrected to use the precise density and temperature power spectra
  from \cite{NB05} (long-dashed curves), and from \cite{NB08} the same
  calculation with a cutoff due to individual H~II regions around
  galaxies (solid curves, the lower of each pair), and the full
  calculation (higher solid curve of each pair) which also includes
  the redistribution of photons due to scattering in the wing of the
  \Lya line. Two possible examples are shown for galactic halos, where
  their minimum circular velocity is assumed to be $V_c = 16.5$ km/s
  (left panels, corresponding to atomic cooling) or $V_c = 35.5$ km/s
  (right panels, an example of a case where internal feedback makes
  lower-mass halos inefficient at star formation). Negative portions
  are shown dotted in absolute value. Note that these results assume
  the simple case of a fixed H~II region size around all galaxies;
  more realistically, the small-scale ringing seen in this Figure may
  be smoothed out by a scatter in H~II region sizes, but the overall
  shape and the peak of each curve are more robust predictions. From
  \cite{NB08}.}
\label{f:warp}
\end{figure}

This discovery of \Lya fluctuations has led to a variety of follow-up
work, including more precise analyses of the atomic cascades of Lyman
series photons \cite{Hirata,Jonathan06a}. Also, a significant boost is
predicted in the 21-cm power spectrum from \Lya fluctuations due to
the repeated scattering of the photons from stars on their way to the
hydrogen atoms, out in the wing of the \Lya line
\cite{CZ07,Semelin,NB08} (Figure~\ref{f:warp}). The repeated
scatterings mean that the \Lya photons do not reach as far (in the
fixed time until they redshift into -- and then out of -- the line),
which decreases the overall large-scale smoothing and thus increases
the predicted level of 21-cm fluctuations.  Moreover, the increased
sensitivity to \Lya photons from short distances makes the overall
21-cm power spectrum sensitive to the sizes of H~II regions at this
very early stage in reionization (Figure~\ref{f:warp}). Note that in
addition to direct stellar emission, \Lya photons are also produced in
the IGM from X-ray ionization; however, despite early overestimates
\cite{Jonathan07}, the contribution of these \Lya photons in typical
models is $\sim 1\%$ compared to stellar \Lya photons \cite{Complete}.

\begin{figure}[tbp]
\includegraphics[width=\textwidth]{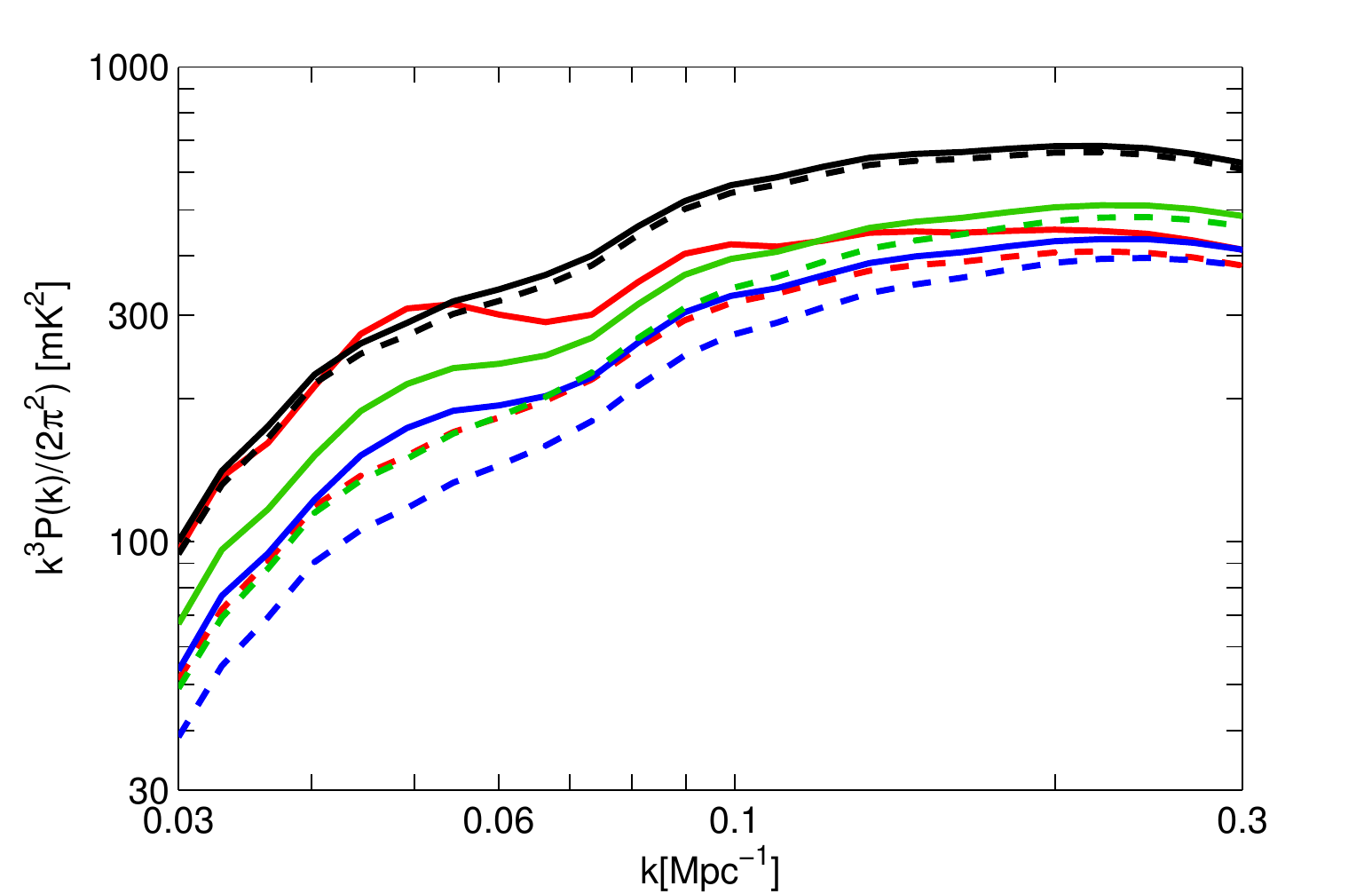}
\caption{Power spectra of the 21-cm brightness temperature for various
  strengths of LW feedback: no feedback (red), relatively weak
  feedback (blue), relatively strong feedback (green) and saturated
  feedback (i.e., no $H_2$ molecules; black); each case is shown
  either with (solid) or without (dashed) the streaming velocity. The
  weak and strong feedback cases roughly bracket current
  uncertainties, although recent simulations \cite{VisbalLW} give some
  support to the strong case. The results in each case are shown at a
  time of particularly strong heating fluctuations, a $\Delta z= 3$
  earlier (i.e., higher $z$) than the moment when the cosmic mean
  21-cm temperature is zero. The streaming velocity increases and
  flattens the large-scale power spectrum, and strengthens the BAOs
  (e.g., at the large-scale peak at $k=0.05$~Mpc$^{-1}$); this effect
  (which is wiped out in the limiting case of saturated feedback) is
  partially suppressed by the LW feedback. This Figure from
  \cite{FialkovLW} assumed the case of early cosmic heating by a soft
  X-ray spectrum (\S~\ref{s:2012}); in the more likely case of late
  heating by a hard X-ray spectrum (\S~\ref{s:late}), the combined
  effect of LW feedback and the streaming velocity would be more
  difficult to observe with heating fluctuations, but would still be
  observable during the somewhat earlier era of \Lya fluctuations.}
\label{f:FialkovLW}
\end{figure}

As discussed in \S~\ref{s:Lya}, LW feedback is an important feedback
effect on early galaxies, as it dissociates molecular hydrogen and
eventually ends star formation driven by molecular cooling
\cite{haiman}. Thus, it affects 21-cm fluctuations indirectly by
changing the amount and distribution of star formation
\cite{Holzbauer:2011}. The effect becomes particularly striking once
the baryon - dark matter streaming velocity (\S~\ref{s:stream}) is
included. Assuming that star formation is dominated by $10^6 M_\odot$
halos at very high redshift, the streaming velocity strongly affects
them and produces a distinctive BAO signature in the 21-cm
fluctuations (\S~\ref{s:2012}). Since LW feedback affects star
formation in precisely the same halos that are affected by the
streaming velocity, the effectiveness of the feedback has a major
effect on 21-cm observations \cite{FialkovLW}
(Figure~\ref{f:FialkovLW}). This is particularly important since there
is a substantial uncertainty in the strength of LW feedback on early
star formation (although this subject has been explored somewhat with
numerical simulations: \S~\ref{s:Lya}); thus, the prospect that 21-cm
observations over a range of redshifts will detect the time evolution
of the LW feedback is quite interesting.

\subsection{Large 21-cm fluctuations from early cosmic heating}

\label{s:2012}

\begin{figure}[tbp]
\includegraphics[width=\textwidth]{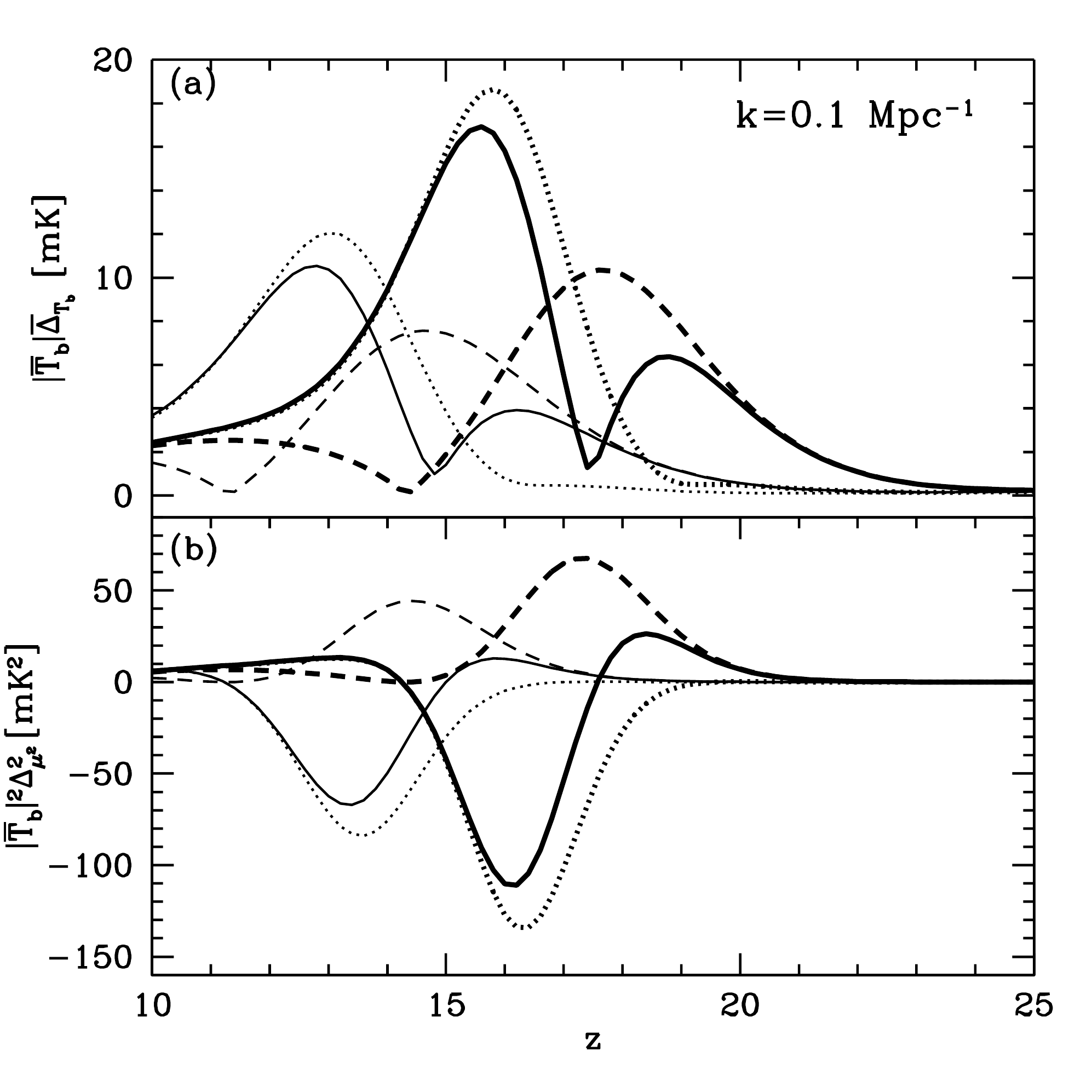}
\caption{Power spectra of 21-cm brightness fluctuations from
  temperature fluctuations during cosmic X-ray heating
  \cite{Jonathan07}. Shown are the total isotropically-averaged
  fluctuation (top panel) and the main anisotropic term $P_{\mu^2}$
  (bottom panel) from the line-of-sight anisotropy of the 21-cm
  fluctuations \cite{BL05a} (\S~\ref{s:anisotropy}). These quantities
  are shown at a wavenumber $k=0.1$~Mpc$^{-1}$, including the effects
  of heating fluctuations only (dotted curves), \Lya fluctuations only
  (dashed curves), or both (solid curves). Two models are shown, one
  corresponding to Pop II stars (thick curves) and the other to Pop
  III stars (thin curves). Note that this Figure from
  \cite{Jonathan07} assumed linear fluctuations, early heating by a
  soft spectrum of X-ray sources, and did not include the boost in the
  \Lya fluctuations by a factor of $\sim 1.5$ (Figure~\ref{f:warp})
  from multiple scattering.}
\label{f:Jonathan07}
\end{figure}

As discussed in detail in \S~\ref{s:heating}, until recently it was
expected that the universe had been pre-heated well before cosmic
reionization.  This early heating was thought to be likely due to the
high heating efficiency of the soft X-ray spectrum that had been
assumed in calculations of cosmic heating. Soft X-rays are absorbed in
the neutral IGM over relatively short distances, making heating a
local phenomenon that can potentially give rise to large temperature
fluctuations in the early IGM. Indeed, when combined with the idea of
unusually large fluctuations in the abundance of early galaxies
(\S~\ref{s:unusually}), the expectation of large-scale fluctuations in
ionization (\S~\ref{s:reion}) and in the \Lya radiation background
(\S~\ref{s:Lya21}) can be extended to the X-ray background. The first
calculation of heating due to an inhomogeneous X-ray background, by
Pritchard \& Furlanetto (2007) \cite{Jonathan07}, applied to X-rays a
similar method as in the \Lya case \cite{BL05b}; integrating the
heating over time to find the distribution of gas temperatures, the
result was the prediction of another era of detectably large 21-cm
fluctuations (Figure~\ref{f:Jonathan07}).

As discussed in \S~\ref{s:numerical}, while numerical simulations are
the best, most accurate method for studying early galaxy formation on
small scales, they are unable to simultaneously cover large volumes.
Simulations that successfully resolve the tiny mini-galaxies that
dominated star formation at early times are limited to $\sim 1$~Mpc
volumes, and cannot explore the large cosmological scales that might
be accessible to 21-cm observations (which are currently limited to
low resolution). On the other hand, analytical calculations are
limited to linear (plus sometimes weakly non-linear) scales, and thus
cannot directly probe the non-linear astrophysics of halo and star
formation. Even if the results of simulations are incorporated within
them, analytical approaches assume small fluctuations and linear bias
(see the end of section~\ref{sec:NL}), assumptions that break down in
the current context, where the stellar density varies by orders of
magnitude on scales of a few Mpc. Even on 100~Mpc scales, fluctuations
in the gas temperature are as large as order unity (see below). Thus,
linear, analytical calculations can only yield rough estimates, even
for large-scale fluctuations.

\begin{figure}[tbp]
\includegraphics[width=\textwidth]{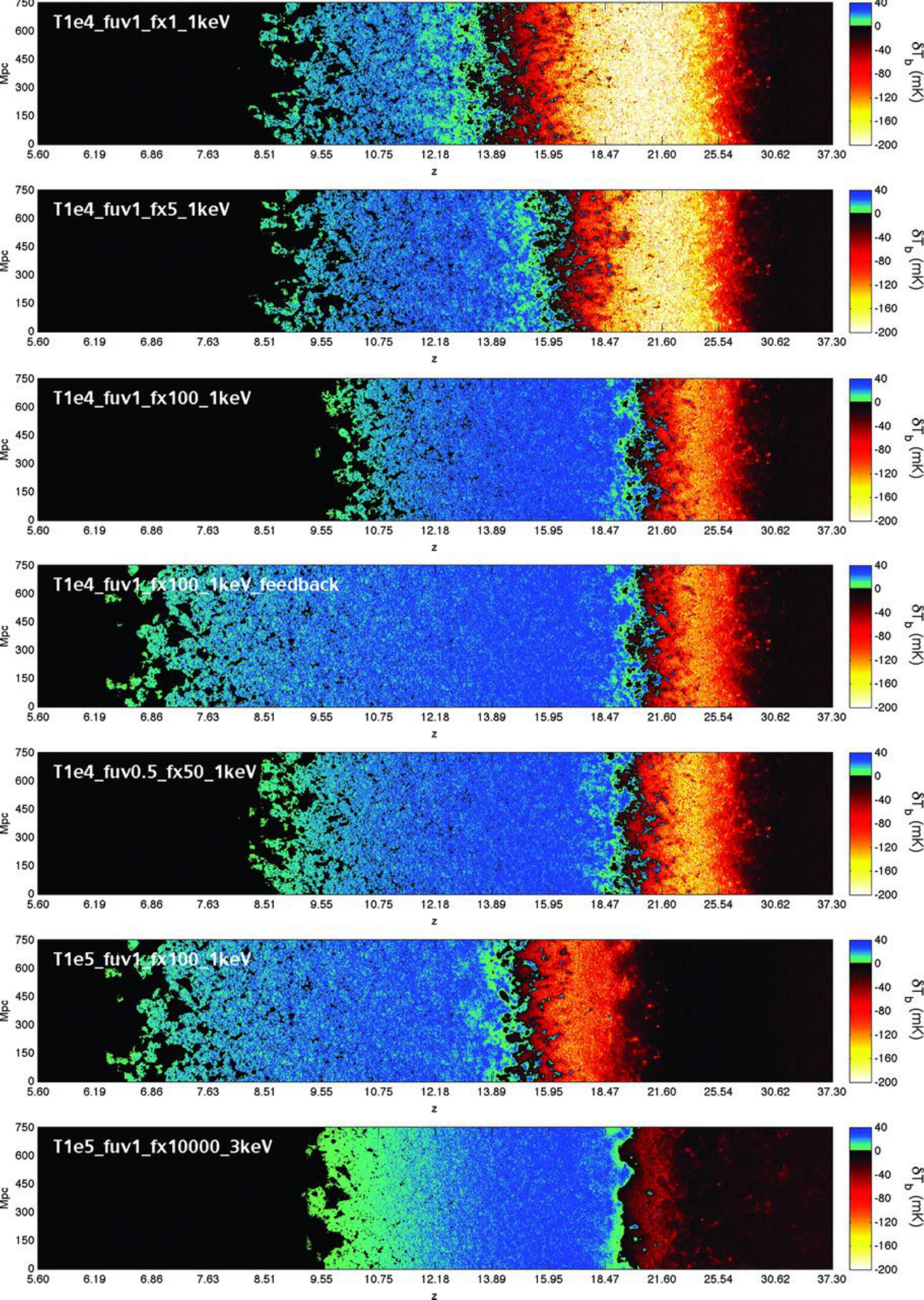}
\caption{Slices through the predicted 21 cm signal for models
  calculated with the semi-numerical code 21CMFAST. The slices show
  the evolution of the 21-cm brightness temperature with distance
  along the light cone, with the redshift indicated on the $x$-axis,
  and the $y$-axis showing spatial structure at each redshift. In the
  model name, T indicates the minimum assumed virial temperature of
  galactic halos ($10^4$~K corresponds to atomic cooling), fuv and fx
  parameterize the ionizing and X-ray efficiency, respectively, and
  the final number indicates the mean X-ray photon energy (``1~keV''
  indicates a soft power-law starting at 0.3~keV, with a mean photon
  energy of 0.9~keV; these quantities are 3 times larger for the
  ``3~keV'' case). These models assume heating via soft X-rays,
  except for the extreme (bottom-most) model in which very intense
  X-rays dominate reionization (not just heating). From
  \cite{Mesinger13}.}
\label{f:21cmfast}
\end{figure}

As a result of these considerations, perhaps the best current method
to generate observable 21-cm predictions from the era of early
galaxies is with a hybrid, semi-numerical code that combines linear
theory and full calculations on large scales with analytical models
and the results of numerical simulations on small scales. Such methods
have been compared with numerical simulations of reionization
\cite{zahn,santos}, and have also been used to predict the effect of
the streaming velocity on high-redshift galaxy formation
\cite{TH10,Dalal}.  Figure~\ref{f:21cmfast} shows a prediction of the
21-cm signatures of X-ray heating made with the semi-numerical code
21CMFAST \cite{Mesinger13}. The light-cone slices show the progression
through cosmic 21-cm history: collisional decoupling during the dark
ages (black, far-right region), \Lya coupling (black to yellow
transition), X-ray heating (yellow to blue), and reionization (blue to
black).

In the case of soft X-ray heating sources, heating fluctuations are
the largest, most promising source of pre-reionization 21-cm
fluctuations, but even in this case there remains a large uncertainty
in predicting the signal. The redshift at which this signal peaks
depends on the overall efficiency of X-ray production, with higher
efficiency leading to an earlier cosmic heating era. This uncertainty
is not too problematic since planned observations will cover a wide
redshift range and find the signal if it is there. Given the correct
redshift, the strength of the signal still depends on the typical mass
of the galactic halos that hosted these sources. The more massive the
halos, the more highly biased (clustered) they are expected to have
been, thus producing a larger 21-cm fluctuation signal. However, the
baryon - dark matter streaming velocity (\S~\ref{s:stream}) greatly
cuts down this uncertainty, as it boosts the expected signal from
low-mass halos nearly to the same level as that from high-mass halos.
Observational predictions that include the streaming velocity were
achieved with a semi-numerical method \cite{nature}.

This approach built upon previous semi-numerical methods used for
high-redshift galaxy formation \cite{TH10,Dalal,21cmfast}. It used the
known statistical properties of the initial density and velocity
perturbations to generate a realistic sample universe on large, linear
scales. This was followed by a calculation of the stellar content of
each pixel on the grid using a model \cite{anastasia} previously
developed to describe the streaming velocity effect on galaxy
formation; this includes analytical models as well as fits to the
results of small-scale numerical simulations. Like other
semi-numerical codes, it assumed standard initial perturbations (e.g.,
from a period of inflation), where the density and velocity components
are Gaussian random fields.

\begin{figure}[tbp]
\includegraphics[width=0.498\textwidth]{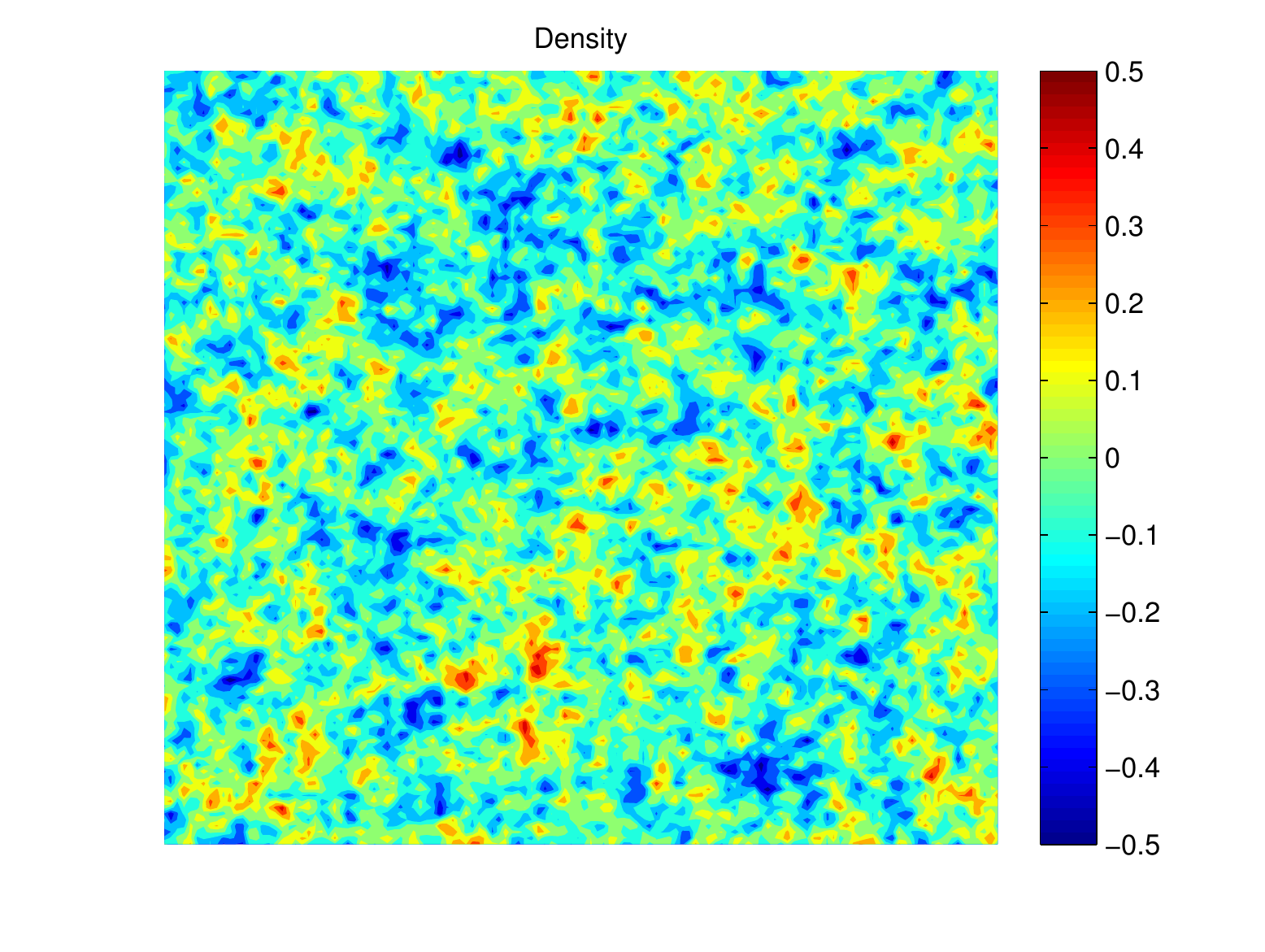}
\includegraphics[width=0.498\textwidth]{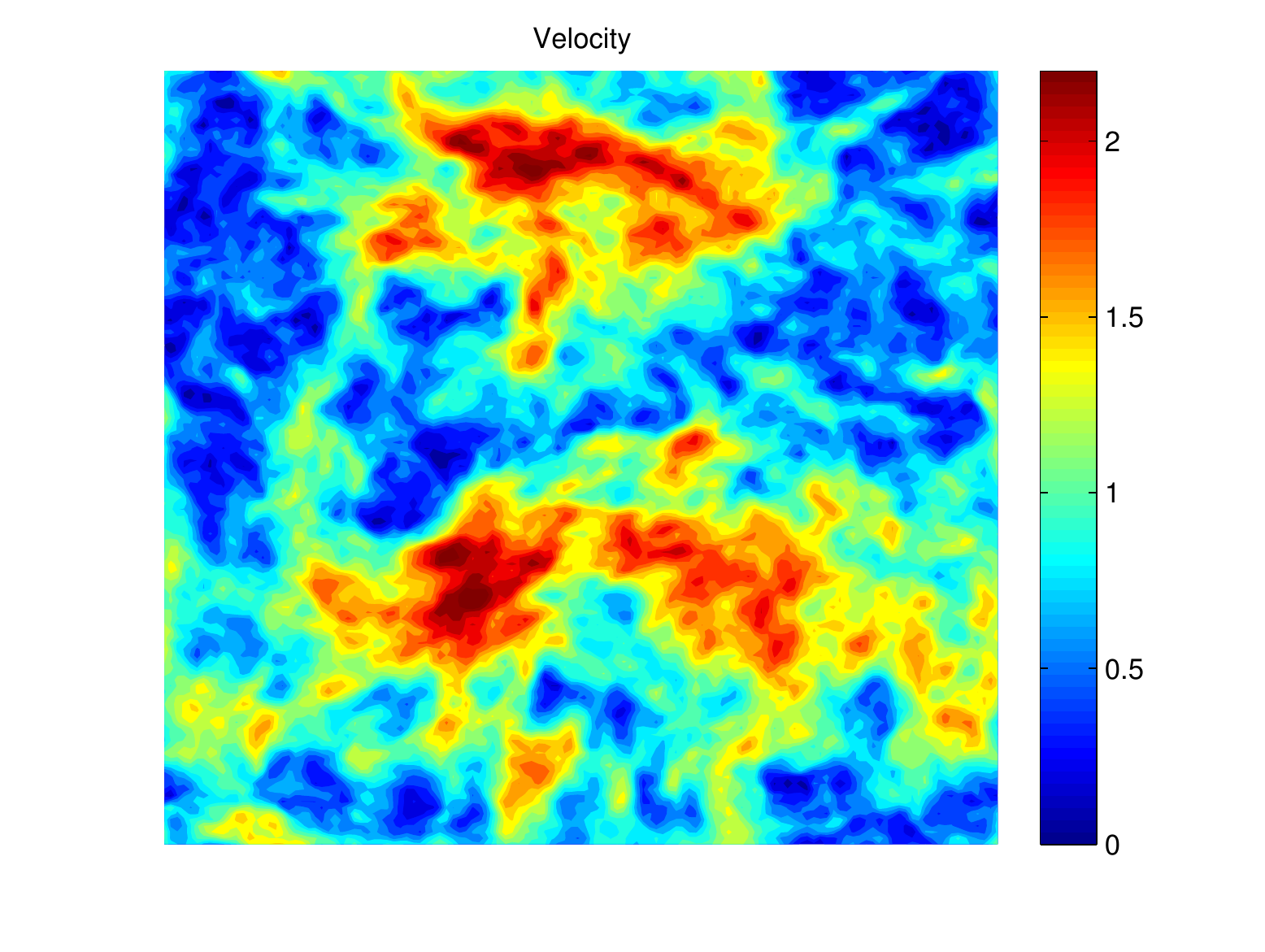}
\includegraphics[width=0.498\textwidth]{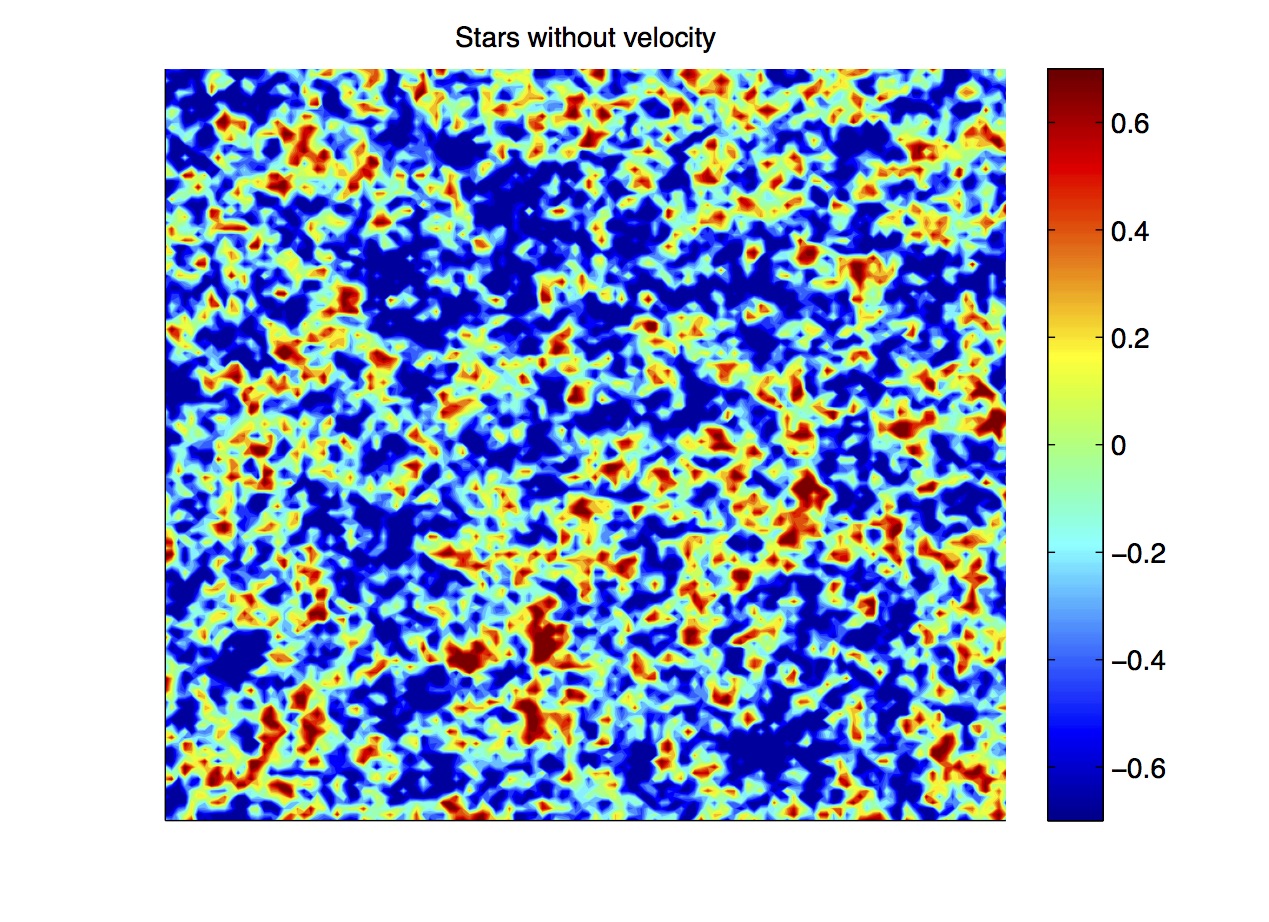}
\includegraphics[width=0.498\textwidth]{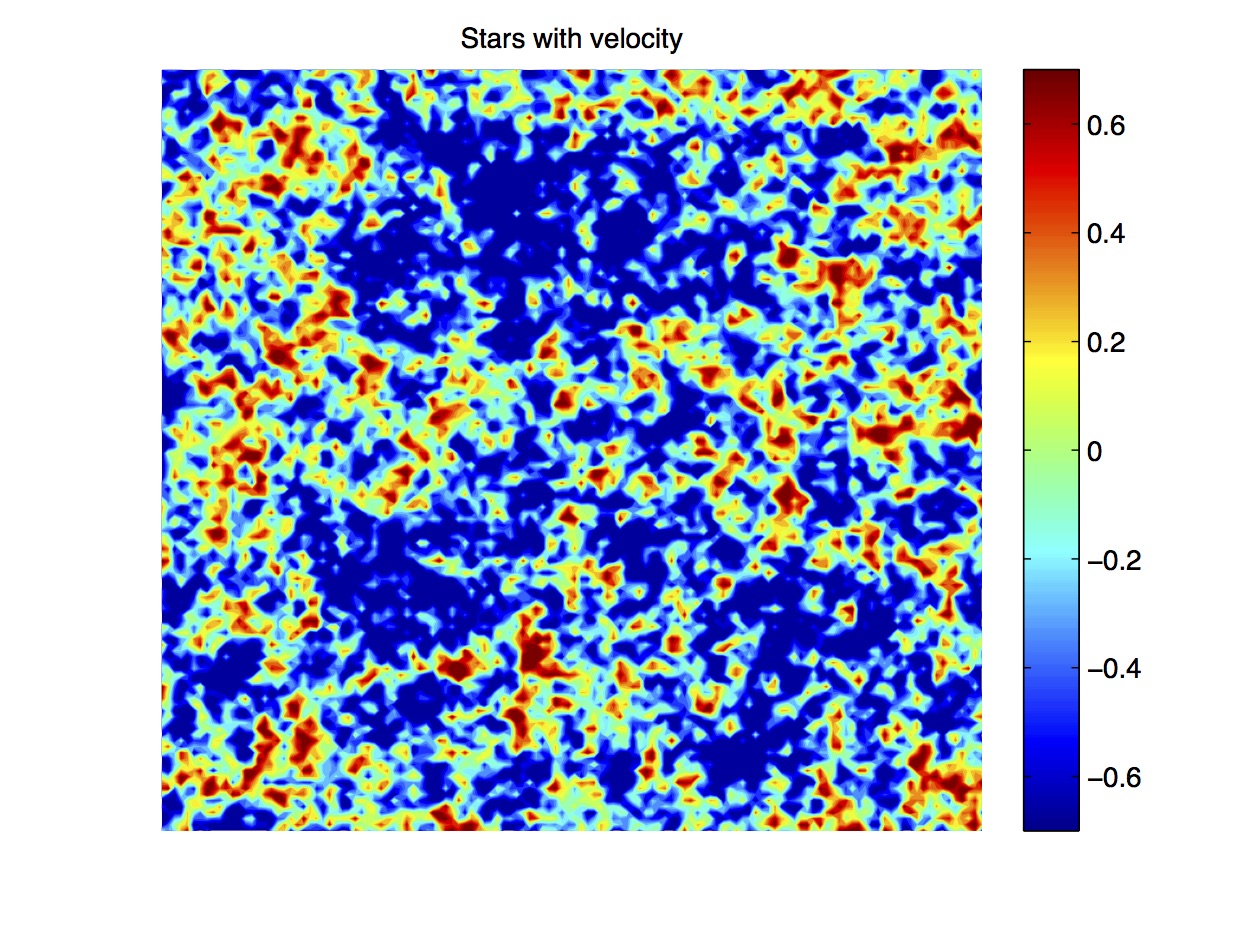}
\caption{Effect of the streaming velocity on the number density of
  stars at redshift 20. The large-scale density (top left panel) and
  velocity (top right panel) fields are shown on the top. For the
  density field, the fractional perturbation relative to the mean is
  shown, at $z=20$; for the velocity field, the magnitude of the
  relative motion in units of the root-mean-square value is shown (the
  map is independent of redshift in these relative units). For the
  same slice, the traditional calculation (lower left panel), which
  includes the effect of density only, is compared to the new
  prediction (lower right panel), which includes the effect of the
  same density field plus that of the streaming velocity. The colors
  in the bottom panels correspond to the logarithm (base 10) of the
  gas fraction in units of its cosmic mean value in each case. The
  panels all show a 3~Mpc thick slice (the pixel size of the grid in
  the semi-numerical code) from a simulated volume 384 Mpc on a side
  (based on \cite{nature}, but taken from a different box from the one
  shown in the Figures in \cite{nature}, i.e., for a different set of
  random initial conditions).}
\label{f:RhoV}
\end{figure}

Velocities are coherent on larger scales than the density, due to the
extra factor of $1/k$ in the velocity from the continuity equation
that relates the two fields. This is clearly apparent in the example
shown in Figure~\ref{f:RhoV} of a thin slice of a simulated volume.
The density field fluctuates on relatively small scales, while the
velocity field shows a larger-scale cosmic web, with coherent
structure on scales of order 100~Mpc. This means that the largest
scales will be dominated by the pattern due to the velocity effect, as
long as the streaming velocity significantly affects star formation.

\begin{figure}[tbp]
\includegraphics[width=0.498\textwidth]{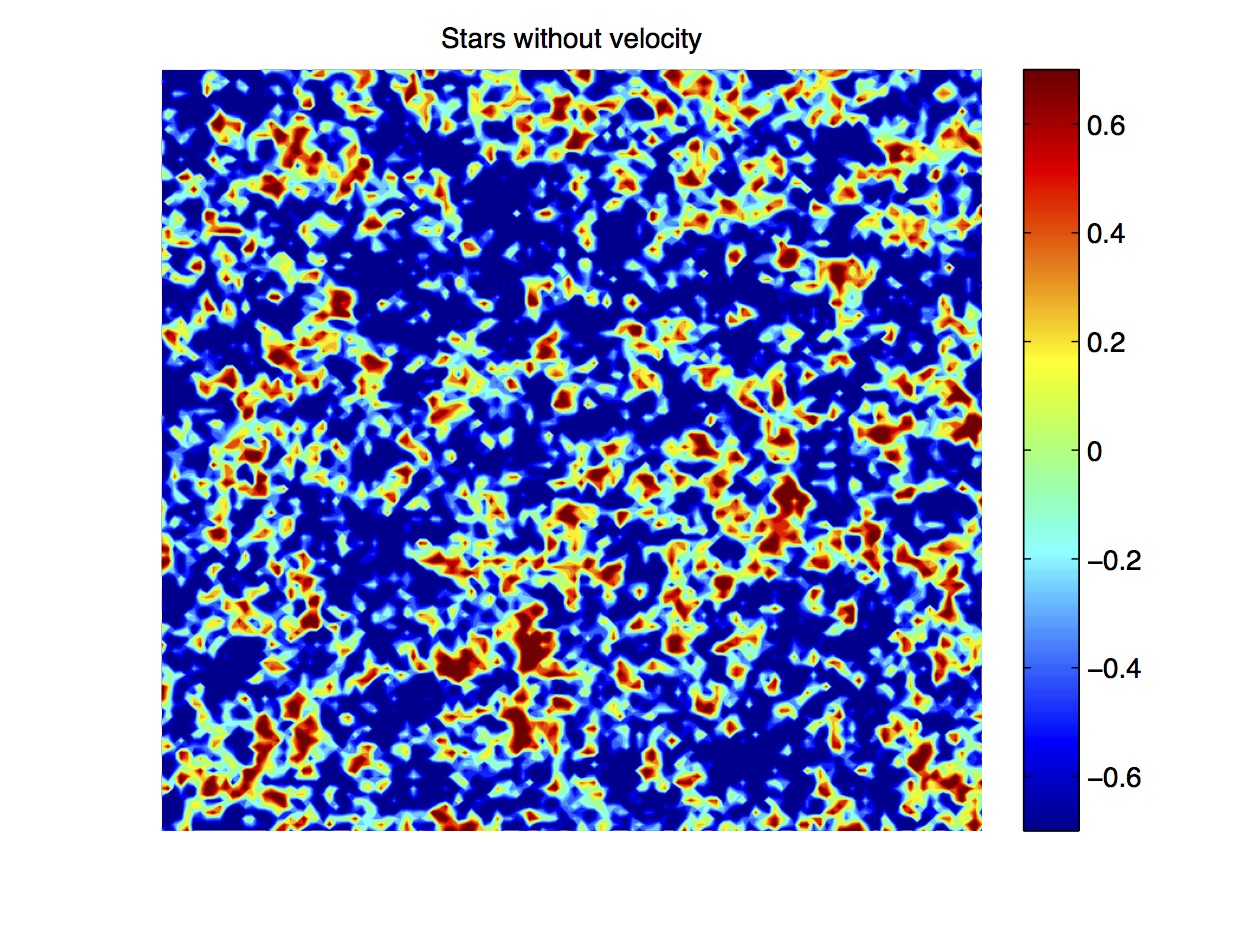}
\includegraphics[width=0.498\textwidth]{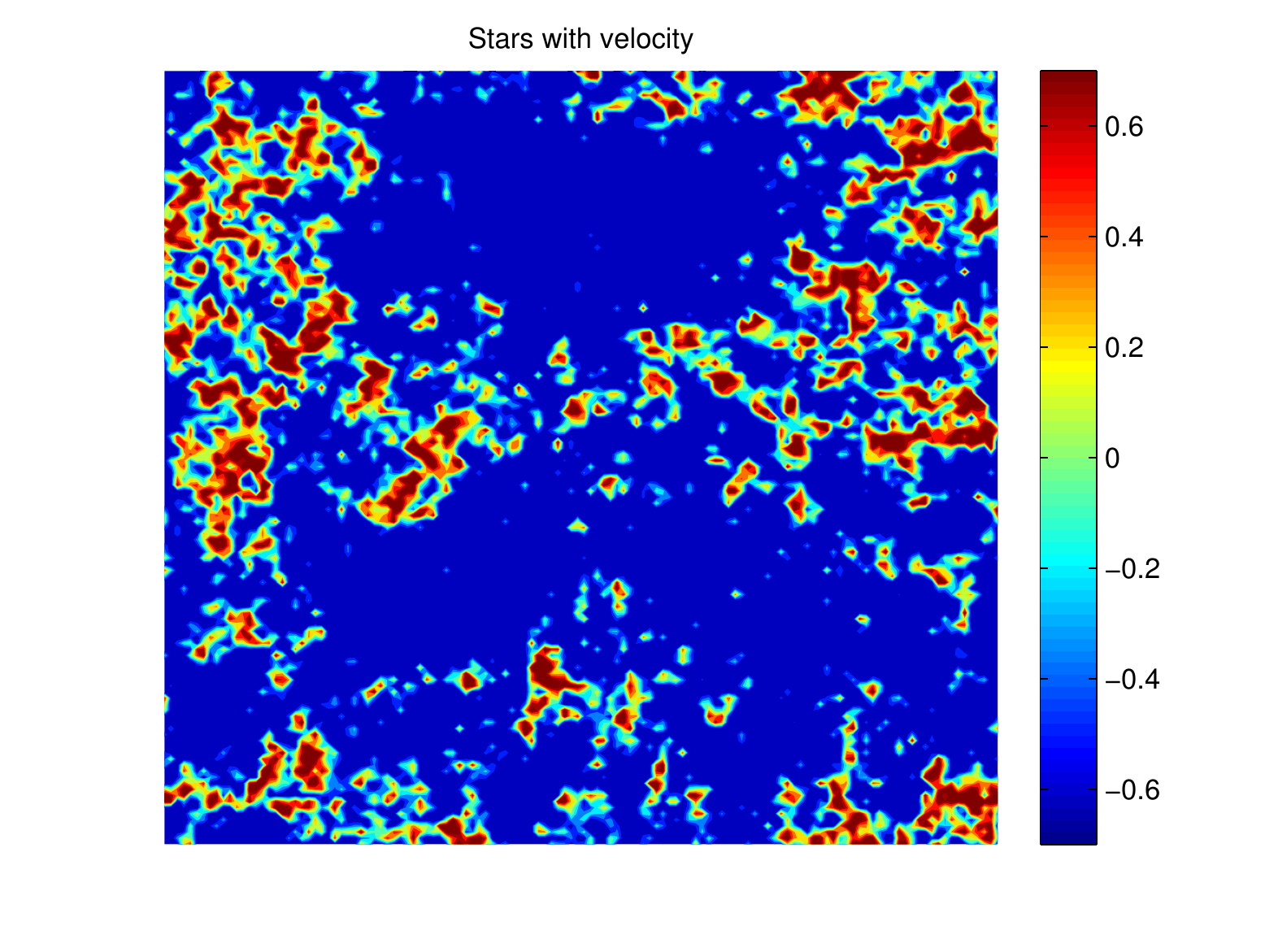}
\caption{Effect of the streaming velocity on the number density of
  stars at redshift 40. For the same slice as in Figure~\ref{f:RhoV},
  we compare the traditional calculation (left panel), which includes
  the effect of density only, to the new prediction (right panel),
  which includes the effect of the same density field plus that of the
  streaming velocity.  The colors correspond to the logarithm of the
  gas fraction in units of its cosmic mean value in each case. The
  color scale spans the same range as in Figure~\ref{f:RhoV} for easy
  comparison.}
\label{f:fgas40}
\end{figure}

The resulting distribution of stellar density at $z=20$ is also shown
in Figure~\ref{f:RhoV}. Note the large biasing (i.e., amplification of
fluctuations) of the stars: density fluctuations ranging up to $\pm
50\%$ yield (without including the streaming velocity) a field of
stellar density that varies by over a factor of 20 (when both fields
are smoothed on a 3~Mpc scale). The velocity effect produces a more
prominent cosmic web on large scales, marked by large coherent regions
that have a low density of stars, separated by ribbons or filaments of
high star formation. The effect is much more striking at higher
redshifts (Figure~\ref{f:fgas40}), and it thus substantially alters
the feedback environment of the very first generations of stars. The
various types of radiation that produce feedback spread out to a
considerable distance from each source, but this distance is typically
not as large as the span of the velocity-induced features.  This means
that regions of low velocity (and thus high star formation) experience
radiative feedback substantially earlier than regions of high velocity
(low star formation). Thus, the substantial effect of the velocities
on early star formation makes early feedback much more inhomogeneous
than previously thought.

Observationally, these degree-scale fluctuations affect various cosmic
radiation backgrounds, and in particular the history of 21-cm emission
and absorption. As noted above, in the presence of soft X-ray heating
sources, the heating fluctuations produce the largest pre-reionization
21-cm fluctuations, typically from sometime after the \Lya coupling
has mostly saturated. As for the LW flux, here we consider the case of
negligible LW feedback (as was assumed in Figures~\ref{f:RhoV} and
\ref{f:fgas40}), but below we bracket the effect of the LW flux by
also considering the opposite limiting case where the LW transition
has already saturated (i.e., completely destroyed hydrogen molecules);
the effect of various strengths of LW feedback was discussed in more
detail in \S~\ref{s:Lya21}.

\begin{figure}[tbp]
\includegraphics[width=0.498\textwidth]{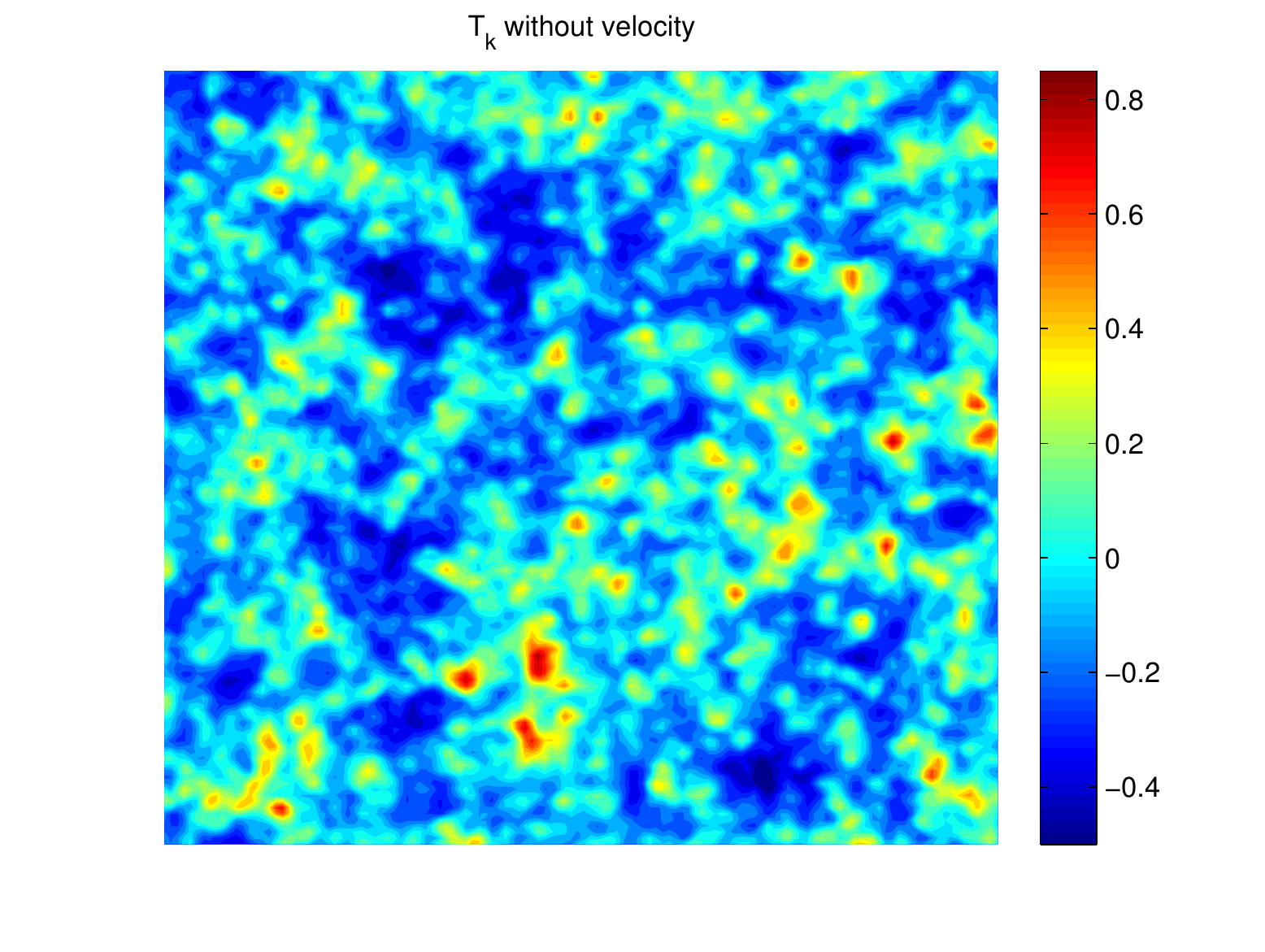}
\includegraphics[width=0.498\textwidth]{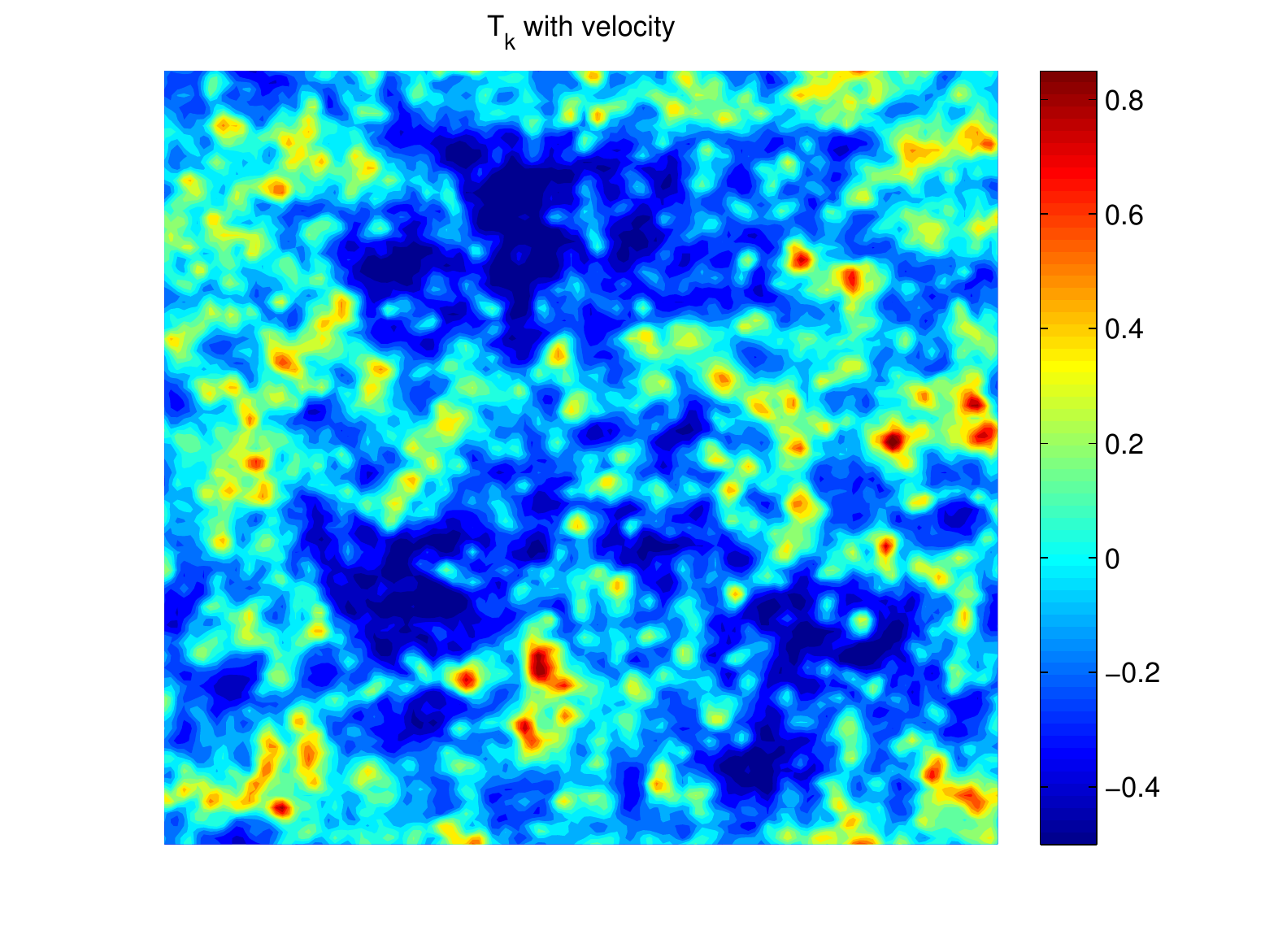}
\includegraphics[width=0.498\textwidth]{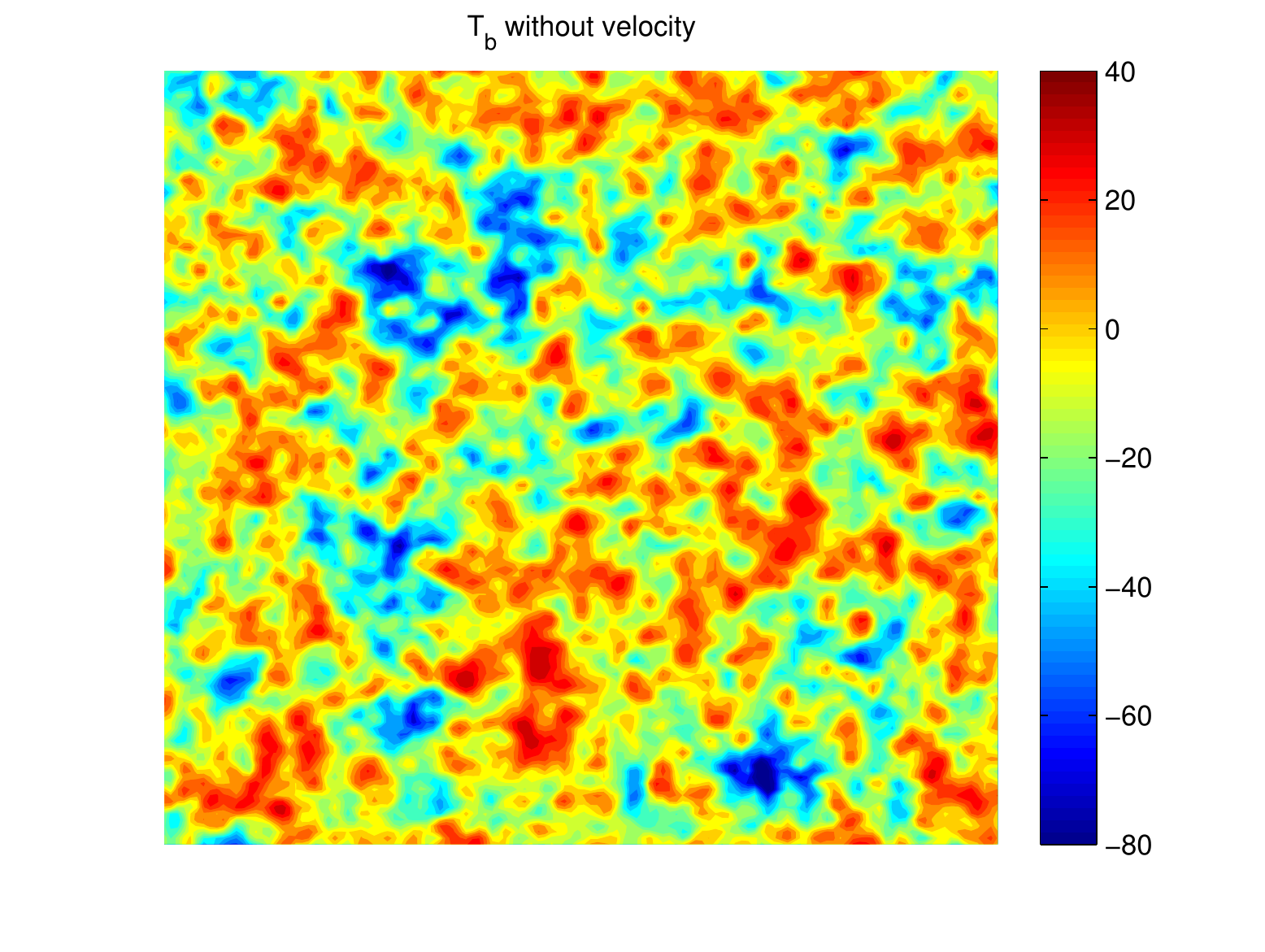}
\includegraphics[width=0.498\textwidth]{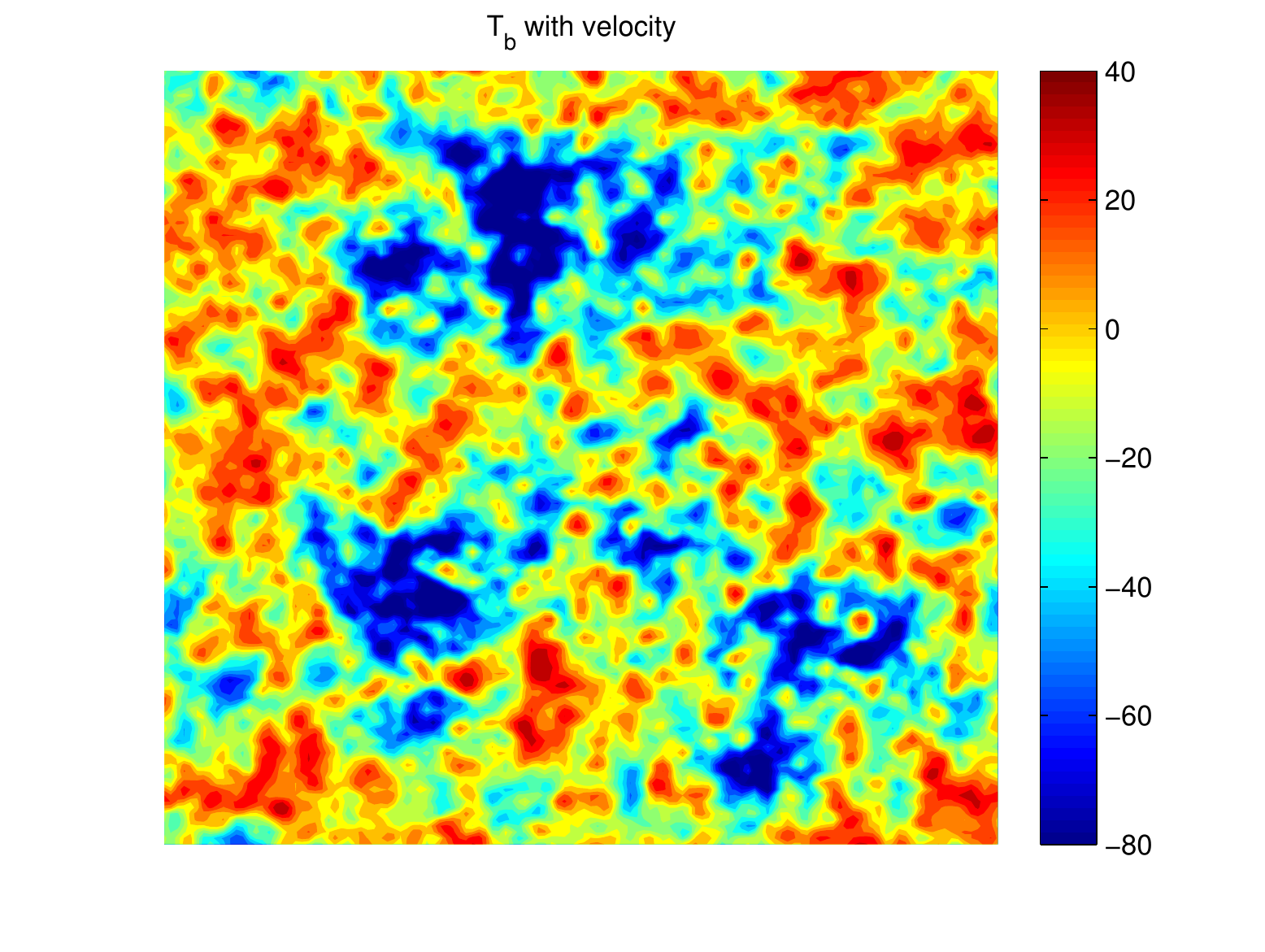}
\caption{Effect of the streaming velocity on the gas temperature $T_k$
  and on the 21-cm brightness temperature $T_{\rm b}$ at redshift 20.
  For the same slice as in Figure~\ref{f:RhoV}, we compare the
  traditional calculations (left-hand panels), which include the
  effect of density only, to the new predictions (right-hand panels),
  which include the effect of density and streaming velocity. The
  colors in the top panels correspond to the logarithm of the gas
  (kinetic) temperature in units of the CMB temperature at $z=20$. The
  colors in the bottom panels correspond to the 21-cm brightness
  temperature in millikelvin units. Note that the observed wavelength
  of this 21-cm radiation is redshifted by the expansion of the
  universe to 4.4 meters (corresponding to a frequency of 68 MHz).}
\label{f:Tk20}
\end{figure}

Figure~\ref{f:Tk20} shows the gas temperature distribution at $z=20$,
assumed to be at the heating transition, i.e., when the mean H~I gas
temperature was equal to that of the CMB. Regions where the gas moved
rapidly with respect to the dark matter (dark red regions, top right
panel of Figure~\ref{f:RhoV}) produced fewer stars (dark blue regions,
bottom right panel of Figure~\ref{f:RhoV}) and thus a lower X-ray
intensity, leaving large regions with gas that is still colder than
the CMB by a factor of several (dark blue regions, top right panel of
Figure~\ref{f:Tk20}). The spatial reach of X-rays results in a gas
temperature distribution that is smoother than the distribution of
stars, and this brings out the effect of large-scale fluctuations and
thus highlights the contrast between the effect of density and
velocity fluctuations.

During the heating transition (\S~\ref{s:heating}), the 21-cm
brightness temperature (shown in the bottom panels of
Figure~\ref{f:Tk20}) mainly measures the gas (kinetic) temperature
$T_K$, although it is also proportional to the gas density (and to the
square root of $1+z$). The form of the dependence, $T_{\rm b} \propto
1 - T_{\rm CMB}/T_K$, makes the 21-cm intensity more sensitive to cold
gas than to hot gas (relative to the CMB temperature). Thus, the large
voids in star formation produced by a high streaming velocity lead to
prominent 21-cm absorption (dark blue regions, bottom right panel of
Figure~\ref{f:Tk20}) seen on top of the pattern from the effect of
density fluctuations. These deep 21-cm cold spots are a major
observable signature of the effect of the streaming velocity on early
galaxies.

\begin{figure}[tbp]
\includegraphics[width=\textwidth]{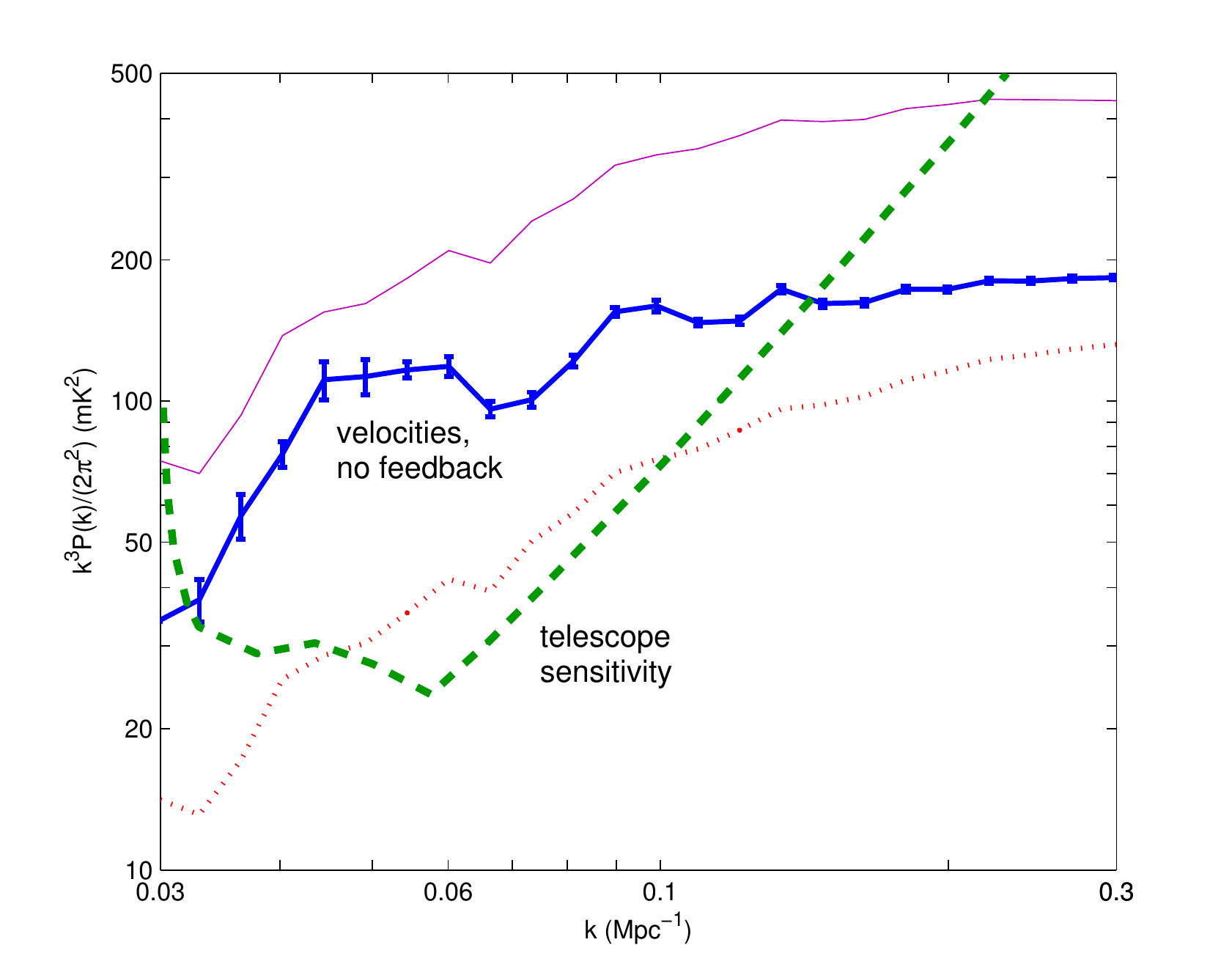}
\caption{Signature of the streaming velocity in the 21-cm power
  spectrum, at the peak of the X-ray heating transition. The
  prediction is shown including the streaming velocity effect (blue
  solid curve) or with the effect of density only (red dotted curve),
  both for the case of a late LW transition for which the LW feedback
  is still negligible at the heating transition. These predictions are
  compared to the projected 1-$\sigma$ telescope sensitivity (green
  dashed curve) based on a 1000-hour observation with an instrument
  like the Murchison Wide-field Array or the Low Frequency Array but
  designed to operate at 50--100 MHz \cite{McQuinn}, including an
  estimated degradation factor due to foreground removal \cite{Liu};
  this sensitivity is defined as the signal that would yield a
  measurement with a signal-to-noise ratio of unity in each $k$ bin of
  size $\Delta k = 0.5 k$ averaged over an 8~MHz frequency band (where
  only thermal noise is included). Future experiments like the Square
  Kilometer Array should reach a better sensitivity by more than an
  order of magnitude \cite{McQuinn}. To allow for the possibility of
  feedback, the prediction is also shown in the opposite limit of
  maximum feedback, i.e., an early LW transition that has already
  saturated (purple solid curve). In this plot, the heating transition
  has been fixed at $z=20$ for easy comparison among the various
  cases. Error bars on the main prediction curve indicate the
  $1-\sigma$ sample variance in one simulation box. From
  \cite{nature}.}
\label{f:nPS}
\end{figure}

While Figure~\ref{f:Tk20} illustrates the detailed pattern that the
streaming velocity imprints on the 21-cm intensity distribution,
upcoming experiments are expected to yield noisy maps that likely must
be analyzed statistically. Figure~\ref{f:nPS} shows the predicted
effect on the power spectrum of the fluctuations in 21-cm intensity
\cite{nature}. The velocities enhance large-scale fluctuations (blue
solid curve compared with red dotted), leading to a flatter power
spectrum with prominent baryon acoustic oscillations (reflecting the
BAO signature in Figure~\ref{f:vbc}). The signal is potentially
observable with a redshift 20 version of current instruments (green
dashed curve). If there is complete LW feedback (solid purple curve),
then the small galaxies that rely on molecular-hydrogen cooling are
unable to form; the larger galaxies that dominate in that case are
almost unaffected by the streaming velocity, so the 21-cm power
spectrum reverts to the density-dominated shape (compare the solid
purple and red dotted curves), but it becomes even higher since more
massive galactic halos are even more strongly biased.

Thus, regardless of the strength of the LW feedback (or other negative
feedback effects on small galaxies), the 21-cm power spectrum at the
peak of the heating transition should feature large fluctuations on
observable scales. Beyond just detection of the signal, only a mild
additional accuracy is necessary in order to determine whether
feedback has suppressed star formation in the smallest halos.  If it
has not, then the velocity effect produces strong BAOs on top of a
flattened power spectrum, in particular raising it by a factor of 4 on
large scales ($k=0.05$ Mpc$^{-1}$, wavelength 130~Mpc, observed angle
2/3 of a degree) where the experimental sensitivity is optimal.  If
this characteristic shape is observed it would confirm that million
mass halos dominated galaxy formation at this early epoch.

While Figure~\ref{f:nPS} considers a single redshift, similar
observations over the full $\Delta z \sim 6$ redshift range of
significant heating fluctuations could actually detect the slow
advance of the LW feedback process, during which the power spectrum is
predicted to continuously change shape, gradually steepening as the
BAO signature weakens towards low redshift (see
Figure~\ref{f:FialkovLW} in \S~\ref{s:Lya21}). This is all the case if
the Universe was heated by soft X-rays. If it was heated by hard
X-rays (see the next subsection), then the heating peak is largely
erased, but similar effects of the streaming velocity are expected on
the 21-cm signal during the $z \sim 25$ fluctuation peak from the
Lyman-$\alpha$ coupling transition (\S~\ref{s:Lya21}).

\subsection{Late heating and reionization}

\label{s:late}

As discussed in \S~\ref{s:heating}, it was recently realized that the
hard X-ray spectrum characteristic of X-ray binaries, the most
plausible source of early cosmic heating, is predicted to have
produced a relatively late heating, possibly encroaching on the
reionization era. The effect of this on the global 21-cm signal is
discussed in \S~\ref{s:global}. Here we discuss the key consequences for
21-cm fluctuations.

A major effect of X-ray heating by a hard spectrum is the suppression
of 21-cm fluctuations due to heating. Under the previously assumed
soft spectra, the short typical distance traveled by the X-ray photons
was found to produce large fluctuations in the gas temperature and
thus in the 21-cm intensity around the time of the heating transition,
regardless of when this transition occurred
\cite{Jonathan07,CLoeb,nature} (\S~\ref{s:2012}). However, the larger
source distances associated with a hard spectrum lead to a much more
uniform heating, with correspondingly low temperature fluctuations
even around the time of the heating transition, when the 21-cm
intensity is quite sensitive to the gas temperature. This trend is
strengthened by late heating, as it occurs at a time when the heating
sources are no longer as rare and strongly biased as they would be in
the case of an earlier heating era. Thus, heating with a hard X-ray
spectrum is predicted to produce a new signature in the 21-cm
fluctuation signal: a deep {\it minimum}\/ during reionization
\cite{Cold}. This results from the low level of gas temperature
fluctuations in combination with a suppression of the 21-cm impact of
other types of fluctuations (i.e., in density and ionization); in
particular, right at the heating transition, the cosmic mean 21-cm
intensity is (very nearly) zero, and thus all fluctuations other than
those in the gas temperature disappear (to linear order) from the
21-cm sky.

\begin{figure}[tbp]
\includegraphics[width=0.498\textwidth]{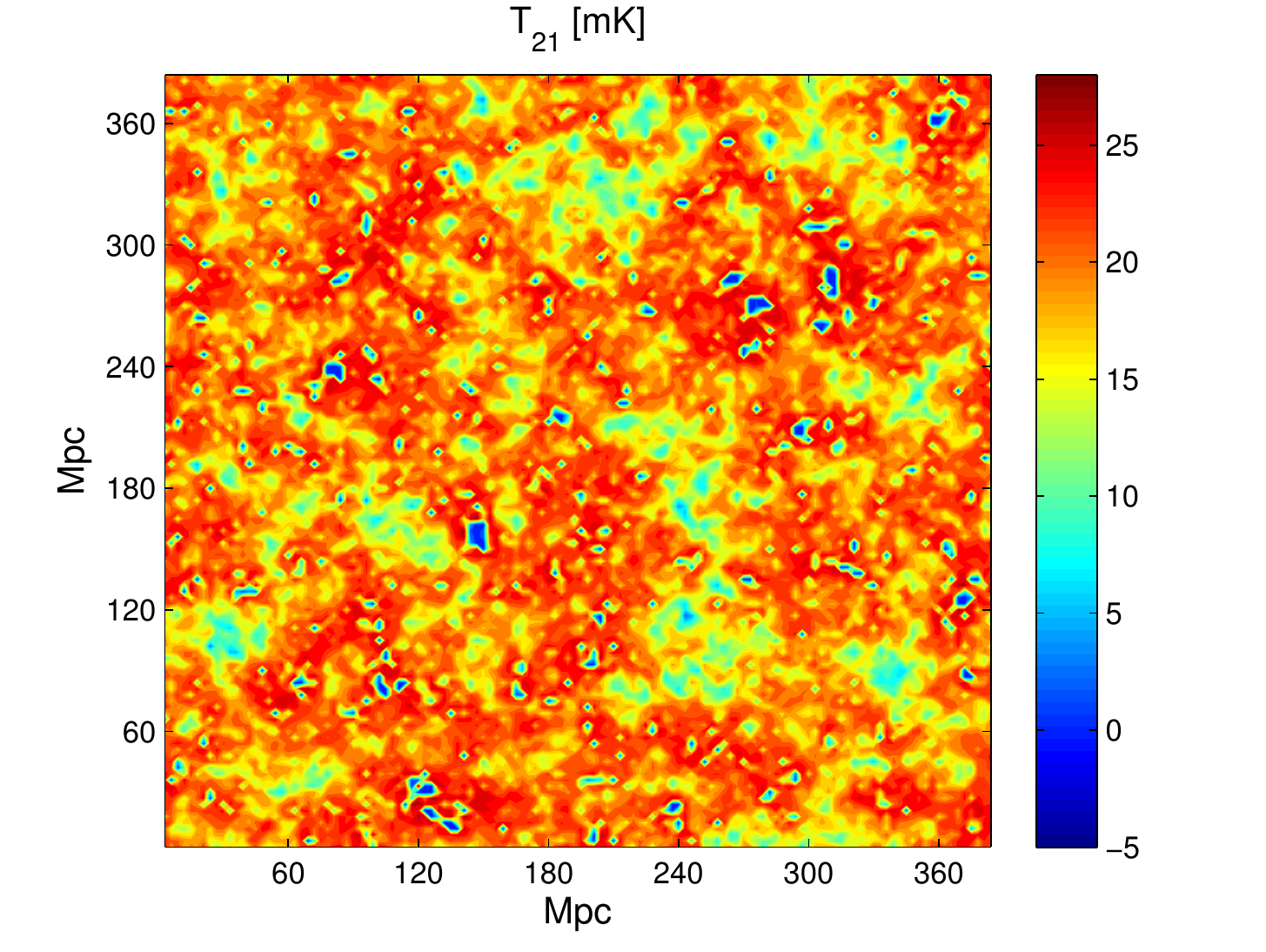}
\includegraphics[width=0.498\textwidth]{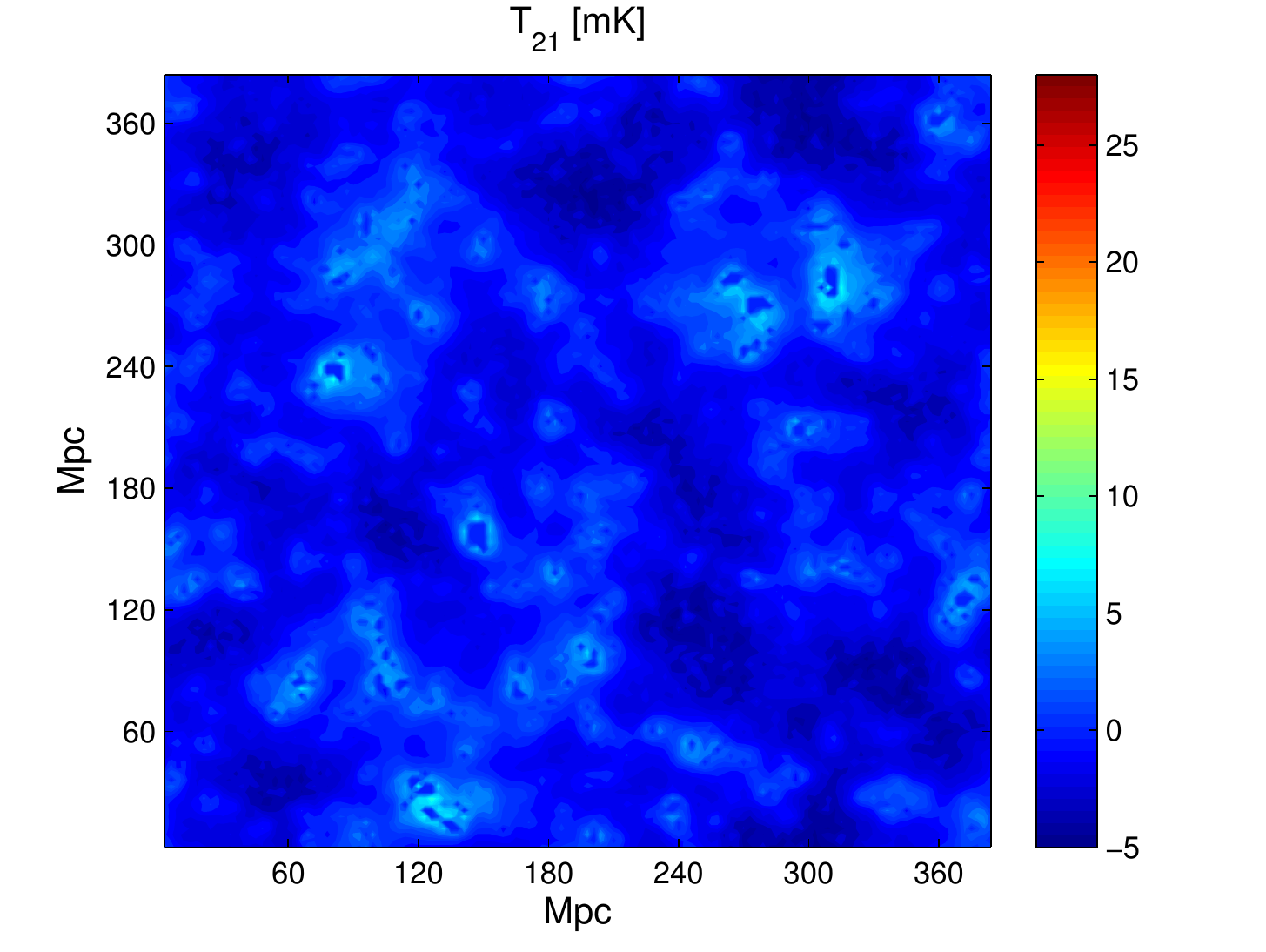}
\caption{Example of a predicted 21-cm map from a semi-numerical code,
  at $z=12.1$, comparing the case of heating sources with a hard X-ray
  spectrum (right panel) and the previously-assumed soft spectrum
  (left panel), shown on a common scale. For the hard spectrum, this
  redshift corresponds to the cosmic heating transition. In this
  comparison, both cases have the same underlying distribution of star
  formation at a given redshift, so they have the same ionized patches
  (at an early stage of reionization, when $14\%$ of the IGM has been
  reionized) and a similar distribution pattern of gas temperature and
  of 21-cm temperature. However, the difference is visually striking,
  in that the map for the hard spectrum is strongly suppressed in
  terms of both the typical value of $T_{\rm b}$ and the typical size
  of its fluctuations. From \cite{complex}.}
\label{f:coldimage}
\end{figure}

\begin{figure}[tbp]
\includegraphics[width=0.498\textwidth]{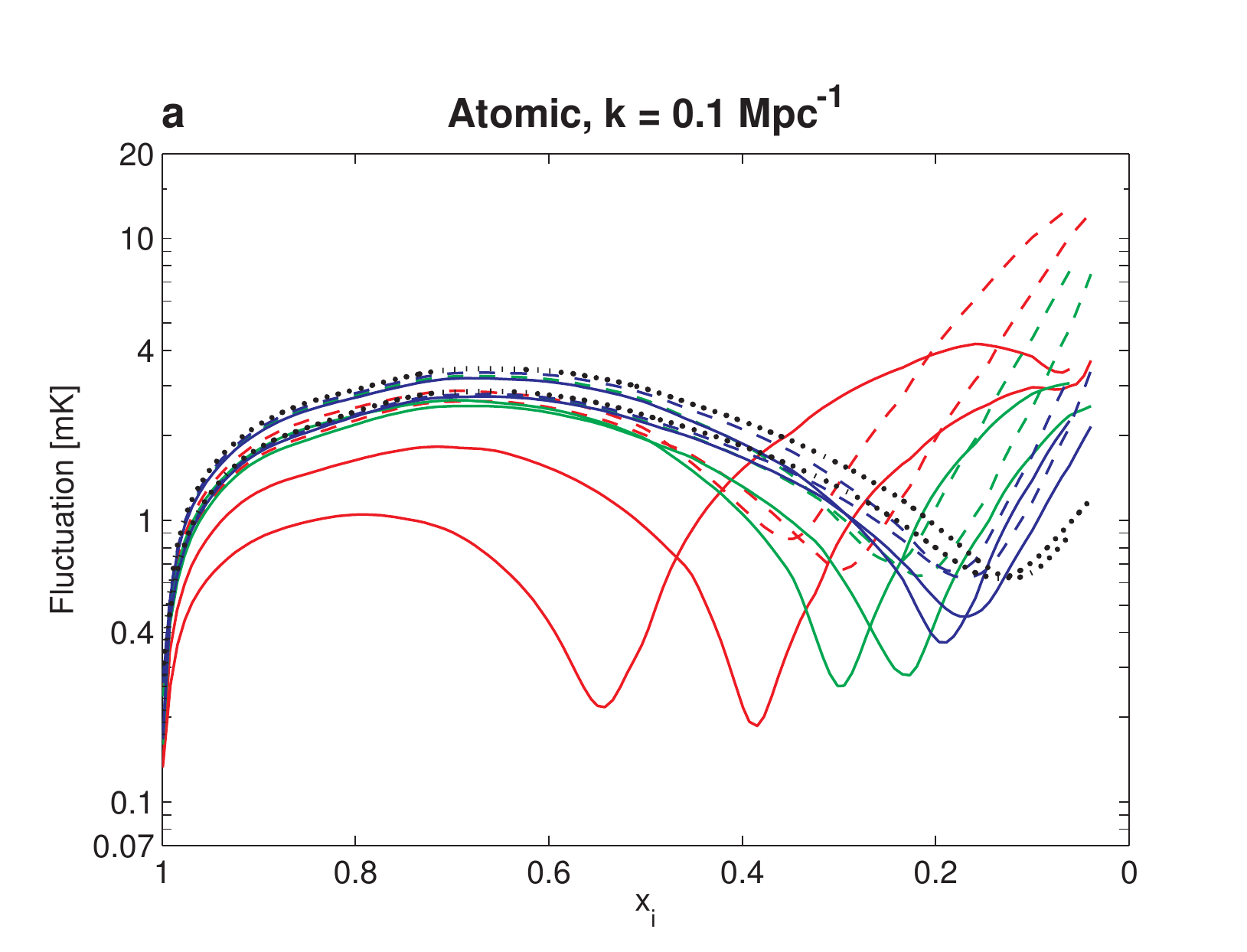}
\includegraphics[width=0.498\textwidth]{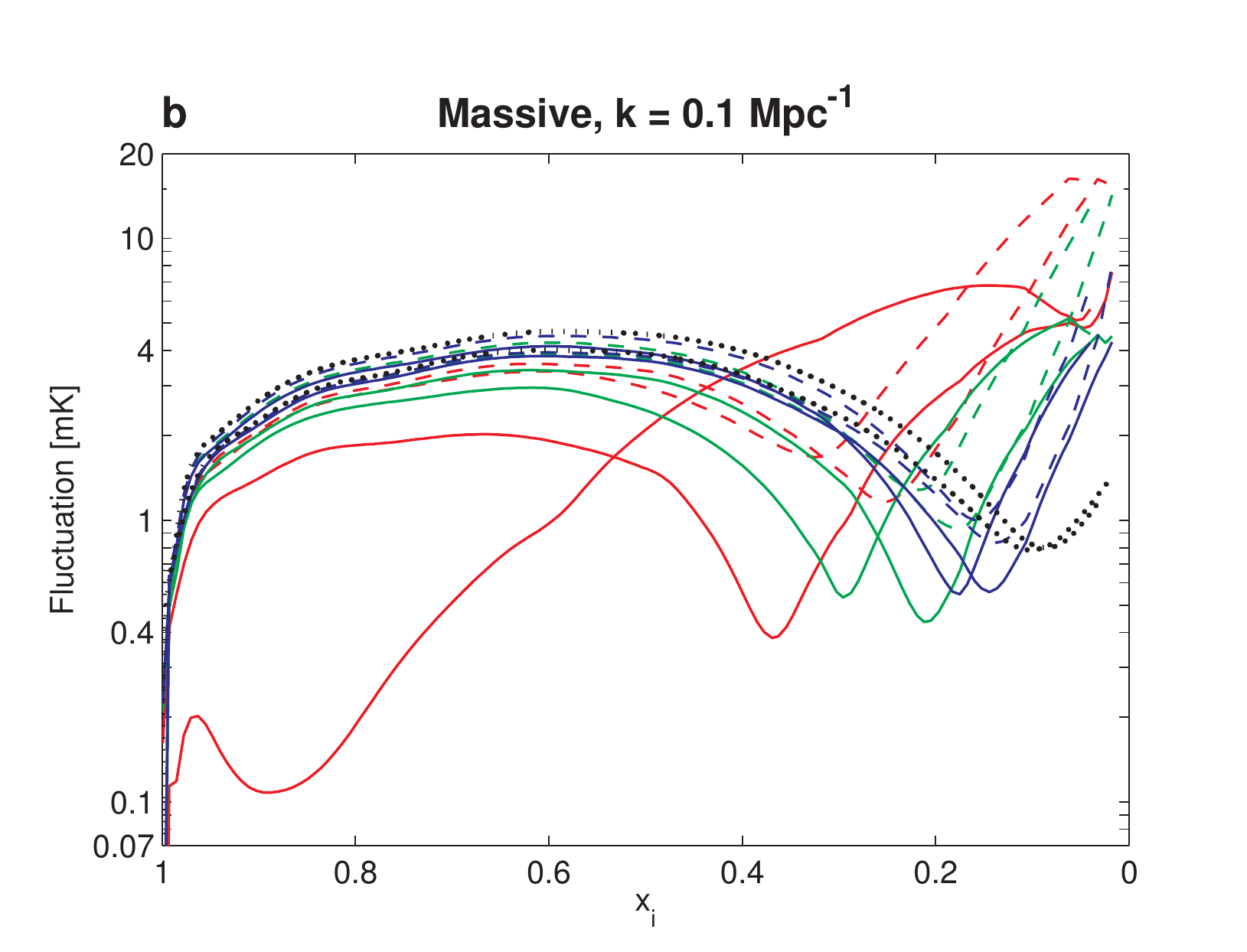}
\includegraphics[width=0.498\textwidth]{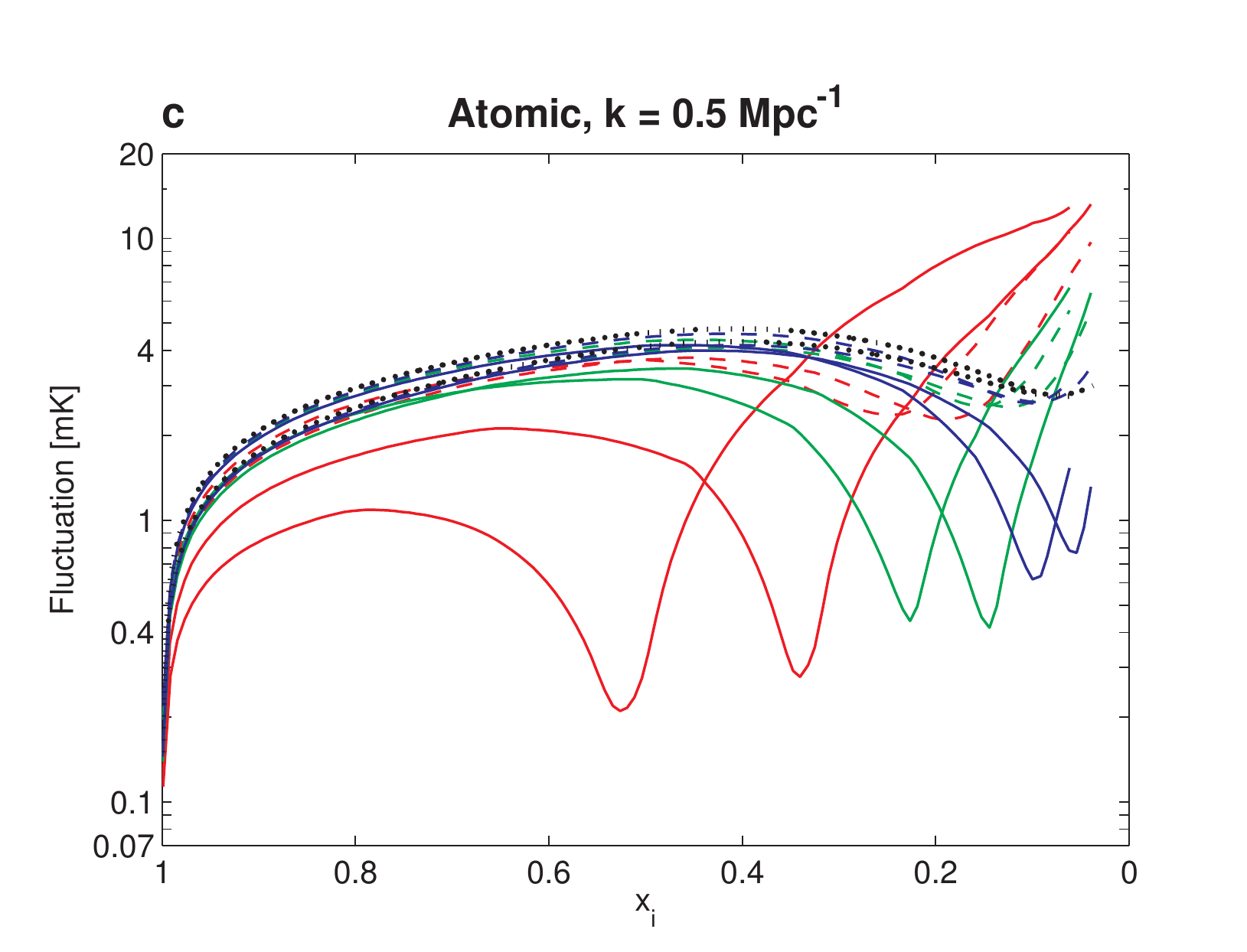}
\includegraphics[width=0.498\textwidth]{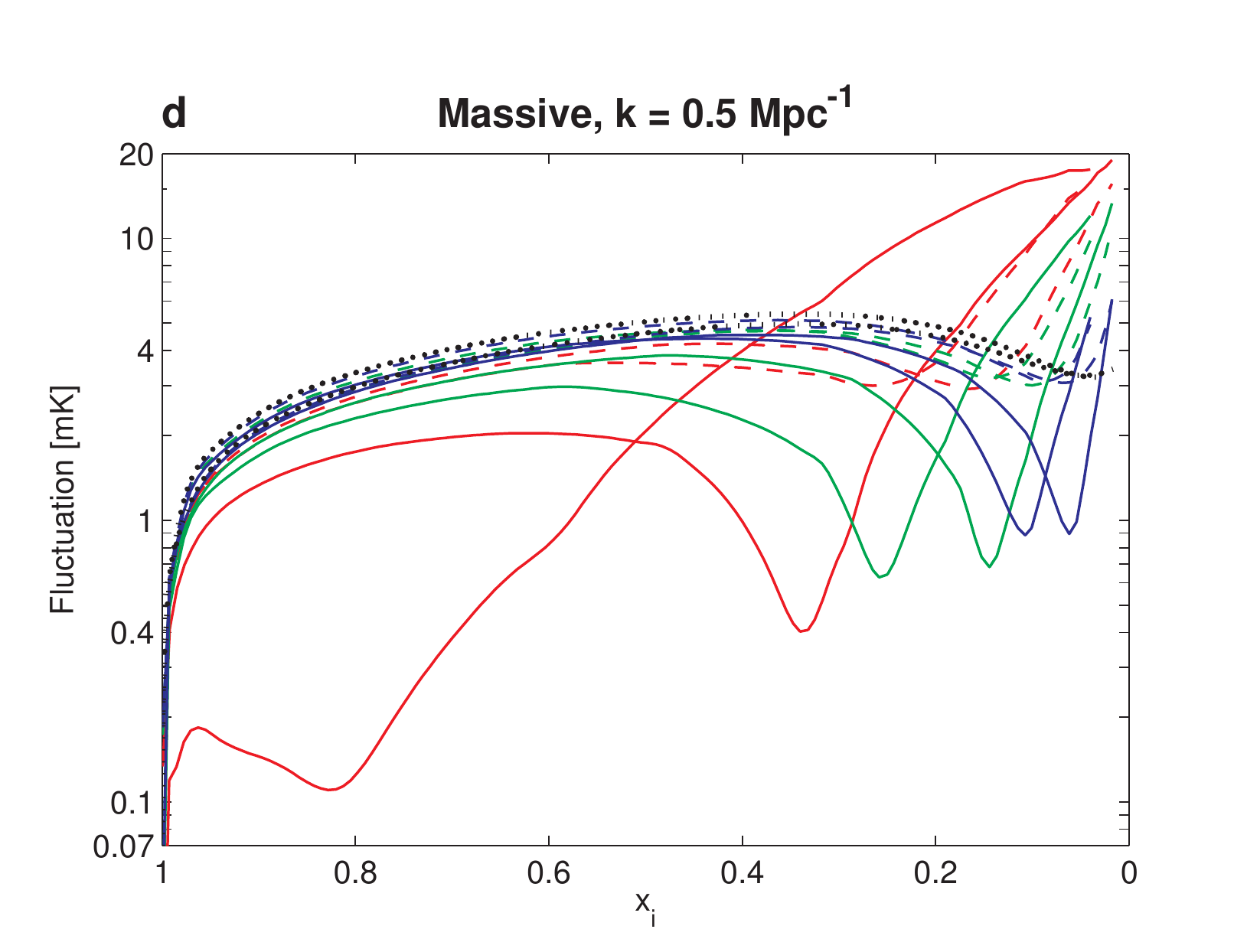}
\caption{The fluctuation level of the 21-cm brightness temperature is
  shown versus the ionized (mass) fraction of the universe $x_i$
  (starting on the right from $z=15$). We compare results obtained
  with a hard X-ray binary spectrum \cite{Frag2} (solid curves) to
  those with the previously-adopted soft spectrum (dashed curves), and
  show the commonly-assumed saturated heating case for reference
  (black dotted curves). The curves show various cases, in order to
  give a reasonable idea of the range of parameter space given current
  uncertainties. Thus, the best guess for the X-ray efficiency (green
  curves) is shown along with an efficiency lower (red curves) or
  higher (blue curves) by a factor of $\sqrt{10}$, each with either
  early or late reionization (given current uncertainties about its
  timing: see \S~\ref{s:reion}). The fluctuation is shown at a
  wavenumber $k=0.1$~Mpc$^{-1}$ (top panels) or $k=0.5$~Mpc$^{-1}$
  (bottom panels), for two possible cases of galactic halos, either a
  minimum halo mass set by atomic cooling (left panels) or halos that
  are ten times more massive (right panels). The lower $k$ value
  roughly tracks large-scale fluctuations (heating early on, and
  ionized bubbles later), while the higher $k$ value corresponds to a
  smaller scale (though one that can still be measured accurately with
  current experiments) and thus tracks more closely the evolution of
  density fluctuations. To illustrate the effect of the X-ray spectrum
  on the results, consider the fluctuation level at $k=0.5$~Mpc$^{-1}$
  at the mid-point of reionization (i.e., $x_i=0.5$); the parameter
  space explored here gives a possible range of 3.6--4.9~mK for the
  soft spectrum, while the hard spectrum gives a much broader range of
  0.3--4.4~mK.  Note also that the latter values are typically much
  lower than the often-assumed limit of saturated heating (which gives
  a corresponding range of 4.1--5.1~mK). From \cite{Cold}.}
\label{f:coldfig3}
\end{figure}

This effect is visually apparent in simulated maps
(Figure~\ref{f:coldimage}). In upcoming observations, it is likely to
be apparent in the measured 21-cm power spectrum
(Figure~\ref{f:coldfig3}). Depending on the parameters, the deep
minimum (reaching below 1~mK) may occur at any time during
reionization, but is likely to occur before its mid-point. Previously,
the fluctuation signal was expected to lie within a narrow,
well-defined range, allowing for a relatively straight-forward
interpretation of the data in terms of the progress of reionization;
the possibility of a hard X-ray spectrum, however, introduces a
variety of possibilities, making it likely that modeling of the 21-cm
data will involve an analysis of the interplay of heating and
reionization.

If a sufficient sensitivity level can be achieved, a low minimum in
the 21-cm power spectrum during reionization would be a clear
signature of late heating due to a hard X-ray spectrum. Indeed, a
clear observational indication that this feature corresponds to a
cosmic milestone is that the minima at all $k>0.5$~Mpc$^{-1}$ should
occur at essentially the same redshift (namely the true redshift of
the heating transition); lower wavenumbers correspond to larger scales
than the typical X-ray mean free path, leading to a more complicated
evolution and to minima delayed to lower redshifts (see also
Figure~\ref{f:complex}). More generally, observations of the 21-cm
power spectrum over a broad range of wavenumbers will clearly probe
the X-ray spectrum of the sources of cosmic heating
\cite{Cold,pacucci,complex,FialkovPRL}.

Beyond reionization, heating by high-energy X-rays removes the
previously expected signal from an early heating transition
(\S~\ref{s:2012}) at $z \sim 15-20$, but leaves in place the similar
$z \sim 20-25$ signal from the Lyman-$\alpha$ coupling transition that
is likely detectable with the Square Kilometre Array
(\S~\ref{s:Lya21}); actually, in this case the \Lya peak is stronger
and more extended in redshift, since it is not cut off by early
heating as in the case of soft X-rays \cite{complex}. It could also
affect other observations of high-redshift galaxies. For example,
since late heating implies weak photoheating feedback during the
cosmic heating era, low-mass halos may continue to produce copious
stars in each region right up to its local reionization; note though
that internal feedback (arising from supernovae or mini-quasars) could
still limit star formation in small halos.

\subsection{The global 21-cm spectrum}

\label{s:global}

This section thus far has focused on 21-cm fluctuations, and in
particular the 21-cm power spectrum. The power spectrum encodes a lot
of information about the various sources of 21-cm fluctuations, and it
is a rich dataset consisting of an entire function of wavenumber at
each redshift, or potentially even much more than that due to the
line-of-sight anisotropy (\S~\ref{s:anisotropy}). This information can
hopefully be extracted from data obtained with radio interferometers,
after dealing with the expected thermal noise and sample variance,
foreground residuals, and artifacts of the imperfectly-known responses
of the radio antennae and receivers.

\begin{figure}[tbp]
\includegraphics[width=0.498\textwidth]{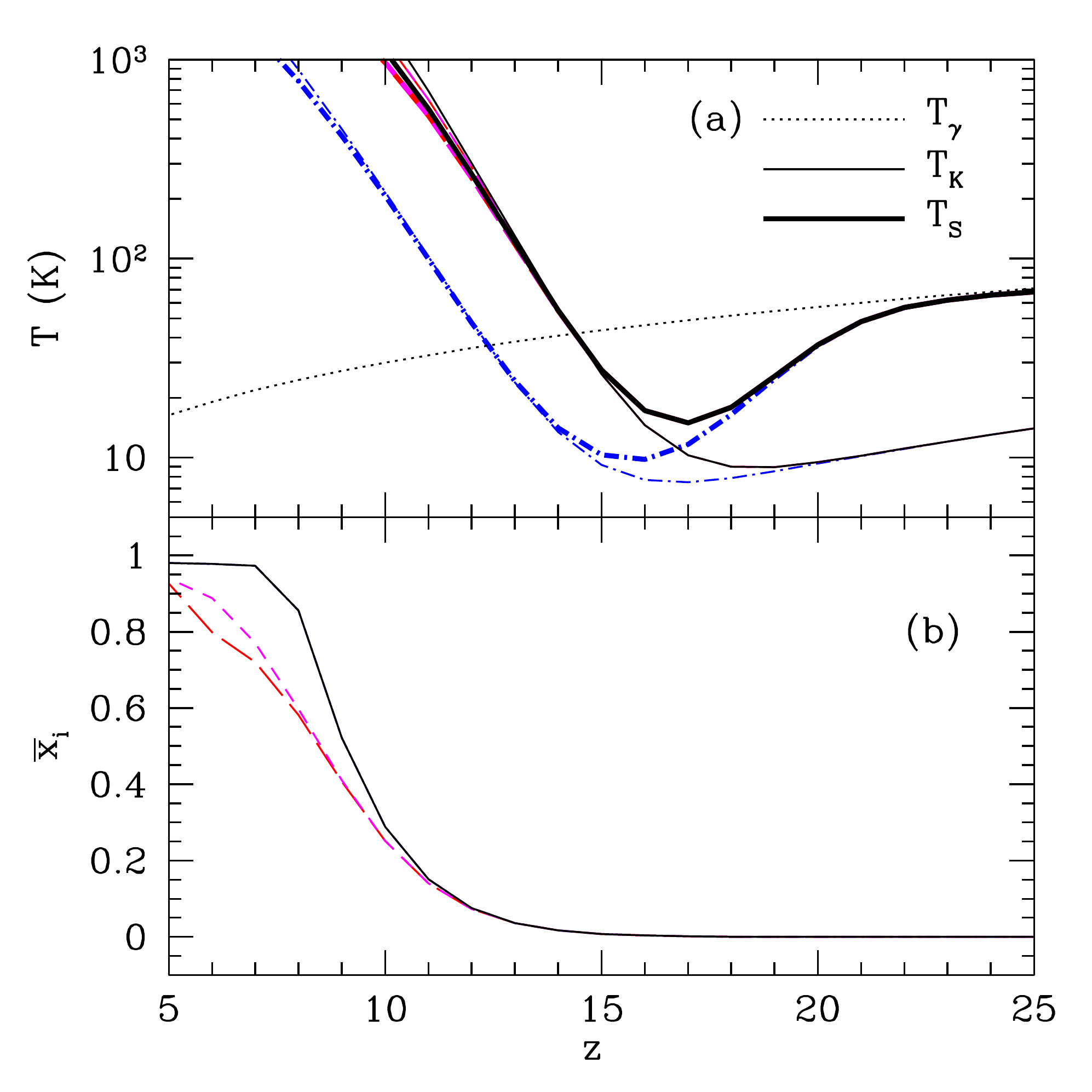}
\includegraphics[width=0.498\textwidth]{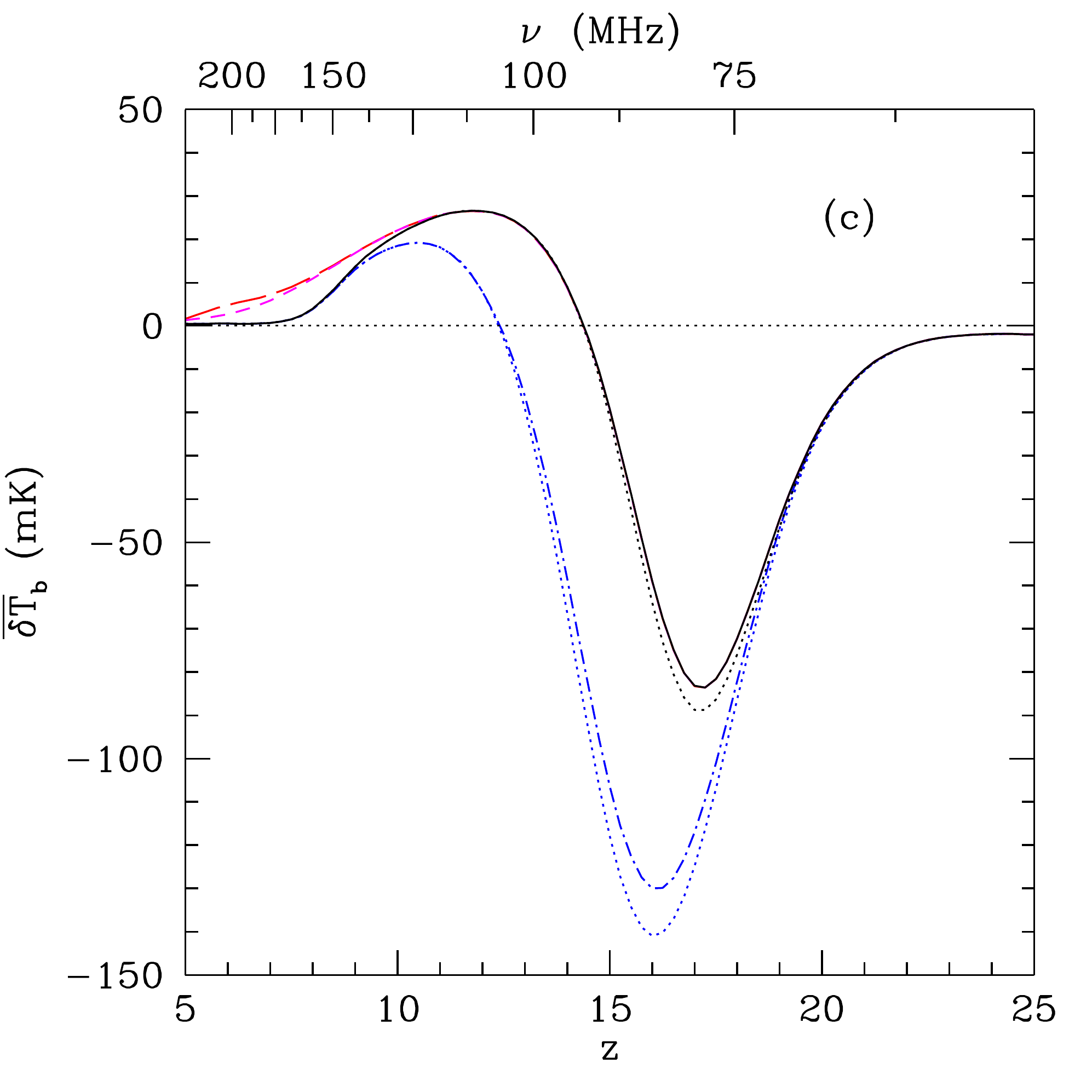}
\caption{Examples of predicted global 21-cm histories, showing how
  they reflect the cosmic history of star formation and its various
  effects on the IGM. A range of parameters are shown in order to
  reflect a reasonable range of uncertainty: the best-guess X-ray
  efficiency (solid curves), a lower efficiency by a factor of 5
  (dot-dashed curves), and the addition of two possible models for
  photoheating feedback (short- and long-dashed curves). Panel (a)
  shows the CMB ($T_\gamma$), gas kinetic ($T_K$) and spin ($T_S$)
  temperatures (dotted, thin, and thick solid curves, respectively).
  Panel (b) shows the progress of reionization, in terms of the cosmic
  mean ionized fraction $\bar{x}_i$. Panel (c) shows the resulting
  global mean 21-cm brightness temperature measured with respect to
  the CMB; in this panel, the two dotted lines show $T_{\rm b}$ if
  shock heating is ignored. Note that this panel shows the observed
  frequency on top in addition to the redshift on the bottom. All
  models here assume Pop~II stars and a soft X-ray spectrum of heating
  sources. From \cite{Fur06}.}
\label{f:global}
\end{figure}

A very different approach is to measure the total sky spectrum and
detect the redshift evolution of the global, cosmic mean 21-cm
intensity. A global experiment requires a simple, relatively cheap
setup (an all-sky antenna) compared to the fluctuation experiments,
and the total sky naturally yields a higher signal-to-noise ratio and
a spectrally smoother foreground than found in small patches (which
are the basic units of the fluctuation experiments). In order to make
success more likely, observations can focus on constraining sharp
frequency features, without attempting to measure the absolute
cosmological 21-cm emission level (which is much harder). During
reionization, there should be a decrease in the global 21-cm emission
due to the overall disappearance of atomic hydrogen
(\S~\ref{s:reion}).  This global step, while not sudden, is still
expected to be fairly sharp in frequency.  At higher redshifts, a
sharp decrease towards negative brightness temperature should occur
due to the rise of the first stars as a result of \Lya coupling of the
cold IGM (\S~\ref{s:Lya}), followed by a sharp rise up to positive
values due to cosmic heating (\S~\ref{s:heating}). Thus, a detection
of the global signal would trace the overall cosmic history of the
first stars through their effect on 21-cm emission
(Figure~\ref{f:global}).  Maximum-likelihood analyses of data fitting
show that global 21-cm measurements during cosmic reionization should
be able to detect a wide range of realistic models and measure the
main features of the reionization history while constraining the key
properties of the ionizing sources; this is true in analyses (that
assumed the saturated heating limit) using a flexible toy model
\cite{PL} or a $\Lambda$CDM-based model \cite{nono}, though the
results are rather sensitive to assumptions on just how difficult it
will be to remove the effect of the foregrounds.

\begin{figure}[tbp]
\includegraphics[width=0.498\textwidth]{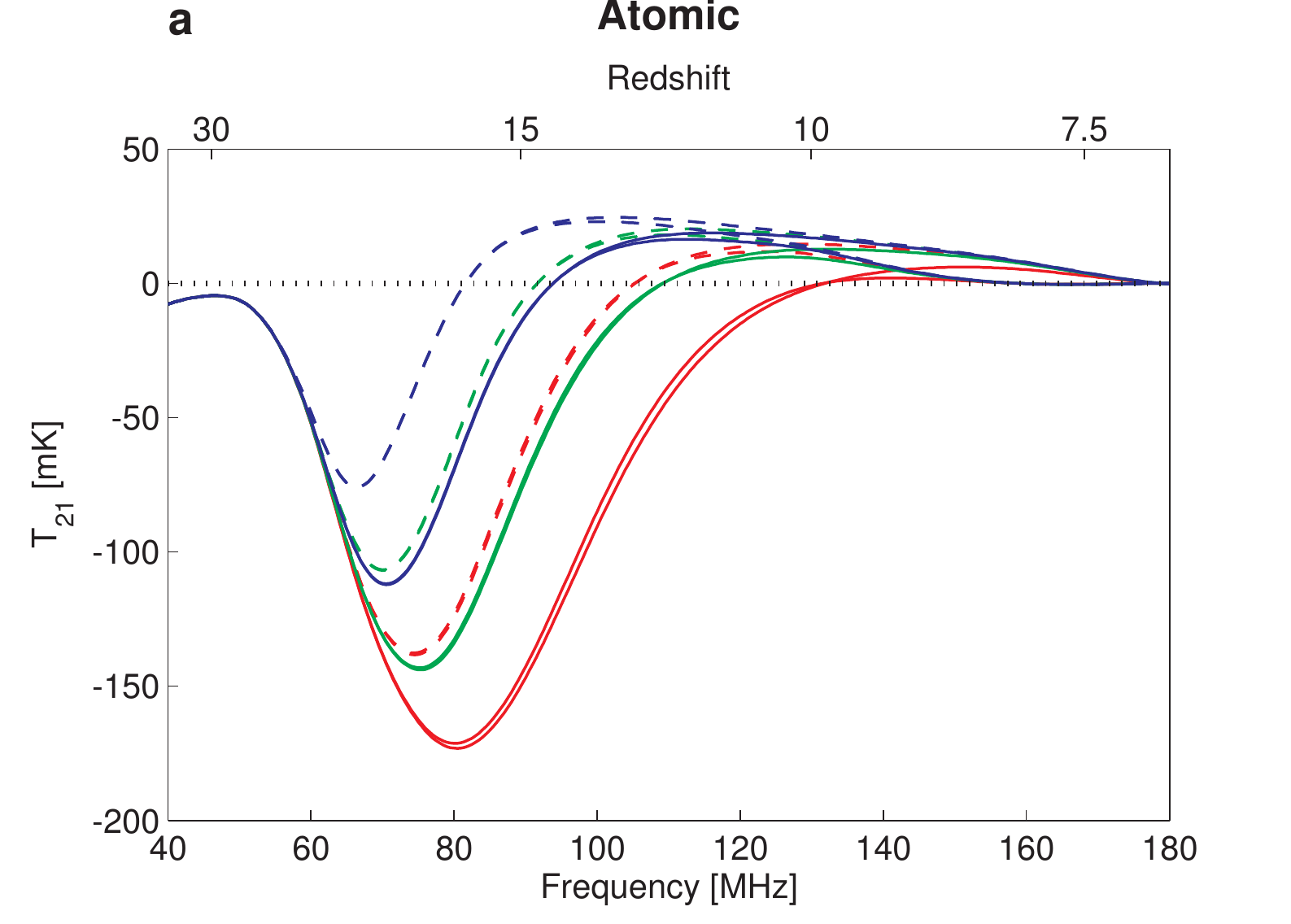}
\includegraphics[width=0.498\textwidth]{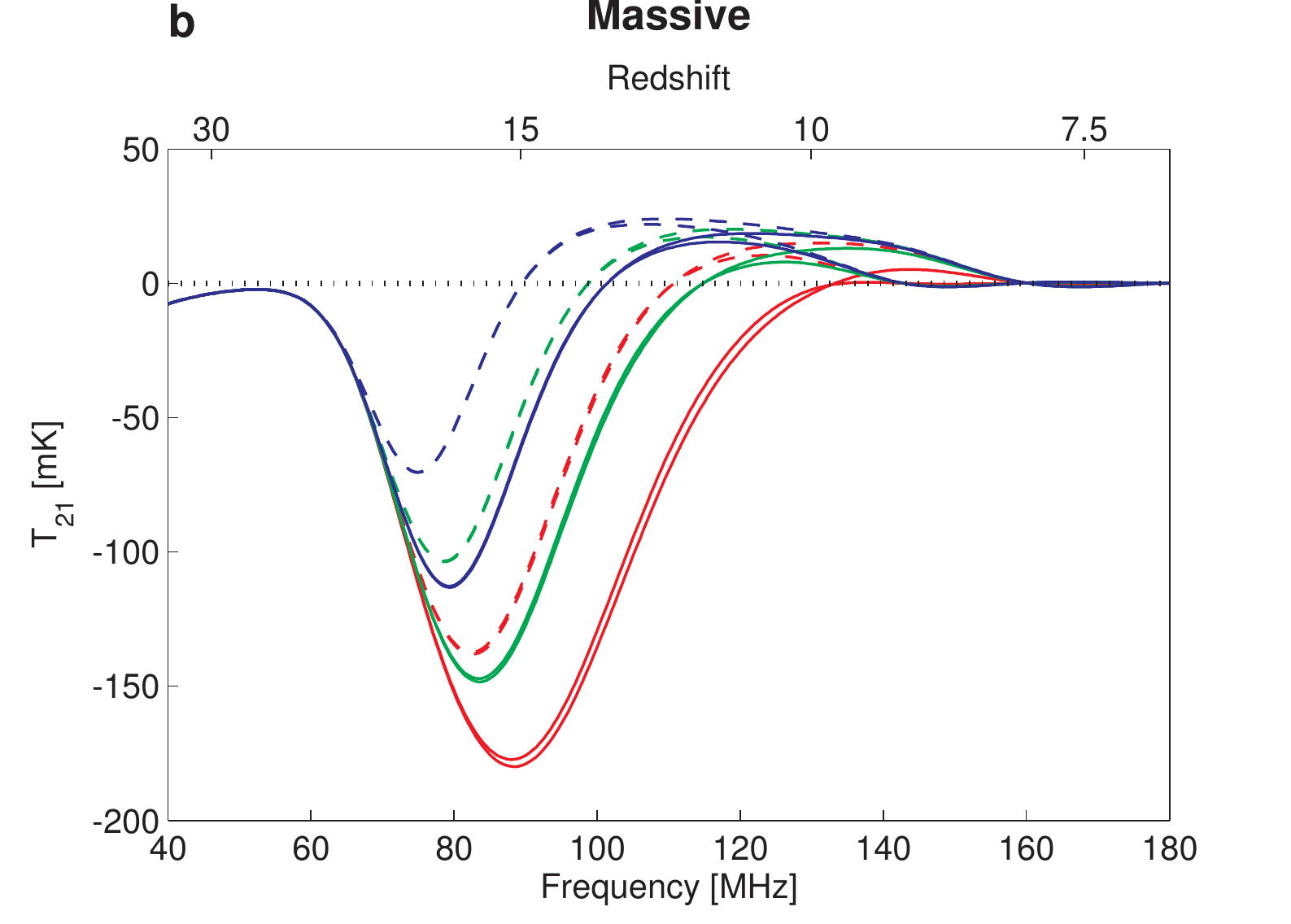}
\caption{The effect of late cosmic heating on the global 21-cm
  spectrum. The cosmic mean 21-cm brightness temperature ($T_{21}$)
  relative to the CMB is shown versus observed frequency (with the
  redshift indicated on top), for the hard X-ray binary spectrum
  \cite{Frag2} (solid curves) and for the previously-adopted soft
  spectrum (dashed curves); note also the fiducial dotted line at
  $T_{21}=0$. Various cases are shown, in order to give a reasonable
  idea of the range of parameter space given current uncertainties;
  the notation matches that in Figure~\ref{f:coldfig3}. From
  \cite{Cold}.}
\label{f:coldfig2}
\end{figure}

The late heating (\S~\ref{s:heating} and \S~\ref{s:late}) expected due
to the hard spectrum of X-ray binaries has a particularly important
effect on the global 21-cm signal. The effect of late heating is to
give the cosmic gas more time to cool adiabatically to well below the
CMB temperature, thus producing mean 21-cm absorption that reaches a
maximum depth in the range $-110$ to $-180$~mK at $z \sim 15-19$
(Figure~\ref{f:coldfig2}). This may make it easier for experiments to
detect the global 21-cm spectrum from before reionization and thus
probe the corresponding early galaxies. Global experiments are most
sensitive to the frequency derivative of the 21-cm brightness
temperature; late heating extends the steep portion of the spectrum to
higher frequencies, moving the maximum positive derivative to a $\sim
10\%$ higher frequency (where the foregrounds are significantly
weaker) while also changing the value of this maximum derivative by
$\pm 10\%$. On the other hand, at lower redshift, late heating
significantly suppresses the global step from reionization, which
suggests that global 21-cm experiments should focus instead on the
earlier eras of \Lya coupling and cosmic heating.

\section{Summary and Conclusions}

\label{s:fin}

The study of the first stars, galaxies, and black holes, and their
effect on the 21-cm sky, is entering a critical stage. While this
subject has been developing theoretically for several decades,
including a great acceleration in recent years, observationally this
field is in its infancy. Thus, we are about to experience that
pinnacle of excitement that comes with the first clash of a scientific
theory with experimental data. In such a collision of theoretical
expectations with reality, there are several possible outcomes. The
predictions can be perfectly verified, an outcome that makes the
successful theorists gleeful and proud, but at the same time is
extremely boring. At the other extreme, the predictions can fail
completely, making the theorists a laughing-stock, but revealing
previously unexpected cosmic events, which makes this possibility the
most exciting one. Neither of these extreme possibilities is expected
in the case of 21-cm cosmology. The shear magnitude of the uncertainty
about high-redshift astrophysics makes the first possibility unlikely,
even in the absence of exotic cosmic events such as dark matter decay.
On the other hand, complete failure is made unlikely by the fact that
the theory is grounded in solid atomic physics as well as models of
galaxy formation that are significantly constrained by observations of
the current Universe, at one end, and the CMB at the other (initial
condition) end. Thus, the most likely outcome is an intermediate one,
where the overall framework of theoretical expectations will be
confirmed, but with some, hopefully interesting and significant,
surprises, such as an unexpected, new class of astrophysical sources
(which will be noticed if it dominated one of the types of radiation
that drove the 21-cm emission). Regardless of the precise outcome, it
is likely that once a clear detection of the 21-cm signal from early
cosmic history is achieved, the field will get a big boost, analogous
to the development of CMB observations and theory after the first
detection of CMB temperature fluctuations by the COBE satellite. This
breakthrough moment for 21-cm cosmology will hopefully occur within
the next few years, and will be followed up with confirmations and
more detailed measurements soon afterwards.

\begin{figure}[tbp]
\includegraphics[width=\textwidth]{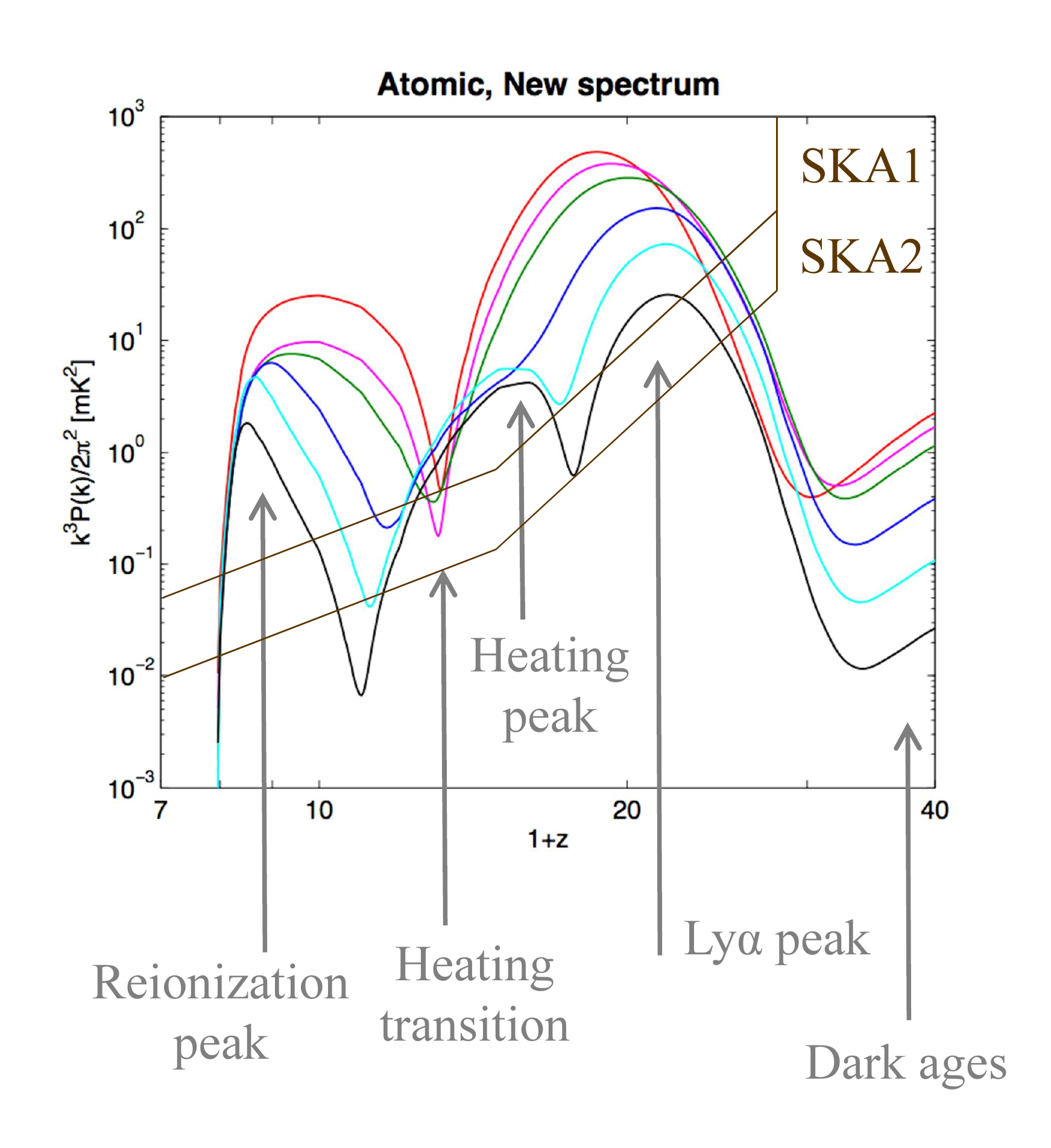}
\caption{A summary view of the rich complexity of even just the
  isotropically-averaged 21-cm power spectrum, shown via the evolution
  with redshift of the squared 21-cm fluctuation at various
  wavenumbers. Going from small to large scales, shown are
  $k=1$~Mpc$^{-1}$ (red), $k=0.5$~Mpc$^{-1}$ (magenta),
  $k=0.3$~Mpc$^{-1}$ (green), $k=0.1$~Mpc$^{-1}$ (blue),
  $k=0.05$~Mpc$^{-1}$ (cyan), and $k=0.03$~Mpc$^{-1}$ (black). The
  particular model shown here assumes cosmic heating by a hard X-ray
  spectrum (\S~\ref{s:heating} and \S~\ref{s:late}), and that stars
  form in all halos in which the gas can cool via atomic cooling. Also
  shown is the approximate observational thermal noise power spectrum
  expected for the SKA phase 1 and phase 2 (at $k=0.1$~Mpc$^{-1}$)
  \cite{Koopmans}. From \cite{complex}, with added labels and
  observational sensitivities.}
\label{f:complex}
\end{figure}

A great wealth of data is potentially available in 21-cm cosmology
(\S~\ref{s:milestone} and \S~\ref{s:Final21cm}). Even just the
isotropically-averaged 21-cm power spectrum, measured as a function of
wavenumber and redshift, is a rich data set that probes many details
of the various cosmological and astrophysical sources of 21-cm
fluctuations (see Figure~\ref{f:complex}). A number of cosmic events
leave clear signatures in the power spectrum, but the redshifts of the
associated features (such as the peaks) vary with scale, since several
different sources of 21-cm fluctuations contribute at any given time,
and these sources vary differently with scale. In the model shown in
Figure~\ref{f:complex}, for which reionization ends at $z \sim 7$, the
reionization peak of fluctuations occurs in the range $z = 7.5 - 9$
depending on wavenumber. While the uncertainties are still large, it
now seems that the IGM was most likely heated by X-ray sources with a
hard spectrum (\S~\ref{s:heating} and \S~\ref{s:late}), a possibility
not considered until recently; in this case, the cosmic heating
transition produces a clear minimum on small scales, but a weak
heating peak remains on the largest scales that are larger than the
typical distance traveled even by hard X-rays. Continuing with
Figure~\ref{f:complex}, the \Lya peak occurs in this example at $z=18
- 20$, and (generally in the case of late heating) it is both the
strongest and highest-redshift signal from the first stars (In the
case of a soft X-ray spectrum, the heating peak is somewhat higher
than the \Lya peak \cite{complex}). We note that additional
theoretical uncertainties result from the complexity of the
astrophysics during early times, including substantial transitions in
the basic character of star formation expected due to various types of
stellar feedback such as supernova outflows, LW radiation, and metal
enrichment. The dark ages, during which 21-cm emission is not
significantly affected by astrophysical sources and becomes a purely
cosmological probe, begin at $z>30$; at this point the predicted
fluctuation signal is quite low, and since the galactic foreground
increases rapidly with redshift (with the brightness temperature of
the sky $\propto (1+z)^{2.6}$ \cite{F06}), observations of this era
lie in the somewhat distant future.

Actually measuring a data set like that shown in
Figure~\ref{f:complex} would obviously constitute an amazing advance
in our understanding of cosmic history. However, it is important to
also look for robust, model-independent signatures that can
convincingly confirm and complement the results obtained from fitting
parameterized models to the (angle-averaged) 21-cm power spectrum.
This is particularly needed in a field looking to probe a new,
unexplored regime of cosmic history. Luckily, the field of 21-cm
cosmology has indeed emerged as a very rich one. For example, the
line-of-sight anisotropy of the power spectrum (\S~\ref{s:anisotropy})
is potentially an immensely important source of additional
information, and it has only begun to be explored. It can provide a
number of model-independent probes of early galaxies that would
complement inferences made based on the angle-averaged power spectrum.
In particular, the dominant $\mu^2$ term of the anisotropy acts as a
cosmic clock, its sign changing as it tracks various cosmic
milestones; for instance, measuring it to be negative during
reionization would directly confirm the inside-out topology of this
transition (i.e., where overdense regions reionize first). Another
example of a possible model-independent signature is the streaming
velocity with its associated strong BAO features. In addition, the
global 21-cm spectrum (\S~\ref{s:global}) is a wonderfully
complementary probe of the same cosmic history. One way to express
this is that the 21-cm fluctuations can be written as a product of the
mean intensity and its relative fluctuations, and information on the
global spectrum helps to separate these two quantities and thus break
a degeneracy.

In this review we have focused on the 21-cm power spectrum (including
its angular anisotropy). There are good reasons for this, even though
it is not an open and shut case as in CMB studies, where the power
spectrum carries the most important cosmological information in the
signal (which is thought to reflect the underlying Gaussian random
field of primordial perturbations). In general, there are two
different modes for studying galaxies: The collective (galaxy
clustering) and the individual (studying individual galaxies). 21-cm
cosmology during cosmic dawn and the EOR will be dominated by the
collective regime.  The 21-cm fluctuations will be dominated by
various radiation fields, and the intensity of those fields at any
point will be made up of the contributions of many individual sources,
except perhaps in a few rare regions. Thus, the structures that will
be seen will be a collective effect, and thus mainly dependent on the
clustering of sources. The power spectrum naturally measures this
clustering. More specifically, the distribution of sources throughout
this era is driven by the underlying density distribution of matter
(except for the additional effect of the streaming velocity). This
density distribution is determined by the power spectrum, and for
linear fluctuations, the 21-cm map is also determined by its power
spectrum (which is the underlying power spectrum times a window
function, corresponding to a convolution in real space that accounts
for the spatial redistribution of photons of the various relevant
frequency regimes). It is true that there are some non-linear
distortions along the way, but still, on the (relatively large) scales
resolvable by upcoming radio arrays, the power spectrum should capture
most of the information available in a full image.  Indeed, as
described throughout this review, the 21-cm power spectrum can be used
to reconstruct the most interesting astrophysical information that we
desire: at what redshifts \Lya coupling, cosmic heating, and
reionization occurred, how fast they progressed, which galactic halos
dominated each era, and what the spectrum was of the sources (e.g.,
the X-ray spectrum in the case of X-ray heating).  The most non-linear
process is reionization (with its sharp edges in the expected scenario
that is dominated by UV photons), but the non-Gaussianity of the
ionization field only reflects the rapid absorption of ionizing
photons, and may not probe much interesting physics beyond that.
Also, in the near future the power spectrum is likely to be the main
available observable from the least explored, and thus most exciting,
high-redshift regime of the pre-EOR cosmic dawn; imaging from such an
early time will be difficult even for the SKA.

That said, the non-Gaussianity of 21-cm fluctuations \cite{BA05} does
make other statistics beyond the power spectrum interesting, including
the bispectrum \cite{BiP,SBP06}, the 21-cm PDF (probability
distribution function, i.e., histogram of values of the 21-cm
brightness temperature)
\cite{CM03,Fur04b,mellema,Ichikawa,IlievPDF,PritchardPDF}, and the
difference PDF (i.e., histogram of $T_b$ differences between pixel
pairs) \cite{BL08,Vera}; some of the additional information available
in the PDF can be captured by its skewness \cite{WM07,Harker}. Also,
while in this review we have focused heavily on the emerging field of
21-cm cosmology, other cosmological probes are making rapid advances
and should explore complementary aspects of high-redshift galaxies.
The James Webb Space Telescope (JWST; http://www.jwst.nasa.gov/)
should discover at least the largest galaxies at early times, as well
as rare bright objects (such as supernovae or gamma-ray bursts) in
more typical galaxies. The planned generation of larger ground-based
optical/IR telescopes, including the Thirty Meter Telescope (TMT;
http://www.tmt.org/), the Giant Magellan Telescope (GMT;
http://www.gmto.org/), and the European Extremely Large Telescope
(E-ELT; http://www.eso.org/public/teles-instr/e-elt/) should give us
detailed, spectroscopic information on some of these objects and their
surrounding IGM. Imaging the 21-cm sky, as planned for the SKA, will
be very interesting around particular bright objects. In another area,
the CMB, in addition to its further development as a cosmological
probe, may allow the detection of the small-scale signature of CMB
scattering by the ionized bubbles during cosmic reionization
\cite{Aghanim96,GruzH98,Santos03,CMBreion}.

We have also discussed in this review the complementary interaction in
this field between numerical simulations, analytical (or
semi-analytical) models, and semi-numerical methods. Each method has
its advantages and disadvantages, and in particular it is important
not to overlook the limitations of numerical simulations
(\S~\ref{s:numerical}).

Another highlight of this work is in pointing out how the idea of
unusually large fluctuations in the number density of high-redshift
galaxies (\S~\ref{s:unusually}) is a common thread that has driven the
whole topic of 21-cm fluctuations, from the understanding of the
character of reionization (\S~\ref{s:reion}) to the first predictions
of large-scale 21-cm fluctuations from the inhomogeneous \Lya
(\S~\ref{s:Lya21}) and X-ray (\S~\ref{s:2012}) backgrounds. It has
recently been joined by an exciting new source of large-scale
fluctuations, the supersonic streaming velocity (\S~\ref{s:stream}).
This new source comes with a strong signature of baryon acoustic
oscillations, making it a potential tool for identifying the presence
of tiny, million solar mass halos at very early times. The streaming
velocity certainly had a major effect on the first generation of
stars, and it may also have had a great significance at redshifts that
are observable with 21-cm experiments (\S~\ref{s:2012}), though this
depends on just how efficiently such small halos were able to form
stars. We have also highlighted a different issue, one of basic 21-cm
physics: we have helped clarify the literature regarding the
low-temperature corrections to the basic expressions of 21-cm
cosmology (\S~\ref{s:lowT}).

In this review we have focused on the astrophysical era of 21-cm
cosmology that is accessible to upcoming experiments. However, it is
also important to keep in mind the great long-term promise of the
development of 21-cm cosmology. When 21-cm measurements reach small
spatial scales, this will open up a variety of new probes and
applications, especially in the dark ages during which 21-cm cosmology
will be a clean cosmological probe. For example, 21-cm fluctuations
should be present down to much smaller scales than CMB fluctuations
(which are cut off by the combination of Silk damping and the width of
the surface of last scattering). This implies a far greater potential
sensitivity of 21-cm measurements to a small primordial
non-Gaussianity \cite{Loeb04,Cooray}. Measuring the primordial power
spectrum on small scales will also probe the tilt of the power
spectrum and could potentially uncover a cutoff due to dark matter
properties (such as in the warm dark matter or fuzzy dark matter
\cite{fdm} models). Also, the gas temperature can in principle be
mapped through its effect on the small-scale power spectrum (i.e., the
filtering mass discussed in \S~\ref{sec:bary}) as well as more
directly through the anisotropic effect of the thermal smoothing of
the 21-cm power spectrum \cite{NB05}; e.g., if the cosmic gas is
radiatively heated to $10^3$~K, then the smoothing is expected on a
scale of $\sim 20$~kpc.  On small scales, the supersonic streaming
velocity also has a significant effect on the 21-cm power spectrum
\cite{Yacine}. Further back in time, a 21-cm signal is expected from
the cosmological epoch of recombination \cite{FL}.

We would like to end in the same way that the author concluded a
review written a decade ago \cite{BSc}, with the sincere hope of not
having to write this again a decade from now: Astronomers are eager to
start tuning into the cosmic radio channels of 21-cm cosmology.

\section{Acknowledgments}

I would like to thank my collaborators over the years on the reviewed
work, especially Avi Loeb, my former students Smadar Naoz and
Anastasia Fialkov, my current student Aviad Cohen, and Eli Visbal.
This work was supported by Israel Science Foundation grant 823/09 and
the Ministry of Science and Technology, Israel. This work was partly
done within the Labex Institut Lagrange de Paris (ILP, reference
ANR-10-LABX-63) part of the Idex SUPER, and received financial state
aid managed by the Agence Nationale de la Recherche, as part of the
programme Investissements d'avenir under the reference
ANR-11-IDEX-0004-02. I also acknowledge a Leverhulme Trust Visiting
Professorship at the University of Oxford. This research was supported
in part by the Perimeter Institute for Theoretical Physics. Research
at Perimeter Institute is supported by the Government of Canada
through Industry Canada and by the Province of Ontario through the
Ministry of Economic Development \& Innovation.


\begin{thebibliography}{14}

\bibitem{sdss} H.\ Aihara, et al., Astroph.\ J.\ Supp.\ 193 (2011) 29;
erratum -- 195 26.

\bibitem{2df}
M.\ Colless, et al., Mon.\ Not.\ R.\ Astron.\ Soc.\ 328 (2001) 1039.

\bibitem{mill} V.\ Springel, C.~S.\ Frenk, S.~D.~M.\ White, Nature
  440 (2006) 1137.

\bibitem{bond}  
J.~R.\ Bond, L.\ Kofman, D.\ Pogosyan, Nature 380 (1996) 603.

\bibitem{z11p1} P.~A.\ Oesch, G.\ Brammer, P.~G.\ van Dokkum, et al.,
  ApJ 819 (2016) 129

\bibitem{Bromm} V.\ Bromm, P.~S.\ Coppi, R.~B.\ Larson, Astroph.\ J.\
  527 (1999) L5.

\bibitem{Abel}  
T.\ Abel, G.~L.\ Bryan, M.~L.\ Norman, Science 295 (2002) 93.

\bibitem{first} S.\ Naoz, S.\ Noter, R.\ Barkana, Mon.\ Not.\ R.\
  Astron.\ Soc.\ 373 (2006) L98.

\bibitem{anastasia} A.\ Fialkov, R.\ Barkana, D.\ Tseliakhovich, C.\
  Hirata, Mon.\ Not.\ R.\ Astron.\ Soc.\ 424 (2012) 1335.

\bibitem{hogan} C.~J.\ Hogan, M.~J.\ Rees, Mon.\ Not.\ R.\
  Astron.\ Soc.\ 188 (1979) 791.

\bibitem{Madau} P.\ Madau, A.\ Meiksin, M.~J.\ Rees, Astroph.\ J.\ 475
  (1997) 429.

\bibitem{Loeb04} A.\ Loeb, M.\ Zaldarriaga, Phys. Rev. Lett.\ 92
  (2004) 211301.

\bibitem{BL05a} R.\ Barkana, A.\ Loeb, Astroph.\ J.\ 624 (2005) 65.

\bibitem{Nusser} A.\ Nusser, Mon.\ Not.\ R.\ Astron.\ Soc.\ 364 (2005)
  743.

\bibitem{APindian} S.~S.\ Ali, S.\ Bharadwaj, B.\ Pandey, Mon.\ Not.\
  R.\ Astron.\ Soc.\ 363 (2005) 251.

\bibitem{MeAP} R.\ Barkana, Mon.\ Not.\ R.\ Astron.\ Soc.\ 372 (2006)
  259.

\bibitem{scott} D.\ Scott, M.~J.\ Rees, Mon.\ Not.\ R.\ Astron.\ Soc.\
  247 (1990) 510.

\bibitem{Wout} S.~A.\ Wouthuysen, Astron.\ J.\ 57 (1952) 31.

\bibitem{Field} G.~B.\ Field, Proc.\ IRE 46 (1958) 240.

\bibitem{haiman} Z.\ Haiman, M.~J.\ Rees, A.\ Loeb, Astroph.\ J.\ 476
  (1997) 458; erratum -- 484 985.

\bibitem{BSc} R.\ Barkana, Science 313 (2006) 931.

\bibitem{GP} J.~E.\ Gunn, B.~A.\ Peterson, Astroph.\ J.\ 142 (1965)
  1633.

\bibitem{Barkana:2001} R.\ Barkana, A.\ Loeb, Phys.\ Rep.\ 349 (2001)
  125.

\bibitem{BL04} 
R.\ Barkana, A.\ Loeb, Astroph.\ J.\ 609 (2004) 474.

\bibitem{fzh04} S.~R.\ Furlanetto, M.\ Zaldarriaga, L.\ Hernquist,
  Astroph.\ J.\ 613 (2004) 1.

\bibitem{Iliev} I.~T.\ Iliev, G.\ Mellema, U.-L.\ Pen, et al., 
  Mon.\ Not.\ R.\ Astron.\ Soc.\ 369 (2006) 1625.

\bibitem{tozzi} P.\ Tozzi, P.\ Madau, A.\ Meiksin, M.~J.\ Rees, 
  Astrophys.\ J.\ 528 (2000) 597.

\bibitem{BL05b} R.\ Barkana, A.\ Loeb, Astroph.\ J.\ 626 (2005) 1.

\bibitem{Jonathan07} J.~R.\ Pritchard, S.~R.\ Furlanetto, Mon.\
  Not.\ R.\ Astron.\ Soc.\ 376 (2007) 1680.

\bibitem{MWAref} J.~D.\ Bowman, M.~F.\ Morales, J.~N.\ Hewitt,
  Astroph.\ J.\ 695 (2009) 183.

\bibitem{LOFARref}
G.\ Harker, Mon.\ Not.\ R.\ Astron.\ Soc.\ 405 (2010) 2492.

\bibitem{GMRT} G.\ Paciga, T.-C.\ Chang, Y.\ Gupta, et al., 
  Mon.\ Not.\ R.\ Astron.\ Soc.\ 413 (2011) 1174.

\bibitem{PAPER} Z.~S.\ Ali, A.~R.\ Parsons, H.\ Zheng, et al.,
  Astroph.\ J.\ 809 (2015) 61.

\bibitem{Bowman} J.~D.\ Bowman, M.~F.\ Morales, J.~N.\ Hewitt,
  Astroph.\ J.\ 661 (2007) 1.

\bibitem{McQuinn} M.\ McQuinn, O.\ Zahn, M.\ Zaldarriaga, L.\
  Hernquist, S.~R.\ Furlanetto, Astroph.\ J.\ 653 (2006) 815.

\bibitem{Sources} R.\ Barkana, Mon.\ Not.\ R.\ Astron.\ Soc.\ 397
  (2009) 1454.

\bibitem{BiP} S.\ Bharadwaj, S.~K.\ Pandey, Mon.\ Not.\ R.\ Astron.\
  Soc.\ 358 (2005) 968.

\bibitem{CM03} B.\ Ciardi, P.\ Madau, Astroph.\ J.\ 596 (2003) 1.

\bibitem{Fur04b} S.\ R.\ Furlanetto, M.\ Zaldarriaga, L.\ Hernquist,
  Astroph.\ J.\ 613 (2004) 16.

\bibitem{Ichikawa} K.\ Ichikawa, R.\ Barkana, I.~T.\ Iliev, G.\
  Mellema, P.~R.\ Shapiro, Mon.\ Not.\ R.\ Astron.\ Soc.\ 406 (2010)
  2521.

\bibitem{Fur06} S.~R.\ Furlanetto, Mon.\ Not.\ R.\ Astron.\
  Soc.\ 371 (2006) 867.

\bibitem{PL} J.~R.\ Pritchard, A.\ Loeb, Phys.\ Rev.\ D 82 (2010)
  023006.

\bibitem{nono} A.\ Morandi, R.\ Barkana, Mon.\ Not.\ R.\ Astron.\
  Soc.\ 424 (2012) 2551.

\bibitem{bowman2} J.~D.\ Bowman, A.~E.~E.\ Rogers, Nature 468 (2010)
  796.

\bibitem{TH10} D.\ Tseliakhovich, C.~M.\ Hirata, Phys.\ Rev.\ D 82
  (2010) 083520.

\bibitem{LoebBook} A.\ Loeb, S.\ R.\ Furlanetto, The First Galaxies in
  the Universe, Princeton University Press, Princeton, 2013.

\bibitem{Weinberg1972} S.\ Weinberg, Gravitation and Cosmology, Wiley,
  New York, 1972.

\bibitem{KT1990} E.~W.\ Kolb, M.~S.\ Turner, The Early Universe,
Addison-Wesley, Redwood City, CA, 1990.

\bibitem{Planck} Planck Collaboration, P.~A.~R.\ Ade, N.\ Aghanim, et
  al.\, Astron.\ \& Astroph.\ 566 (2014) A54.

\bibitem{Peebles1980} P.~J.~E.\ Peebles, The Large-Scale Structure
  of the Universe, Princeton University Press, Princeton, 1980.

\bibitem{Peebles1993} P.~J.~E.\ Peebles, Principles of Physical
  Cosmology, Princeton University Press, Princeton, 1993.

\bibitem{NFW} J.~F.\ Navarro, C.~S.\ Frenk, S.~D.~M.\ White,
  Astrophys.\ J.\ 490 (1997) 493.

\bibitem{ps74} W.~H.\ Press, P.\ Schechter, Astroph.\ J.\ 187 (1974)
  425.

\bibitem{bond91} J.~R.\ Bond, S.\ Cole, G.\ Efstathiou, N.\ Kaiser,
  Astroph.\ J.\ 379 (1991) 440.

\bibitem{shetht99} R.~K.\ Sheth, G.\ Tormen, Mon.\ Not.\ R.\
  Astron.\ Soc.\ 308 (1999) 119.

\bibitem{shetht02} R.~K.\ Sheth, G.\ Tormen, Mon.\ Not.\ R.\
  Astron.\ Soc.\ 329 (2002) 61.

\bibitem{jenkins} A.\ Jenkins, et al., Mon.\ Not.\ R.\ Astron.\ Soc.\
  321 (2001) 372.

\bibitem{k84} N.\ Kaiser, Astroph.\ J.\ 284 (1984) L9.

\bibitem{b86} J.~M.\ Bardeen, J.~R.\ Bond, N.\ Kaiser, A.~S.\ Szalay,
  Astroph.\ J., 304 (1986) 15.

\bibitem{ck89} S.\ Cole, N.\ Kaiser, Mon.\ Not.\ R.\ Astron.\ Soc.\
  237 (1989) 1127.

\bibitem{mw96} H.~J.\ Mo, S.~D.~M.\ White, Mon.\ Not.\ R.\ Astron.\
  Soc.\ 282 (1996) 347.

\bibitem{NBhalos} S.\ Naoz, R.\ Barkana, Mon.\ Not.\ R.\ Astron.\
  Soc.\ 377 (2007) 667.

\bibitem{BL05c} R.\ Barkana, A.\ Loeb, Mon.\ Not.\ R.\ Astron.\ Soc.\
  363 (2005) L36.

\bibitem{BL11} R.\ Barkana, A.\ Loeb, Mon.\ Not.\ R.\ Astron.\ Soc.\
  415 (2011) 3113.

\bibitem{Maayane} M.~T.\ Soumagnac, R.\ Barkana, C.~G.\ Sabiu, et al.,
  2016, arXiv:1602.01839

\bibitem{MaBert1995} C.\ Ma, E.\ Bertschinger, Astroph.\ J.\ 455
  (1995) 7.

\bibitem{CMBFAST} U.\ Seljak, M.\ Zaldarriaga, Astroph.\ J.\ 469
  (1996) 437.

\bibitem{Yamamoto1} K.\ Yamamoto, N.\ Sugiyama, H.\ Sato, Phys.\ Rev.\
  D, 56 (1997) 7566.

\bibitem{Yamamoto2} K.\ Yamamoto, N.\ Sugiyama, H.\ Sato, Astroph.\
  J.\ 501 (1998) 442.

\bibitem{NB05} S.\ Naoz, R.\ Barkana, Mon.\ Not.\ R.\ Astron.\ Soc.\
  362 (2005) 1047.

\bibitem{GH98} N.~Y.\ Gnedin, L.\ Hui, Mon.\ Not.\ R.\ Astron.\ Soc.\
  296 (1998) 44.

\bibitem{G00b} N.~Y.\ Gnedin, Astroph.\ J.\ 542 (2000) 535.

\bibitem{Hoeft} M.\ Hoeft, G.\ Yepes, S.\ Gottl{\"o}ber, V.\ Springel,
  Mon.\ Not.\ R.\ Astron.\ Soc.\ 371 (2006) 401.

\bibitem{Okamoto} T.\ Okamoto, L.\ Gao, T.\ Theuns, Mon.\ Not.\ R.\
  Astron.\ Soc.\ 390 (2008) 920.

\bibitem{NaozMF1} S.\ Naoz, R.\ Barkana, A.\ Mesinger, 
Mon.\ Not.\ R.\ Astron.\ Soc.\ 399 (2009) 369.

\bibitem{NaozMF2} S.\ Naoz, N.\ Yoshida, R.\ Barkana, Mon.\
  Not.\ R.\ Astron.\ Soc.\ 416 (2011) 232.

\bibitem{MesingerMF} E.\ Sobacchi, A.\ Mesinger, Mon.\ Not.\ R.\
  Astron.\ Soc.\ 432 (2013) L51.

\bibitem{McQuinnMF} Y.\ Noh, M.\ McQuinn, Mon.\ Not.\ R.\ Astron.\
  Soc.\ 444 (2014) 503.

\bibitem{1967H2} W.~C.\ Saslaw, D.\ Zipoy, Nature 216 (1967) 976.

\bibitem{yoshida} N.\ Yoshida, A.\ Sokasian, L.\ Hernquist, V.\
  Springel, Astroph.\ J.\ 598 (2003) 73.

\bibitem{abn02} T.\ Abel, G.\ L.\ Bryan, M.\ L.\ Norman, Science 295
  (2002) 93.

\bibitem{avihabit} A.\ Loeb, International Journal of Astrobiology 13
  (2014) 337.

\bibitem{miralda03} J.\ Miralda-Escud\'e, Science 300 (2003) 1904.

\bibitem{wise05} J.~H.\ Wise, T.\ Abel, Astroph.\ J.\ 629 (2005) 615.

\bibitem{casas02} R.\ Casas-Miranda,H.~J.\ Mo, R.~K.\ Sheth, G.\
  B\"{o}rner, Mon.\ Not.\ R.\ Astron.\ Soc.\ 333 (2002) 730.

\bibitem{Spergel07} D.~N.\ Spergel, R.\ Bean, O.\ Dor{\'e}, et al.,
  Astroph.\ J.\ 170 (2007) 377.

\bibitem{Viel06} M.\ Viel, M.~G.\ Haehnelt, A.\ Lewis, Mon.\ Not.\ R.\
  Astron.\ Soc.\ 370 (2006) L51.

\bibitem{F06} S.~R.\ Furlanetto, S.~P.\ Oh, F.~H.\ Briggs, Phys.\
  Rep.\ 433 (2006) 181.

\bibitem{converge} V.\ Springel, L.\ Hernquist, Mon.\ Not.\ R.\
  Astron.\ Soc.\ 339 (2003) 312.

\bibitem{WL04} J.~S.~B.\ Wyithe, A.\ Loeb, Nature 432 (2004) 194.

\bibitem{21cmfast} A.\ Mesinger, S.\ Furlanetto, R.\ Cen, Mon.\
  Not.\ R.\ Astron.\ Soc.\ 411 (2011) 955.

\bibitem{Cold} A.\ Fialkov, R.\ Barkana, E.\ Visbal, Nature 506 (2014)
  197.

\bibitem{BrommRev} V.\ Bromm, Rep.\ Prog.\ Phys.\ 76 (2013) 112901.

\bibitem{BrommRev2} T.\ Karlsson, V.\ Bromm, J.\ Bland-Hawthorn, 
  Rev.\ Mod.\ Phys.\ 85 (2013) 809.

\bibitem{Bfield1} R.\ Durrer, A.\ Neronov, Astron.\ Astroph.\ Rev.\ 21
  (2013) 62.

\bibitem{Bfield2} L.~M.\ Widrow, D.\ Ryu, D.~R.~G.\ Schleicher, et
  al., Space Science Rev.\ 166 (2012) 37.

\bibitem{NaozB} S.\ Naoz, R.\ Narayan, Phys.\ Rev.\ Lett.\ 111 (2013)
  051303.

\bibitem{H2a} Z.\ Haiman, A.\ A.\ Thoul, A.\ Loeb, Astrophys.\ J.\ 464
  (1996) 523.

\bibitem{H2b} T.\ Abel, P.\ Anninos, Y.\ Zhang, M.\ L.\ Norman, 
  New Astron.\ 2 (1997) 181.

\bibitem{H2c} D.\ Galli, F.\ Palla, Astron.\ Astrophys.\ 335 (1998)
  403.

\bibitem{H2d} S.~C.~O.\ Glover, T.\ Abel, Mon.\ Not.\ R.\ Astron.\
  Soc.\ 388 (2008) 1627.

\bibitem{H2e} M.\ Tegmark, J.\ Silk, M.~J.\ Rees, A.\ Blanchard, 
  T.\ Abel, F.\ Palla, Astrophys.\ J.\ 474 (1997) 1.

\bibitem{AbelBinary} M.~J.\ Turk, T.\ Abel, B.\ O'Shea, Science 325
  (2009) 601.

\bibitem{PopIIIa} J.~C.\ Tan, C.~F.\ McKee, Astrophys.\ J.\ 603 (2004)
  383

\bibitem{PopIIIb} A.\ Stacy, T.~H.\ Greif, V.\ Bromm, Mon.\ Not.\ R.\
  Astron.\ Soc.\ 403 (2010) 45.

\bibitem{PopIIIc} P.~C.\ Clark, S.~C.~O.\ Glover, R.~J.\ Smith, T.~H.\
  Greif, R.~S.\ Klessen, V.\ Bromm, Science 331 (2011) 1040.

\bibitem{PopIIId} T.~H.\ Greif, V.\ Bromm, P.~C.\ Clark, S.~C.~O.\
  Glover, R.\ J.\ Smith, R.~S.\ Klessen, N.\ Yoshida, V.\ Springel, 
  Mon.\ Not.\ R.\ Astron.\ Soc.\ 424 (2012) 399.

\bibitem{YoshidaStars} S.\ Hirano, T.\ Hosokawa, N.\ Yoshida, et al.\
  Astroph.\ J.\ 781 (2014) 60.

\bibitem{YStars2} S.\ Hirano, T.\ Hosokawa, N.\ Yoshida, K.\ Omukai,
  H.~W.\ Yorke, Mon.\ Not.\ R.\ Astron.\ Soc.\ 448 (2015) 568.

\bibitem{MW10} M.~F.\ Morales, J.~S.~B.\ Wyithe, Ann.\ Rev.\ Astron.\
  Astroph.\ 48 (2010) 127.

\bibitem{PL12} J.~R.\ Pritchard A.\ Loeb, Rep.\ Prog.\ Phys.\ 75
  (2012) 086901.

\bibitem{purcell} E.~M.\ Purcell, G.~B.\ Field, Astroph.\ J.\ 124
  (1956) 542.

\bibitem{Field58} G.~B.\ Field, Proc.\ IRE, 46 (1958) 240.

\bibitem{Rybicki} G.~B.\ Rybicki, Astroph.\ J.\ 647 (2006) 709.

\bibitem{AD} A.~C.\ Allison, A.\ Dalgarno, Astroph.\ J.\ 158 (1969)
  423.

\bibitem{Zyg} B.\ Zygelman, Astroph.\ J.\ 622 (2005) 1356.

\bibitem{Field59} G.~B.\ Field, Astroph.\ J.\ 129 (1959) 551.

\bibitem{ChSh06} L.\ Chuzhoy, P.~R.\ Shapiro, Astroph.\ J.\ 651 (2006)
  1.

\bibitem{Grachev} S.~I.\ Grachev, Astrofizika, 30 (1989) 347.

\bibitem{Miralda} X.\ Chen, J.\ Miralda-Escud\'e, Astroph.\ J.\ 602
  (2004) 1.

\bibitem{Basko} M.~M.\ Basko, Astrophysics 17 (1981) 69.

\bibitem{Ryb94} G.~B.\ Rybicki, I.~P.\ dell'Antonio, Astroph.\ J.\ 
  427 (1994) 603.

\bibitem{Hirata} C.~M.\ Hirata, Mon.\ Not.\ R.\ Astron.\ Soc.\ 367
  (2006) 259.

\bibitem{Jonathan06b} J.~R.\ Pritchard, S.~R.\ Furlanetto, Mon.\
  Not.\ R.\ Astron.\ Soc.\ 372 (2006) 1093.

\bibitem{Kaiser87} N.\ Kaiser, Mon.\ Not.\ R.\ Astron.\ Soc.\ 227
  (1987) 1.

\bibitem{BA04} S.\ Bharadwaj, S.~S.\ Ali, Mon.\ Not.\ R.\ Astron.\
  Soc.\ 352 (2004) 142.

\bibitem{MaoY} Y.\ Mao, P.~R.\ Shapiro, G.\ Mellema, et al., 
  Mon.\ Not.\ R.\ Astron.\ Soc.\ 422 (2012) 926.

\bibitem{Jensen} H.\ Jensen, K.~K.\ Datta, G.\ Mellema, et al., 
  Mon.\ Not.\ R.\ Astron.\ Soc.\ 435 (2013) 460.

\bibitem{ShapiroPRL} P.~R.\ Shapiro, Y.\ Mao, I.~T.\ Iliev, et al.,
  Phys.\ Rev.\ Lett.\ 110 (2013) 151301.

\bibitem{FialkovPRL} A.\ Fialkov, R.\ Barkana, A.\ Cohen,
  Phys.\ Rev.\ Lett.\ 114 (2015) 101303.

\bibitem{Beutler} F.\ Beutler, S.\ Saito, H.-J.\ Seo, et al., 
  Mon.\ Not.\ R.\ Astron.\ Soc.\ 443 (2014) 1065.

\bibitem{Seljak} U.\ Seljak, J.\ Cosmo.\ Astropart.\ Phys.\ 3 (2012)
  004.

\bibitem{BL06} R.\ Barkana, A.\ Loeb, Mon.\ Not.\ R.\ Astron.\
  Soc.\ 372 (2006) 43.

\bibitem{Datta12} K.~K.\ Datta, G.\ Mellema, Y.\ Mao, et al.,
  Mon.\ Not.\ R.\ Astron.\ Soc.\ 424 (2012) 1877.

\bibitem{Zawada14} K.\ Zawada, B.\ Semelin, P.\ Vonlanthen, S.\ Baek,
  Y.\ Revaz, Mon.\ Not.\ R.\ Astron.\ Soc.\ 439 (2014) 1615.

\bibitem{LaPlante14} P.\ La Plante, N.\ Battaglia, A.\ Natarajan, et
  al., Astroph.\ J.\ 789 (2014) 31.

\bibitem{Datta14} K.~K.\ Datta, H.\ Jensen, S.\ Majumdar, et al.,
  Mon.\ Not.\ R.\ Astron.\ Soc.\ 442 (2014) 1491.

\bibitem{AP} C.\ Alcock, B.\ Paczynski, Nature 281 (1979) 358.

\bibitem{shaver} P.~A.\ Shaver, R.~A.\ Windhorst, P.\ Madau, A.~G.\ de
  Bruyn, Astron. \& Astroph.\ 345 (1999) 380.

\bibitem{Koopmans} L.\ Koopmans, J.\ Pritchard, G.\ Mellema, J.\
  Aguirre, K.\ Ahn, R.\ Barkana, et al., Advancing Astrophysics with
  the Square Kilometre Array (AASKA14), 2015, p.\ 1.

\bibitem{Mellema13} G.\ Mellema, L.~V.~E.\ Koopmans, F.~A.\ Abdalla,
  et al., Experimental Astronomy 36 (2013) 235.

\bibitem{Fialkov} A.\ Fialkov, Intern.\ J.\ Mod.\ Phys.\ D 23 (2014)
  30017.

\bibitem{Dalal} N.\ Dalal, U.-L.\ Pen, U.\ Seljak, J.\ Cosmo.\
  Astropart.\ Phys.\ 11 (2010) 7.

\bibitem{Us}  D.\ Tseliakhovich, R.\ Barkana, C.\ Hirata, 
Mon.\ Not.\ R.\ Astron.\ Soc.\ 418 (2011) 906.

\bibitem{Maio:2011} U.\ Maio, L.~V.~E.\ Koopmans, B.\ Ciardi, 
  Mon.\ Not.\ R.\ Astron.\ Soc.\ 412 (2011) L40.

\bibitem{Stacy:2011} A.\ Stacy, V.\ Bromm, A.\ Loeb, Astroph.\ J.\ 730
  (2011) 1.

\bibitem{Greif:2011} T.\ Greif, S.\ White, R.\ Klessen, V.\ Springel,
  Astroph.\ J.\ 736 (2011) 147.

\bibitem{mcquinn12} R.~M.\ O'Leary, M.\ McQuinn, Astroph.\ J.\ 760
  (2012) 4.

\bibitem{mcquinn12b} M.\ McQuinn, R.~M.\ O'Leary, Astroph.\ J.\ 760
  (2012) 3.

\bibitem{naoz1} S.\ Naoz, N.\ Yoshida, N.~Y.\ Gnedin, Astroph.\ J.\
  747 (2012) 128.

\bibitem{naoz2} S.\ Naoz, N.\ Yoshida, N.~Y.\ Gnedin, Astroph.\ J.\ 
  763 (2013) 27.

\bibitem{UrosBAO1} J.\ Yoo, N.\ Dalal, U.\ Seljak, J.\ Cosmo.\
  Astropart.\ Phys.\ 7 (2011) 18.

\bibitem{UrosBAO2} J.\ Yoo, U.\ Seljak, Phys.\ Rev.\ D 88 (2013)
  103520.

\bibitem{BovyDvorkin} J.\ Bovy, C.\ Dvorkin, Astroph.\ J.\ 768 (2013)
  70.

\bibitem{FialkovLW} A.\ Fialkov, R.\ Barkana, E.\ Visbal, D.\
  Tseliakhovich, C.~M.\ Hirata, Mon.\ Not.\ R.\ Astron.\ Soc.\ 432
  (2013) 2909.

\bibitem{Tanaka} T.~L.\ Tanaka, M.\ Li, Z.\ Haiman, Mon.\ Not.\ R.\
  Astron.\ Soc.\ 435 (2013) 3559.

\bibitem{Tanaka2} T.~L.\ Tanaka, M.\ Li, Mon.\ Not.\ R.\ Astron.\
  Soc.\ 439 (2014) 1092.

\bibitem{GCs} S.\ Naoz, R.\ Narayan, Astroph.\ J.\ 791 (2014) L8.

\bibitem{Yacine} Y.\ Ali-Ha{\"i}moud, P.~D.\ Meerburg, S.\ Yuan, 
  Phys.\ Rev.\ D 89 (2014) 083506.

\bibitem{Blazek} J.~A.\ Blazek, J.~E.\ McEwen, C.~M.\ Hirata, 
  Phys.\ Rev.\ Lett. 116 (2016) 121303.

\bibitem{Ahn} K.\ Ahn 2016, arXiv:1603.09356.

\bibitem{Schmidt} F.\ Schmidt 2016, arXiv:1602.09059.
  
\bibitem{CO93} R.\ Cen, J.~P.\ Ostriker, Astroph.\ J.\ 417 (1993) 404.

\bibitem{GO97} N.~Y.\ Gnedin, J.~P.\ Ostriker, Astrophys.\ J.\ 
  486 (1997) 581.

\bibitem{FK94} M.\ Fukugita, M.\ Kawasaki, Mon.\ Not.\ R.\
  Astron.\ Soc.\ 269 (1994) 563.

\bibitem{SGB94} P.~R.\ Shapiro, M.~L.\ Giroux, A.\ Babul, Astroph.\
  J.\ 427 (1994) 25.

\bibitem{KSS94} M.\ Kamionkowski, D.~N.\ Spergel, N.\ Sugiyama,
  Astroph.\ J.\ 426 (1994) 57.

\bibitem{TSB94} M.\ Tegmark, J.\ Silk, A.\ Blanchard, Astroph.\ J.\
  420 (1994) 484.

\bibitem{HL97} Z.\ Haiman, A.\ Loeb, Astroph.\ J.\ 483 (1997) 21.

\bibitem{VS99} P.\ Valageas, J.\ Silk, Astron. \& Astroph.\ 347 (1999)
  1.

\bibitem{G00a} N.~Y.\ Gnedin, Astroph.\ J.\ 535 (2000) 530.

\bibitem{HS99} Z.\ Haiman, M.\ Spaans, Astroph.\ J.\ 518 (1999) 138.

\bibitem{H02} Z.\ Haiman, Astroph.\ J.\ 576 (2002) L1.

\bibitem{CGO02} C.~L.\ Carilli, N.~Y.\ Gnedin, F.\ Owen, Astroph.\ J.\
  577 (2002) 22.

\bibitem{Aghanim96} N.\ Aghanim, F.~X.\ Desert, J.~L.\ Puget, 
  R.\ Gispert, Astron.\ \& Astroph.\ 311 (1996) 1.

\bibitem{GruzH98} A.\ Gruzinov, W.\ Hu, Astroph.\ J.\ 508 (1998) 435.

\bibitem{Santos03} M.~G.\ Santos, A.\ Cooray, Z.\ Haiman, L.\ Knox, 
  C.-P.\ Ma, Astroph.\ J.\ 598 (2003) 756.

\bibitem{MHR00} J.\ Miralda-Escud{\'e}, M.\ Haehnelt, M.~J.\ Rees, 
  Astroph.\ J.\ 530 (2000) 1.

\bibitem{Sokasian} S.~R.\ Furlanetto, A.\ Sokasian, L.\ Hernquist, 
  Mon.\ Not.\ R.\ Astron.\ Soc.\ 347 (2004) 187.

\bibitem{Ciardi} B.\ Ciardi, F.\ Stoehr, S.~D.~M.\ White, Mon.\ Not.\
  R.\ Astron.\ Soc.\ 343 (2003) 1101.

\bibitem{mellema} G.\ Mellema, I.~T.\ Iliev, U.-L.\ Pen, P.~R.\
  Shapiro, Mon.\ Not.\ R.\ Astron.\ Soc.\ 372 (2006) 679.

\bibitem{B08} R.\ Barkana, Mon.\ Not.\ R.\ Astron.\ Soc.\ 391 (2008)
  727.

\bibitem{zahn} O.\ Zahn, A.\ Lidz, M.\ McQuinn, S.\ Dutta, L.\
  Hernquist, M.\ Zaldarriaga, S.~R.\ Furlanetto, Astroph.\ J.\ 
  654 (2007) 12.

\bibitem{santos} M.~G.\ Santos, A.\ Amblard, J.\ Pritchard, H.\ Trac,
  R.\ Cen, A.\ Cooray, Astroph.\ J.\ 689 (2008) 1.

\bibitem{Planck16} Planck Collaboration, Aghanim, N., Ashdown, M., et
  al.\ 2016, arXiv:1605.02985

\bibitem{Planck16b} Planck Collaboration, Adam, R., Aghanim, N., et
  al.\ 2016, arXiv:1605.03507

\bibitem{D01} S.~G.\ Djorgovski, S.\ Castro, D.\ Stern, A.~A.\
  Mahabal, Astroph.\ J.\ 560 (2001) L5.

\bibitem{Becker01} R.~H.\ Becker, X.\ Fan, R.~L.\ White, et al.,
  Astron.\ J.\ 122 (2001) 2850.

\bibitem{Fan02} X.\ Fan, V.~K.\ Narayanan, M.~A.\ Strauss, et al.,
  Astron.\ J.\ 123 (2002) 1247.

\bibitem{Fan06} X.\ Fan, M.~A.\ Strauss, R.~H.\ Becker, et al., 
  Astron.\ J.\ 132 (2006) 117.

\bibitem{z6} J.\ Schroeder, A.\ Mesinger, Z.\ Haiman, Mon.\ Not.\ R.\
  Astron.\ Soc.\ 428 (2013) 3058.

\bibitem{B02} R.\ Barkana, New Astron.\ 7 (2002) 85.

\bibitem{N02} A.\ Nusser, A.~J.\ Benson, N.\ Sugiyama, C.\ Lacey, 
  Astroph.\ J.\ 580 (2002) L93.

\bibitem{L02} A.\ Lidz, L.\ Hui, M.\ Zaldarriaga, \& R.\ Scoccimarro,
  Astroph.\ J.\ 579 (2002) 491.

\bibitem{MF08} A.\ Mesinger, S.~R.\ Furlanetto, Mon.\ Not.\ R.\
  Astron.\ Soc.\ 385 (2008) 1348.

\bibitem{M10} A.\ Mesinger, Mon.\ Not.\ R.\ Astron.\ Soc.\ 407 (2010)
  1328.

\bibitem{R86} M.~J.\ Rees, Mon.\ Not.\ R.\ Astron.\ Soc.\ 222 (1986)
  27.

\bibitem{E92} G.\ Efstathiou, Mon.\ Not.\ R.\ Astron.\ Soc.\ 256
  (1992) 43.

\bibitem{TW96} A.~A.\ Thoul, D.~H. Weinberg, Astroph.\ J.\ 465 (1996)
  608.

\bibitem{QKE96} T.\ Quinn, N.\ Katz, G.\ Efstathiou, Mon.\ Not.\ R.\
  Astron.\ Soc.\ 278 (1996) L49.

\bibitem{WHK97} D.~H.\ Weinberg, L.\ Hernquist, N.\ Katz, Astroph.\
  J.\ 477 (1997) 8.

\bibitem{MR98} J.\ Miralda-Escud\'e, M.~J.\ Rees, Astroph.\ J.\ 497
  (1998) 21.

\bibitem{KI00} T.\ Kitayama, S.\ Ikeuchi, Astroph.\ J.\ 529 (2000)
  615.

\bibitem{NS00} J.~F.\ Navarro, M.\ Steinmetz, Astroph.\ J.\ 538 (2000)
  477.

\bibitem{BL00} R.\ Barkana, A.\ Loeb, Astroph.\ J.\ 539 (2000) 20.

\bibitem{WL06} J.~S.~B.\ Wyithe, A.\ Loeb, Nature, 441 (2006) 322.

\bibitem{BLhistory} R.\ Barkana, A.\ Loeb, Mon.\ Not.\ R.\ Astron.\
  Soc.\ 371 (2006) 395.

\bibitem{DS86} A.\ Dekel, J.\ Silk, Astroph.\ J.\ 303 (1986) 39.

\bibitem{WL13} J.~S.~B.\ Wyithe, A.\ Loeb, Mon.\ Not.\ R.\ Astron.\
  Soc.\ 428 (2013) 2741.

\bibitem{HAM01} Z.\ Haiman, T.\ Abel, P.\ Madau, Astroph.\ J.\ 551
  (2001) 599.

\bibitem{BL02} R.\ Barkana, A.\ Loeb, Astroph.\ J.\ 578 (2002) 1.

\bibitem{BL99} R.\ Barkana, A.\ Loeb, Astroph.\ J.\ 523 (1999) 54.

\bibitem{SIR04} P.~R.\ Shapiro, I.~T.\ Iliev, A.~C.\ Raga, Mon.\ Not.\
  R.\ Astron.\ Soc.\ 348 (2004) 753.

\bibitem{Ciardi06} B.\ Ciardi, E.\ Scannapieco, F.\ Stoehr, et al.,
  Mon.\ Not.\ R.\ Astron.\ Soc.\ 366 (2006) 689.

\bibitem{Yue09} B.\ Yue, B.\ Ciardi, E.\ Scannapieco, X.\ Chen, 
  Mon.\ Not.\ R.\ Astron.\ Soc.\ 398 (2009) 2122.

\bibitem{Holzbauer:2011} L.~N.\ Holzbauer, S.~R.\ Furlanetto, Mon.\
  Not.\ R.\ Astron.\ Soc.\ 419 (2012) 718.

\bibitem{Complete} A.\ Fialkov, R.\ Barkana, A.\ Pinhas, \& E.\
  Visbal, Mon.\ Not.\ R.\ Astron.\ Soc.\ 437 (2014) L36.

\bibitem{Ahn:2009} K.\ Ahn, P.~R.\ Shapiro, I.~T.\ Iliev, G.\ Mellema,
  U.\ Pen, Astroph.\ J.\ 695 (2009) 1430.

\bibitem{Machacek:2001} M.~E.\ Machacek, G.~L.\ Bryan, T.\ Abel, 
  Astroph.\ J.\ 548 (2001) 509.

\bibitem{Wise:2007} J.~H.\ Wise, T.\ Abel, Astroph.\ J.\ 671 (2007)
  1559.

\bibitem{O'Shea:2008} B. W.\ O'Shea, M.~L.\ Norman, Astroph.\ J.\ 673,
  (2008) 14.

\bibitem{VisbalLW} E.\ Visbal, Z.\ Haiman, B.\ Terrazas, G.~L.\ Bryan,
  R.\ Barkana Mon.\ Not.\ R.\ Astron.\ Soc.\ 445 (2014) 107.

\bibitem{Bromm:2001} V.\ Bromm, R.~P.\ Kudritzki, A.\ Loeb, 
  Astroph.\ J.\ 552 (2001) 464.

\bibitem{Shull} J.~M.\ Shull, M.~E.\ van Steenberg, Astroph.\ J.,
  298 (1985) 268.

\bibitem{ChenK} X.\ Chen, M.\ Kamionkowski, Phys.\ Rev.\ D 70 (2004)
  043502.

\bibitem{Stoever} S.~R.\ Furlanetto, S.~J.\ Stoever, Mon.\ Not.\ R.\
  Astron.\ Soc.\ 404 (2010) 1869.

\bibitem{S08} J.~D.\ Silverman, P.~J.\ Green, W.~A.\ Barkhouse, et
  al., Astroph.\ J.\ 679 (2008) 118.

\bibitem{chuzhoy07} L.\ Chuzhoy, P.~R.\ Shapiro, Astroph.\ J.\ 655
  (2007) 843.

\bibitem{ciardiLya} B.\ Ciardi, R.\ Salvaterra, T.\ Di Matteo, Mon.\
  Not.\ R.\ Astron.\ Soc.\ 401 (2010) 2635.

\bibitem{Mirabel0} I.~F.\ Mirabel, Proc.\ IAU Symp.\ 275, Jets at
  all Scales, Cambridge University Press, pages 2-8. ed. G. E. Romero,
  R. A. Sunyaev, T. Belloni [arXiv:1012.4944v1], 2010.

\bibitem{Frag2} T.\ Fragos, B.~D.\ Lehmer, S.\ Naoz, A.\ Zezas, A.\
  Basu-Zych, Astroph.\ J.\ 776 (2013) 31.

\bibitem{Mirabel} I.~F.\ Mirabel, M.\ Dijkstra, P.\ Laurent, A.\ Loeb,
  J.~R.\ Pritchard, Astron.\ \& Astroph.\ 528 (2011) A149.

\bibitem{Basu} A.~R.\ Basu-Zych, et al., Astroph.\ J.\ 762 (2013) 45.

\bibitem{Basu2} A.~R.\ Basu-Zych, et al., Astroph.\ J.\ 774 (2013)
  152.

\bibitem{Lofar14} A.~H.\ Patil, S.\ Zaroubi, E.\ Chapman, et al.,
  Mon.\ Not.\ R.\ Astron.\ Soc.\ 443 (2014) 1113.

\bibitem{CLoeb} P.\ Christian, A.\ Loeb, J.\ Cosmo.\ Astropart.\
  Phys.\ 09 (2013) 014.

\bibitem{Mesinger13} A.\ Mesinger, A.\ Ferrara, D.~S.\ Spiegel, 
  Mon.\ Not.\ R.\ Astron.\ Soc.\ 431 (2013) 621.

\bibitem{Frag1} T.\ Fragos, et al., Astroph.\ J.\ 764 (2013) 41.

\bibitem{spectra} J.~E.\ McClintock, R.~A.\ Remillard, In: Compact
  stellar X-ray sources. Edited by Walter Lewin \& Michiel van der
  Klis, Cambridge Astrophysics Series, 157, 2006.

\bibitem{tamura} M.\ Tamura, et al., Astroph.\ J.\ 753 (2012) 65.

\bibitem{mineo} S.\ Mineo, M.\ Gilfanov, R.\ Sunyaev, Mon.\ Not.\ R.\
  Astron.\ Soc.\ 426 (2012) 1870.

\bibitem{oh01} S.~P.\ Oh, Astroph.\ J. 553 (2001) 499.

\bibitem{shocks} S.~R.\ Furlanetto, A.\ Loeb, Astroph.\ J.\ 611 (2004)
  642.

\bibitem{shocks2} P.~R.\ Shapiro, K.\ Ahn, M.~A.\ Alvarez, et al.,
  Astroph.\ J.\ 646 (2006) 681.

\bibitem{vasudevan} R.~V.\ Vasudevan, R.~F.\ Mushotzky, P.\ Gandhi,
  Astroph.\ J.\ 770 (2013) L37.

\bibitem{BL03} R.\ Barkana, A.\ Loeb, Nature 421 (2003) 341.

\bibitem{imbh} N.\ L{\"u}tzgendorf, M.\ Kissler-Patig, N.\ Neumayer,
  et al., Astron. \& Astroph.\ 555 (2013) A26.

\bibitem{tanakaX} T.\ Tanaka, R.\ Perna, Z.\ Haiman, Mon.\ Not.\ R.\
  Astron.\ Soc.\ 425 (2012) 2974.

\bibitem{WL03} J.~S.~B.\ Wyithe, A.\ Loeb, Astroph.\ J.\ 595 (2003) 614.

\bibitem{ss} N.~I.\ Shakura, R.~A.\ Sunyaev, Astron. \& Astroph.\ 24
  (1973) 337.

\bibitem{Evan} E.\ Scannapieco, R.\ Barkana, Astroph.\ J.\ 571 (2002)
  585.

\bibitem{B07} R.\ Barkana, Mon.\ Not.\ R.\ Astron.\ Soc.\ 376 (2007)
  1784.

\bibitem{21CMMC} B.\ Greig, A.\ Mesinger, Mon.\ Not.\ R.\ Astron.\
  Soc.\ 449 (2015) 4246.

\bibitem{Jonathan06a} J.~R.\ Pritchard, S.~R.\ Furlanetto, Mon.\ Not.\
  R.\ Astron.\ Soc.\ 367 (2006) 1057.

\bibitem{CZ07} L.\ Chuzhoy, Z.\ Zheng, Astroph.\ J.\ 670 (2007) 912.

\bibitem{Semelin} B.\ Semelin, F.\ Combes, S.\ Baek, Astron.\ \&
  Astroph.\ 474 (2007) 365.

\bibitem{NB08} S.\ Naoz, R.\ Barkana, Mon.\ Not.\ R.\ Astron.\ Soc.\
  385 (2008) 63.

\bibitem{nature} E.\ Visbal, R.\ Barkana, A.\ Fialkov, D.\
  Tseliakhovich, C.~M.\ Hirata, Nature 487 (2012) 70.

\bibitem{Liu} A.\ Liu, M.\ Tegmark, Mon.\ Not.\ R.\ Astron.\ Soc.\ 
419 (2012) 3491.

\bibitem{pacucci} F.\ Pacucci, A.\ Mesinger, S.\ Mineo, A.\ Ferrara,
  Mon.\ Not.\ R.\ Astron.\ Soc.\ 443 (2014) 678.

\bibitem{complex} A.\ Fialkov, R.\ Barkana, Mon.\ Not.\ R.\ Astron.\
  Soc.\ 445 (2014) 213.

\bibitem{BA05} S.\ Bharadwaj, S.~S.\ Ali, Mon.\ Not.\ R.\ Astron.\
  Soc.\ 356 (2005) 1519.

\bibitem{SBP06} S.\ Saiyad Ali, S.\ Bharadwaj, S.~K.\ Pandey, 
  Mon.\ Not.\ R.\ Astron.\ Soc.\ 366 (2006) 213.

\bibitem{IlievPDF} I.~T.\ Iliev, G.\ Mellema, P.~R.\ Shapiro, et al.,
  Mon.\ Not.\ R.\ Astron.\ Soc.\ 423 (2012) 2222.

\bibitem{PritchardPDF} C.~A.\ Watkinson, J.~R.\ Pritchard, 
  Mon.\ Not.\ R.\ Astron.\ Soc.\ 443 (2014) 3090.

\bibitem{BL08} R.\ Barkana, A.\ Loeb, Mon.\ Not.\ R.\ Astron.\ Soc.\
  384 (2008) 1069.

\bibitem{Vera} V.\ Gluscevic, R.\ Barkana, Mon.\ Not.\ R.\ Astron.\
  Soc.\ 408 (2010) 2373.

\bibitem{WM07} S.\ Wyithe, M.\ Morales, Mon.\ Not.\ R.\ Astron.\
  Soc.\ 379 (2007) 1647.

\bibitem{Harker} G.~J.~A.\ Harker, et al., Mon.\ Not.\ R.\ Astron.\
  Soc.\ 393 (2009) 1449.

\bibitem{CMBreion} E.\ Calabrese, R.\ Hlo{\v z}ek, N.\ Battaglia, et
  al., J.\ Cosmo.\ Astropart.\ Phys.\ 8 (2014) 10.

\bibitem{Cooray} A.\ Cooray, Phys.\ Rev.\ Lett.\ 97 (2006) 261301.

\bibitem{fdm} W.\ Hu, R.\ Barkana, A.\ Gruzinov, Phys.\ Rev.\ Lett.\
  85 (2000) 1158.

\bibitem{FL} A.\ Fialkov, A.\ Loeb, J.\ Cosmo.\ Astropart.\ Phys.\ 11
  (2013) 66.

\end{thebibliography}
\end{document}